\documentclass[a4paper,11pt]{article}
\pdfoutput=1 
\usepackage{jheppub}
\usepackage{tikz}

\usepackage{subfig}
\usepackage{enumitem}
\usepackage[T1]{fontenc} 
\usepackage{floatrow}
\usepackage{appendix}

\allowdisplaybreaks

\usepackage{epsf}
\usepackage{amsmath}
\usepackage{amsfonts}
\usepackage{amssymb}
\usepackage{psfrag,epsfig,graphicx,graphics}
\graphicspath{{./images/}}
\newcommand\numberthis[1][]{%
    \refstepcounter{equation}%
    \ifx#1\empty\else\label{eq:#1}\fi%
    \tag{\theequation}%
}

\providecommand{\U}[1]{\protect\rule{.1in}{.1in}}

\def\slashchar#1{\setbox0=\hbox{$#1$}
   \dimen0=\wd0
   \setbox1=\hbox{/} \dimen1=\wd1
   \ifdim\dimen0>\dimen1
      \rlap{\hbox to \dimen0{\hfil/\hfil}}
      #1
   \else
      \rlap{\hbox to \dimen1{\hfil$#1$\hfil}}
      /
   \fi}

\def\tr{{\rm tr}}

\def\bei{\begin{itemize}}
\def\ei{\end{itemize}}

\def\beeq{\begin{eqnarray}} 
\def\beqa{\begin{eqnarray}}
\def\bea{\begin{eqnarray}}

\def\eea{\end{eqnarray}}
\def\eqa{\end{eqnarray}}
\def\eeeq{\end{eqnarray}}

\def\eqar{\end{array}}
\def\beqar{\begin{array}}

\def\beas{\begin{eqnarray*}}
\def\beqas{\begin{eqnarray*}}

\def\eqas{\end{eqnarray*}}
\def\eeas{\end{eqnarray*}}

\def\beq{\begin{equation}} 
\def\be{\begin{equation}}

\def\ee{\end{equation}}
\def\eq{\end{equation}}
\def\eeq{\end{equation}}

\def\beqd{\begin{displaymath}}
\def\eeqd{\end{displaymath}}
\def\eqd{\end{displaymath}}

\def\beeq{\begin{eqnarray}} \def\eeeq{\end{eqnarray}}

\newcommand{\fin}{\end{document}}

\newcommand{\Arrow}[1]{%
\parbox{#1}{\tikz{\draw[->](0,0)--(#1,0);}}
}

\title{\boldmath NLO computation of diffractive di-hadron production in a saturation framework }

\author[a,b,c,1]{Michael Fucilla, \note{Corresponding author.}}
\author[d,e]{Andrey Grabovsky,}
\author[c]{Emilie Li,}
\author[f]{Lech Szymanowski,}
\author[c]{Samuel Wallon}

\affiliation[a]{Dipartimento di Fisica, Università della Calabria, I-87036 Arcavacata di Rende, Cosenza, Italy}
\affiliation[b]{Istituto Nazionale di Fisica Nucleare, Gruppo collegato di Cosenza,-87036 Arcavacata di Rende, Cosenza, Italy}
\affiliation[c]{Universit\'e Paris-Saclay, CNRS/IN2P3, IJCLab, 91405, Orsay, France}
\affiliation[d]{Budker Institute of Nuclear Physics, 11, Lavrenteva avenue, 630090, Novosibirsk, Russia}
\affiliation[e]{Novosibirsk State University, 630090, 2, Pirogova street, Novosibirsk, Russia}
\affiliation[f]{National Centre for Nuclear Research (NCBJ),Pasteura 7, 02-093 Warsaw,  Poland}

\emailAdd{Michael.Fucilla@unical.it}
\emailAdd{A.V.Grabovsky@inp.nsk.su}
\emailAdd{Emilie.Li@ijclab.in2p3.fr}
\emailAdd{Lech.Szymanowski@ncbj.gov.pl}
\emailAdd{Samuel.Wallon@ijclab.in2p3.fr}

\abstract{The cross-sections of diffractive double hadron photo- or electroproduction with large $p_T$, on a nucleon or a nucleus, are calculated to NLO accuracy. A hybrid formalism mixing collinear factorization and high energy small-$x$ factorization, more precisely the shockwave formalism for the latter, is used to derive the results.
The cancellation of divergences is explicitly shown, and the finite parts of the NLO differential cross-sections are found. We work in  arbitrary kinematics such that both photoproduction and leptoproduction are considered. The results are therefore usable, to detect saturation effects, at both the future EIC or already at LHC, using Ultra Peripheral Collisions.}

\begin{document} 
\maketitle
\flushbottom

\section{Introduction}
\label{sec:intro}

One of the most intriguing phenomena of strong interaction is the existence of diffractive processes, which have been revealed experimentally by H1 and ZEUS experiments at HERA, for the first time in the semi-hard regime in which a hard scale allows one to describe such processes from first principles, relying on QCD. Besides,
due to the parametrically large center-of-mass energy, these processes provide an access to the regime of very high gluon densities~\cite{Wusthoff:1999cr,Wolf:2009jm}.
Indeed, HERA showed that almost 10~\%  of the $\gamma^* p \to X$ deep inelastic scattering (DIS) events present
 a rapidity gap between the proton remnants 
and the hadrons 
coming from the fragmentation region of the initial virtual photon. These
events are called diffractive deep inelastic scattering (DDIS), and look like
$\gamma^* p \to X  \, Y$~\cite{Aktas:2006hx,Aktas:2006hy,Chekanov:2004hy,Chekanov:2005vv,Aaron:2010aa,Aaron:2012ad,Chekanov:2008fh,Aaron:2012hua}, where $Y$ is either the outgoing proton or one of its low-mass excited states, and $X$ is the diffractive final state. Besides inclusive DDIS, one can select specific diffractive states, like jet(s) production, exclusive meson production, or hadrons.

The existence of a rapidity gap between $X$ and $Y$ leads naturally to describe diffraction through a Pomeron exchange in the $t-$channel between these $X$ and $Y$ states. 

A QCD factorization theorem was derived~\cite{Collins:1997sr} in the collinear framework, justified by the existence of a hard scale, the photon virtuality $Q^2$ of DIS, involving a  convolution of
a coefficient function with diffractive parton distributions, describing the partonic content of the Pomeron. 

Besides this collinear treatment, at high energies, it is natural to model the diffractive events by a {\em direct} Pomeron contribution involving the coupling of a Pomeron with the diffractive state $X$ of invariant mass $M.$ For low values of $M^2$, $X$ can be modeled by a  $q \bar{q}$ pair, while for larger values of $M^2,$
the cross section with an additional produced gluon, i.e. $X=q \bar{q} g,$
is enhanced.

In the present article, we extend the series of works by some of us, devoted to a complete Next-to-Leading Order (NLO) description of the direct coupling of the Pomeron
to several kinds of  diffractive $X$ states, namely exclusive diffractive dijet production~\cite{Boussarie:2014lxa, Boussarie:2016ogo, Boussarie:2019ero} and exclusive meson production~\cite{Boussarie:2016bkq}, here for the case of double hadron production at large $p_T.$
Such a study is strongly motivated by present and future possibilities of accessing gluonic saturation through large-$p_T$ dihadron production. In photoproduction and leptoproduction, this could be studied at the future EIC, in particular on a nucleus target in order to have a saturation scale $Q_s$ large enough to be clearly in the perturbative regime, since $Q_s^2 \simeq (A/x)^{1/3}.$ Here, the hard scale is provided by the large virtuality $Q^2$ of the virtual photon and/or the large $p_T$ of the produced hadron. In photoproduction, this could be achieved already now at the LHC in $pA$ and $AA$ scattering, using Ultra Peripheral Collisions (UPC). Indeed, heavy ions produce very high (real) photon fluxes, and when scattered  off a nucleon or a nucleus far away in impact parameter, the Coulomb peak dominates over any hadronic exchange, so that these photons can be used as rather clean probes of the nucleon/nucleus. In particular, this could be studied at LHCb which is very well equipped for identifying and  reconstructing large-$p_T$ hadron (when saying large $p_T$, we mean here large enough so that it provides the required hard scale). Therefore, in view of future possible phenomenological applications, we will keep in the present work very general kinematics, meaning both $Q^2$ and $p_T$ arbitrary.

The "Pomeron" is presently a color singlet QCD shockwave, either built from  Balitsky's high energy operator expansion~\cite{Balitsky:1995ub, Balitsky:1998kc, Balitsky:1998ya, Balitsky:2001re}, 
or from the color glass condensate formulation~\cite{JalilianMarian:1997jx,JalilianMarian:1997gr,JalilianMarian:1997dw,JalilianMarian:1998cb,Kovner:2000pt,Weigert:2000gi,Iancu:2000hn,Iancu:2001ad,Ferreiro:2001qy}, which satisfies the Balitsky-Jalilian Marian-Iancu-McLerran-Weigert-Leonidov-Kovner (B-JIMWLK) evolution equation.
The results we derive are obtained in the QCD shockwave approach and depend on the total available center-of-mass energy. This framework
is rather general and can have many applications. Restricting to the non saturated regime, one might describe 
the $t-$channel exchanged state in the linear Balitsky-Fadin-Kuraev-Lipatov (BFKL) regime~\cite{Fadin:1975cb, Kuraev:1976ge, Kuraev:1977fs, Balitsky:1978ic}, with next-to-leading logarithmic (NLL) precision~\cite{Fadin:1998py,Ciafaloni:1998gs,Fadin:2004zq,Fadin:2005zj} to be consistent with our computation of impact factor at NLO\footnote{For a recent review on tests of BFKL through semi-hard processes involving jets and hadrons, see Ref.~\cite{Celiberto:2020wpk}.}.
At higher energies, i.e. when the gluonic saturation settles in, 
the Wilson-line operators, describing the $t-$channel exchanged state,  
 evolve with respect to rapidity according to the Balitsky hierarchy. Restricting to the case of a dipole operator, it reduces to the Balitsky-Kovchegov (BK) 
equation~\cite{Balitsky:1995ub, Balitsky:1998kc, Balitsky:1998ya, Balitsky:2001re, Kovchegov:1999yj, Kovchegov:1999ua} in the large $N_c$ limit. 

In the present paper, we calculate the matrix element for the $\gamma^{(*)}\to h_1 \, h_2 \, X$ transition ($h_1$ and $h_2$ being two identified hadrons) in the shockwave background of
the target. It depends on the target via the matrix elements of two Wilson line operators $\tr(U_1 U_2^\dag)$ and
$\tr(U_1 U_3^\dag) \, \tr(U_3 U_2^\dag) - N_c \tr(U_1 U_2^\dag)$ between the $in$ and $out$ target states. The Wilson lines are functions of the rapidity
 that separates the {\it quantum} gluons belonging to the impact factor and the {\it classical} gluons from the Wilson lines. For hadron targets, these matrix
elements are to be described by some models, or 
 can be calculated as solutions of the NLO BK and the LO double dipole evolution
equations, using initial conditions at the rapidity of the target. In the linear limit (BFKL) for forward scattering, these solutions are known analytically with a running coupling \cite{Chirilli:2013kca, Grabovsky:2013gta}. In the low-density regime, the second Wilson line operator, namely the double dipole operator, can always be linearized so that
the cross-section can be written in terms of matrix elements of color dipoles only. 

We will particularly focus on the way the various kinds of divergences will compensate each other since one of the technical difficulties in this framework is to prove explicitly that the various infrared (IR) and ultraviolet (UV) singularities cancel properly.

First, we will very briefly present the two kinds of factorization we will be dealing with. Then, we will present the 
Leading Order (LO) results. Next, we will compute explicitly the various real and virtual contributions at NLO, and will show how different types of divergences cancel properly. Finally, we will give our main results, i.e. the finite cross-sections at NLO for any kind of initial photon polarization. 

\section{Theoretical framework}
\label{sec: framework}
\subsection{Hybrid collinear/high-energy factorization}
\begin{figure}[!htb]
   \centering
    \begin{minipage}{.5\textwidth}
        \centering
        \includegraphics[scale=0.36]{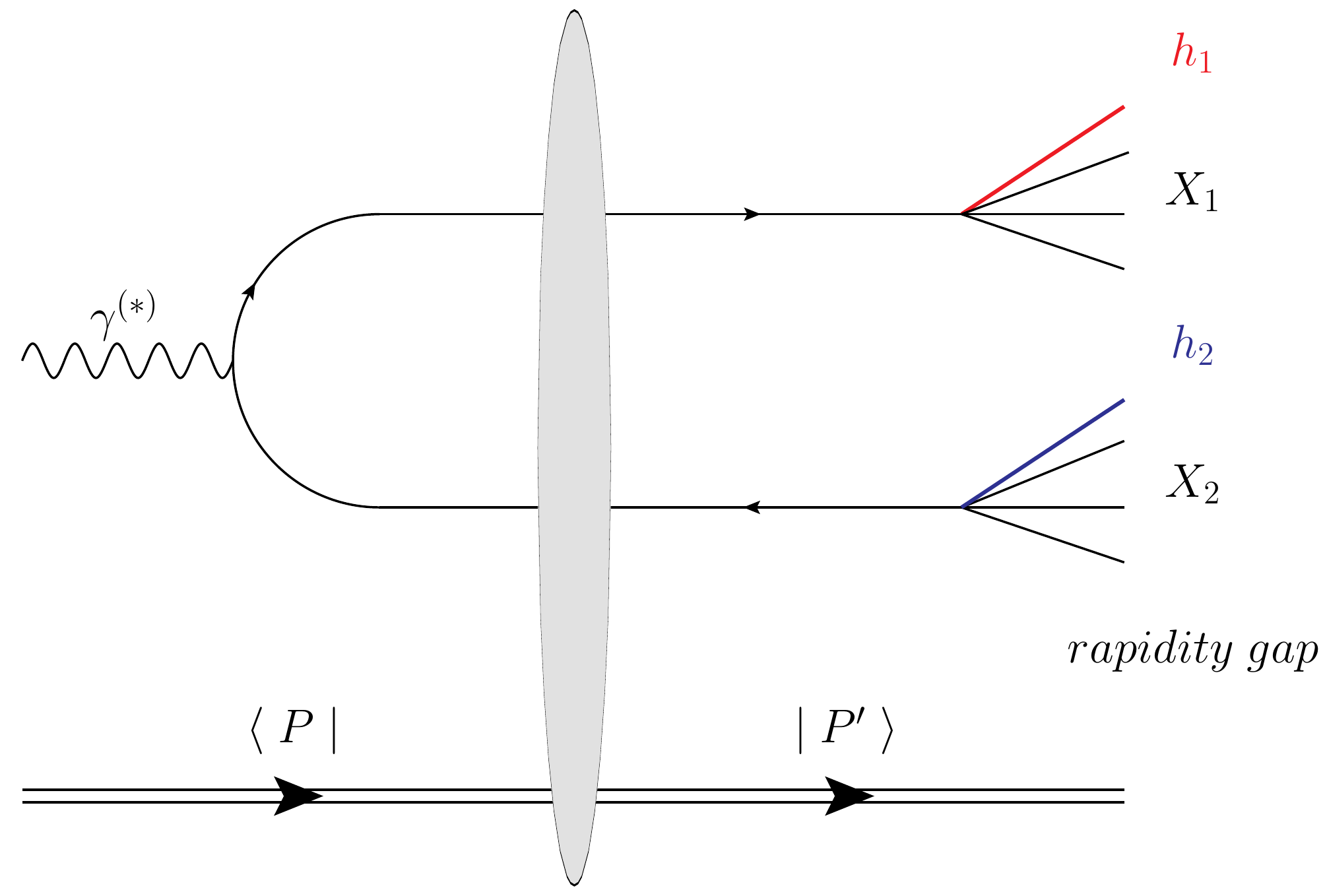}
    \end{minipage}%
    \begin{minipage}{0.5\textwidth}
        \centering
        \includegraphics[scale=0.36]{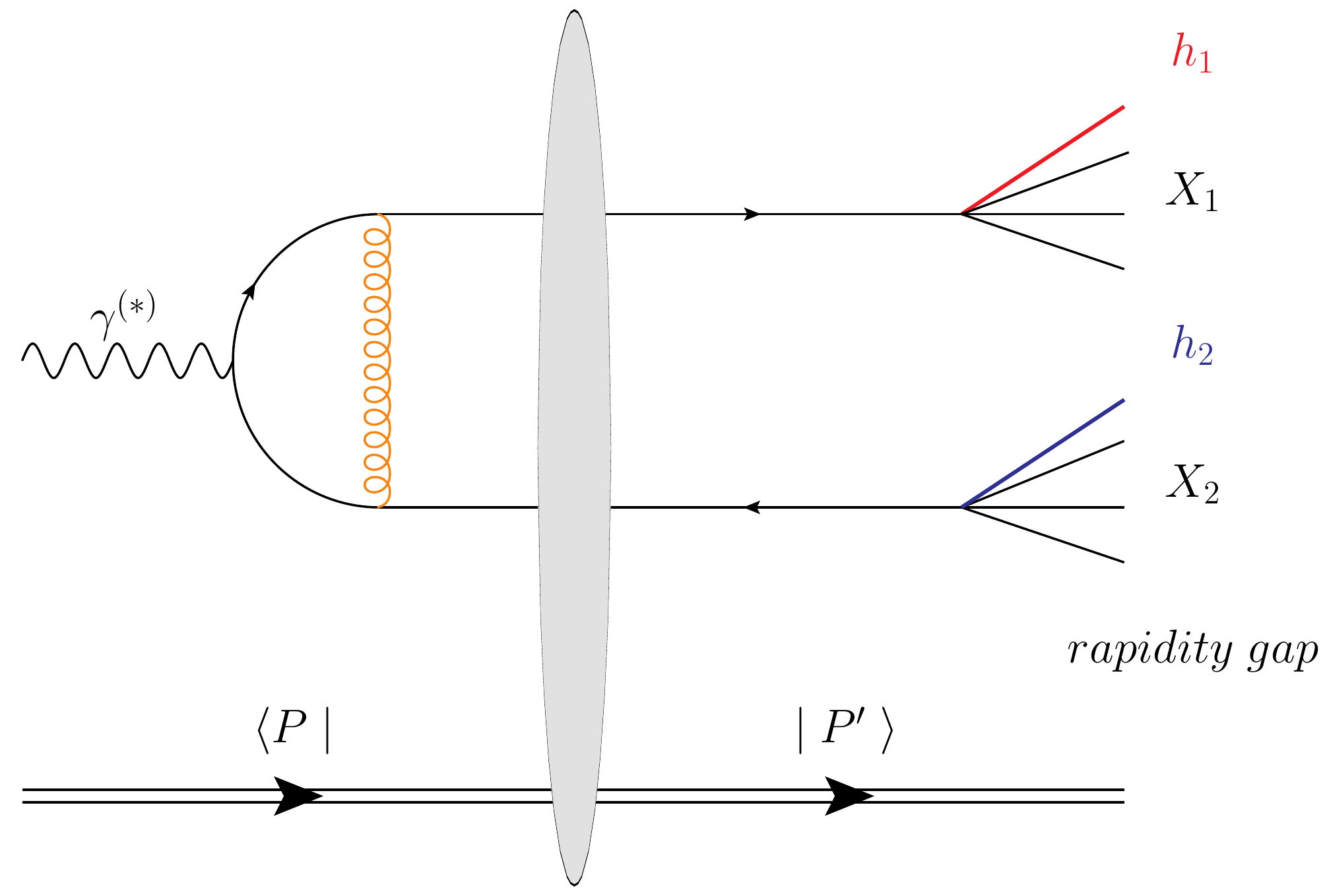}
    \end{minipage}
    \caption{Left: Amplitude of the process (\ref{process}) at LO. Right: An example of amplitude contributing to the process (\ref{process}) at NLO. The grey blob symbolizes the QCD shockwave. The double line symbolizes the target, which remains intact in the figure, but could just as well break. The quark and antiquark fragment into the systems $(h_1 X_1)$ and $(h_2 X_2)$. The two tagged hadrons $h_1$ and $h_2$ are drawn in red and blue.}
    \label{fig:process}
\end{figure}

The aim of this paper is to perform a full NLO computation  of the semi-inclusive diffractive di-hadron production in the high energy limit: 
\begin{equation}
\label{process}
    \gamma^{(*)}(p_\gamma) + P(p_0) \rightarrow h_1(p_{h_1}) + h_2(p_{h_2})  + X + P'(p'_0)
\end{equation}
where $P$ is a nucleon or a nucleus target, generically called proton in the following. The initial photon plays the role of a probe (also named projectile). Our computation applies both to the photoproduction case (including ultraperipheral collisions) and to the electroproduction case (e.g. at EIC). A gap in rapidity is assumed between the outgoing nucleon/nucleus and the diffractive system $(X h_1 h_2)$. This is illustrated by Fig.~\ref{fig:process}.

We will be working in a combination of collinear factorization and small-$x$ factorization, more precisely in the shockwave formalism for the latter. 

\subsubsection*{Kinematics}

We introduce a light-cone basis composed of $n_1 $ and $n_2$, with $n_1 \cdot n_2 = 1$ defining the $+/-$ direction.
We write the Sudakov decomposition for any vector as 
\begin{equation}
p^\mu = p^+ n_1^\mu + p^- n_2^\mu + p_\perp^\mu
\end{equation}
and the scalar product of two vectors as 

\begin{equation}
    \begin{aligned}
    p \cdot q &= p^+ q^- + p^- q^+ + p_\perp \cdot q_\perp \\ 
    &= p^+ q^- + p^- q^+ -\vec{p} \cdot \vec{q}\,.
    \end{aligned} 
\end{equation}
We work in a reference frame such that the target moves ultra-relativistically and such that 
$s = (p_\gamma + p_0)^2 \sim 2 p_\gamma^+ p_0^- \gg \Lambda_{\text{QCD}}^2$, $s$ also being larger than any other scale.  Particles on the projectile side are moving in the $n_1$ (i.e. $+$) direction while particles on the target side have a large component along $n_2$ (i.e. $-$ direction).

We will use kinematics such that the photon with virtuality $Q$ is forward, and thus it does not carry any transverse momentum\footnote{Any transverse momentum in Euclidean space will be denoted with an arrow, while a $\perp$ index will be used in Minkowski space.}:
\begin{equation}
\vec{p}_{\gamma}=0,\quad p_{\gamma}^{\mu}=p_{\gamma}^{+}n_{1}^{\mu}+\frac{p_{\gamma}^{2}}{2p_{\gamma}^{+}}n_{2}^{\mu},\quad-p_{\gamma}^{2}\equiv Q^{2}\geq 0. \label{photonk}
\end{equation}
We will denote its transverse polarization $\varepsilon_T$. Its longitudinal polarization vector reads
\begin{equation}
\varepsilon_{L}^{\alpha}=\frac{1}{\sqrt{-p_{\gamma}^{2}}}\left(  p_{\gamma
}^{+}n_{1}^{\alpha}-\frac{p_{\gamma}^{2}}{2p_{\gamma}^{+}}n_{2}^{\alpha
}\right)  ,\quad\varepsilon_{L}^{+}=\frac{p_{\gamma}^{+}}{Q},\quad
\varepsilon_{L}^{-}=\frac{Q}{2p_{\gamma}^{+}}.
\end{equation}
We write the momentum of the produced hadrons as
\beqa
\label{ph}
p^\mu_{h_i}=p^+_{h_i} n_1^\mu + \frac{m_{h_i}^2 + \vec{p}_{h_i}^{\,2}}{2 p^+_{h_i}} n_2^\mu + p^\mu_{h_i\perp}\quad (i=1,2) \,.
\eqa
The momenta of the fragmenting quark of virtuality $p_q^2$ reads
\beqa
\label{pq}
p^\mu_q=p^+_q n_1^\mu + \frac{p_q^2+\vec{p}_{q}^{\,2}}{2 p^+_{q}} n_2^\mu + p^\mu_{q\perp}\,
\eqa
and similarly for an antiquark of virtuality $p_{\bar{q}}^2$ 
\beqa
\label{pqbar}
p^\mu_{\bar{q}}=p^+_{\bar{q}} n_1^\mu + \frac{p^2_{\bar{q}}+\vec{p}_{\bar{q}}^{\,2}}{2 p^+_{\bar{q}}} n_2^\mu + p^\mu_{\bar{q}\perp}\,.
\eqa
From now, we will use the notation $p_{ij}=p_i-p_j.$

\subsubsection*{Collinear factorization}

The kinematical region considered here is such that $\vec{p}_{h_1}^{\,2} \sim \vec{p}_{h_2}^{\,2} \gg \Lambda_{\text{QCD}}^2$. The hadron momenta are the hard scale, making the use of perturbative QCD and collinear factorization possible. 
The constraint $\vec{p}^{\, 2} \gg \vec{p}_{h_{1,2}}^{\,2} $, with $\vec{p}$, the relative transverse momentum of the two hadrons has also been considered. This means that the two produced hadrons have a large enough separation angle (or in other words a large enough invariant mass) so that it will not be necessary to consider the di-hadron unpolarized fragmentation functions: each hadron, typically pion, can be produced by two well-separated fragmentation cascades.
The quark and antiquark in the hard part after collinear factorization will be treated as on-shell particles. For further use, we introduce the longitudinal momentum fraction $x_q$ and $x_{\bar{q}}$ as
\beqa
\label{xq-xqbar}
p_q^+ = x_q p^+_{\gamma} \quad \hbox{ and } \quad p_{\bar{q}}^+ = x_{\bar{q}} p^+_{\gamma}\,.
\eqa
Similarly, we denote
\beqa
\label{xh}
p_{h_i}^+ = x_{h_i} p^+_{\gamma} \,.
\eqa

\subsubsection*{Shockwave approach}

Shockwave formalism is an effective approach to deal with gluonic saturation. We briefly give here a survey of the required technicalities to be used in the present paper.

When considering the photon impact factor, the gluonic field $A$ is separated into external background fields $b$ (resp. internal fields $\mathcal{A}$) depending on whether their $+$-momentum is below (resp. above) the arbitrary rapidity cut-off $e^\eta p_\gamma^+$, with $\eta < 0$. The light-cone gauge $n_2 \cdot A = $ is used.
The external fields, after being highly boosted from the target rest frame to this frame, take the form 
\begin{equation}
    b^\mu (x) = b^-(x_\perp) \delta (x^+) n_2^\mu \,.
\end{equation}

The resummation of all order interactions with those fields leads to a high-energy Wilson line, that represents the shockwave and is located exactly at $x^+ =0$:
\begin{equation}
    U_{\vec{z}} = \mathcal{P} \exp \left(i g \int d z^+ b^-(z)\right)\,,
\end{equation}
where $\mathcal{P}$ is the usual path ordering operator.

The small-$x$ factorization applies here and the scattering amplitude is the convolution of the projectile impact factor and the non-perturbative matrix element of operators from the Wilson lines operators on the target states. 
One of such operators is the dipole operator, which in the fundamental representation of $SU(N_c)$ takes the form:
\begin{equation}
\left[\operatorname{Tr} \left(U_1 U_2^\dag\right)-N_c\right]\left(\vec{p_1},\vec{p}_2\right) = \int d^d \vec{z}_{1} d^d \vec{z}_{2\perp} e^{- i \vec{p}_1 \cdot \vec{z}_1} e^{- i \vec{p}_2 \cdot \vec{z}_2} \left[\operatorname{Tr} \left(U_{\vec{z}_1} U_{\vec{z}_2}^\dag\right)-N_c\right]\,,
\end{equation}
where
$\vec{z}_{1,2}$ are the transverse positions of the $q,\bar{q}$ coming from the photon and $\vec{p}_{1,2}$ their respective transverse momentums kicks from the shockwave.

The proton matrix element should be parameterized. This can be done  through a generic function $F$, following the definition of ref.~\cite{Boussarie:2016ogo} 
\begin{eqnarray}
\left\langle P^{\prime}\left(p_{0}^{\prime}\right)\left|T\left(\operatorname{Tr}\left(U_{\frac{z_{\perp}}{2}} U_{-\frac{z_{\perp}}{2}}^{\dagger}\right)-N_{c}\right)\right| P\left(p_{0}\right)\right\rangle
& \equiv & 2 \pi \delta\left(p_{00^{\prime}}^{-}\right) F_{p_{0 \perp} p_{0 \perp}^{\prime}}\left(z_{\perp}\right) \nonumber \\
   & \equiv & 2 \pi \delta\left(p_{00^{\prime}}^{-}\right) F\left(z_{\perp}\right) 
\end{eqnarray}
and its Fourier Transform (FT) is
\begin{equation}
\label{eq: FT F}
\int d^{d} z_{\perp} e^{i\left(z_{\perp} \cdot p_{\perp}\right)} F\left(z_{\perp}\right) \equiv \mathbf{F}\left(p_{\perp}\right).
\end{equation}

Similar definitions exist for the double dipole operator and its action on proton states, as can be seen with eqs.~(5.3) and (5.6) in \cite{Boussarie:2016ogo}, with
\begin{eqnarray}
    && \left\langle P^{\prime}\left(p_0^{\prime}\right)\left|\left(\operatorname{Tr}\left(U_{\frac{z}{2}} U_x^{\dagger}\right) \operatorname{Tr}\left(U_x U_{-\frac{z}{2}}^{\dagger}\right)-N_c \operatorname{Tr}\left(U_{\frac{z}{2}} U_{-\frac{z}{2}}^{\dagger}\right)\right)\right| P\left(p_0\right)\right\rangle  \nonumber \\
     &&\equiv 2 \pi \delta\left(p_{00^{\prime}}^{-}\right) \tilde{F}_{p_{0 \perp} p_{0 \perp}^{\prime}}\left(z_{\perp}, x_{\perp}\right) \equiv 2 \pi \delta\left(p_{00^{\prime}}^{-}\right) \tilde{F}\left(z_{\perp}, x_{\perp}\right)
\end{eqnarray}
and its FT is
\begin{equation}
\int d^d z_{\perp} d^d x_{\perp} e^{i\left(p_{\perp} \cdot x_{\perp}\right)+i\left(z_{\perp} \cdot q_{\perp}\right)} \tilde{F}\left(z_{\perp}, x_{\perp}\right) \equiv \tilde{\mathbf{F}}\left(q_{\perp}, p_{\perp}\right).
\end{equation}

In this paper, dimensional regularization will be used with $D=2 + d$, where $d= 2+2\epsilon$ is the transverse dimension.

\subsection{LO order}
\begin{figure}[h!]
\begin{picture}(430,70)
\put(100,1){\includegraphics[scale=0.35]{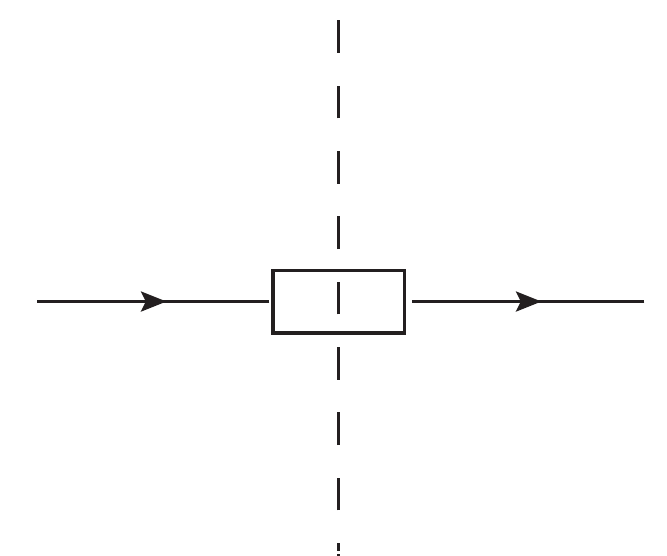}}
\put(182,24){$\equiv$}
\put(200,0){\includegraphics[scale=0.35]{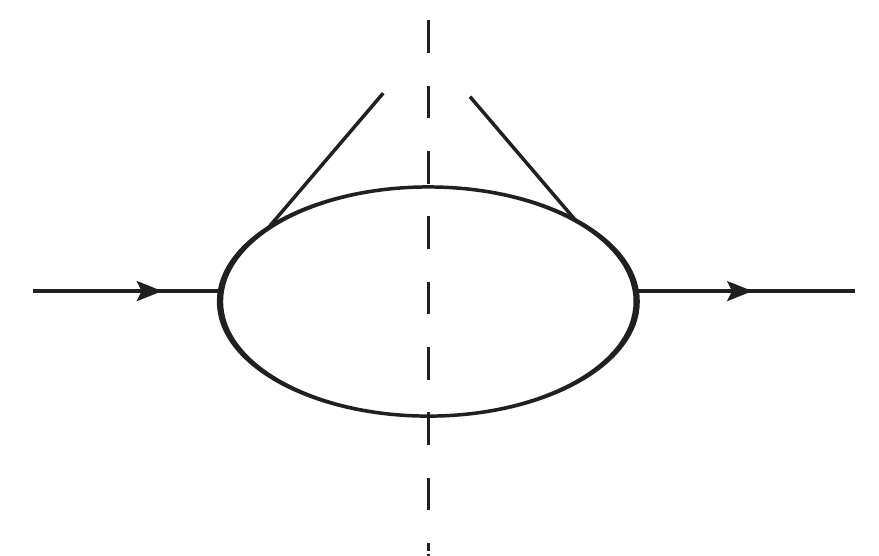}}
\put(225,45){$h$}
\end{picture} 
 \caption{Graphical convention for the fragmentation function of a parton (here a quark for illustration) to a hadron $h$ plus spectators. In the rest of this article, we will use the left-hand side of this drawing.}
    \label{fig:fragmentation}
\end{figure}

QCD collinear factorization stipulates that the total cross-section, at leading-twist and LO, reads, see ref~\cite{Collins:2011zzd} (chap.~12) and ref.~\cite{Altarelli:1979kv}  
\begin{equation}
\label{eq: coll facto}
\frac{d \sigma_{0JI}^{h_1 h_2}}{d x_{h_1} d x_{h_2}}= \sum_{q} \int_{x_{h_1}}^1 \frac{d x_q}{x_q} \int_{x_{h_2}}^1 \frac{d x_{\bar{q}}}{x_{\bar{q}}} D_q^{h_1}\left(\frac{x_{h_1}}{x_q},\mu_F\right) D_{\bar{q}}^{h_2}\left(\frac{x_{h_2}}{x_{\bar{q}}}, \mu_F\right) \frac{d\hat{\sigma}_{JI}}{d x_q d x_{\bar{q}}} + (h_1 \leftrightarrow h_2) \; ,
\end{equation}
where $q$ specifies the quark flavor types ($q=u,d,s,c,b$), and $J,I=L,T$ specify the photon polarization since we deal here with a modulus square amplitude ($J$ labels the photon polarization in the complex conjugated amplitude and $I$ in the amplitude). Here  $x_q$ and $x_{\bar{q}}$ are the longitudinal fractions of the photon momentum carried by the fragmenting partons, $x_ {h_1,h_2}$ are the longitudinal fraction of the photon momentum carried by the produced hadrons, $\mu_F$ is the factorization scale, $D_{q(\bar{q})}^{h}$ denotes the quark (antiquark) Fragmentation Function (FF) and $d\hat{\sigma}$ is the partonic cross-section, i.e. the cross-section for the subprocess 
\begin{equation}
\label{partonic_LO}
     \gamma^{(*)}(p_\gamma) + P(p_0) \rightarrow q (p_q) + \bar{q}(p_{\bar{q}})  + P'(p'_0)\,.
\end{equation}
The graphical convention used in the present article for any fragmentation function is given in fig.~\ref{fig:fragmentation}.

All detailed computations will be done considering only the first term in \eqref{eq: coll facto}, remembering the second term is just simply obtained by doing $h_1 \leftrightarrow h_2$. 

\begin{figure}[h!]
\centering
 \includegraphics[scale=0.35]{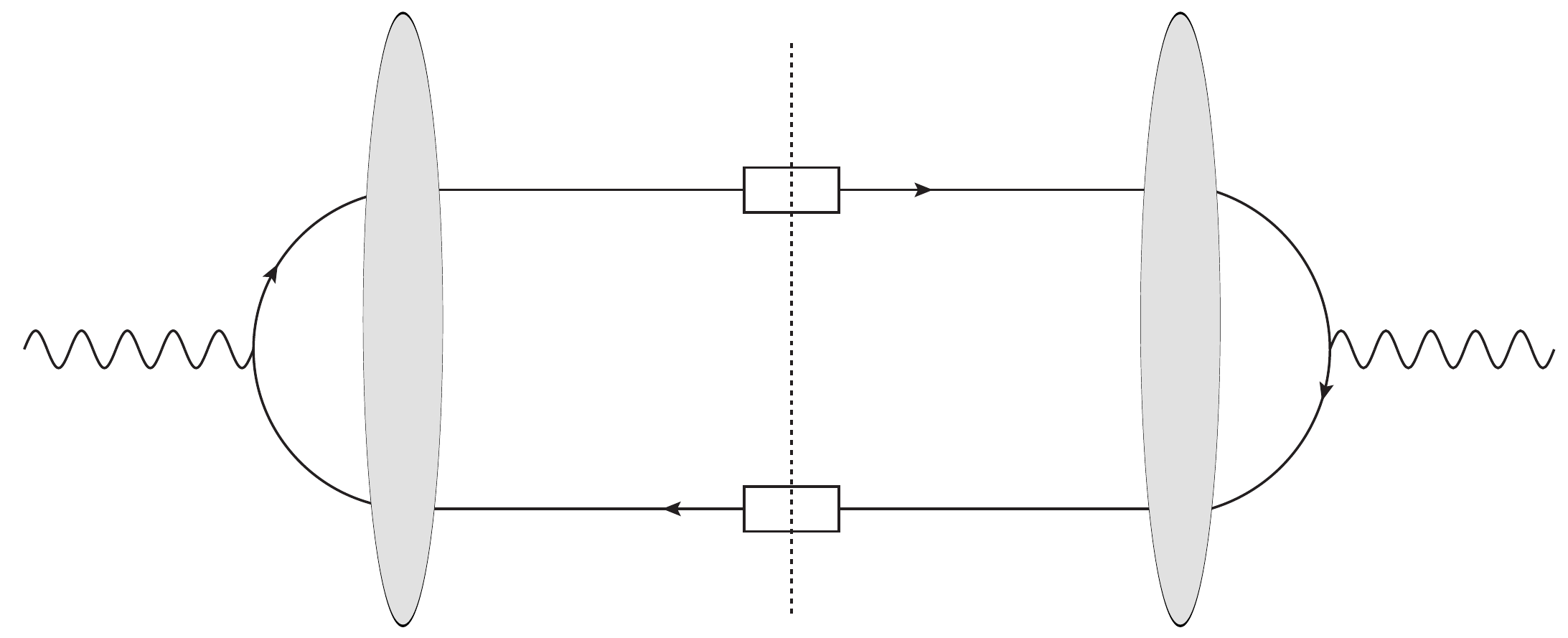}
 \caption{Diagram of the LO process at cross-section level. The blob is the shockwave (we do not draw the coupling with the target for clarity) and the squares the FFs, see fig.~\ref{fig:fragmentation}. The dashed line is to represent the integration over phase space.}
    \label{fig:LO}
\end{figure}

The partonic cross-section (\ref{partonic_LO}) has been computed in the shockwave framework. The structure of the result for the whole process \ref{process} at LO is illustrated in Fig.~\ref{fig:LO}.

Collinear factorization means that the produced hadrons should fly collinearly to the fragmenting partons. This means here that the following constraints should be fulfilled 
\beqa
\label{constraint-collinear-q}
p^+_q &=& \frac{x_q}{x_{h_1}} p^+_{h_1},    \quad \vec{p}_q = \frac{x_q}{x_{h_1}} \vec{p}_{h_1}\,,
\\
\label{constraint-collinear-qbar}
p^+_{\bar{q}} &=& \frac{x_{\bar{q}}}{x_{h_2}} p^+_{h_2}, \quad \vec{p}_{\bar{q}}= \frac{x_{\bar{q}}}{x_{h_2}}\vec{p}_{h_2}\,.
\eqa

Since the photon in the initial state can appear with different polarizations, we construct the density matrix 
\begin{equation}
\label{eq:density_matrix}
d\sigma_{JI}=
\begin{pmatrix}
d\sigma_{LL} & d\sigma_{LT}\\
d\sigma_{TL} & d\sigma_{TT}
\end{pmatrix}
,\qquad d\sigma_{TL}=d\sigma_{LT}^{\ast}.
\end{equation}
Each element of this matrix has a LO contribution $d\sigma_{0}$.
This Born order result, see eq.~(5.14) of ref.~\cite{Boussarie:2016ogo}, has the following structure:
\begin{align}  
\label{dsigma0}
d\sigma_{0JI}  & =  \frac{\alpha_{\mathrm{em}}Q_{q}^{2}}{\left(2\pi\right)^{4\left(d-1\right)}N_{c}}\frac{\left(p_{0}^{-}\right)^{2}}{2x_q x_{\bar{q}} s^{2}}d x_q d x_{\bar{q}} d^{d}p_{q\perp}d^{d}p_{\bar{q}\perp}\delta\left(1-x_q-x_{\bar{q}} \right)\left(\varepsilon_{I\beta}\varepsilon_{J\gamma}^\ast\right)\nonumber \\
& \quad  \times  \int d^{d}p_{1\perp}d^{d}p_{2\perp}d^{d}p_{1^{\prime}\perp}d^{d}p_{2^{\prime}\perp}\delta\left(p_{q1\perp}+p_{\bar{q}2\perp}\right)\delta\left(p_{11^{\prime}\perp}+p_{22^{\prime}\perp}\right)\nonumber \\
& \quad  \times  \sum_{\lambda_q,\lambda_{\bar{q}}}\Phi_{0}^{\beta}\left(p_{1\perp},\, p_{2\perp}\right)\Phi_{0}^{\gamma*}\left(p_{1^{\prime}\perp},\, p_{2^{\prime}\perp}\right)\mathbf{F}\left(\frac{p_{12\perp}}{2}\right)\mathbf{F^{*}}\left(\frac{p_{1^{\prime}2^{\prime}\perp}}{2}\right) .
\end{align}
Using the explicit expressions of the product $\Phi_{0}^{\beta}\Phi_{0}^{\gamma*}$, see eq.~(5.18-20) of ref.~\cite{Boussarie:2016ogo}, as well as  eq.~\eqref{eq: coll facto} 
the LO cross-sections are obtained and read for $LL$:

\begin{eqnarray}
\label{eq:LL-LO}
&&  \hspace{-1cm}  \frac{d \sigma_{0 L L}^{h_1 h_2}}{d x_{h_1}d x_{h_2}d^d p_{h_1 \perp}d^d p_{h_2 \perp}}  = \frac{4 \alpha_{\mathrm{em}} Q^2 }{(2\pi)^{4(d-1)}N_c} \sum_{q} \int_{x_{h_1}}^1 d x_q \int_{x_{h_2}}^1 d x_{\bar{q}} \;  x_q x_{\bar{q}} \left(\frac{x_q}{x_{h_1}}\right)^d \left(\frac{x_{\bar{q}}}{x_{h_2}}\right)^d \nonumber  \\
& \times &  \delta (1-x_q - x_{\bar{q}})  Q_q^2 D_q^{h_1}\left(\frac{x_{h_1}}{x_q} \right) D_{\bar{q}}^{h_2}\left(\frac{x_{h_2}}{x_{\bar{q}}}\right)  {\cal F}_{LL}  + (h_1 \leftrightarrow h_2)\,,
\end{eqnarray}
where
\begin{equation}
\label{F-LL}
{\cal F}_{LL} =  \left|\int d^{d} p_{2 \perp} \frac{\mathbf{F}\left(\frac{x_q}{2x_{h_1}}  p_{h_1\perp} + \frac{x_{\bar{q}} }{2 x_{h_2}}p_{h_2\perp} -p_{2\perp}\right)}{\left(\frac{x_{\bar{q}}}{x_{h_2}}\vec{p}_{h_2}- \vec{p}_{2}\right)^{2}+x_q x_{\bar{q}} Q^{2}} \right|^{2}   \,.
\end{equation}

This LO cross-section can be written differently, using transverse momentum conservation, see eq.~\eqref{dsigma0}
\begin{eqnarray}
\label{eq:LL-LO-minus}
&&  \hspace{-1cm}  \frac{d \sigma_{0 L L}^{h_1 h_2}}{d x_{h_1}d x_{h_2}d^d p_{h_1 \perp}d^d p_{h_2 \perp}}  = \frac{4 \alpha_{\mathrm{em}} Q^2 }{(2\pi)^{4(d-1)}N_c} \sum_{q} \int_{x_{h_1}}^1 d x_q \int_{x_{h_2}}^1 d x_{\bar{q}} \; x_q x_{\bar{q}} \left(\frac{x_q}{x_{h_1}}\right)^d \left(\frac{x_{\bar{q}}}{x_{h_2}}\right)^d \nonumber  \\
& \times &  \delta (1-x_q - x_{\bar{q}})  Q_q^2 D_q^{h_1}\left(\frac{x_{h_1}}{x_q} \right) D_{\bar{q}}^{h_2}\left(\frac{x_{h_2}}{x_{\bar{q}}}\right)   \tilde{\cal F}_{LL}  + (h_1 \leftrightarrow h_2)\,,
\end{eqnarray}
where
\begin{equation}
\label{Ftilde-LL}
{\cal \tilde{F}}_{LL} =  \left|\int d^{d} p_{1 \perp} \frac{\mathbf{F}\left(-\left(\frac{x_q}{2x_{h_1}}  p_{h_1\perp} + \frac{x_{\bar{q}} }{2 x_{h_2}}p_{h_2\perp} -p_{1\perp}\right)\right)}{\left(\frac{x_q}{x_{h_1}}\vec{p}_{h_1}- \vec{p}_{1}\right)^{2}+x_q x_{\bar{q}} Q^{2}} \right|^{2}   \,.
\end{equation}

Both forms can be used interchangeably in NLO cross-sections that are proportional to the LO cross-sections, i.e when dealing with the soft, virtual, and counter-term contribution to the NLO cross-sections. For the collinear quark-gluon contribution, eq.~\eqref{eq:LL-LO} will be used while for the collinear anti-quark-gluon contribution, eq.~\eqref{eq:LL-LO-minus} will be used.

Similarly, the non-diagonal interference term $TL$ can be written in the two equivalent forms
\begin{eqnarray}
\label{eq:TL-LO}
&&  \hspace{-1cm}  \frac{d \sigma_{0 T L}^{h_1 h_2}}{d x_{h_1}d x_{h_2}d^d p_{h_1 \perp}d^d p_{h_2 \perp}}  = \frac{2 \alpha_{\mathrm{em}} Q }{(2\pi)^{4(d-1)}N_c}  \sum_{q} \int_{x_{h_1}}^1 d x_q \int_{x_{h_2}}^1 d x_{\bar{q}}  \left(\frac{x_q}{x_{h_1}}\right)^d \left(\frac{x_{\bar{q}}}{x_{h_2}}\right)^d \nonumber  \\
    & \times &  (x_{\bar{q}}-x_q) \delta (1-x_q - x_{\bar{q}}) Q_q^2 D_q^{h_1}\left(\frac{x_{h_1}}{x_q}\right) D_{\bar{q}}^{h_2}\left(\frac{x_{h_2}}{x_{\bar{q}}}\right)  {\cal F}_{TL}  + (h_1 \leftrightarrow h_2)\,,
\end{eqnarray}
where
\begin{equation}
\begin{aligned}
{\cal F}_{TL} & =  \int d^d p_{2\perp} \frac{\mathbf{F}\left( \frac{x_q}{2 x_{h_1}} p_{h_1 \perp} + \frac{x_{\bar{q}}}{2 x_{h_2}} p_{h_2 \perp} - p_{2 \perp}
\right)}{ \left( \frac{x_{\bar{q}}}{x_{h_2}} \vec{p}_{h_2} -\vec{p}_2  \right)^2 + x_q x_{\bar{q}} Q^2} \\
&  \times \left[\int d^d p_{2'\perp} \frac{\mathbf{F}\left( \frac{x_q}{2 x_{h_1}} p_{h_1 \perp} + \frac{x_{\bar{q}}}{2 x_{h_2}} p_{h_2 \perp} - p_{2' \perp}
\right)}{ \left( \frac{x_{\bar{q}}}{x_{h_2}} \vec{p}_{h_2} -\vec{p}_{2'}  \right)^2 + x_q x_{\bar{q}} Q^2} \left( \frac{x_{\bar{q}}}{x_{h_2}} \vec{p}_{h_2}  - \vec{p}_{2'}\right) \cdot \vec{\varepsilon}_T  \right]^*
\end{aligned}
\end{equation}
and
\begin{eqnarray}
\label{eq:TL-LO-minus}
&&  \hspace{-1cm}  \frac{d \sigma_{0 T L}^{h_1 h_2}}{d x_{h_1}d x_{h_2}d^d p_{h_1 \perp}d^d p_{h_2 \perp}}  = \frac{2 \alpha_{\mathrm{em}} Q }{(2\pi)^{4(d-1)}N_c} \sum_{q} \int_{x_{h_1}}^1 d x_q \int_{x_{h_2}}^1 d x_{\bar{q}}  \left(\frac{x_q}{x_{h_1}}\right)^d \left(\frac{x_{\bar{q}}}{x_{h_2}}\right)^d \nonumber  \\
    & \times & (x_{\bar{q}}-x_q)  \delta (1-x_q - x_{\bar{q}})  Q_q^2 D_q^{h_1}\left(\frac{x_{h_1}}{x_q}\right) D_{\bar{q}}^{h_2}\left(\frac{x_{h_2}}{x_{\bar{q}}}\right)  {\cal \tilde{F}}_{TL}  + (h_1 \leftrightarrow h_2)\,,
\end{eqnarray}
where
\begin{equation}
\begin{aligned}
{\cal \tilde{F}}_{TL} & =  \int d^d p_{1\perp} \frac{\mathbf{F}\left( - \left(\frac{x_q}{2 x_{h_1}} p_{h_1 \perp} + \frac{x_{\bar{q}}}{2 x_{h_2}} p_{h_2 \perp} - p_{1 \perp}
\right)\right)}{ \left( \frac{x_q}{x_{h_1}} \vec{p}_{h_1} -\vec{p}_1  \right)^2 + x_q x_{\bar{q}} Q^2} \\
&  \times \left[\int d^d p_{1'\perp} \frac{\mathbf{F} \left( - \left(\frac{x_q}{2 x_{h_1}} p_{h_1 \perp} + \frac{x_{\bar{q}}}{2 x_{h_2}} p_{h_2 \perp} - p_{1' \perp}
\right) \right)}{ \left( \frac{x_q}{x_{h_1}}  \vec{p}_{h_1} -\vec{p}_{1'}  \right)^2 + x_q x_{\bar{q}} Q^2} \left( \frac{x_q}{x_{h_1}} p_{h_1 }  - p_{1'}\right) \cdot \varepsilon_{T}  \right]^*.
\end{aligned}
\end{equation}
The most complicated contribution $TT$ reads
\begin{eqnarray}
\label{eq:TT-LO}
&&  \hspace{-1cm}  \frac{d \sigma_{0 T T}^{h_1 h_2}}{d x_{h_1}d x_{h_2}d^d p_{h_1 \perp}d^d p_{h_2 \perp}}  = \frac{\alpha_{\mathrm{em}} }{(2\pi)^{4(d-1)}N_c} \sum_{q} \int_{x_{h_1}}^1 \frac{d x_q}{x_q} \int_{x_{h_2}}^1 \frac{d x_{\bar{q}} }{x_{\bar{q}}}  \left(\frac{x_q}{x_{h_1}}\right)^d \left(\frac{x_{\bar{q}}}{x_{h_2}}\right)^d \nonumber  \\
    & \times &  \delta (1-x_q - x_{\bar{q}})  Q_q^2 D_q^{h_1}\left(\frac{x_{h_1}}{x_q}\right) D_{\bar{q}}^{h_2}\left(\frac{x_{h_2}}{x_{\bar{q}}}\right)  {\cal F}_{TT}  + (h_1 \leftrightarrow h_2)\,,
\end{eqnarray}
where
\begin{equation}
\begin{aligned}
{\cal F}_{TT} 
& =   \left[ (x_{\bar{q}} -x_q )^2 g_{\perp}^{ri} g_\perp^{lk} - g_\perp^{rk} g_\perp^{li} + g_{\perp}^{rl} g_\perp^{ik} \right] \\
& \times  \int d^d p_{2\perp} \frac{\mathbf{F}\left( \frac{x_q}{2 x_{h_1}} p_{h_1 \perp} + \frac{x_{\bar{q}}}{2 x_{h_2}} p_{h_2 \perp} - p_{2 \perp}
\right)}{ \left( \frac{x_{\bar{q}}}{x_{h_2}} \vec{p}_{h_2} -\vec{p}_2  \right)^2 + x_q x_{\bar{q}} Q^2} \left( \frac{x_{\bar{q}}}{x_{h_2}} p_{h_2 }  - p_{2}\right)_{r}  \varepsilon_{T i}  \\
&  \times  \left[\int d^d p_{2'\perp} \frac{\mathbf{F}\left( \frac{x_q}{2 x_{h_1}} p_{h_1 \perp} + \frac{x_{\bar{q}}}{2 x_{h_2}} p_{h_2 \perp} - p_{2' \perp}
\right)}{ \left( \frac{x_{\bar{q}}}{x_{h_2}} \vec{p}_{h_2} -\vec{p}_{2'}  \right)^2 + x_q x_{\bar{q}} Q^2} \left( \frac{x_{\bar{q}}}{x_{h_2}} p_{h_2}  - p_{2'}\right)_l  \varepsilon_{T k} \right]^*
\end{aligned}
\end{equation}
or equivalently
\begin{eqnarray}
\label{eq:TT-LO-minus}
&&  \hspace{-1cm}  \frac{d \sigma_{0 T T}^{h_1 h_2}}{d x_{h_1}d x_{h_2}d^d p_{h_1 \perp}d^d p_{h_2 \perp}}  = \frac{\alpha_{\mathrm{em}} }{(2\pi)^{4(d-1)}N_c} \sum_{q} \int_{x_{h_1}}^1 \frac{d x_q}{x_q} \int_{x_{h_2}}^1 \frac{d x_{\bar{q}} }{x_{\bar{q}}}  \left(\frac{x_q}{x_{h_1}}\right)^d \left(\frac{x_{\bar{q}}}{x_{h_2}}\right)^d \nonumber  \\
    & \times &  \delta (1-x_q - x_{\bar{q}})  Q_q^2 D_q^{h_1}\left(\frac{x_{h_1}}{x_q}\right) D_{\bar{q}}^{h_2}\left(\frac{x_{h_2}}{x_{\bar{q}}}\right)  {\cal \tilde{F}}_{TT}  + (h_1 \leftrightarrow h_2)\,,
\end{eqnarray}
where 
\begin{equation}
\begin{aligned}
{\cal \tilde{F}}_{TT} 
& =   \left[ (x_{\bar{q}} -x_q )^2 g_{\perp}^{ri} g_\perp^{lk} - g_\perp^{rk} g_\perp^{li} + g_{\perp}^{rl} g_\perp^{ik} \right] \\
& \times  \int d^d p_{1\perp} \frac{\mathbf{F}\left( -\left( \frac{x_q}{2 x_{h_1}} p_{h_1 \perp} + \frac{x_{\bar{q}}}{2 x_{h_2}} p_{h_2 \perp} - p_{1 \perp}
\right) \right)}{ \left( \frac{x_q}{x_{h_1}} \vec{p}_{h_1} -\vec{p}_1 \right)^2 + x_q x_{\bar{q}} Q^2} \left( \frac{x_q}{x_{h_1}} p_{h_1 }  - p_{1}\right)_{r}  \varepsilon_{T i}  \\
&  \times  \left[\int d^d p_{1'\perp} \frac{\mathbf{F}\left( - \left(\frac{x_q}{2 x_{h_1}} p_{h_1 \perp} + \frac{x_{\bar{q}}}{2 x_{h_2}} p_{h_2 \perp} - p_{1' \perp}
\right) \right)}{ \left( \frac{x_q}{x_{h_1}} \vec{p}_{h_1} -\vec{p}_{1'}  \right)^2 + x_q x_{\bar{q}} Q^2} \left( \frac{x_q}{x_{h_1}} p_{h_1}  - p_{1'}\right)_l \varepsilon_{T k} \right]^*.
\end{aligned}
\end{equation}

Compared to the $LL$ cross-section, the $TL$ cross-section has the same form up to a factor of 
$$\frac{1}{Q} \frac{x_{\bar{q}}-x_q}{2 x_{\bar{q}} x_q } \left( \frac{x_{\bar{q}}}{x_{h_2}} \vec{p}_{h_2}  - \vec{p}_{2'}\right) \cdot \vec{\varepsilon}^{\,*}_T $$
or $$ \frac{1}{Q} \frac{x_{\bar{q}}-x_q}{2 x_{\bar{q}} x_q }  \left( \frac{x_q}{x_{h_1}} p_{h_1}  - p_{1'}\right) \cdot \varepsilon_{T}^{\,*} \,.$$
The $TT$ cross-section differs from the $LL$ cross-section by a factor of 
$$\frac{1}{Q^2} \frac{1}{4 x_q^2 x_{\bar{q}}^2 } \left[ (x_{\bar{q}} -x_q )^2 g_{\perp}^{ri} g_\perp^{lk} - g_\perp^{rk} g_\perp^{li} + g_{\perp}^{rl} g_\perp^{ik} \right] \left( \frac{x_q}{x_{h_1}} p_{h_1 }  - p_{1}\right)_{r}  \varepsilon_{T i}  \left( \frac{x_q}{x_{h_1}} p_{h_1}  - p_{1'}\right)_l \varepsilon_{T k}^{\,*} $$
or 
$$
\frac{1}{Q^2} \frac{1}{4 x_q^2 x_{\bar{q}}^2 } \left[ (x_{\bar{q}} -x_q )^2 g_{\perp}^{ri} g_\perp^{lk} - g_\perp^{rk} g_\perp^{li} + g_{\perp}^{rl} g_\perp^{ik} \right]
\left( \frac{x_{\bar{q}}}{x_{h_2}} p_{h_2 }  - p_{2}\right)_{r}  \varepsilon_{T i} \left( \frac{x_{\bar{q}}}{x_{h_2}} p_{h_2}  - p_{2'}\right)_l  \varepsilon_{T k}^{\,*} \,.
$$

The factors of $1/Q$ and $1/Q^2$ come from the photon polarization while the other modifications come from the expression of the squared of the impact factors.   
Those modifications and additional factors between $TL$ and $TT$ cross-section wrt $LL$ will remain true when going to NLO, for what concerns the extraction of divergences. This means that no additional detailed calculations are needed for those cases.  

\subsection{NLO computations in a nutshell}

\subsubsection*{Different types of contributions in the dipole picture}

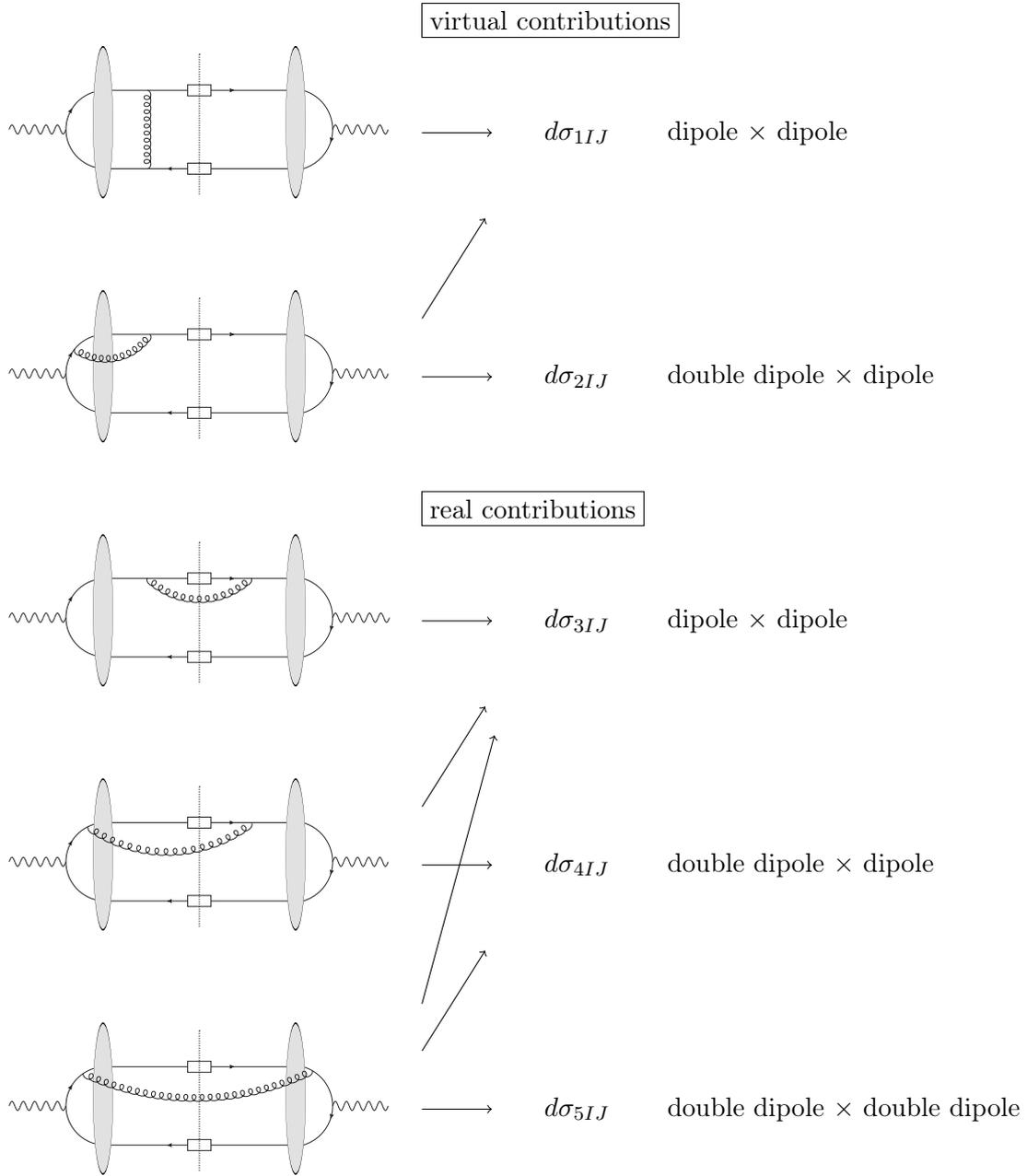
\begin{figure}
\begin{picture}(430,470)

\put(180,470){\fbox{virtual contributions}}

\put(10,400){\includegraphics[scale=0.25]{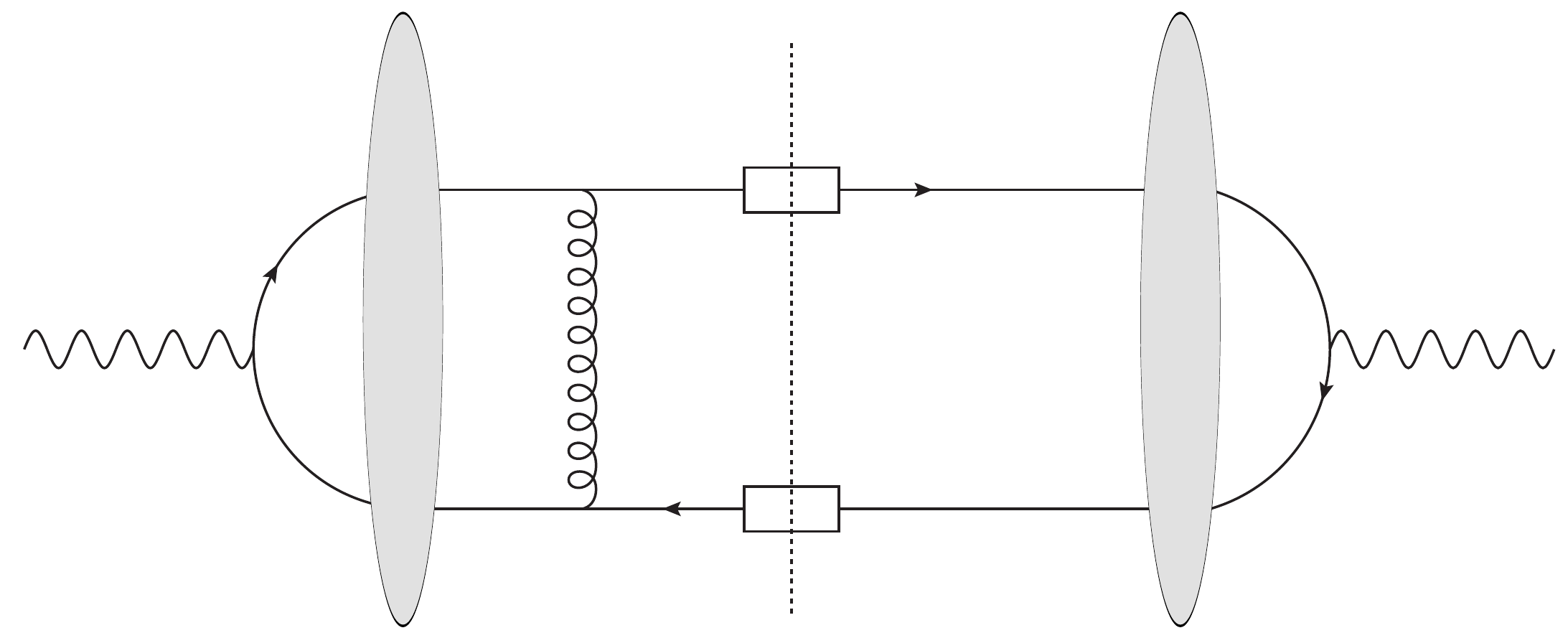}}
\put(180,425){$\Arrow{1cm}$}
\put(230,425){$d\sigma_{1IJ}$}
\put(280,425){dipole $\times$ dipole}

\put(10,300){\includegraphics[scale=0.25]{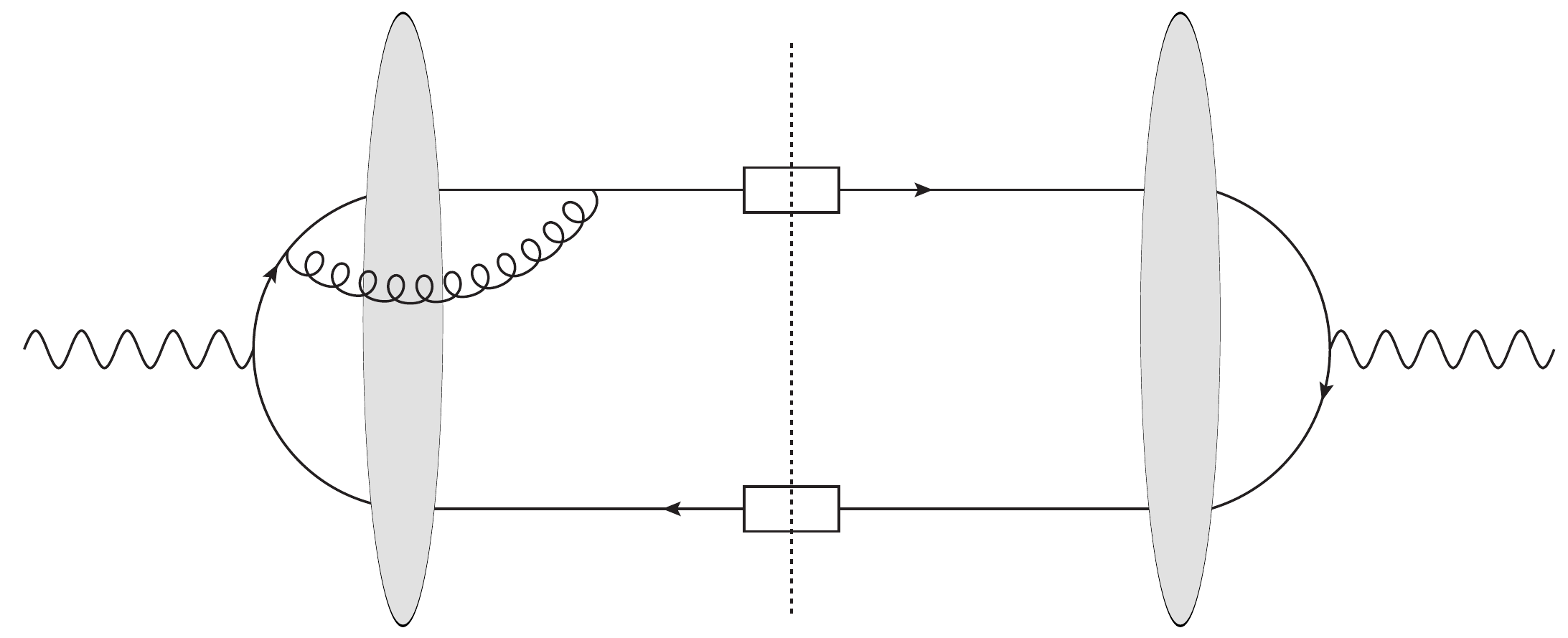}}
\put(180,350){\rotatebox{58}{$\Arrow{1.7cm}$}}
\put(180,325){$\Arrow{1cm}$}
\put(230,325){$d\sigma_{2IJ}$}
\put(280,325){double dipole $\times$ dipole}

\put(180,270){\fbox{real contributions}}

\put(10,200){\includegraphics[scale=0.25]{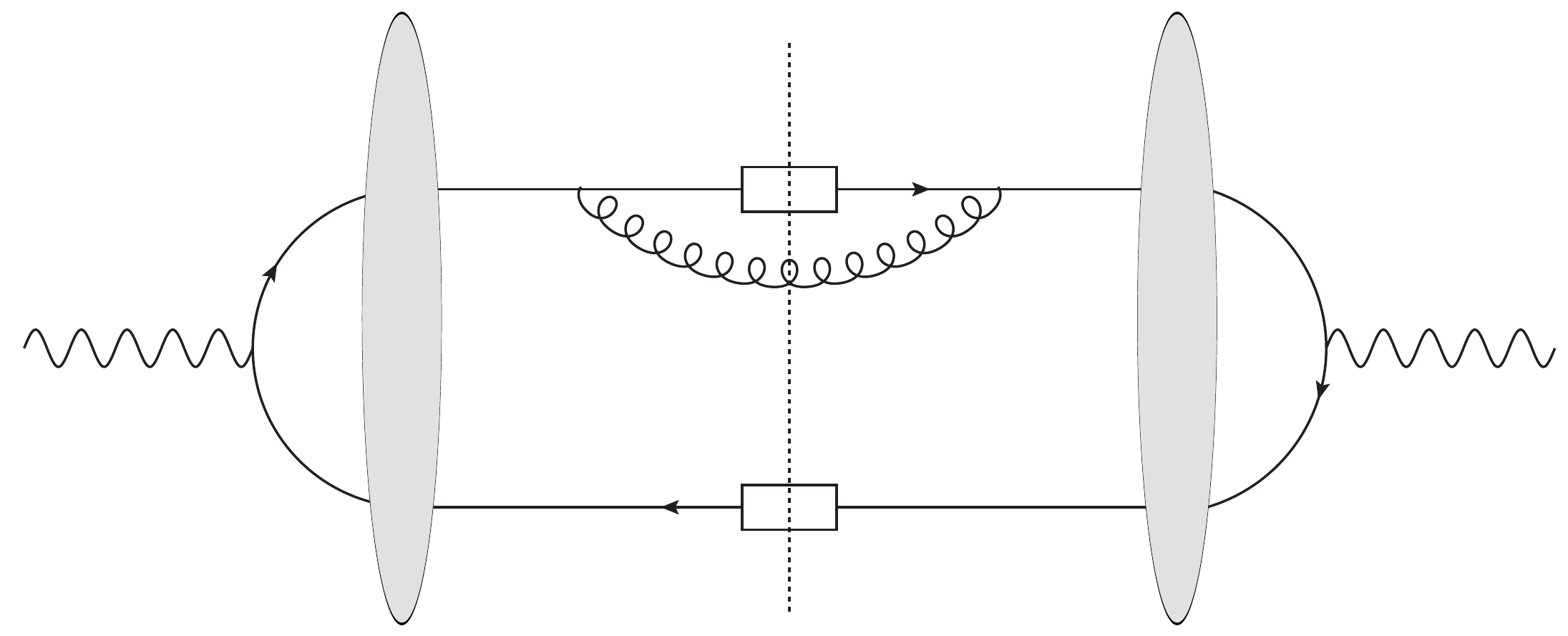}}
\put(180,225){$\Arrow{1cm}$}
\put(230,225){$d\sigma_{3IJ}$}
\put(280,225){dipole $\times$ dipole}

\put(10,100){\includegraphics[scale=0.25]{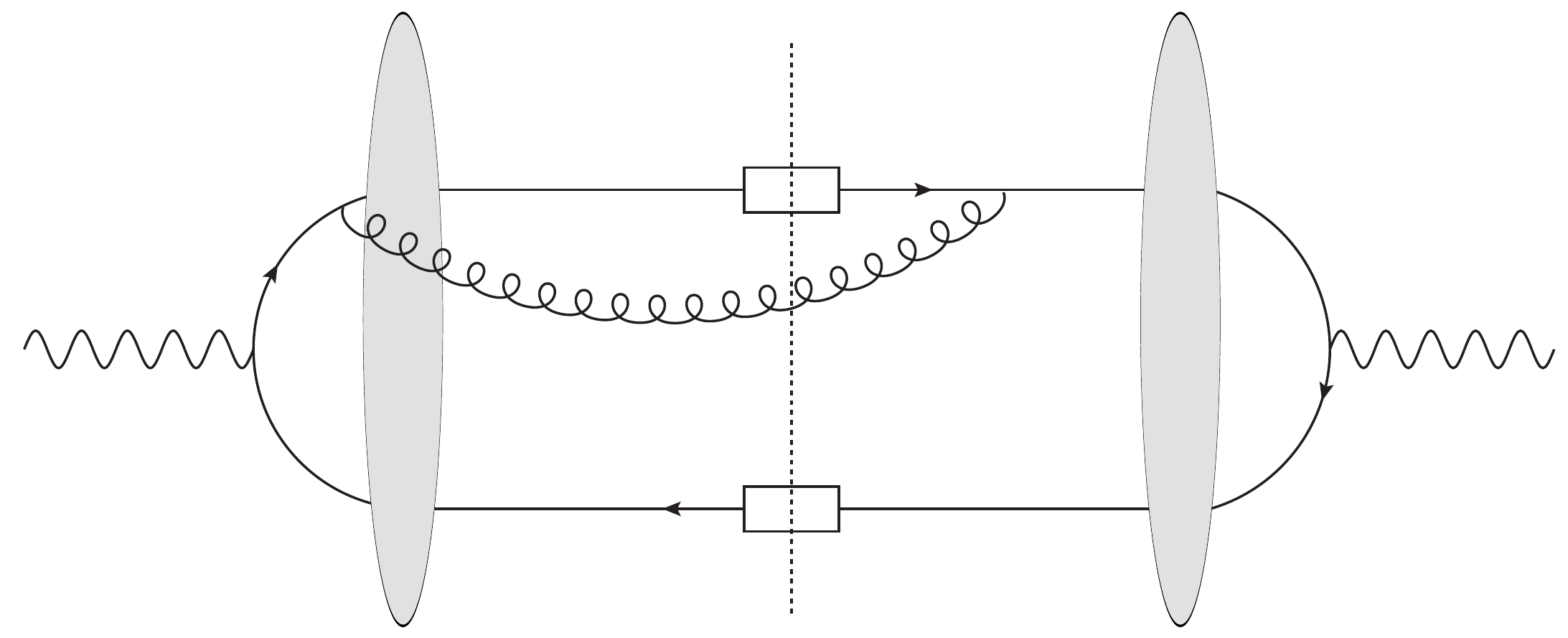}}
\put(180,125){$\Arrow{1cm}$}
\put(180,150){\rotatebox{58}{$\Arrow{1.7cm}$}}
\put(230,125){$d\sigma_{4IJ}$}
\put(280,125){double dipole $\times$ dipole}

\put(10,0){\includegraphics[scale=0.25]{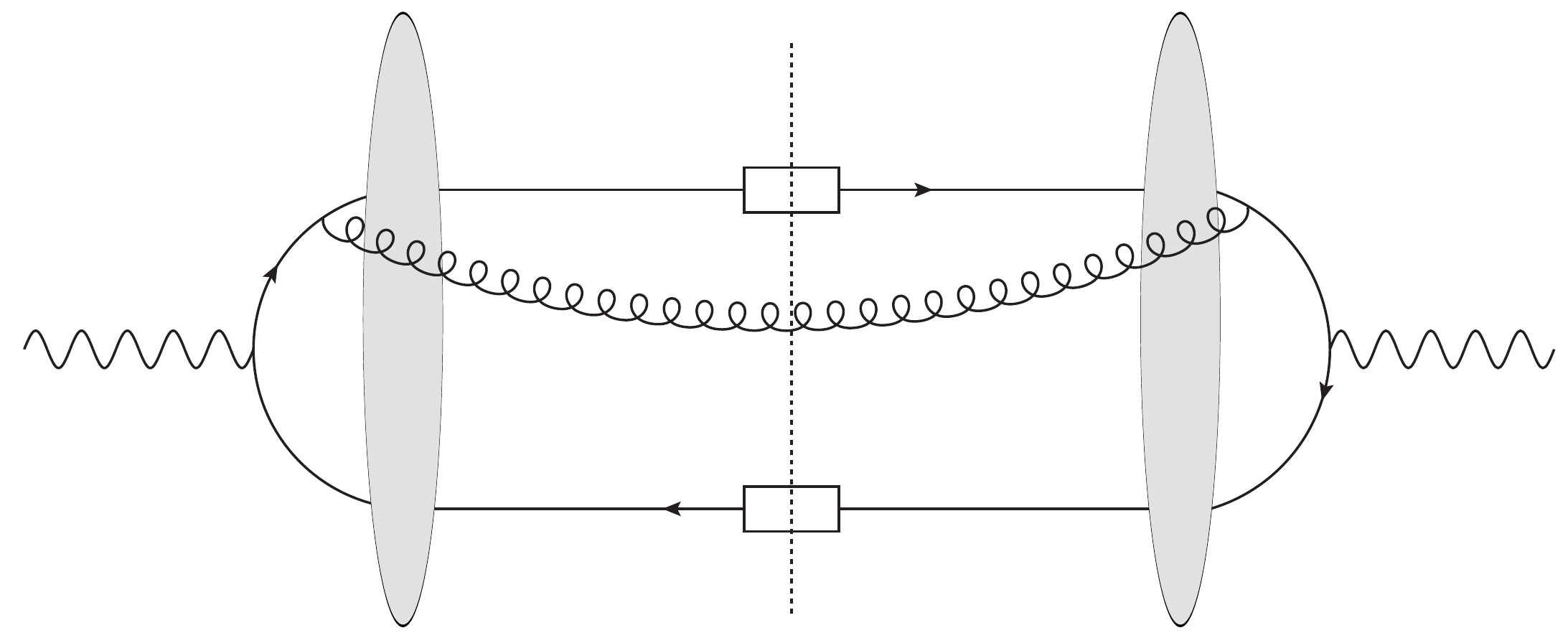}}
\put(180,25){$\Arrow{1cm}$}
\put(180,50){\rotatebox{58}{$\Arrow{1.7cm}$}}
\put(180,70){\rotatebox{75}{$\Arrow{4cm}$}}
\put(230,25){$d\sigma_{5IJ}$}
\put(280,25){double dipole $\times$ double dipole}

\end{picture}
\vspace{.1cm}
\caption{Illustration of the 5 kinds of  contributions to the NLO cross-section from the dipole point of view. Arrows show to which combination of dipole structures each type of diagrams contributes.}
  \label{fig:sigma-NLO-dipole}
\end{figure}

At NLO, since we rely on the shockwave approach, it is convenient to separate the various contributions from the dipole point of view, as illustrated in Fig.~\ref{fig:sigma-NLO-dipole}. In this figure, we exhibit a few examples of diagrams, either virtual or real, as a representative of each 5 classes of diagrams. There are indeed 5 classes of contributions from the dipole point of view, namely $d \sigma_{iJI}\, (i=1,\cdots 5)$, so that the NLO density matrix can be written as
\begin{equation}
d\sigma_{JI}=d\sigma_{0JI}+d\sigma_{1JI}+d\sigma_{2JI}+d\sigma_{3JI}+d\sigma_{4JI}+d\sigma_{5JI}. 
\label{sigmaNLO}
\end{equation}
Now, we will shortly discuss each of these 5 NLO corrections.

For the virtual diagrams, there are two classes of diagrams: the diagrams in which the virtual gluon does not cross the shockwave, thus contributing to $d \sigma_{1IJ}$, purely made of dipole $\times$ dipole terms; the diagrams in which the virtual gluon does cross the shockwave, contributing both to $d \sigma_{1IJ}$, made of dipole $\times$ dipole terms as well as to $d \sigma_{2IJ}$, made of double dipole $\times$ dipole  (and dipole $\times$ double-dipole) terms.

For the real diagrams, there are three classes of diagrams: the diagrams in which the real gluon does not cross the shockwave, thus contributing to $d \sigma_{3IJ}$, purely made of dipole $\times$ dipole terms; the diagrams in which the real gluon crosses  exactly once the shockwave, contributing both to $d \sigma_{3IJ}$, made of dipole $\times$ dipole terms as well as to $d \sigma_{4IJ}$, made of double dipole $\times$ dipole  (and dipole $\times$ double-dipole) terms; the diagrams in which the real gluon crosses  exactly twice the shockwave, contributing to $d \sigma_{3IJ}$, made of dipole $\times$ dipole terms, to $d \sigma_{4IJ}$, made of double dipole $\times$ dipole  (and dipole $\times$ double-dipole) terms, and to $d \sigma_{5IJ}$, made of double dipole $\times$ double dipole terms.

\subsubsection*{Overview of cancellation of divergences}
Before providing technical details, let us sketch the way the computation will be done, putting emphasis on the infrared (IR) sector.

When generically decomposing any on-shell parton momentum in the Sudakov basis as\footnote{Here $p^+$ is a large fixed momentum, eg $p_\gamma^+$ in our present case.}
\begin{equation}
\label{p-sudakov}
p^\mu = z p^+ n_1^\mu + \frac{\vec{p}^{\,2}}{2 z p^+} n_2^\mu + p_\perp^\mu\,,
\end{equation} 
in the IR sector, we face three kinds of divergences:
\begin{itemize}
    \item Rapidity:
$z$ goes to 0 and $p_{\perp}$ arbitrary.
    \item Soft:
any component of the gluon momentum goes linearly to 0 (obtained with both $z$ and $p_\perp = z \tilde{p}_\perp \sim z$ going to 0).
    \item Collinear:
parton's $p_\perp$ goes to zero, $z$ being arbitrary.
\end{itemize} 

Technically, since the $z$ integration is regulated through a lower cut-off (named $\alpha$), one should be careful with the fact that the appearance of 
$\ln \alpha$ may have originated from both rapidity or soft divergences.  

The calculation goes as follows. First, the rapidity divergences, appearing only in the virtual corrections in the present computation, are taken care of at the amplitude level by absorbing them in the shockwave through one step of B-JIMWLK evolution. This removes part of terms with $\ln \alpha$ related to pure rapidity divergences.

Next, at the level of cross-section, we separate the soft divergent contribution from the non-soft divergent terms. Combining real and virtual contributions, these soft divergent terms will disappear as guaranteed by the Kinoshita-Lee-Nauenberg theorem.

Finally, the remaining type of divergences, which are of purely collinear nature, will be absorbed into the fragmentation functions through one step of the Dokshitzer-Gribov-Lipatov-Altarelli-Parisi (DGLAP) evolution equation~\cite{Gribov:1972ri, Lipatov:1974qm, Altarelli:1977zs, Dokshitzer:1977sg}.

\subsubsection*{Different fragmentation contributions to the NLO cross-section}

\def\sca{.27}
\begin{figure}[htbp]
\begin{picture}(430,400)
\put(10,300){\includegraphics[scale=\sca]{FF_dihadron_LO_box.pdf}}
\put(90,370){NLO}
\put(188,328){$=$}
%1-a
\put(202,300){\includegraphics[scale=\sca]{FF_dihadron_LO_box.pdf}}
\put(250,370){1-loop}
\put(380,328){+ c.c}
\put(282,275){(a)}
%%%%%%%%%%%%%%%%%%%%%%%%%%%%%%%%%%%%%%%
\put(0,200){+}
%2-b
\put(10,170){\includegraphics[scale=\sca]{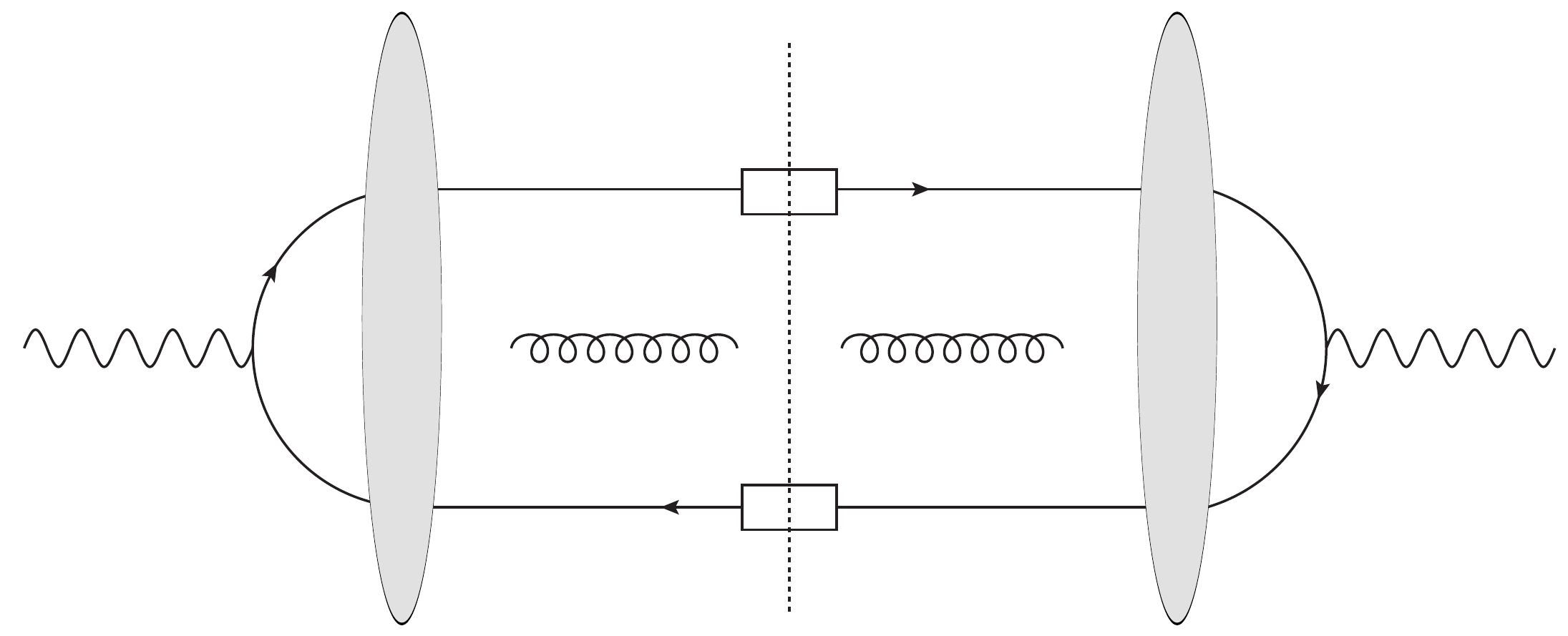}}
\put(188,198){+}
%3-c
\put(202,170){\includegraphics[scale=\sca]{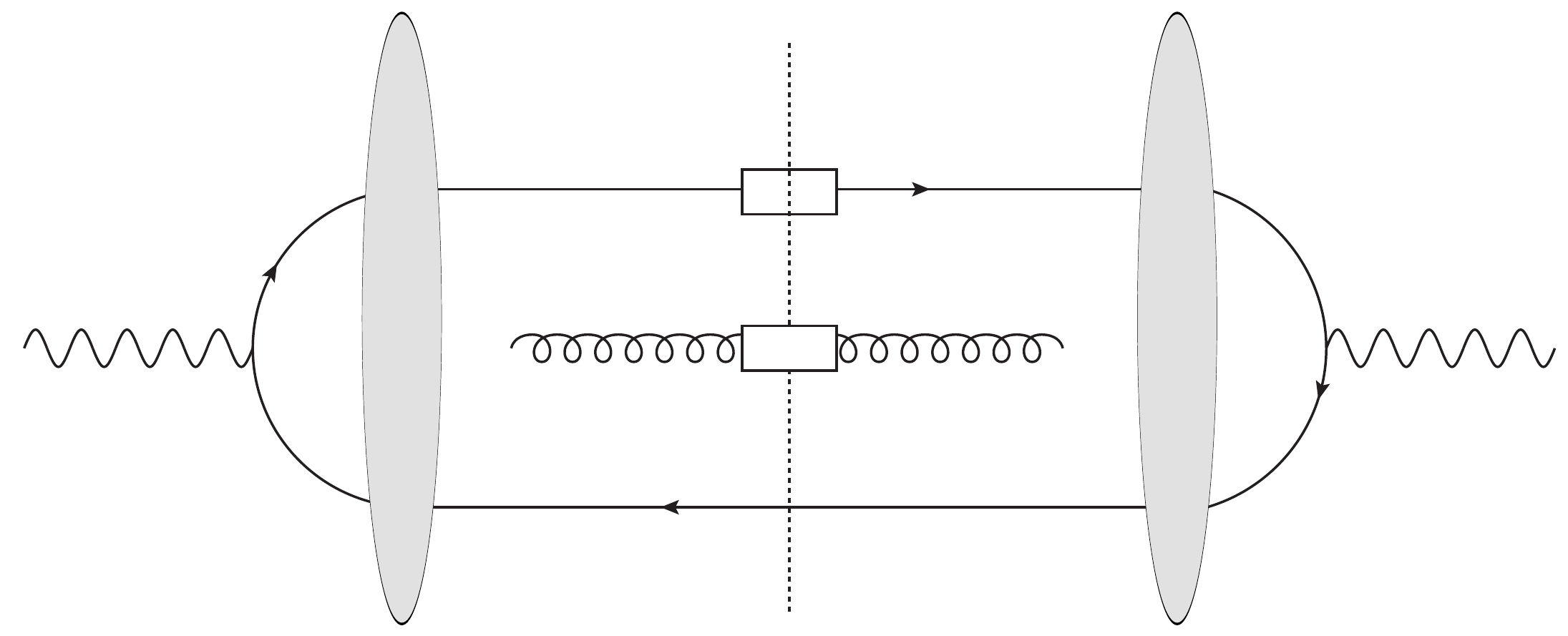}}
\put(82,145){(b)}
\put(282,145){(c)}
%%%%%%%%%%%%%%%%%%%%%%%%%%%%%%%%%%%%%%%
\put(0,70){+}
%4-d
\put(10,40){\includegraphics[scale=\sca]{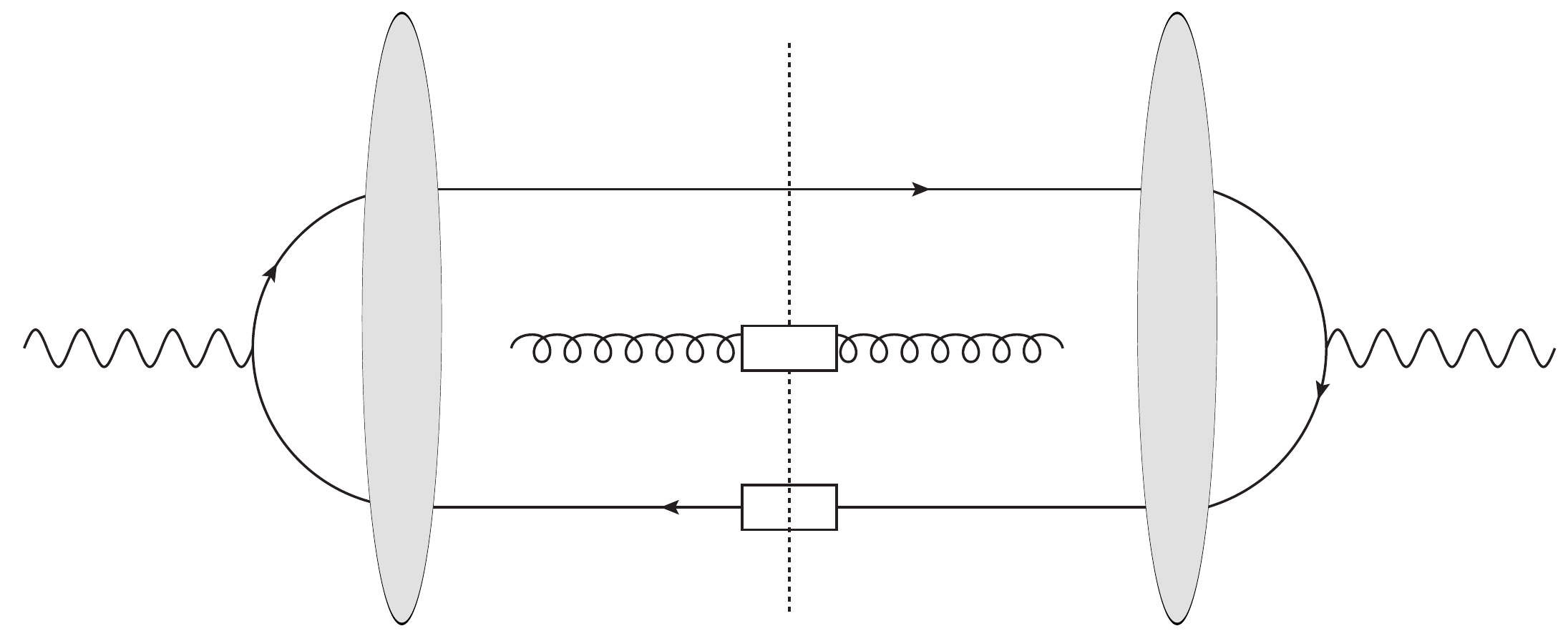}}
\put(188,68){+}
%5-e
\put(205,40){\includegraphics[scale=\sca]{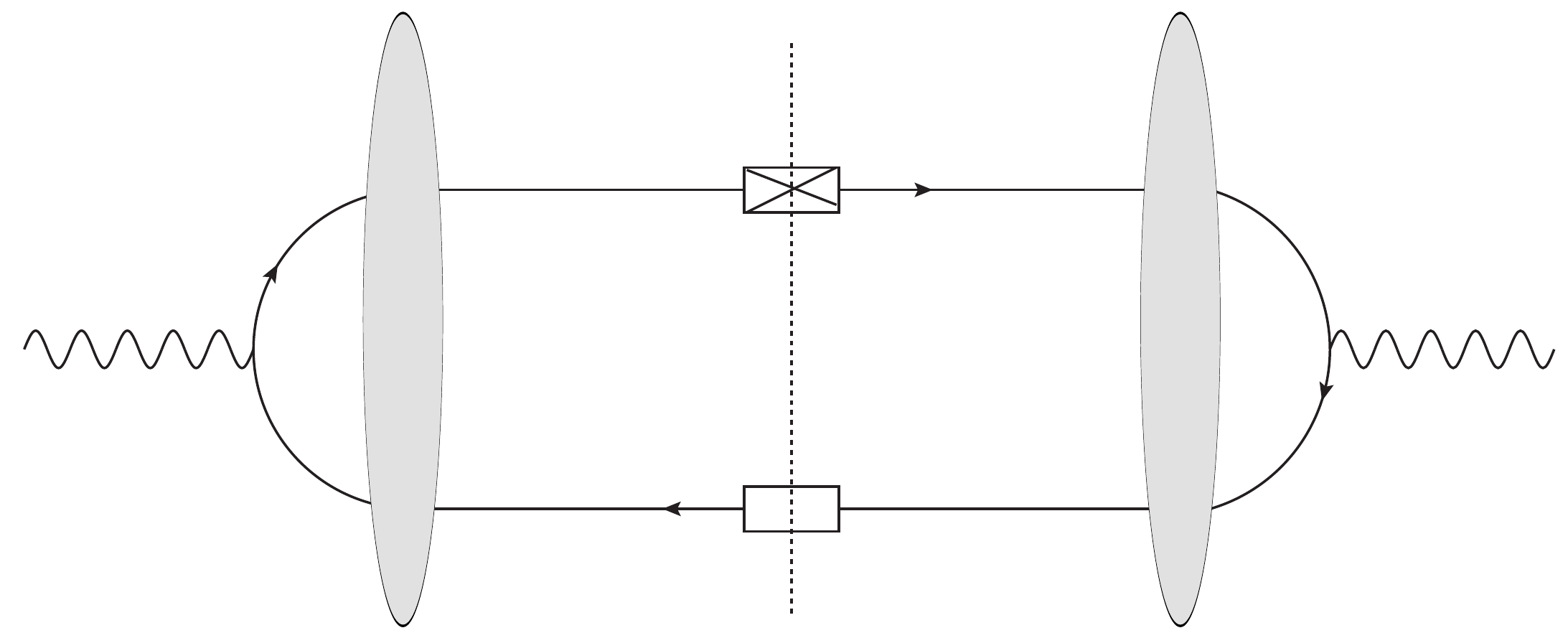}}
\put(197,68){\scalebox{2}{\Bigg ( }}
\put(380,68){$+ \, q \leftrightarrow \bar{q}$ \scalebox{2}{\Bigg )}}
\put(82,15){(d)}
\put(282,15){(e)}
\end{picture}
  \caption{The 5 kinds of  contributions to the NLO cross-section.}
  \label{fig:sigma-NLO}
\end{figure}

At NLO, we have to deal with 5 kinds of contributions to the cross-section, illustrated in fig.~\ref{fig:sigma-NLO}:
\begin{enumerate}[label=(\alph*)]
    \item $\gamma^{*} + P \rightarrow h_1 + h_2 + X + P$ cross-section at one-loop (i.e. virtual contributions),
    \item $\gamma^{*} + P \rightarrow h_1 + h_2 + g + X + P$ cross-section at Born level (i.e. real contributions),
    \item $\gamma^{*} + P \rightarrow h_1 + h_2 + \bar{q} + X + P$ cross-section at Born level (i.e. real contributions),
    \item $\gamma^{*} + P \rightarrow h_1 + h_2 + q + X + P$ cross-section at Born level (i.e. real contributions),
    \item FFs counterterms,
\end{enumerate}
where $X$ denotes the remnants of the fragmentation.

%%%%%%%%%%%%%%%%%%%%%% b %%%%
\begin{figure}[htbp]
\begin{picture}(420,410)
\put(100,340){\includegraphics[scale=\sca]{FF_dihadron_NLOqqbarg_box.pdf}}
\put(180,315){(b)}
%%%%%%%%%%%%%%%%%%%%%%%%%%%%%%%%%%%%%%%%%
\put(0,240){=}
%1
\put(10,210){\includegraphics[scale=\sca]{FF_dihadron_soft_1.pdf}}
\put(195,240){+}
%2
\put(210,210){\includegraphics[scale=\sca]{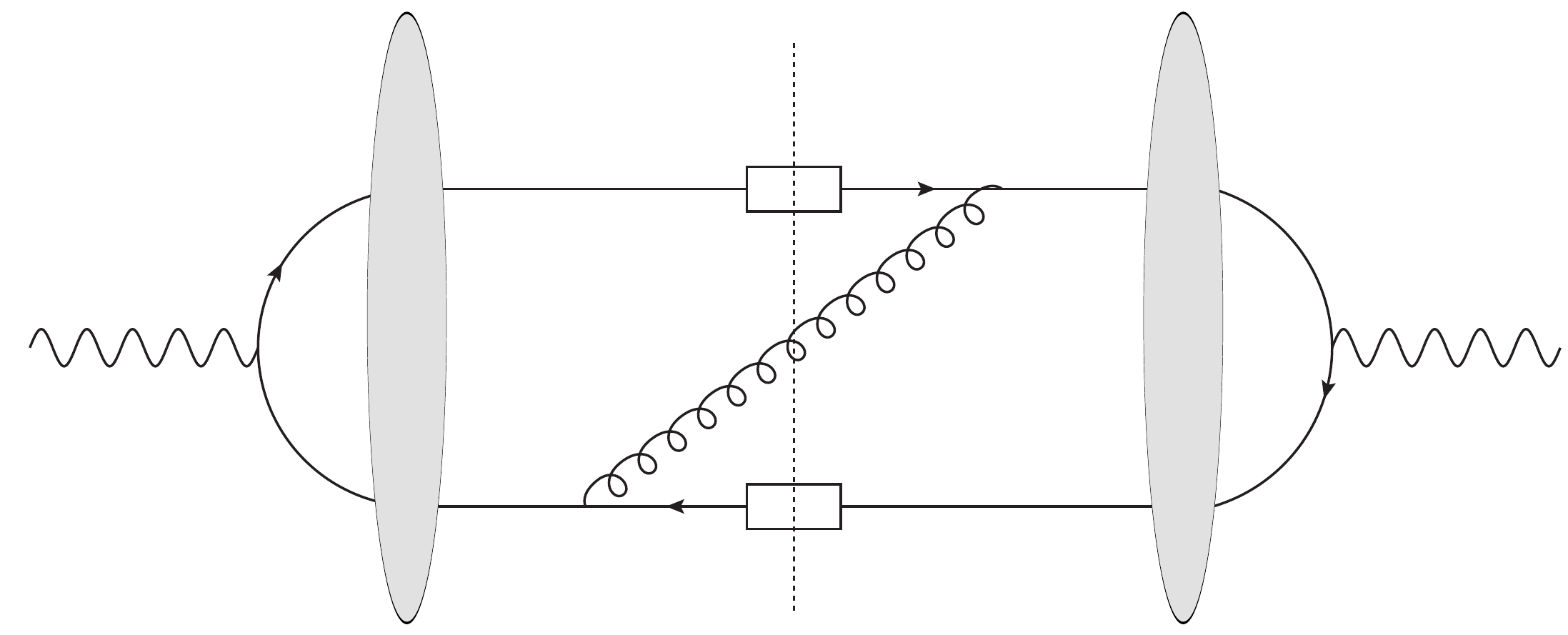}}
\put(92,185){(1)}
\put(292,185){(2)}
%%%%%%%%%%%%%%%%%%%%%%%%%%%%%%%%%%%%%%%
\put(0,112){+}
%4
\put(10,80){\includegraphics[scale=\sca]{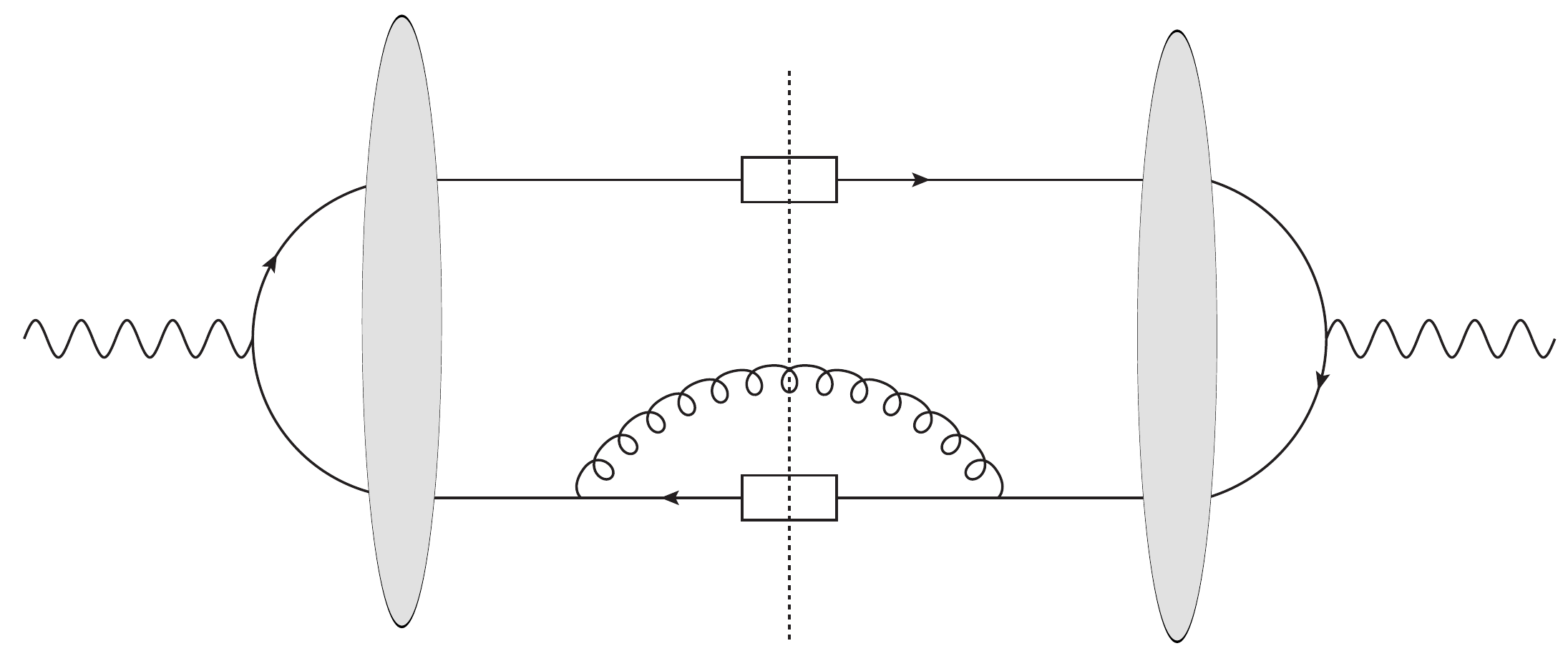}}
\put(195,112){+}
%5
\put(210,80){\includegraphics[scale=\sca]{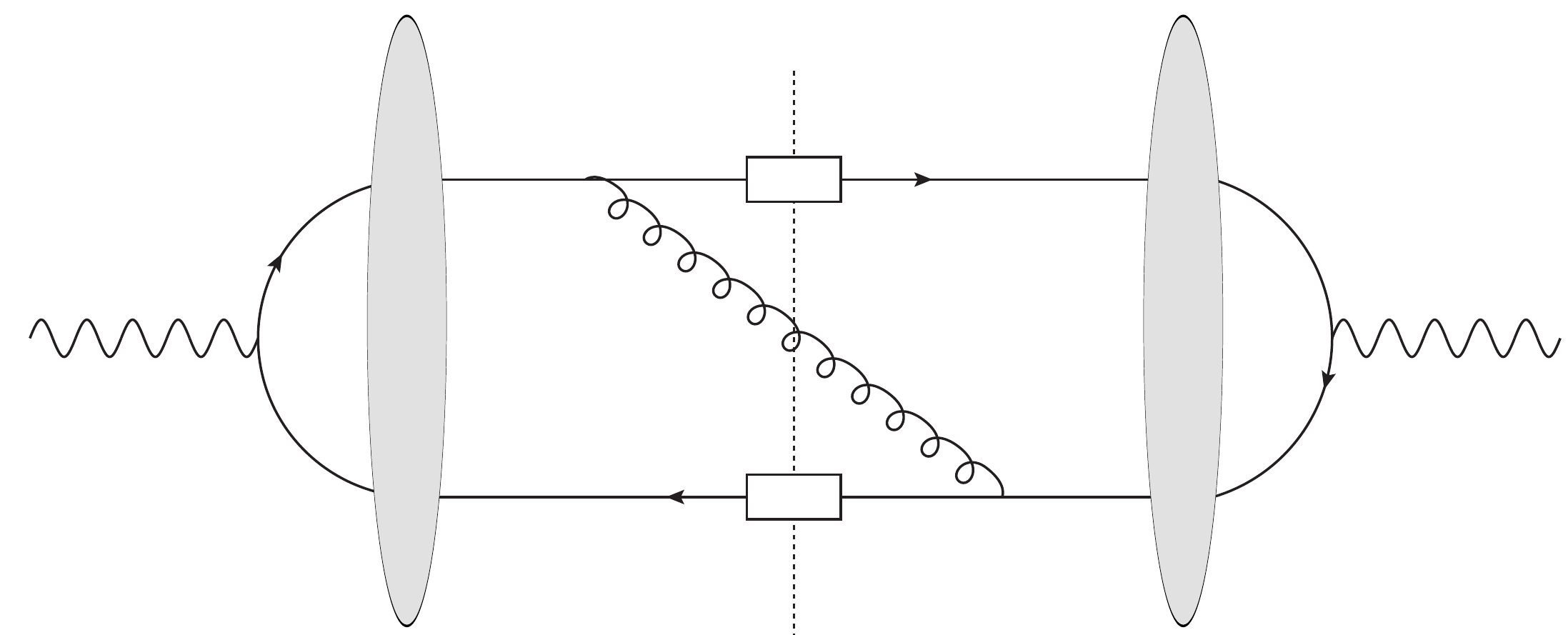}}
\put(92,45){(3)}
\put(292,45){(4)}
%%%%%%%%%%%%%%%%%%%%%%%%%%
\put(0,0){+ finite contributions}
\end{picture}
\vspace{.5cm}
\caption{NLO cross-section in the case of fragmentation from the quark and the antiquark. We explicitly isolate the diagrams which contain divergences.
  Diagram (1) contains a collinear divergence between the fragmenting quark and the gluon as well as a soft gluon divergence. Diagram (2) contains a soft gluon divergence. Diagram (3) contains a collinear divergence between the fragmenting antiquark and the gluon as well as a soft gluon divergence. Diagram (4) contains a soft gluon divergence. By "finite terms", here, we mean all diagrams in which the gluon crosses the shockwave at least once.}
  \label{fig:NLO-b-div}
\end{figure}
Contributions (a) and (e) are easy to treat since (a) is simply the convolution of a known one-loop result with fragmentation functions, while (e) is obtained from the Born result when one renormalizes the fragmentation functions. We just split them into finite and divergent parts. 

For the real contributions (b), (c), (d), the treatment is less straightforward even if the partonic real corrections are also already known.

Contribution (b) is the most complicated one, it contains both soft and collinear divergences. When we square the amplitude contributing to (b), see fig.~\ref{fig:NLO-b-div}, there is a series of finite contributions plus one, represented by the sum of contributions (1), (2), (3), (4) of fig.~\ref{fig:NLO-b-div}, that contains all divergences, and that belongs only to the dipole-dipole contribution. 
We  add and subtract to the latter its soft limit to obtain the following structure
\begin{eqnarray}
\label{sigmatilde_q-qbar}
\tilde{\sigma}_{(b)div} &=&  \sum_{\lambda_q,\lambda_g, \lambda_{\bar{q}}}|A_{q\bar{q},sing-dipole}  |^2_{div}= \tilde{\sigma}_{(b)div,1} + \tilde{\sigma}_{(b)div,2} + \tilde{\sigma}_{(b)div,3} + \tilde{\sigma}_{(b)div,4} 
\nonumber \\
&=& \tilde{\sigma}_{(b)div}^{soft}
+ (\tilde{\sigma}_{(b)div,1}-\tilde{\sigma}_{(b)div,1}^{soft})
+
(\tilde{\sigma}_{(b)div,2}-\tilde{\sigma}_{(b)div,2}^{soft})
+
(\tilde{\sigma}_{(b)div,3}-\tilde{\sigma}_{(b)div,3}^{soft}) \nonumber \\
&&+
(\tilde{\sigma}_{(b)div,4}-\tilde{\sigma}_{(b)div,4}^{soft})
\end{eqnarray}
with the meaning that each of these four contributions is in one-to-one correspondence to the four diagrams (1), (2), (3), (4) in fig.~\ref{fig:NLO-b-div}.
Among these various terms, the terms $(\tilde{\sigma}_{(b)div,1}-\tilde{\sigma}_{(b)div,1}^{soft})$ and $(\tilde{\sigma}_{(b)div,3}-\tilde{\sigma}_{(b)div,3}^{soft})$ are collinearly divergent while the terms $(\tilde{\sigma}_{(b)div,2}-\tilde{\sigma}_{(b)div,2}^{soft})$ and $(\tilde{\sigma}_{(b)div,4}-\tilde{\sigma}_{(b)div,4}^{soft})$ are finite.

%%%%%%%%%%%%%%%%%%%  c %%%%%%%%%%%%%%%%%%%%%%%
\begin{figure}[h!]
\begin{picture}(420,140)
\put(10,70){\includegraphics[scale=\sca]{diagram4a_box.pdf}} \put(195,100){=}
\put(210,67){\includegraphics[scale=\sca]{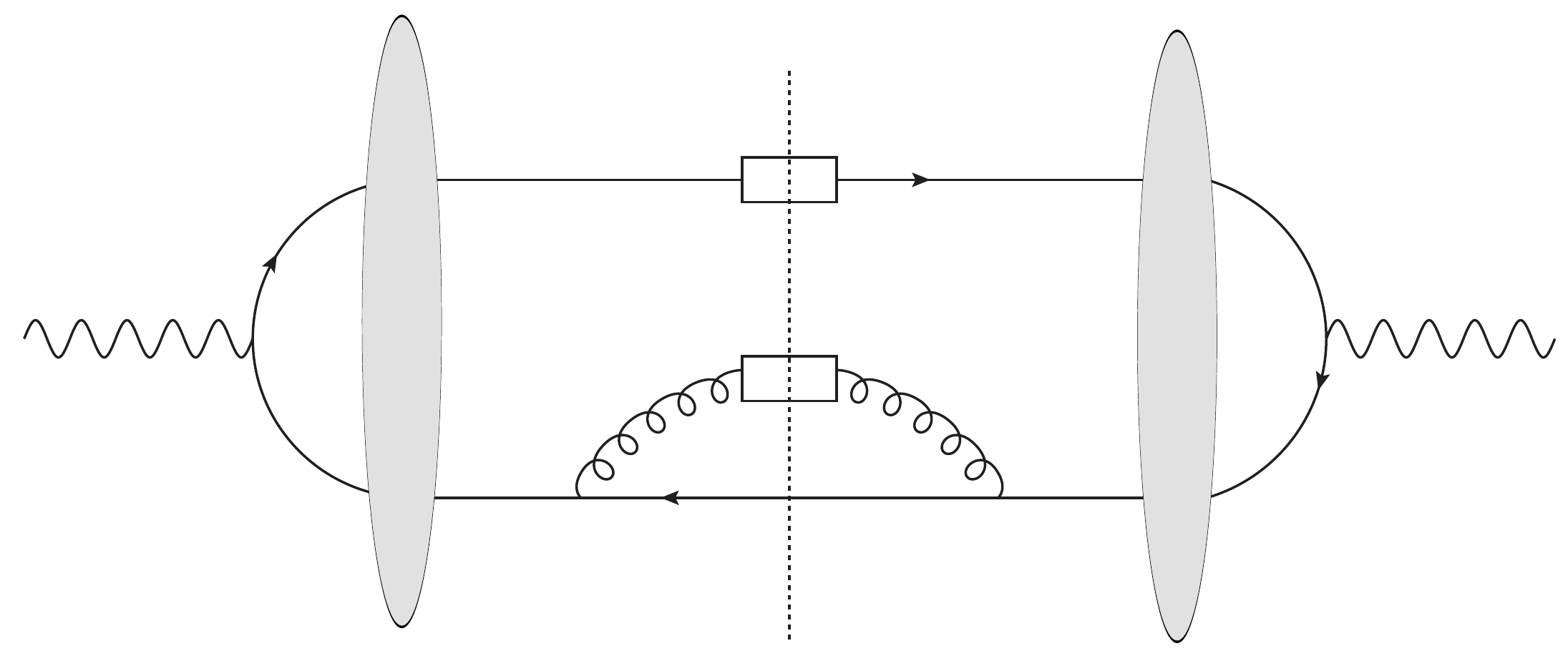}}
\put(0,0){+ finite contributions.}
\put(90,45){(c)}
\put(290,45){(3)}
\end{picture}
\vspace{.5cm}

\caption{NLO cross-section in the case of fragmentation from the gluon and the quark. We explicitly isolate the diagram which contains divergences, namely a collinear divergence between the fragmenting gluon and the antiquark.}
\label{fig:NLO-c-div}
\end{figure}
In fig.~\ref{fig:NLO-c-div}, the contribution with fragmentation from quark and gluon is considered. Again, we have
\begin{eqnarray}
\label{sigmatilde_q-gluon}
\tilde{\sigma}_{(c)div} &=&  \sum_{\lambda_q,\lambda_g, \lambda_{\bar{q}}}|A_{qg,sing-dipole}  |^2_{div}= 
\tilde{\sigma}_{(c)div,3}   
\,.
\end{eqnarray}
Here, one does not encounter any soft divergence. The only divergence comes from the contribution $\tilde{\sigma}_{(c)div,3}$ which has a collinear divergence when the fragmenting gluon and the antiquark are collinear.

\begin{figure}[h!]
\begin{picture}(420,150)
\put(10,70){\includegraphics[scale=\sca]{diagram3a_box.pdf}} \put(195,100){=}
\put(210,70){\includegraphics[scale=\sca]{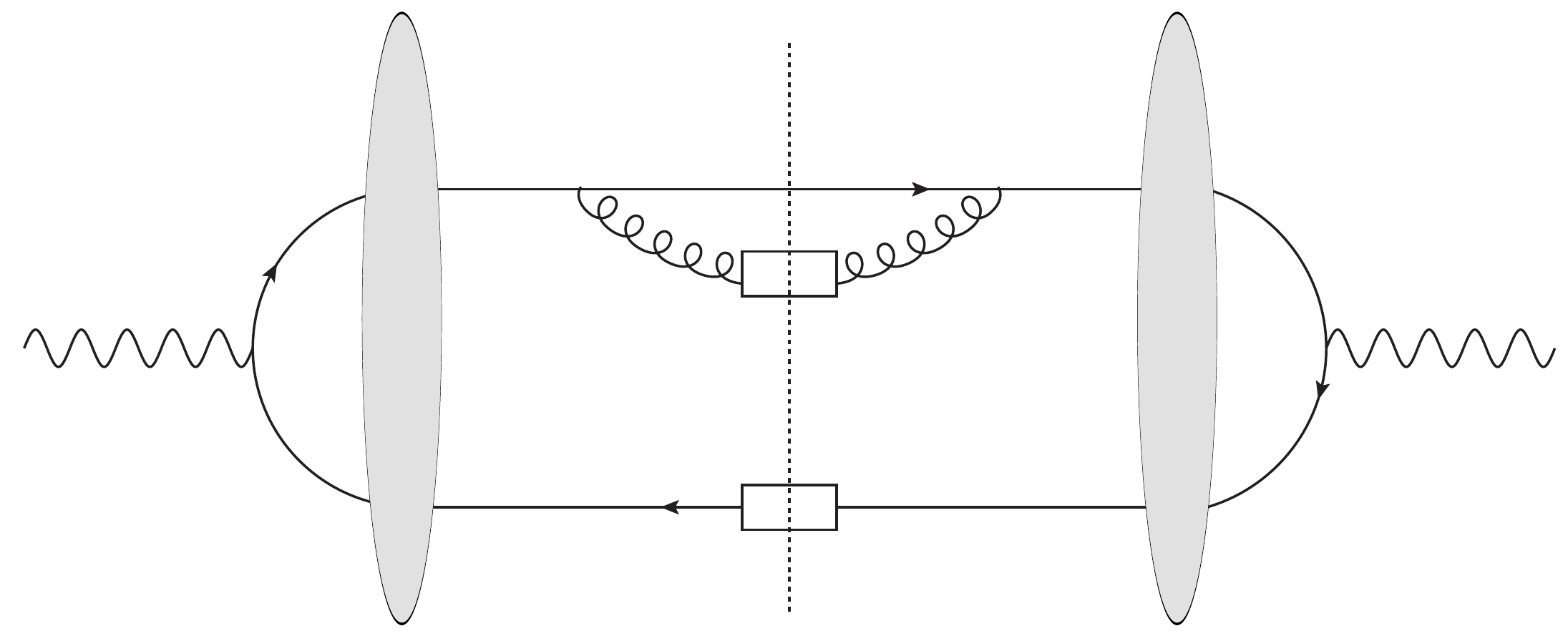}}
\put(0,0){+ finite contributions.}
\put(90,45){(d)}
\put(290,45){(1)}
\end{picture}
\vspace{.5cm}

\caption{NLO cross-section in the case of fragmentation from the gluon and the antiquark. We explicitly isolate the diagram which contains divergences, namely a collinear divergence between the fragmenting gluon and the quark.}
\label{fig:NLO-d-div}
\end{figure}

The discussion for the fourth case, see fig.~\ref{fig:NLO-d-div}, involving the fragmentation from antiquark and gluon goes along the same line: the only divergence comes from the contribution $\tilde{\sigma}_{(d)div,1}$ which has a collinear divergence when the fragmenting gluon and the quark are collinear.

%%%%%%%%%%%%%%%%%%%  d   %%%%%%%%%%%%%%%%%%%%%%%

\section{Counterterms from FFs renormalization and evolution}
\label{sec:CounterTerms}

The renormalization and evolution equations of FFs express the bare FFs in terms of dressed ones. In $\overline{\text{MS}}$ scheme, at factorization scale $\mu_F$, they take the form, following notations of ref.~\cite{Ivanov:2012iv} 
\begin{equation}
\label{eq: FF evolution} 
\begin{aligned}
    D_{q}^{h}(x)& =D_{q}^{h}\left(x, \mu_{F}\right)-\frac{\alpha_{s}}{2 \pi}\left(\frac{1}{\hat{\epsilon}}+\ln \frac{\mu_{F}^{2}}{\mu^{2}}\right) \int_{x}^{1} \frac{d z}{z}\left[D_{q}^{h}\left(\frac{x}{z}, \mu_{F}\right) P_{q q}(z)+D_{g}^{h}\left(\frac{x}{z}, \mu_{F}\right) P_{gq}(z)\right], \\
    D_g^h (x) &= D_g^h (x, \mu_F) - \frac{\alpha_s}{2 \pi} \!\left( \frac{1}{\hat{\epsilon}}+\ln \frac{\mu_{F}^{2}}{\mu^{2}}\right)\! \!\int_{x}^{1} \!\frac{d z}{z} \! \left[ \sum_{q,\bar{q}} D_{q}^{h}\left(\frac{x}{z}, \mu_{F}\right) \! P_{qg}(z)+D_{g}^{h}\left(\frac{x}{z}, \mu_{F}\!\right)\! P_{gg}(z)\right]\!,
\end{aligned}
\end{equation}
where $\frac{1}{\hat{\epsilon}} = \frac{\Gamma (1- \epsilon)}{\epsilon (4 \pi )^\epsilon} \sim \frac{1}{\epsilon} + \gamma_E - \ln (4 \pi)$ and $\mu$ is an arbitrary parameter introduced by dimensional regularization. 
The LO splitting functions are given by  
\begin{eqnarray}
    P_{qq}(z) &=& C_F \left[ \frac{1 + z^2}{(1-z)_+} + \frac{3}{2} \delta (1-z) \right], \\
    P_{gq}(z) &=& C_F \frac{1 + (1-z)^2}{z}\,, \\
    P_{qg}(z) &=& T_R \left[z^2 + (1-z)^2 \right] \ \text{ with } \ T_R = \frac{1}{2} \,,\\\
    P_{gg}(z) &=& 2 C_A \left[ \frac{1}{(1-z)_+} + \frac{1}{z} - 2 + z(1-z) \right] + \left( \frac{11}{6}C_A - \frac{n_f}{3} \right) \delta (1-z)\, ,
\end{eqnarray}
where the + prescription is defined as 
\begin{equation}
\label{eq: plus prescription}
    \int_a^1 d \beta \frac{F(\beta)}{(1-\beta)_+} = \int_a^1 d \beta \frac{F(\beta)- F(1)}{(1-\beta)} - \int_0^{a} d\beta \frac{F(1)}{1-\beta} \,.\\ 
\end{equation}
The collinear counterterms due to the renormalization of the bare FFs are calculated by inserting eq.~\eqref{eq: FF evolution} in the contributions~(\ref{eq:LL-LO}, \ref{eq:TL-LO}, \ref{eq:TT-LO}) to the LO cross-section. This corresponds to the contribution  (e) in figure \ref{fig:sigma-NLO}. 

For $LL$ cross-section, this counterterm takes the form  
    \begin{align*}
   & \frac{d \sigma_{L L}^{q\bar{q} \rightarrow h_1 h_2}}{d x_{h_1}d_{h_2}d^d p_{h_1 \perp}d^d p_{h_2 \perp}} \bigg |_{\text{ct}}  \\
   & = \frac{4 \alpha_{\mathrm{em}} Q^2 }{(2\pi)^{4(d-1)}N_c} \sum_{q} \int_{x_{h_1}}^1 \!\!\!\! d x_q \int_{x_{h_2}}^1 \!\!\!\! d x_{\bar{q}} \; x_q x_{\bar{q}}  \left( \frac{x_q}{x_{h_1}}\right)^d \! \left(\frac{x_{\bar{q}}}{x_{h_2}}\right)^d \! \delta (1-x_q - x_{\bar{q}}) \\
     & \times  {\cal F}_{LL} \left(- \frac{\alpha_s}{2\pi}\right) 
     \left(\frac{1}{\hat{\epsilon}}+\ln \frac{\mu_{F}^{2}}{\mu^{2}}\right)
      Q_q^2 \left\{ \int_{\frac{x_{h_1}}{x_q}}^1 \frac{d \beta_1}{\beta_1} \left[ P_{qq}(\beta_1) D_q^{h_1}\left(\frac{x_{h_1}}{\beta_1 x_q},\mu_F\right)D_{\bar{q}}^{h_2}\left(\frac{x_{h_2}}{x_{\bar{q}}},\mu_F\right)  \right. \right. \\
    & \left. + P_{gq}(\beta_1) D_g^{h_1}\left(\frac{x_{h_1}}{\beta_1 x_q}, \mu_F\right) D_{\bar{q}}^{h_2}\left(\frac{x_{h_2}}{x_{\bar{q}}},\mu_F\right) \right]\\
    & + \int_{\frac{x_{h_2}}{x_{\bar{q}}}}^1 \frac{d \beta_2}{\beta_2} \left[ P_{qq}(\beta_2) D_q^{h_1}\left(\frac{x_{h_1}}{x_q}, \mu_F\right)D_{\bar{q}}^{h_2}\left(\frac{x_{h_2}}{\beta_2 x_{\bar{q}}}, \mu_F\right) \right. \\
    & \left.  \left. + P_{gq}(\beta_2) D_q^{h_1}\left(\frac{x_{h_1}}{x_q}, \mu_F\right)D_{g}^{h_2}\left(\frac{x_{h_2}}{\beta_2 x_{\bar{q}}}, \mu_F\right) \right] \right\} + (h_1 \leftrightarrow h_2)  \\ 
    &= \frac{d \sigma_{L L}^{h_1 h_2}}{d x_{h_1}d_{h_2}d^d p_{h_1 \perp}d^d p_{h_2 \perp}} \bigg |_{\text{ct div}} +  \frac{d \sigma_{L L}^{h_1 h_2}}{d x_{h_1}d_{h_2}d^d p_{h_1 \perp}d^d p_{h_2 \perp}} \bigg |_{\text{ct fin}} .
    \numberthis[ct_LL]
   \end{align*}
We stress here that, for any separate term in the curly bracket, one can indifferently use $\mathcal{F}_{LL}$ or $\tilde{\mathcal{F}}_{LL}$. In particular, for the first two terms we can use $\mathcal{F}_{LL}$ and for the last two $\tilde{\mathcal{F}}_{LL}$. This simple observation is useful when observing the cancellation of divergences at the level of integrands. The same remark applies also for other transitions.  \\
For cross-sections involving other combinations of polarizations, we have
\begin{equation}
    \begin{aligned}
   & \frac{d \sigma_{TL}^{q\bar{q} \rightarrow h_1 h_2}}{d x_{h_1}d_{h_2}d^d p_{h_1 \perp}d^d p_{h_2 \perp}} \bigg |_{\text{ct}}  \\
   & = \frac{2 \alpha_{\mathrm{em}} Q}{(2\pi)^{4(d-1)}N_c} \sum_{q}  \int_{x_{h_1}}^1 \!\!\!\! d x_q \int_{x_{h_2}}^1 \!\!\!\! d x_{\bar{q}}  \left( \frac{x_q}{x_{h_1}}\right)^d \! \left(\frac{x_{\bar{q}}}{x_{h_2}}\right)^d \! (x_{\bar{q}}-x_q ) \; \delta (1-x_q - x_{\bar{q}}) \\
     & \times  {\cal F}_{TL} \left(- \frac{\alpha_s}{2\pi}\right)
     \left(\frac{1}{\hat{\epsilon}}+\ln \frac{\mu_{F}^{2}}{\mu^{2}}\right)
     Q_q^2 \left\{ \int_{\frac{x_{h_1}}{x_q}}^1 \frac{d \beta_1}{\beta_1} \left[ P_{qq} (\beta_1) D_q^{h_1}\left(\frac{x_{h_1}}{\beta_1 x_q},\mu_F\right)D_{\bar{q}}^{h_2}\left(\frac{x_{h_2}}{x_{\bar{q}}},\mu_F\right)  \right. \right. \\
    & \left. + P_{gq}(\beta_1) D_g^{h_1}\left(\frac{x_{h_1}}{\beta_1 x_q}, \mu_F\right) D_{\bar{q}}^{h_2}\left(\frac{x_{h_2}}{x_{\bar{q}}},\mu_F\right) \right] \\
    & + \int_{\frac{x_{h_2}}{x_{\bar{q}}}}^1 \frac{d \beta_2}{\beta_2} \left[ P_{qq}(\beta_2) D_q^{h_1}\left(\frac{x_{h_1}}{x_q}, \mu_F\right)D_{\bar{q}}^{h_2}\left(\frac{x_{h_2}}{\beta_2 x_{\bar{q}}}, \mu_F\right) \right.  \\
    & \left. \left. + P_{gq}(\beta_2) D_q^{h_1}\left(\frac{x_{h_1}}{x_q}, \mu_F\right)D_{g}^{h_2}\left(\frac{x_{h_2}}{\beta_2 x_{\bar{q}}}, \mu_F\right) \right]
    \right\} + (h_1 \leftrightarrow h_2) \\ 
    &= \frac{d \sigma_{T L}^{h_1 h_2}}{d x_{h_1}d_{h_2}d^d p_{h_1 \perp}d^d p_{h_2 \perp}} \bigg |_{\text{ct div}} +  \frac{d \sigma_{T L}^{h_1 h_2}}{d x_{h_1}d_{h_2}d^d p_{h_1 \perp}d^d p_{h_2 \perp}} \bigg |_{\text{ct fin }} ,
   \end{aligned}
\label{eq:ct TL }   
\end{equation}
and
\begin{equation}
\label{eq:ct TT }
    \begin{aligned}
   & \frac{d \sigma_{TT}^{q\bar{q} \rightarrow h_1 h_2}}{d x_{h_1}d_{h_2}d^d p_{h_1 \perp}d^d p_{h_2 \perp}} \bigg |_{\text{ct}} \\
   & = \frac{ \alpha_{\mathrm{em}}  }{(2\pi)^{4(d-1)}N_c} \sum_{q} \int_{x_{h_1}}^1 \!\!\!\! \frac{d x_q}{x_q} \int_{x_{h_2}}^1 \!\!\!\! \frac{d x_{\bar{q}} }{x_{\bar{q}}}  \left( \frac{x_q}{x_{h_1}}\right)^d \! \left(\frac{x_{\bar{q}}}{x_{h_2}}\right)^d \! \delta (1-x_q - x_{\bar{q}}) \\
     & \times  {\cal F}_{TT} \left(- \frac{\alpha_s}{2\pi}\right) 
     \left(\frac{1}{\hat{\epsilon}}+\ln \frac{\mu_{F}^{2}}{\mu^{2}}\right)
      Q_q^2 \left\{ \int_{\frac{x_{h_1}}{x_q}}^1 \frac{d \beta_1}{\beta_1} \left[  P_{qq}(\beta_1) D_q^{h_1}\left(\frac{x_{h_1}}{\beta_1 x_q},\mu_F\right)D_{\bar{q}}^{h_2}\left(\frac{x_{h_2}}{x_{\bar{q}}},\mu_F\right)  \right. \right.  \\
    & \left. + P_{gq}(\beta_1) D_g^{h_1}\left(\frac{x_{h_1}}{\beta_1 x_q}, \mu_F\right) D_{\bar{q}}^{h_2}\left(\frac{x_{h_2}}{x_{\bar{q}}},\mu_F\right) \right] \\
    & + \int_{\frac{x_{h_2}}{x_{\bar{q}}}}^1 \frac{d \beta_2}{\beta_2} \left[ P_{qq}(\beta_2) D_q^{h_1}\left(\frac{x_{h_1}}{x_q}, \mu_F\right)D_{\bar{q}}^{h_2}\left(\frac{x_{h_2}}{\beta_2 x_{\bar{q}}}, \mu_F\right) \right.  \\
    & \left. \left.  + P_{gq}(\beta_2) D_q^{h_1}\left(\frac{x_{h_1}}{x_q}, \mu_F\right)D_{g}^{h_2}\left(\frac{x_{h_2}}{\beta_2 x_{\bar{q}}}, \mu_F\right) \right] \right\} + (h_1 \leftrightarrow h_2) \\ 
    &= \frac{d \sigma_{TT}^{h_1 h_2}}{d x_{h_1}d_{h_2}d^d p_{h_1 \perp}d^d p_{h_2 \perp}} \bigg |_{\text{ct div}} +  \frac{d \sigma_{TT}^{h_1 h_2}}{d x_{h_1}d_{h_2}d^d p_{h_1 \perp}d^d p_{h_2 \perp}} \bigg |_{\text{ct fin }} .
   \end{aligned}
\end{equation}
The divergent parts are the ones containing $1/\hat{\epsilon}$ and the finite terms are the ones with $\ln (\mu_F^2/ \mu^2) $. The dependence on the arbitrary parameter $\mu$ disappears at the end when all finite terms are put together.

\section{NLO cross-section: Virtual corrections}
\label{sec: VirtualDiv}

Here, we compute 1-loop virtual corrections to the leading order cross-section. 
For sake of comprehension, we report the  dipole-dipole virtual corrections to the $\gamma^{*} \rightarrow q \bar{q}$ cross-section, as presented in (5.24) of \cite{Boussarie:2016ogo}, which we refer to as our partonic cross section. Adapting to our notation, we have
\begin{equation}
\label{eq: VirtBeg}
\begin{aligned}
d \hat{\sigma}_{1 L L}  &= \frac{\alpha_{s}}{2\pi} \frac{\Gamma(1-\epsilon)}{(4 \pi)^{\epsilon}} C_F \left( \frac{S_{V}+S_{V}^{*}}{2} \right) d \hat{\sigma}_{0 L L} \\
&  + \frac{\alpha_{s} Q^{2}}{4 \pi}\left(\frac{N_{c}^{2}-1}{N_{c}}\right) \frac{\alpha_{\mathrm{em}} Q_{q}^{2}}{(2 \pi)^{4} N_{c}} d x_q d x_{\bar{q}} d^{2} p_{q \perp} d^{2} p_{\bar{q} \perp} \delta(1-x_q-x_{\bar{q}}) \\
 & \times \int d^{2} p_{1 \perp} d^{2} p_{2 \perp} d^{2} p_{1 \perp}^{\prime} d^{2} p_{2 \perp}^{\prime} \delta\left(p_{q 1 \perp}+p_{\bar{q} 2 \perp}\right) \frac{\delta\left(p_{11^{\prime} \perp}+p_{22^{\prime} \perp}\right)}{\vec{p}_{q 1^{\prime}}^{\,2}+x_q x_{\bar{q}} Q^{2}} \mathbf{F}\left(\frac{p_{12 \perp}}{2}\right) \mathbf{F}^{*}\left(\frac{p_{1^{\prime} 2^{\prime} \perp}}{2}\right) \\
 &  \times {\left[\frac{6 x_q^{2} x_{\bar{q}}^{2}}{\vec{p}_{q 1}^{\,2}+x_q x_{\bar{q}} Q^{2}} \ln \left(\frac{x_q^{2} x_{\bar{q}}^{2} \mu^{4} Q^{2}}{\left(x_q \vec{p}_{\bar{q}}-x_{\bar{q}} \vec{p}_{q}\right)^{2}\left(\vec{p}_{q 1}^{\,2}+x_q x_{\bar{q}} Q^{2}\right)^{2}}\right)\right.} \\
& + \left.\frac{\left(p_{0}^{-}\right)^{2}}{s^{2} p_{\gamma}^{+}} \operatorname{tr}\left(\left(C_{\|}^{4}+C_{1 \|}^{5}+C_{1 \|}^{6}\right) \hat{p}_{\bar{q}} \gamma^{+} \hat{p}_{q}\right)\right]+ h.c . 
\end{aligned}
\end{equation} 
in which the divergences are inside the contribution
\beqa
\label{SV}
&&\frac{S_V + S_V^*}{2}  =  \frac{1}{\epsilon} \Bigg [ - 4 \epsilon \ln (\alpha) \ln \left(\frac{x_q ^2 x_{\bar{q}}^2 \mu^{2}}{\left(x_q \vec{p}_{\bar{q}}-x_{\bar{q}} \vec{p}_{q}\right)^{2}}\right) + 4 \ln (\alpha) + 4 \epsilon \ln^2(\alpha)  - 2 \ln (x_q x_{\bar{q}})+ 3 \nonumber \\ 
&& \hspace{- 1 cm}+  2 \epsilon \ln \left(\frac{x_q x_{\bar{q}} \mu^{2}}{\left(x_q \vec{p}_{\bar{q}}-x_{\bar{q}}\vec{p}_{q}\right)^{2}}\right) \ln (x_q x_{\bar{q}}) + \epsilon \ln^2 (x_q x_{\bar{q}})  - 3 \epsilon \ln \left(\frac{x_q x_{\bar{q}} \mu^{2}}{\left(x_q \vec{p}_{\bar{q}}-x_{\bar{q}} \vec{p}_{q}\right)^{2}}\right)  - \frac{\pi^2}{3}\epsilon + 6 \epsilon \Bigg ] , \,
\eqa
where $\alpha$ is an infra-red cut-off imposed on the longitudinal fraction of gluon momenta in order to regularize rapidity divergences. 

In eq.~\eqref{eq: VirtBeg} the 
$C$ functions have been parametrized using (5.25) and (5.26) in ref.~\cite{Boussarie:2016ogo} as
\begin{align}
\frac{(p_{0}^{-})^{2}}{s^{2}p_{\gamma}^{+}}tr(C_{||}^{4}\hat{p}_{\bar{q}}\gamma^{+}\hat{p}_{q})=\int_{0}^{x}dz\left[(\phi_4)_{LL}\right]_+ +(q\leftrightarrow\bar{q}) \, , \label{phi1LL}
\end{align}
and
\begin{align}
\frac{(p_{0}^{-})^{2}}{s^{2}p_{\gamma}^{+}}tr(C_{1||}^{n}\hat{p}_{\bar{q}}\gamma^{+}\hat{p}_{q})=\int_{0}^{x}dz\left[(\phi_n)_{LL}\right]_+|_{\vec{p}_3=\vec{0}} +(q\leftrightarrow\bar{q}) \, ,
\end{align}
where $n=5 \, \, \mathrm{or} \, \, 6$, and $(q\leftrightarrow\bar{q})$ stands for $p_{q}\leftrightarrow p_{\bar{q}},\,p_{1}^{(\prime)}\leftrightarrow p_{2}^{(\prime)},\,x_{q}\leftrightarrow
x_{\bar{q}}.$ The expressions for $(\phi_n)_{LL}$ are given in appendix~\ref{AppendixB}.

By using a factorization formula analogous to eq.~\eqref{eq: coll facto} and the expression  \eqref{eq: VirtBeg}, as well as the collinear constraints (\ref{constraint-collinear-q}) and (\ref{constraint-collinear-qbar}), we obtain the dipole-dipole virtual corrections to the full cross-section. We split it into divergent part,
\begin{equation}
\label{eq:virtual div LL}
    \begin{aligned}
& \frac{d \sigma_{1 L L}^{q\bar{q} \rightarrow  h_1 h_2}}{d x_{h_1}d_{h_2}d^d p_{h_1 \perp}d^d p_{h_2 \perp}}\bigg |_{\text{div}} \\
&=  \frac{4 \alpha_{\mathrm{em}} Q^2}{( 2 \pi)^{4(d-1)} N_c}  \sum_{q}   \int_{x_{h_1}}^1 d x_q  \int_{x_{h_2}}^1 d x_{\bar{q}} \;  x_q x_{\bar{q}} \left( \frac{x_q}{x_{h_1}}\right)^d \left( \frac{x_{\bar{q}}}{x_{h_2}}\right)^d  \delta(1-x_q-x_{\bar{q}})\\
& \times Q_q^2 D_q^{h_1}\left(\frac{x_{h_1}}{x_q}, \mu_F\right) D_{\bar{q}}^{h_2}\left(\frac{x_{h_2}}{x_{\bar{q}}}, \mu_F\right)  {\cal F}_{LL} \\
& \times \frac{\alpha_s}{2 \pi}C_F  \frac{1}{\hat{\epsilon}}  \left[ - 4 \epsilon \ln (\alpha) \ln \left(\frac{  \mu^{2}}{\left( \frac{\vec{p}_{h_2}}{x_{h_2}}- \frac{\vec{p}_{h_1}}{x_{h_1}} \right)^{2}}\right) + 4 \ln (\alpha)  \right. \\
& + \left.   4 \epsilon \ln^2(\alpha)   - 2 \ln (x_q x_{\bar{q}})+ 3 \right] + (h_1 \leftrightarrow h_2)
    \end{aligned}
\end{equation}
and finite part,
\begin{align*}
& \frac{d \sigma_{1 L L}^{q\bar{q} \rightarrow  h_1 h_2}}{d x_{h_1}d_{h_2}d^d p_{h_1 \perp}d^d p_{h_2 \perp}}\bigg |_{\text{fin}} \\
& =  \frac{4 \alpha_{\mathrm{em}} Q^2}{( 2 \pi)^{4(d-1)} N_c}   \sum_{q} \int_{x_{h_1}}^1 d x_q  \int_{x_{h_2}}^1 d x_{\bar{q}} \;   x_q x_{\bar{q}} \left( \frac{x_q}{x_{h_1}}\right)^d \left( \frac{x_{\bar{q}}}{x_{h_2}}\right)^d  \delta(1-x_q-x_{\bar{q}})\\
& \times  Q_q^2 D_q^{h_1}\left(\frac{x_{h_1}}{x_q}, \mu_F\right) D_{\bar{q}}^{h_2}\left(\frac{x_{h_2}}{x_{\bar{q}}}, \mu_F\right)  {\mathcal{F}}_{LL} \\
& \times \frac{\alpha_s}{2 \pi}C_F  \frac{1}{\hat{\epsilon}} \left[2 \epsilon \ln \left(\frac{ \mu^{2}}{ x_q x_{\bar{q}} \left( \frac{\vec{p}_{h_2}}{x_{h_2}}- \frac{\vec{p}_{h_1}}{x_{h_1}}\right)^{2}}\right)  \ln (x_q x_{\bar{q}}) \right. \\
& \left. + \epsilon \ln^2 (x_q x_{\bar{q}})  - 3 \epsilon \ln \left(\frac{ \mu^{2}}{ x_q x_{\bar{q}} \left( \frac{\vec{p}_{h_2}}{x_{h_2}}- \frac{\vec{p}_{h_1}}{x_{h_1}}\right)^{2}}\right) - \frac{\pi^2}{3}\epsilon + 6 \epsilon\right] \\
& + \frac{\alpha_s Q^2}{4\pi}\left(\frac{N_c^2-1}{N_c}\right) \frac{\alpha_{\mathrm{em}}}{(2 \pi)^{4} N_{c}} \sum_{q} \int_{x_{h_1}}^{1} \frac{d x_q}{x_q} \int_{x_{h_2}}^1  \frac{d x_{\bar{q}}}{x_{\bar{q}}}  \delta(1-x_q-x_{\bar{q}}) \\
& \times \left(\frac{x_{q}}{x_{h_1}}\right)^d \left(\frac{x_{\bar{q}}}{x_{h_2}}\right)^d  Q_{q}^{2} D_{q}^{h_1}\left(\frac{x_{h_1}}{x_q},\mu_F\right) D_{\bar{q}}^{h_2}\left(\frac{x_{h_2}}{x_{\bar{q}}} ,\mu_F\right)  \\ 
 & \times \int d^{d} p_{2 \perp} \frac{\mathbf{F}\left(\frac{x_q}{2x_{h_1}}  p_{h_1\perp} + \frac{x_{\bar{q}} }{2 x_{h_2}}p_{h_2\perp} -p_{2\perp}\right)}{\left(\frac{x_{\bar{q}}}{x_{h_2}}\vec{p}_{h_2}- \vec{p}_{2}\right)^{2}+x_q x_{\bar{q}} Q^{2}} \int d^{d} p_{2' \perp} \frac{\mathbf{F}^{*}\left(\frac{x_q}{2x_{h_1}}  p_{h_1\perp} + \frac{x_{\bar{q}} }{2 x_{h_2}}p_{h_2\perp} -p_{2'\perp}\right)}{\left(\frac{x_{\bar{q}}}{x_{h_2}}\vec{p}_{h_2}- \vec{p}_{2'} \right)^{2}+x_q x_{\bar{q}} Q^{2}} \\ & \times \left[ 6 x_q^{2} x_{\bar{q}}^{2}  
  \ln \left( \frac{\mu^{4} Q^{2}}{\left( \frac{\vec{p}_{h_2}}{x_{h_2}}- \frac{\vec{p}_{h_1}}{x_{h_1}}\right)^{2} \left(\left(\frac{x_{\bar{q}}}{x_{h_2}} \vec{p}_{h_2}- \vec{p}_2 \right)^{2}+x_q x_{\bar{q}} Q^{2}\right)^{2}}\right) +  \left(  \int_0^{x_q} dz \left[(\phi_4)_{LL}\right]_+   \right. \right.  \\
 & \left. \left. + \sum_{n=5,6} \left[(\phi_n)_{LL}\right]_+ \bigg |_{\vec{p}_3 = \vec{0}}  + (q \leftrightarrow \bar{q}) \right) \left(\left(\frac{x_{\bar{q}}}{x_{h_2}}\vec{p}_{h_2}-\vec{p}_{2}\right)^2 + x_q x_{\bar{q}}Q^2 \right) \right] + (h_1 \leftrightarrow h_2) \; .
\numberthis[virtual finite LL]
\end{align*}

For the $d \sigma_{TL}$ element of the matrix \eqref{eq:density_matrix}, we get from (5.28) of \cite{Boussarie:2016ogo}
    \begin{align*}
& \frac{d \sigma_{1 T L}^{q\bar{q} \rightarrow  h_1 h_2}}{d x_{h_1}d_{h_2}d^d p_{h_1 \perp}d^d p_{h_2 \perp}}\bigg |_{\text{div}} \\
&=  \frac{2 \alpha_{\mathrm{em}} Q}{( 2 \pi)^{4(d-1)} N_c}   \sum_{q}  \int_{x_{h_1}}^1 d x_q  \int_{x_{h_2}}^1 d x_{\bar{q}}   \left( \frac{x_q}{x_{h_1}}\right)^d \left( \frac{x_{\bar{q}}}{x_{h_2}}\right)^d (x_{\bar{q}}-x_q ) \\
& \times \delta(1-x_q-x_{\bar{q}}) Q_q^2 D_q^{h_1}\left(\frac{x_{h_1}}{x_q}, \mu_F\right) D_{\bar{q}}^{h_2}\left(\frac{x_{h_2}}{x_{\bar{q}}}, \mu_F\right)  {\cal F}_{TL} \\
& \times \frac{\alpha_s}{2 \pi}C_F  \frac{1}{\hat{\epsilon}}  \left[ - 4 \epsilon \ln (\alpha) \ln \left(\frac{  \mu^{2}}{\left( \frac{\vec{p}_{h_2}}{x_{h_2}}- \frac{\vec{p}_{h_1}}{x_{h_1}} \right)^{2}}\right) + 4 \ln (\alpha)   \right. \\
& \left. + 4 \epsilon \ln^2(\alpha)  - 2 \ln (x_q x_{\bar{q}})+ 3 \right] + (h_1 \leftrightarrow h_2) \; ,  
\numberthis[virtual div TL]
\end{align*}
and
{\allowdisplaybreaks
\begin{align*}
& \frac{d \sigma_{1T L}^{q\bar{q} \rightarrow h_1 h_2}}{d x_{h_1}d_{h_2}d^d p_{h_1 \perp}d^d p_{h_2 \perp}}\bigg |_{\text{fin}} \\
& =  \frac{2 \alpha_{\mathrm{em}} Q}{( 2 \pi)^{4(d-1)} N_c}   \sum_{q} \int_{x_{h_1}}^1 d x_q  \int_{x_{h_2}}^1 d x_{\bar{q}}  \left( \frac{x_q}{x_{h_1}}\right)^d \left( \frac{x_{\bar{q}}}{x_{h_2}}\right)^d (x_{\bar{q}}-x_q ) \\
& \times \delta(1-x_q-x_{\bar{q}})  Q_q^2 D_q^{h_1}\left(\frac{x_{h_1}}{x_q}, \mu_F\right) D_{\bar{q}}^{h_2}\left(\frac{x_{h_2}}{x_{\bar{q}}}, \mu_F\right)  {\cal F}_{TL} \\
& \times \frac{\alpha_s}{2 \pi}C_F  \frac{1}{\hat{\epsilon}} \left[2 \epsilon \ln \left(\frac{ \mu^{2}}{ x_q x_{\bar{q}} \left( \frac{\vec{p}_{h_2}}{x_{h_2}}- \frac{\vec{p}_{h_1}}{x_{h_1}}\right)^{2}}\right) \ln (x_q x_{\bar{q}}) + \epsilon \ln^2 (x_q x_{\bar{q}}) \right. \\
& \left. - 3 \epsilon \ln \left(\frac{ \mu^{2}}{ x_q x_{\bar{q}} \left( \frac{\vec{p}_{h_2}}{x_{h_2}}- \frac{\vec{p}_{h_1}}{x_{h_1}}\right)^{2}}\right)  - \frac{\pi^2}{3}\epsilon + 6 \epsilon\right] \\
& + \frac{\alpha_s Q}{4\pi}\left(\frac{N_c^2-1}{N_c}\right) \frac{\alpha_{\mathrm{em}}}{(2 \pi)^{4} N_{c}} \sum_{q}  \int_{x_{h_1}}^{1} \frac{d x_q}{x_q} \int_{x_{h_2}}^1  \frac{d x_{\bar{q}}}{x_q} \;  \delta(1-x_q-x_{\bar{q}})  \\
& \times \left(\frac{x_{q}}{x_{h_1}}\right)^d \left(\frac{x_{\bar{q}}}{x_{h_2}}\right)^d  Q_{q}^{2} D_{q}^{h_1}\left(\frac{x_{h_1}}{x_q},\mu_F\right) D_{\bar{q}}^{h_2}\left(\frac{x_{h_1}}{x_q},\mu_F\right)\\ 
& \times \int d^d p_{1 \perp} d^d p_{2 \perp} \mathbf{F}\left(\frac{p_{12\perp}}{2}\right) \delta \left( \frac{x_q}{x_{h_1}} p_{h_1 \perp} -p_{1\perp} + \frac{x_{\bar{q}}}{x_{h_2}} p_{h_2 \perp} - p_{2 \perp} \right) \\
& \times  \int d^d p_{1' \perp} d^d p_{2' \perp} \mathbf{F}^{*} \left(\frac{p_{1'2'\perp}}{2}\right) \delta \left( \frac{x_q}{x_{h_1}} p_{h_1 \perp} -p_{1'\perp} + \frac{x_{\bar{q}}}{x_{h_2}} p_{h_2 \perp} - p_{2' \perp} \right) \varepsilon_{Ti}^* \\
& \times   \left[ \frac{1}{\left(\frac{x_q}{x_{h_1}} \vec{p}_{h_1} - \vec{p}_1\right)^2 + x_q x_{\bar{q}}Q^2 }  \left( \int_0^{x_q} \left[(\phi_4^i)_{TL}\right]_+ + \sum_{n=5,6} \int_0^{x_q} \left[(\phi_n^i)_{TL}\right]_+ \bigg |_{\vec{p}_3 = \vec{0}} + (q \leftrightarrow \bar{q} ) \right)^\dag  \right. \\
& + \frac{3 x_q x_{\bar{q}} (x_{\bar{q}} - x_q) \left(\frac{x_q}{x_{h_1}} p_{h_1 \perp} - p_{1' \perp}\right)^i }{\left(\left(\frac{x_q}{x_{h_1}} \vec{p}_{h_1} - \vec{p}_1\right)^2 + x_q x_{\bar{q}}Q^2 \right) \left(\left(\frac{x_q}{x_{h_1}} \vec{p}_{h_1} - \vec{p}_{1'}\right)^2 + x_q x_{\bar{q}}Q^2 \right)} \\
& \times \left( \ln \left( \frac{ Q^2 \mu^8 \left(\frac{\vec{p}_{h_2}}{x_{h_2}} - \frac{\vec{p}_{h_1}}{x_{h_1}}\right)^{-4}}{x_q x_{\bar{q}} \left(\left(\frac{x_q}{x_{h_1}} \vec{p}_{h_1} - \vec{p}_1\right)^2 + x_q x_{\bar{q}}Q^2 \right) \left(\left(\frac{x_q}{x_{h_1}} \vec{p}_{h_1} - \vec{p}_{1'}\right)^2 + x_q x_{\bar{q}}Q^2 \right)} \right)\right. \\ 
& \left. \hspace{-0.1 cm} - \frac{x_q x_{\bar{q}} Q^2}{\left(\frac{x_q}{x_{h_1}} \vec{p}_{h_1} - \vec{p}_{1'} \right)^2} \ln \left( \frac{x_q x_{\bar{q}} Q^2}{\left(\frac{x_q}{x_{h_1}} \vec{p}_{h_1} - \vec{p}_{1'} \right)^2 + x_q x_{\bar{q}}Q^2 }\right) \hspace{-0.1 cm} \right) \hspace{-0.1 cm} + \frac{1}{2 x_q x_{\bar{q}} \left(\left(\frac{x_q}{x_{h_1}} \vec{p}_{h_1} - \vec{p}_{1'} \right)^2 + x_q x_{\bar{q}} Q^2 \right)} \\
& \times  \left.   \left( \int_0^{x_q} \left[(\phi_4^i)_{LT}\right]_+ + \sum_{n=5,6} \int_0^{x_q} \left[(\phi_n^i)_{LT}\right]_+ \bigg |_{\vec{p}_3 = \vec{0}}  +   (q \leftrightarrow \bar{q} ) \right)\right]  + (h_1 \leftrightarrow h_2) \,. \numberthis[virtual finite TL]
\end{align*}}
In the case of the $TT$ transition, the result was obtained in ref.~\cite{Boussarie:2016ogo} eq.~(5.35) and the divergent part is
{\allowdisplaybreaks
    \begin{align*}
& \frac{d \sigma_{1T T}^{q\bar{q} \rightarrow  h_1 h_2}}{d x_{h_1}d_{h_2}d^d p_{h_1 \perp}d^d p_{h_2 \perp}}\bigg |_{\text{div}} \\
&=  \frac{ \alpha_{\mathrm{em}} }{( 2 \pi)^{4(d-1)} N_c}  \sum_{q}  \int_{x_{h_1}}^1 \frac{d x_q}{x_q}  \int_{x_{h_2}}^1 \frac{d x_{\bar{q}} }{x_{\bar{q}}}   \left( \frac{x_q}{x_{h_1}}\right)^d \left( \frac{x_{\bar{q}}}{x_{h_2}}\right)^d \delta(1-x_q -x_{\bar{q}}) \\
& \times  Q_q^2 D_q^{h_1}\left(\frac{x_{h_1}}{x_q}, \mu_F\right) D_{\bar{q}}^{h_2}\left(\frac{x_{h_2}}{x_{\bar{q}}}, \mu_F\right)  {\cal F}_{TT} \\
& \times \frac{\alpha_s}{2 \pi}C_F  \frac{1}{\hat{\epsilon}}  \left[ - 4 \epsilon \ln (\alpha) \ln \left(\frac{  \mu^{2}}{\left( \frac{\vec{p}_{h_2}}{x_{h_2}}- \frac{\vec{p}_{h_1}}{x_{h_1}} \right)^{2}}\right) + 4 \ln (\alpha)  \right. \\
& \left.  + 4 \epsilon \ln^2(\alpha) - 2 \ln (x_q x_{\bar{q}})+ 3 \right] + (h_1 \leftrightarrow h_2)  \numberthis[virtual div TT]
    \end{align*}}
while the finite part reads
{\allowdisplaybreaks
\begin{align*}
& \frac{d \sigma_{1T T}^{q\bar{q} \rightarrow h_1 h_2}}{d x_{h_1}d_{h_2}d^d p_{h_1 \perp}d^d p_{h_2 \perp}} \bigg |_{\text{fin}} \\
& =  \frac{\alpha_{\mathrm{em}} }{( 2 \pi)^{4(d-1)} N_c}   \sum_{q} \int_{x_{h_1}}^1 \frac{d x_q }{x_q} \int_{x_{h_2}}^1 \frac{d x_{\bar{q}}}{x_{\bar{q}}}  \left( \frac{x_q}{x_{h_1}}\right)^d \left( \frac{x_{\bar{q}}}{x_{h_2}}\right)^d  \delta(1-x_q -x_{\bar{q}}) \\
& \times  Q_q^2 D_q^{h_1}\left(\frac{x_{h_1}}{x_q}, \mu_F\right) D_{\bar{q}}^{h_2}\left(\frac{x_{h_2}}{x_{\bar{q}}}, \mu_F\right)  {\cal F}_{TT} \\
& \times \frac{\alpha_s}{2 \pi}C_F  \frac{1}{\hat{\epsilon}} \left[2 \epsilon\ln \left(\frac{ \mu^{2}}{ x_q x_{\bar{q}} \left( \frac{\vec{p}_{h_2}}{x_{h_2}}- \frac{\vec{p}_{h_1}}{x_{h_1}}\right)^{2}}\right)  \ln (x_q x_{\bar{q}}) + \epsilon \ln^2 (x_q x_{\bar{q}}) \right. \\
& \left. - 3 \epsilon \ln \left(\frac{ \mu^{2}}{ x_q x_{\bar{q}} \left( \frac{\vec{p}_{h_2}}{x_{h_2}}- \frac{\vec{p}_{h_1}}{x_{h_1}}\right)^{2}}\right) - \frac{\pi^2}{3}\epsilon + 6 \epsilon\right] \\
& + \frac{\alpha_s }{4 \pi} \left(\frac{N_c^2 -1 }{N_c}\right) \frac{\alpha_{\mathrm{em}}}{(2\pi)^4 N_c} \sum_{q} \int_{x_{h_1}}^1 \frac{ d x_q }{x_q} \int_{h_2}^1 \frac{ d x_{\bar{q}}}{x_{\bar{q}}} \; \delta (1-x_q -x_{\bar{q}} ) \\ 
& \times \left(\frac{x_q}{x_{h_1}}\right)^d \left(\frac{x_{\bar{q}}}{x_{h_2}}\right)^d   Q_q^2 D_q^{h_1}\left(\frac{x_{h_1}}{x_q}, \mu_F\right) D_{\bar{q}}^{h_2}\left(\frac{x_{h_2}}{x_{\bar{q}}}, \mu_F\right) \\
&\times  \int d^d p_{1 \perp} d^d p_{2 \perp} \mathbf{F}\left(\frac{p_{12\perp}}{2}\right) \delta \left( \frac{x_q}{x_{h_1}} p_{h_1 \perp} -p_{1\perp} + \frac{x_{\bar{q}}}{x_{h_2}} p_{h_2 \perp} - p_{2 \perp} \right) \\
& \times \int d^d p_{1' \perp} d^d p_{2' \perp} \mathbf{F}^{*} \left(\frac{p_{1'2'\perp}}{2}\right) \delta \left( \frac{x_q}{x_{h_1}} p_{h_1 \perp} -p_{1'\perp} + \frac{x_{\bar{q}}}{x_{h_2}} p_{h_2 \perp} - p_{2' \perp} \right) \\
& \times  \varepsilon_{T i} \varepsilon_{T k}^* \left\{ \frac{3}{2} \frac{\left(\frac{x_q}{x_{h_1}} p_{h_1 \perp}- p_{1 \perp}\right)_r \left(\frac{x_q}{x_{h_1}} p_{h_1 \perp}- p_{1'\perp }\right)_l}{\left(\left(\frac{x_q}{x_{h_1}} \vec{p}_{h_1}- \vec{p}_{1 }\right)^2 + x_q x_{\bar{q} } Q^2 \right) \left(\left(\frac{x_q}{x_{h_1}} \vec{p}_{h_1}- \vec{p}_{1'}\right)^2 + x_q x_{\bar{q} } Q^2 \right)} \right. \\ 
&  \times \left[(x_{\bar{q}} -x_q)^2 g_{\perp}^{ri} g_{\perp}^{lk} - g_{\perp}^{rk} g_{\perp}^{li} + g_{\perp}^{rl} g_{\perp}^{ik}\right] \\
& \times  \left[ \ln \left( \frac{\mu^4}{x_q x_{\bar{q}} \left(\frac{\vec{p}_{h_2}}{x_{h_2}} - \frac{\vec{p}_{h_1}}{x_{h_1}}\right)^2 \left(\left(\frac{x_q}{x_{h_1}} \vec{p}_{h_1}- \vec{p}_{1 }\right)^2 + x_q x_{\bar{q} } Q^2 \right)  }\right) \right. \\ 
& \left. - \frac{x_q x_{\bar{q}}Q^2}{\left(\frac{x_q}{x_{h_1}} \vec{p}_{h_1}- \vec{p}_{1 }\right)^2  } \ln \left( \frac{x_q x_{\bar{q}} Q^2 }{\left(\frac{x_q}{x_{h_1}} \vec{p}_{h_1}- \vec{p}_{1 }\right)^2 + x_q x_{\bar{q} } Q^2  }\right) \right] +  \frac{1}{\left(\left(\frac{x_q}{x_{h_1}} \vec{p}_{h_1}- \vec{p}_{1 '}\right)^2 + x_q x_{\bar{q} } Q^2 \right) x_q x_{\bar{q}}} \\ 
& \times  \left(\int_{0}^{x_q} \left[(\phi_4)^{ik}_{TT}\right]_+ + \sum_{n=5,6} \int_{0}^{x_q} \left[(\phi_n)^{ij}_{TT}\right]_+ \bigg |_{\vec{p}_3 = \vec{0} } + (q \leftrightarrow \bar{q}) \right)   \left. + h.c. \bigg  |_{1\leftrightarrow 1', i \leftrightarrow k} \right\}  \\
& + (h_1 \leftrightarrow h_2) \numberthis[virtual finite TT] \,.
\end{align*}}

\section{NLO cross-section: Real corrections}
\label{sec: RealDiv}

In this section, we will discuss the real corrections. Since, as explained above, the calculation is almost completely identical in the $LL$, $TL$, and $TT$ cases (apart from factors that do not affect the general strategy), we will show the details of the $LL$ case only. For the others, we will just report the final results. \\
The dipole-dipole partonic cross-section is given by eq.~(6.6) of ref.~\cite{Boussarie:2016ogo}: 
\begin{equation}
\begin{aligned}
    d \hat{\sigma}_{3JI} &  = \frac{\alpha_s}{\mu^{2\epsilon}} \left( \frac{N_c^2 -1}{N_c}\right) \frac{\alpha_{\mathrm{em}}Q_q^2}{(2\pi)^{4(d-1)}N_c} \frac{(p_0^-)^2}{s^2 x_q'x_{\bar{q}}'} \varepsilon_{I\alpha} \varepsilon_{J\beta}^*  d x_q' d x_{\bar{q}}'   \delta (1-x_q'-x_{\bar{q}}'-x_g) d^d p_{q\perp}  d^d p_{\bar{q}\perp} \\
    & \times  \frac{d x_g  d^d p_{g\perp}}{x_g (2\pi)^d} \int d^d p_{1\perp} d^d p_{2\perp} \mathbf{F} \left(\frac{p_{12\perp}}{2}\right)\delta (p_{q1\perp} + p_{\bar{q}2\perp} + p_{g\perp}) \\
    & \times  \int d^d p_{1'\perp} d^d p_{2'\perp} \mathbf{F}^*\left(\frac{p_{1'2'\perp}}{2}\right)  \delta (p_{q1'\perp} + p_{\bar{q}2'\perp} + p_{g\perp}) \\
   &  \times \Phi_3^\alpha (p_{1\perp}, p_{2\perp}) \Phi_3^{\beta*} (p_{1'\perp}, p_{2'\perp})\,. 
\end{aligned}
\end{equation}
where we introduce shorthand notation by suppressing summation over helicities of partons
\begin{equation}
    \Phi_3^\alpha (p_{1\perp}, p_{2\perp}) \Phi_3^{\beta*} (p_{1'\perp}, p_{2'\perp}) \equiv \sum_{\lambda_q, \lambda_g, \lambda_{\bar{q}} } \Phi_3^\alpha (p_{1\perp}, p_{2\perp}) \Phi_3^{\beta*} (p_{1'\perp}, p_{2'\perp}) \; .
    \label{eq:ShortHand}
\end{equation}
The impact factor has the form $\Phi_3^\alpha = \Phi_4^\alpha |_{\vec{p}_3 = 0} + \Tilde{\Phi}_3^\alpha$. Only the square of $\Tilde{\Phi}_3^\alpha$ provides divergences in the cross-section and it is given by (B.3) in ref.~\cite{Boussarie:2016ogo}. The $LL$ contribution reads
\begin{equation}
\label{eq: div real impact factor}
\begin{aligned}
   &    \Tilde{\Phi}_3^+(\vec{p}_1, \vec{p}_2) \Tilde{\Phi}_3^{+*}(\vec{p}_{1'}, \vec{p}_{2'}) \\
   &= \frac{8 x_q' x_{\bar{q}}' (p_\gamma^+)^4 \left( d x_g^2 + 4 x_q' (x_q' + x_g) \right)}{\left(Q^2 + \frac{\vec{p}_{\bar{q}2}^{\,2}}{x_{\bar{q}}'(1-x_{\bar{q}}')} \right) \left(Q^2 + \frac{\vec{p}_{\bar{q}2'}^{\,2}}{x_{\bar{q}}'(1-x_{\bar{q}}')} \right) (x_q' \vec{p}_g -x_g \vec{p}_q)^2 } \\ 
   & - \frac{8 x_q' x_{\bar{q}}'(p_\gamma^+)^4 \left(2 x_g -d x_g^2 + 4 x_q' x_{\bar{q}}' \right) \left(x_q' \vec{p}_g - x_g \vec{p}_q \right) \cdot \left( x_{\bar{q}}' \vec{p}_g - x_g \vec{p}_{\bar{q}} \right)}{\left(Q^2 + \frac{\vec{p}_{\bar{q}2'}^{\,2}}{x_{\bar{q}}'(1-x_{\bar{q}}')} \right) \left(Q^2 + \frac{\vec{p}_{q1}^{\,2}}{x_q' (1-x_q')} \right)  \left(x_q' \vec{p}_g - x_g \vec{p}_q \right)^2  \left( x_{\bar{q}}' \vec{p}_g - x_g \vec{p}_{\bar{q}} \right)^2 }
   \\
   & +  \frac{8 x_q' x_{\bar{q}}' (p_\gamma^+)^4 \left( d x_g^2 + 4 x_{\bar{q}}' (x_{\bar{q}}' + x_g) \right)}{\left(Q^2 + \frac{\vec{p}_{q1}^{\,2}}{x_q' (1-x_q')} \right) \left(Q^2 + \frac{\vec{p}_{q1'}^{\,2}}{x_q' (1-x_q')} \right) (x_{\bar{q}}' \vec{p}_g -x_g \vec{p}_{\bar{q}})^2 } \\ 
   & - \frac{8 x_q' x_{\bar{q}}' (p_\gamma^+)^4 \left(2 x_g -d x_g^2 + 4 x_q' x_{\bar{q}}' \right) \left(x_q'  \vec{p}_g - x_g \vec{p}_q \right) \cdot \left( x_{\bar{q}}' \vec{p}_g - x_g \vec{p}_{\bar{q}} \right)}{\left(Q^2 + \frac{\vec{p}_{q1'}^{\,2}}{x_q' (1-x_q')} \right) \left(Q^2 + \frac{\vec{p}_{\bar{q}2}^{\,2}}{x_{\bar{q}}' (1-x_{\bar{q}}')} \right)  \left(x_q' \vec{p}_g - x_g \vec{p}_q \right)^2  \left( x_{\bar{q}}' \vec{p}_g - x_g \vec{p}_{\bar{q}} \right)^2 }\,.
\end{aligned}
\end{equation}
The $TL$ contribution is
{\allowdisplaybreaks
\begin{align*}
&  \tilde{\Phi}_3^{+}(\vec{p}_{1},\vec{p}_{2}) \tilde{\Phi}_3^{i*}(\vec{p}_{1'},\vec{p}_{2'}) \\*
& = \frac{4 x_q'\left(p_\gamma^{+}\right)^3}{\left(x_q' +x_g\right)\left(Q^2+\frac{\vec{p}_{\bar{q} 2'}^{\,2}}{x_{\bar{q}}'\left(1-x_{\bar{q}}'\right)}\right)\left(Q^2+\frac{\vec{p}_{q1}^{\,2}}{x_q'\left(1-x_q'\right)}\right)}\left(\frac{ \left(x_q' p_{g\perp} - x_g p_{q\perp}\right)_\mu \left(x_{\bar{q}}' p_{g\perp} - x_g p_{\bar{q}\perp}\right)_\nu}{\left(x_q' \vec{p}_g - x_g \vec{p}_q\right)^2 \left(x_{\bar{q}}' \vec{p}_g - x_g \vec{p}_{\bar{q}}\right)^2}\right) \\ 
& \times \left[x_g\left(4 x_{\bar{q}}' +x_g d-2\right)\left(p_{\bar{q} 2' \perp}^\mu g_{\perp}^{i \nu}-p_{\bar{q} 2' \perp}^\nu g_{\perp}^{\mu i}\right)-\left(2 x_{\bar{q}}'-1\right)\left(4 x_q' x_{\bar{q}}'+x_g\left(2-x_g d\right)\right) g_{\perp}^{\mu \nu} p_{\bar{q} 2' \perp}^i\right] \\
& -\frac{4 x_q ' \left(p_\gamma^{+}\right)^3\left(2 x_{\bar{q}}' -1\right)\left(x_g^2 d+4 x_q' \left(x_q' +x_g\right)\right) p_{\bar{q} 2' \perp}^i}{\left(x_q'+x_g\right)\left(Q^2+\frac{\vec{p}_{\bar{q} 2'}^{\,2}}{x_{\bar{q}}'\left(1-x_{\bar{q}}'\right)}\right)\left(Q^2+\frac{\vec{p}_{\bar{q} 2}^{\,2}}{x_{\bar{q}}'\left(1-x_{\bar{q}}'\right)}\right) \left(x_q' \vec{p}_g- x_g \vec{p}_q\right)^2}
+(q \leftrightarrow \bar{q}) \,,
\numberthis[impact_factor_TL]
\end{align*}}
and finally, the $TT$ contribution reads
{\allowdisplaybreaks
\begin{align*}
&  \tilde{\Phi}_3^i(\vec{p}_1,\vec{p}_2) \tilde{\Phi}_3^{k*}(\vec{p}_{1'},\vec{p}_{2'}) \\
& =\frac{-2\left(p_\gamma^{+}\right)^2}{\left(x_q'+x_g\right)\left(x_{\bar{q}}'+x_g\right)\left(Q^2+\frac{\vec{p}_{\bar{q} 2}^{\,2}}{x_{\bar{q}}'\left(1-x_{\bar{q}}'\right)}\right)\left(Q^2+\frac{\vec{p}_{q 1^{\prime}}^{\,2}}{x_q'\left(1-x_q'\right)}\right)} \\
& \times  \left(\frac{ \left(x_q' p_{g\perp} - x_g p_{q\perp}\right)_\mu \left(x_{\bar{q}}' p_{g\perp} - x_g p_{\bar{q}\perp}\right)_\nu}{\left(x_q' \vec{p}_g - x_g \vec{p}_q\right)^2  \left(x_{\bar{q}}' \vec{p}_g - x_g \vec{p}_{\bar{q}}\right)^2}\right) \left\{ x_g ((d-4)) x_g -2) \left[p_{q 1^{\prime} \perp}^\nu\left(p_{\bar{q} 2 \perp}^\mu g_{\perp}^{i k}+p_{\bar{q} 2 \perp}^k g_{\perp}^{\mu i}\right) \right. \right. \\
& \left. +g_{\perp}^{\mu \nu}\left(\left(\vec{p}_{q 1^{\prime}} \cdot \vec{p}_{\bar{q} 2}\right) g_{\perp}^{i k}+p_{q 1^{\prime} \perp}^i p_{\bar{q} 2 \perp}^k\right) -g_{\perp}^{\nu k} p_{q 1^{\prime} \perp}^i p_{\bar{q} 2 \perp}^\mu -g_{\perp}^{\mu i} g_{\perp}^{\nu k}\left(\vec{p}_{q 1^{\prime}} \cdot \vec{p}_{\bar{q} 2}\right) \right] -g_{\perp}^{\mu \nu} \\
& \times  \left[ \left(2x_q' -1 \right) \left(2 x_{\bar{q}}' - 1\right) p_{q1'\perp}^k p_{\bar{q}2\perp}^i \left( 4 x_q' x_{\bar{q}}' + x_g (2 - x_g d)\right)  + 4 x_q' x_{\bar{q}}' ((\vec{p}_{q1'} \cdot \vec{p}_{\bar{q}2})g_\perp^{ik} + p_{q1'\perp}^i p_{\bar{q}2\perp}^k  )\right] \\
& + \left( p_{q1'\perp}^\mu p_{\bar{q}2\perp}^\nu g_\perp^{ik} - p_{q1'\perp}^\mu p_{\bar{q}2\perp}^k g_\perp^{\nu i } - p_{q1'\perp}^i p_{\bar{q}2\perp}^\nu g_\perp^{\mu k } - g_\perp^{\mu k } g_\perp^{\nu i } (\vec{p}_{q1'} \cdot \vec{p}_{\bar{q}2} ) \right) \\ 
& \times x_g ((d-4)x_g + 2) + x_g (2x_{\bar{q}}' - 1 ) (x_g d + 4 x_q' -2 ) \left( g_\perp^{\mu k } p_{q1'\perp}^\nu - g_\perp^{\nu k} p_{q1'\perp}^\mu \right) p_{\bar{q}2\perp}^i \\
& \left.  + x_g (2 x_q' -1 ) p_{q1'\perp}^k (4 x_{\bar{q}}' + x_g d -2) \left( g_\perp^{\nu i } p_{\bar{q}2\perp}^\mu -g_\perp^{\mu i } p_{\bar{q}2\perp}^\nu \right) \right\} \\
& - \frac{2 x_q' (p_\gamma^+)^2 (x_g^2 d + 4x_q'(x_q'+ x_g)) \left( (\vec{p}_{\bar{q}2} \cdot \vec{p}_{\bar{q}2'}) g_\perp^{ik}-(1-2x_{\bar{q}}')^2 p_{\bar{q}2\perp}^i p_{\bar{q}2'\perp}^k + p_{\bar{q}2'\perp}^i p_{\bar{q}2\perp}^k\right) }{x_{\bar{q}}' (x_q' + x_g)^2 \left(Q^2 + \frac{\vec{p}_{\bar{q}2}^{\,2} }{x_{\bar{q}}' (1 - x_{\bar{q}}')} \right) \left(Q^2 + \frac{\vec{p}_{\bar{q}2'}^{\,2}}{x_{\bar{q}}' (1 - x_{\bar{q}}')} \right) \left(x_q' \vec{p}_{g} - x_g \vec{p}_q \right)^2 }    \\
& + (q \leftrightarrow \bar{q}) \numberthis[impact_factor_TT] \,.
\end{align*}} 
The divergent part of the $LL$ partonic cross-section from real emission is given by
%
%\label{eq:real_div_LL}
\begin{align*}
   \left. d \hat{\sigma}_{3LL}\right|_{div} &=  \frac{4 \alpha_{\mathrm{em}} Q^2}{(2\pi)^{4(d-1)}N_c} Q_q^2 d x_q' d x_{\bar{q}}' \delta(1-x_q'-x_{\bar{q}}' -x_g) d^d p_{q\perp} d^d p_{\bar{q}\perp}  \frac{\alpha_s C_F}{\mu^{2\epsilon}} \frac{d x_g}{x_g} \frac{d^d p_{g\perp}}{(2\pi)^d} \\
    & \times  \int d^d p_{1\perp} d^d p_{2\perp}  \delta (p_{q1\perp} + p_{\bar{q}2\perp} + p_{g\perp})  \; \mathbf{F} \left(\frac{p_{12\perp}}{2}\right)  \\
    & \times \int d^d p_{1'\perp} d^d p_{2'\perp} \delta (p_{q1'\perp} + p_{\bar{q}2'\perp} + p_{g\perp}) \; \mathbf{F}^*\left(\frac{p_{1'2'\perp}}{2}\right)   \\
    & \times  \left\{   \frac{ d x_g^2 + 4 x_q' (x_q' + x_g) }{\left(Q^2 + \frac{\vec{p}_{\bar{q}2}^{\,2}}{x_{\bar{q}}'(1-x_{\bar{q}}')} \right) \left(Q^2 + \frac{\vec{p}_{\bar{q}2'}^{\,2}}{x_{\bar{q}}'(1-x_{\bar{q}}')} \right) (x_q' \vec{p}_g -x_g \vec{p}_q)^2 } \right. \\ 
   & - \frac{ \left(2 x_g -d x_g^2 + 4 x_q' x_{\bar{q}}'\right) \left(x_q' \vec{p}_g - x_g \vec{p}_q \right) \cdot \left( x_{\bar{q}}' \vec{p}_g - x_g \vec{p}_{\bar{q}} \right)}{\left(Q^2 + \frac{\vec{p}_{\bar{q}2'}^{\,2}}{x_{\bar{q}}'(1-x_{\bar{q}}')} \right) \left(Q^2 + \frac{\vec{p}_{q1}^{\,2}}{x_q' (1-x_q')} \right)  \left(x_q' \vec{p}_g - x_g \vec{p}_q \right)^2  \left( x_{\bar{q}}' \vec{p}_g - x_g \vec{p}_{\bar{q}} \right)^2 }
   \\
   & +  \frac{ d x_g^2 + 4 x_{\bar{q}}' (x_{\bar{q}}' + x_g) }{\left(Q^2 + \frac{\vec{p}_{q1}^{\,2}}{x_q' (1-x_q')} \right) \left(Q^2 + \frac{\vec{p}_{q1'}^{\,2}}{x_q' (1-x_q')} \right) (x_{\bar{q}}' \vec{p}_g -x_g \vec{p}_{\bar{q}})^2 } \\ 
   & \left. - \frac{ \left(2 x_g -d x_g^2 + 4 x_q' x_{\bar{q}}'\right) \left(x_q' \vec{p}_g - x_g \vec{p}_q \right) \cdot \left( x_{\bar{q}}' \vec{p}_g - x_g \vec{p}_{\bar{q}} \right)}{\left(Q^2 + \frac{\vec{p}_{q1'}^{\,2}}{x_q' (1-x_q')} \right) \left(Q^2 + \frac{\vec{p}_{\bar{q}2}^{\,2}}{x_{\bar{q}}' (1-x_{\bar{q}}')} \right)  \left(x_q' \vec{p}_g - x_g \vec{p}_q \right)^2  \left( x_{\bar{q}}' \vec{p}_g - x_g \vec{p}_{\bar{q}} \right)^2 } \right \} \; .
   \\  \numberthis[real_div_LL]
\end{align*}

When two partons labeled $i$ and $j$ become collinear, the variable
\beqa
\vec{A}_{ij} = x_i \vec{p}_j - x_j \vec{p}_i
\eqa
vanishes. In the present case, 
the first term in the bracket of eq.~\eqref{eq:real_div_LL} gives the collinear divergences ($\vec{A}_{qg}^2 \to 0$, i.e. quark-gluon channel)  and the third ($\vec{A}_{\bar{q}g}^2 \to 0$, i.e. antiquark-gluon channel). 

For $LT$, the relevant divergent squared impact factor is \eqref{eq:impact_factor_TL} and for $TT$, it is \eqref{eq:impact_factor_TT}.

\subsection{Fragmentation from quark and anti-quark}
\label{sec:qqbarfrag}

As explained above, there are several contributions to the final cross-section that contain divergences. In this section, we deal with extracting the soft and collinear divergences associated with the contribution (b) in Fig. \ref{fig:sigma-NLO}. This contribution corresponds to the situation in which the quark and the anti-quark fragment and there is an additional emission of a gluon with respect to the LO case. Below,
\begin{itemize}
    \item[\textbullet] We compute the collinear divergence of the diagram (1) of Fig. \ref{fig:NLO-b-div} and show that it is removed by the + prescription part of the first term of eq. \eqref{eq:ct_LL}. 
    \item[\textbullet] Similarly, we calculate the collinear divergence of the diagram (3) of Fig. \ref{fig:NLO-b-div} and show that it is removed by the + prescription part of the third term of eq. \eqref{eq:ct_LL}.
    \item[\textbullet] We extract the soft divergences of diagrams (1), (2), (3), (4) of Fig. \ref{fig:NLO-b-div} and discuss the complete cancellation of divergences of this contribution.
\end{itemize}
The calculations of the collinear divergences is done by Fourier transforming the $\mathbf{F} (\frac{p_{12\perp}}{2})$, as defined in \eqref{eq: FT F}, and by using the identity  (see eg ref.~\cite{Chirilli:2012jd}):
\begin{equation}
\label{eq:expo}
\frac{1}{\mu^{2\epsilon}} \int d^{2+2 \epsilon} q_\perp e^{-i q_\perp \cdot r_\perp} \frac{1}{q_\perp^2} = \pi \left( \frac{4\pi}{\mu^2  r_\perp^2} \right)^{\epsilon} \Gamma (\epsilon) \,.
\end{equation}
We also have to change variables from the usual fraction of longitudinal photon momentum of the partons $(x_i',x_g)$ with $i= q,\bar{q}$, in the spirit of the definition \eqref{xq-xqbar}, as used in eq.~\eqref{eq:real_div_LL}, to the variable of the fraction of longitudinal photon momentum of the parent parton $x_i$ and longitudinal fraction $\beta$  with respect to the parent parton and not with respect to the photon anymore:
\beqa
(x_i',x_g) \rightarrow (x_i,\beta) \qquad \hbox{ with } \qquad x_i' = \beta x_i \,.
\eqa
Changing variables is necessary to be able to compare with \eqref{eq:ct_LL}, \eqref{eq:ct TL }, and \eqref{eq:ct TT }. Note that to make notations lighter, we will remove the ' when one particle is a spectator, see figs.~\ref{fig:kinematics_qg} and~\ref{fig:kinematics_qbarg}.

The Fourier transform of $\mathbf{F} \left(\frac{p_{12\perp}}{2}\right)$ is necessary in order to be able to integrate over the transverse momentum of the spectator parton, as it allows for the complete factorization of this momentum from this non-perturbative function. 

\subsubsection{Collinear contributions: $q$-$g$ splitting}
\label{sec:qqbarfragColl-qg}

\begin{figure}[h!]
\begin{picture}(420,160)
\put(-50,0){\includegraphics[scale=0.5]{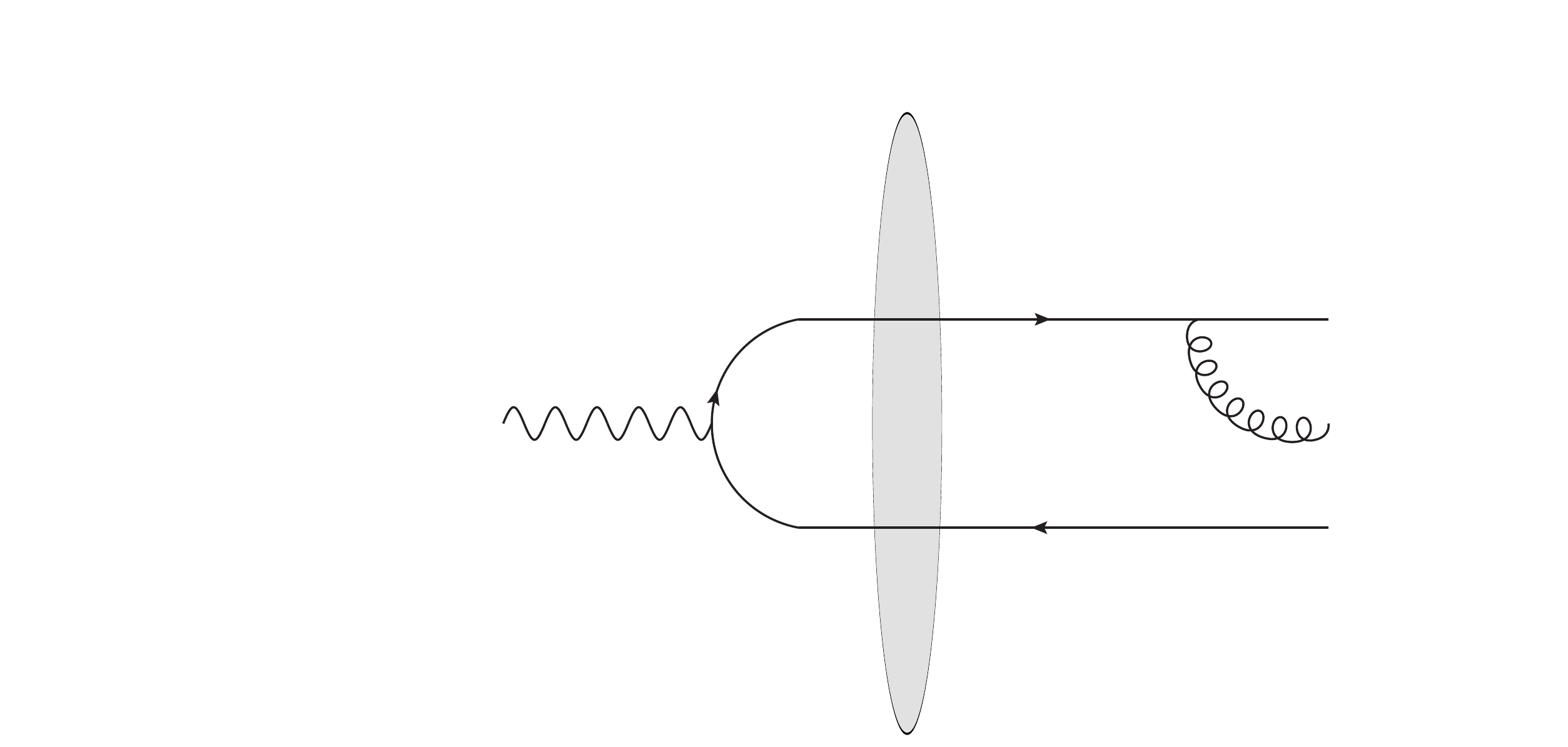}}
\put(170,110){$x_q, \vec{p}_q+\vec{p}_g$}
\put(270,100){$x'_q = \beta_1 x_q, \vec{p}_q$}
\put(270,75){$x_g=(1-\beta_1) x_q, \vec{p}_g$}
\put(270,50){$x_{\bar{q}}, \vec{p}_{\bar{q}}$}
\end{picture}
\caption{Kinematics for the $q-g$ splitting contribution. We indicate the longitudinal fraction of momentum carried by the partons as well as their transverse momenta.}
\label{fig:kinematics_qg}
\end{figure}
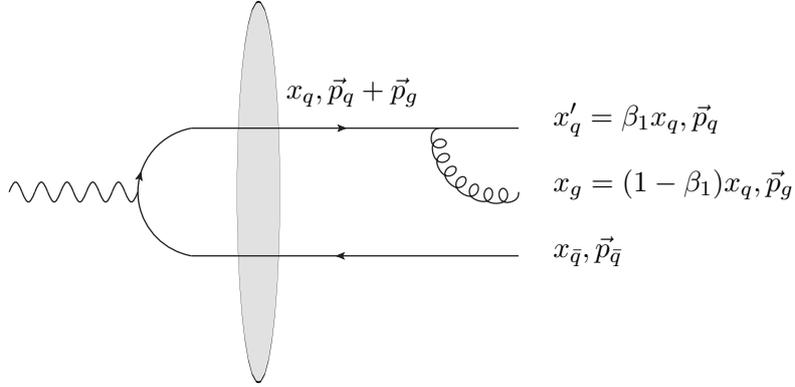
We use the kinematics illustrated in fig.~\ref{fig:kinematics_qg}.
The term in \eqref{eq: div real impact factor} considered for this collinear contribution is the first one in the bracket. 
{\allowdisplaybreaks
\begin{align*}
 &  d \sigma_{3LL}^{q \bar{q} \rightarrow h_1 h_2}|_{\text{coll. qg}} \\
    & = d x_{h_1} d x_{h_2} \frac{4  \alpha_{\mathrm{em}} Q^2}{(2\pi)^{4(d-1)} N_c} \sum_{q}  \int_{x_{h_1}}^1 \frac{d x_q'}{x_q'} \int_{x_{h_2}}^1 \frac{d x_{\bar{q}}}{x_{\bar{q}}} \int_{\alpha}^1 \frac{d x_g}{x_g} \delta(1-x_q'-x_{\bar{q}}-x_g) \\
    & \times Q_q^2 D_q^{h_1}\left(\frac{x_{h_1}}{x_q'}, \mu_F\right) D_{\bar{q}}^{h_2}\left(\frac{x_{h_2}}{x_{\bar{q}}}, \mu_F\right) d^d p_{q\perp} d^d p_{\bar{q}\perp}   \frac{\alpha_s}{\mu^{2\epsilon}}C_F  \frac{d^d p_{g\perp}}{(2\pi)^d}   \\ 
     & \times \int d^d p_{1\perp}  d^d p_{2\perp}    \delta(p_{q1\perp} + p_{\bar{q}2\perp} + p_{g\perp}) \; \mathbf{F} \left(\frac{p_{12\perp}}{2}\right)  \\
     & \times \int d^d p_{1'\perp}   d^d p_{2'\perp} \delta(p_{q1'\perp} + p_{\bar{q}2'\perp}  + p_{g\perp}) \; \mathbf{F}^*\left(\frac{p_{1'2'\perp}}{2}\right)  \\
    & \times \frac{(d x_g^2 + 4 x_q' (x_q' + x_g)) x_{\bar{q}}^2 (1-x_{\bar{q}})^2}{\left( x_{\bar{q}} (1-x_{\bar{q}})Q^2 + \vec{p}_{\bar{q}2}^{\,2}\right) \left( x_{\bar{q}} (1-x_{\bar{q}})Q^2 + \vec{p}_{\bar{q}2'}^{\,2}\right) (x_q' \vec{p}_g - x_g \vec{p}_q)^2}  + (h_1 \leftrightarrow h_2) \\ 
    & = d x_{h_1} d x_{h_2} d^d p_{h_1 \perp} d^d p_{h_2\perp} \frac{4  \alpha_{\mathrm{em}} Q^2}{(2\pi)^{4(d-1)} N_c} \sum_{q} \int_{x_{h_1}}^1 \frac{d x_q'}{x_q'} \int_{\alpha}^1 \frac{d x_g}{x_g} \int_{x_{h_2}}^1 \frac{d x_{\bar{q}}}{x_{\bar{q}}} \delta(1-x_q'-x_{\bar{q}}-x_g) \\
    & \times   \left(\frac{x_q'}{x_{h_1}}\right)^d \left(\frac{x_{\bar{q}}}{x_{h_2}}\right)^d  Q_q^2 D_q^{h_1}\left(\frac{x_{h_1}}{x_q'}, \mu_F\right) D_{\bar{q}}^{h_2}\left(\frac{x_{h_2}}{x_{\bar{q}}}, \mu_F\right)  \frac{\alpha_s}{\mu^{2\epsilon}} C_F  \frac{d^d p_{g\perp}}{(2\pi)^d} \\
     & \times \int d^d p_{1\perp}  d^d p_{2\perp}    \delta\left(\frac{x_q'}{x_{h_1}} p_{h_1\perp} -p_{1\perp} + \frac{x_{\bar{q}}}{x_{h_2}} p_{h_2 \perp} -p_{2\perp} + p_{g\perp}\right)  \mathbf{F}\left(\frac{p_{12\perp}}{2}\right) \\
     & \times \int d^d p_{1'\perp}  d^d p_{2'\perp}    \delta \left(\frac{x_q'}{x_{h_1}} p_{h_1\perp} -p_{1'\perp} + \frac{x_{\bar{q}}}{x_{h_2}}p_{h_2 \perp} -p_{2'\perp} + p_{g\perp}\right)  \mathbf{F}^*\left(\frac{p_{1'2'\perp}}{2}\right) \\
     & \times \frac{(d x_g^2 + 4 x_q' (x_q' + x_g)) x_{\bar{q}}^2 (1-x_{\bar{q}})^2}{\left( x_{\bar{q}} (1-x_{\bar{q}})Q^2 \hspace{-0.1 cm} + \hspace{-0.05 cm} \left(\frac{x_{\bar{q}}}{x_{h_2}} \vec{p}_{h_2}-\vec{p}_{2} \hspace{-0.05 cm} \right)^2 \hspace{-0.05 cm} \right) \left( x_{\bar{q}} (1-x_{\bar{q}})Q^2 \hspace{-0.1 cm} + \hspace{-0.05 cm} \left(\frac{x_{\bar{q}}}{x_{h_2}} \vec{p}_{h_2}-\vec{p}_{2'} \hspace{-0.05 cm} \right)^2 \hspace{-0.05 cm} \right) \left(x_q' \vec{p}_g - x_g \frac{x_{q}'}{x_{h_1}}\vec{p}_{h_1} \hspace{-0.05 cm} \right)^2} \\
     & + (h_1 \leftrightarrow h_2)\\ 
      & = d x_{h_1} d x_{h_2} d^d p_{h_1 \perp} d^d p_{h_2\perp} \frac{4  \alpha_{\mathrm{em}} Q^2}{(2\pi)^{4(d-1)} N_c} \sum_{q} \int_{x_{h_1}}^1 \frac{d x_q'}{x_q'} \int_{\alpha}^1 \frac{d x_g}{x_g} \int_{x_{h_2}}^1 \frac{d x_{\bar{q}}}{x_{\bar{q}}} \delta(1-x_q'-x_{\bar{q}}-x_g) \\
    & \times   \left(\frac{x_q'}{x_{h_1}}\right)^d \left(\frac{x_{\bar{q}}}{x_{h_2}}\right)^d  Q_q^2 D_q^{h_1}\left(\frac{x_{h_1}}{x_q'}, \mu_F\right) D_{\bar{q}}^{h_2}\left(\frac{x_{h_2}}{x_{\bar{q}}}, \mu_F\right)  \frac{\alpha_s}{\mu^{2\epsilon}} C_F  \frac{d^d p_{g\perp}}{(2\pi)^d} \\
     & \times \int d^d p_{2\perp}   \;  \mathbf{F} \left(\frac{x_q'}{2 x_{h_1}} p_{h_1\perp} + \frac{x_{\bar{q}}}{2 x_{h_2}} p_{h_2 \perp} -p_{2\perp} + \frac{p_{g\perp}}{2}\right)  \\
     & \times \int  d^d p_{2'\perp}  \;   \mathbf{F}^* \left(\frac{x_q'}{2x_{h_1}} p_{h_1\perp} + \frac{x_{\bar{q}}}{ 2 x_{h_2}}p_{h_2 \perp} -p_{2'\perp} + \frac{p_{g\perp}}{2}\right)  \\
     & \times \hspace{-0.1 cm} \frac{(d x_g^2 + 4 x_q' (x_q' + x_g)) x_{\bar{q}}^2 (1-x_{\bar{q}})^2}{\left( x_{\bar{q}} (1-x_{\bar{q}})Q^2 + \hspace{-0.1 cm} \left(\frac{x_{\bar{q}}}{x_{h_2}} \vec{p}_{h_2}-\vec{p}_{2}\right)^2 \hspace{-0.05 cm} \right) \hspace{-0.1 cm} \left( x_{\bar{q}} (1-x_{\bar{q}})Q^2 \hspace{-0.1 cm} +\left(\frac{x_{\bar{q}}}{x_{h_2}} \vec{p}_{h_2}-\vec{p}_{2'}\right)^2 \hspace{-0.05 cm} \right) \hspace{-0.1 cm} \left(x_q' \vec{p}_g - x_g \frac{x_{q}'}{x_{h_1}}\vec{p}_{h_1} \hspace{-0.05 cm} \right)^2} \\
     & + (h_1 \leftrightarrow h_2) \; .  
\end{align*}}
After performing the change of variable
\begin{eqnarray}
        x_q' &=& \beta_1 x_q \nonumber \\ 
        x_g &=& (1-\beta_1) x_q 
\label{eq:Transbeta}    
\end{eqnarray}
and using the
Jacobian $d x_q' d x_g = x_q \,d x_q d \beta_1 $
we can rewrite the longitudinal integration in the symbolic form
    \begin{align*}
         & \int_{x_{h_1}}^1 \frac{d x_q'}{x_q'} \int_{x_{h_2}}^1 \frac{d x_{\bar{q}} }{x_{\bar{q}}} \int_{\alpha}^1 \frac{d x_g}{x_g} \delta(1-x_q'-x_{\bar{q}}-x_g) \\
        & = \int_{x_{h_1}}^1 \frac{d x_q'}{x_q'} \int_{\alpha}^1 \frac{d x_g}{x_g} \int_{- \infty}^{+\infty} \frac{d x_{\bar{q}}}{x_{\bar{q}}} \theta(x_{\bar{q}}-x_{h_2}) \theta(1- x_{\bar{q}}) \delta(1-x_q'-x_{\bar{q}}-x_g) \\
        &= \int_{x_{h_1}}^1 \frac{d x_q'}{x_q'} \int_{\alpha}^1 \frac{d x_g}{x_g}  \theta(1-x_q'-x_g-x_{h_2}) \theta(x_q'+x_g) \frac{1}{1-x_q'-x_g} \\
        &= \int_{x_{h_1}}^{1-x_{h_2}} \frac{d x_q}{x_q} \frac{1}{1-x_q} \int_{\frac{x_{h_1}}{x_q}}^{1-\frac{\alpha}{x_q}}\frac{d\beta_1}{ \beta_1  (1-\beta_1)} \,.\numberthis[change_variable_integral]
    \end{align*}
After this manipulation, we obtain
\begin{align*}
& \frac{d \sigma_{3LL}^{q \bar{q} \rightarrow h_1 h_2}}{ d x_{h_1} d x_{h_2} d^d p_{h_1 \perp} d^d p_{h_2\perp} } \Bigg |_{\text{coll qg}} \\
&=  \frac{4  \alpha_{\mathrm{em}} Q^2}{(2\pi)^{4(d-1)} N_c} \sum_{q} \int_{x_{h_1}}^{1-x_{h_2}} d x_q x_q (1-x_q) \left(\frac{ x_q}{x_{h_1}}\right)^d \left(\frac{1-x_q}{x_{h_2}}\right)^d  \\
&\times \int_{\frac{x_{h_1}}{x_q}}^{1-\frac{\alpha}{x_q}}\frac{d\beta_1}{\beta_1}    Q_q^2 D_q^{h_1}\left(\frac{x_{h_1}}{\beta_1 x_q}, \mu_F\right) D_{\bar{q}}^{h_2}\left(\frac{x_{h_2}}{1-x_q}, \mu_F\right)  \\
 & \times \int   d^d p_{2\perp}  \int d^d z_{1\perp}   \frac{e^{i z_{1\perp}\cdot \left(\frac{\beta_1 x_q}{2 x_{h_1}} p_{h_1\perp} + \frac{1-x_q}{2 x_{h_2}} p_{h_2 \perp} -p_{2\perp} \right)} F(z_{1\perp})}{x_q (1-x_q)Q^2 +\left(\frac{1-x_q}{x_{h_2}} \vec{p}_{h_2}-\vec{p}_{2}\right)^2} \\
& \times \int   d^d p_{2'\perp}  \int d^d z_{2\perp}   \frac{e^{-i z_{2\perp}\cdot \left(\frac{\beta_1 x_q}{2 x_{h_1}} p_{h_1\perp} + \frac{1-x_q}{2 x_{h_2}} p_{h_2 \perp} -p_{2'\perp} \right)} F^*(z_{2\perp})}{x_q (1-x_q)Q^2 +\left(\frac{1-x_q}{x_{h_2}} \vec{p}_{h_2}-\vec{p}_{2'}\right)^2} \\
&  \times \frac{2(1+\beta_1^2)+ 2 \epsilon (1-\beta_1)^2 + 4 \epsilon (1+\beta_1^2)  \ln \beta_1  }{ 1-\beta_1}  \\
& \times e^{i \left(\frac{z_{1\perp}-z_{2\perp}}{2}\right)\cdot \frac{(1-\beta_1)x_q}{x_{h_1}} p_{h_1\perp}}   \frac{\alpha_s}{\mu^{2\epsilon}} C_F  \int \frac{d^d p_{g\perp}}{(2\pi)^d} \frac{e^{i \left(\frac{z_{1\perp}-z_{2\perp}}{2}\right)\cdot p_{g\perp}}}{\left(\vec{p}_g  \right)^2} + (h_1 \leftrightarrow h_2 ).
\numberthis[dsigmaLL-qg-collinear]
    \end{align*}    

\noindent
The integral over $p_{g\perp}$ gives, using eq.~\eqref{eq:expo},
\begin{align*}
\mu^{-2\epsilon}\int \frac{d^d p_{g\perp}}{(2\pi)^d} \frac{e^{i \left(\frac{z_{1\perp}-z_{2\perp}}{2}\right)\cdot p_{g\perp}}}{\left(\vec{p}_g  \right)^2} &  = \frac{1}{(2\pi)^d} \pi \mu^{-2\epsilon} \left[ \frac{\left( \frac{z_{1\perp}-z_{2\perp}}{2}\right)^2}{4\pi}\right]^{-\epsilon} \Gamma (\epsilon) \\*
& = \frac{1}{4\pi} \left( \frac{1}{\hat{\epsilon}} + \ln \left( \frac{c_0^2}{\left(\frac{z_{1\perp} - z_{2\perp}}{2}\right)^2 \mu^2}\right) \right) + O(\epsilon) \\
& = \frac{1}{4\pi} \frac{1}{\hat{\epsilon}} \left(\frac{c_0^2}{\left(\frac{z_{1\perp}-z_{2\perp}}{2}\right)^2 \mu^2}\right)^\epsilon + O(\epsilon)
\numberthis[integration-over-pgperp]
\end{align*}
where $c_0 = 2 e^{-\gamma_E}.$ This leads to
\begin{align*}
& \frac{d \sigma_{3LL}^{q \bar{q} \rightarrow h_1 h_2}}{ d x_{h_1} d x_{h_2} d^d p_{h_1 \perp} d^d p_{h_2\perp} } \Bigg |_{\text{coll qg.}} \\
& =  \frac{4  \alpha_{\mathrm{em}} Q^2}{(2\pi)^{4(d-1)} N_c}  \sum_{q} \int_{x_{h_1}}^{1} d x_q  \int_{x_{h_2}}^{1} d x_{\bar{q}} x_q x_{\bar{q}} \delta(1-x_q -x_{\bar{q}}) \left(\frac{x_q}{x_{h_1}}\right)^d \left(\frac{x_{\bar{q}}}{x_{h_2}}\right)^d \\
& \times \int   d^d p_{2\perp}  \int d^d z_{1\perp}   \frac{e^{i z_{1\perp}\cdot \left(\frac{x_q}{2 x_{h_1}} p_{h_1\perp} + \frac{1-x_q}{2 x_{h_2}} p_{h_2 \perp} -p_{2\perp} \right)} F(z_{1\perp})}{x_q (1-x_q)Q^2 +\left(\frac{1-x_q}{x_{h_2}} \vec{p}_{h_2}-\vec{p}_{2}\right)^2} \\
& \times \int   d^d p_{2'\perp}  \int d^d z_{2\perp}   \frac{e^{-i z_{2\perp}\cdot \left(\frac{x_q}{2 x_{h_1}} p_{h_1\perp} + \frac{1-x_q}{2 x_{h_2}} p_{h_2 \perp} -p_{2'\perp} \right)} F^*(z_{2\perp})}{x_q (1-x_q)Q^2 +\left(\frac{1-x_q}{x_{h_2}} \vec{p}_{h_2}-\vec{p}_{2'}\right)^2} \\
& \times  \int_{\frac{x_{h_1}}{x_q}}^{1-\frac{\alpha}{x_q}}\frac{d\beta_1}{\beta_1}  Q_q^2 D_q^{h_1}\left(\frac{x_{h_1}}{\beta_1 x_q}, \mu_F\right) D_{\bar{q}}^{h_2}\left(\frac{x_{h_2}}{1-x_q}, \mu_F\right) \\
& \times \frac{\alpha_s C_F}{2\pi} \left[ \frac{1}{\hat{\epsilon}} \left(\frac{c_0^2}{\left(\frac{z_{1\perp}-z_{2\perp}}{2}\right)^2 \mu^2}\right)^\epsilon \frac{1+ \beta_1^2}{1-\beta_1}  + \frac{(1-\beta_1)^2 + 2 (1+ \beta_1^2) \ln \beta_1 }{(1-\beta_1)} \right] + (h_1 \leftrightarrow h_2 ). \numberthis[coll_qg_qqbar_FF_1]
\end{align*}
Here in eq.~\eqref{eq:coll_qg_qqbar_FF_1}
we have put back the integral in $x_{\bar{q}} $ using
\begin{equation}
 \label{eq: xq xbarq }
\begin{aligned}
\int_{x_{h_1}}^1 d x_q \int_{x_{h_2}}^1 d x_{\bar{q}}\, \delta(1-x_q-x_{\bar{q}}) & =  \int_{x_{h_1}}^1 d x_q \int_{-\infty}^{+\infty} d x_{\bar{q}} \, \delta(1-x_q-x_{\bar{q}}) \theta(1-x_{\bar{q}}) \theta(x_{\bar{q}}-x_{h_2}) \\
&= \int_{x_{h_1}}^{1-x_{h_2}} d x_q  
\end{aligned}
\end{equation}
in order to have the same form as in the LO cross-section \eqref{eq:LL-LO}.
  
Now, to separate the collinear and soft contribution we introduce the plus prescription, as defined in eq.~\eqref{eq: plus prescription}, and after, we expand the factor $ \frac{1}{\epsilon} \left(\frac{c_0^2}{\left(\frac{z_{1\perp}-z_{2\perp}}{2}\right)^2 \mu^2}\right)^\epsilon$ within accuracy of order $\epsilon^0$, only in those terms whose integrand is  safe in the limit $\beta_1 \rightarrow 1$.  

\begin{align*}
& \frac{d \sigma_{3LL}^{q \bar{q} \rightarrow h_1 h_2}}{ d x_{h_1} d x_{h_2} d^d p_{h_1 \perp} d^d p_{h_2\perp} } \Bigg |_{\text{coll qg.}} \\
& =  \frac{4  \alpha_{\mathrm{em}} Q^2}{(2\pi)^{4(d-1)} N_c}  \sum_{q} \int_{x_{h_1}}^{1} d x_q  \int_{x_{h_2}}^{1} d x_{\bar{q}} \; x_q x_{\bar{q}} \delta(1-x_q -x_{\bar{q}}) \left(\frac{x_q}{x_{h_1}}\right)^d \left(\frac{x_{\bar{q}}}{x_{h_2}}\right)^d \\
& \times \int   d^d p_{2\perp}  \int d^d z_{1\perp}   \frac{e^{i z_{1\perp}\cdot \left(\frac{x_q}{2 x_{h_1}} p_{h_1\perp} + \frac{x_{\bar{q}}}{2 x_{h_2}} p_{h_2 \perp} -p_{2\perp} \right)} F(z_{1\perp})}{x_q x_{\bar{q}}Q^2 +\left(\frac{x_{\bar{q}}}{x_{h_2}} \vec{p}_{h_2}-\vec{p}_{2}\right)^2} \\
& \times \int   d^d p_{2'\perp}  \int d^d z_{2\perp}   \frac{e^{-i z_{2\perp}\cdot \left(\frac{x_q}{2 x_{h_1}} p_{h_1\perp} + \frac{x_{\bar{q}}}{2 x_{h_2}} p_{h_2 \perp} -p_{2'\perp} \right)} F^*(z_{2\perp})}{x_q x_{\bar{q}}Q^2 +\left(\frac{x_{\bar{q}}}{x_{h_2}} \vec{p}_{h_2}-\vec{p}_{2'}\right)^2} \\
& \times \left \{ \int_{\frac{x_{h_1}}{x_q}}^{1}\frac{d\beta_1}{\beta_1}  Q_q^2 D_q^{h_1}\left(\frac{x_{h_1}}{\beta_1 x_q}, \mu_F\right) D_{\bar{q}}^{h_2}\left(\frac{x_{h_2}}{x_{\bar{q}}}, \mu_F\right) \frac{\alpha_s C_F}{2\pi} \frac{1}{\hat{\epsilon}}  \frac{1+ \beta_1^2}{(1-\beta_1)_+} \right. \\
& + \int_{\frac{x_{h_1}}{x_q}}^{1-\frac{\alpha}{x_q}} d\beta_1  Q_q^2 D_q^{h_1}\left(\frac{x_{h_1}}{ x_q}, \mu_F\right) D_{\bar{q}}^{h_2}\left(\frac{x_{h_2}}{x_{\bar{q}}}, \mu_F\right) \frac{\alpha_s C_F}{2\pi} \frac{1}{\hat{\epsilon}} \left(\frac{c_0^2}{\left(\frac{z_{1\perp}-z_{2\perp}}{2}\right)^2 \mu^2}\right)^\epsilon \frac{2}{1-\beta_1}  \\
& - Q_q^2 D_q^{h_1}\left(\frac{x_{h_1}}{x_q},\mu_F\right) D_{\bar{q}}^{h_2}\left(\frac{x_{h_2}}{x_{\bar{q}}},\mu_F \right) \frac{\alpha_s C_F}{2\pi} \frac{1}{\hat{\epsilon}} 2 \ln \left(1-\frac{x_{h_1}}{x_q}\right) \\
& + \int_{\frac{x_{h_1}}{x_q}}^{1}\frac{d\beta_1}{\beta_1}  Q_q^2 D_q^{h_1}\left(\frac{x_{h_1}}{\beta_1 x_q}, \mu_F\right) D_{\bar{q}}^{h_2}\left(\frac{x_{h_2}}{x_{\bar{q}}}, \mu_F\right) \frac{\alpha_s C_F}{2\pi} \left[ \ln \left( \frac{c_0^2}{\left(\frac{z_{1\perp}-z_{2\perp}}{2}\right)^2 \mu^2} \right) \frac{1+ \beta_1^2}{(1-\beta_1)_+} \right. \\ & \left. 
+ \frac{(1-\beta_1)^2 + 2 (1+ \beta_1^2) \ln \beta_1 }{(1-\beta_1)} \right] - Q_q^2 D_q^{h_1}\left(\frac{x_{h_1}}{x_q},\mu_F\right) D_{\bar{q}}^{h_2}\left(\frac{x_{h_2}}{x_{\bar{q}}},\mu_F \right) \\
& \left. \times \frac{\alpha_s C_F}{2\pi} 2 \ln \left(1-\frac{x_{h_1}}{x_q}\right) \ln \left(\frac{c_0^2}{\left(\frac{z_{1\perp}-z_{2\perp}}{2}\right)^2 \mu^2}\right) \right \} + (h_1 \leftrightarrow h_2 ) \\
&= \frac{d \sigma_{3LL}^{q \bar{q} \rightarrow h_1 h_2}}{ d x_{h_1} d x_{h_2} d p_{h_1 \perp} d^d p_{h_2\perp} } \bigg |_{\text{coll. qg div}}     + \frac{d \sigma_{3LL}^{q \bar{q} \rightarrow h_1 h_2}}{ d x_{h_1} d x_{h_2} d p_{h_1 \perp} d^d p_{h_2\perp} } \bigg |_{\text{coll. qg fin}}   \numberthis[coll_qg_qqbar_FF]  \;,
    \end{align*}
where the term denoted by the label "coll. qg div" corresponds to the sum of the first three terms in the curly bracket, whereas the remaining terms in the curly bracket contribute to the term denoted "coll. qg fin".

This gives the following expression for the divergent part: 
\begin{align*}
& \frac{d \sigma_{3LL}^{q \bar{q} \rightarrow h_1 h_2}}{ d x_{h_1} d x_{h_2} d p_{h_1 \perp} d^d p_{h_2\perp} } \bigg |_{\text{coll. qg div}}       \\
&=  \frac{4  \alpha_{\mathrm{em}} Q^2}{(2\pi)^{4(d-1)} N_c} \sum_{q} \int_{x_{h_1}}^{1} d x_q \int_{x_{h_2}}^1 d x_{\bar{q}} \; x_q x_{\bar{q}} \left(\frac{x_q}{x_{h_1}}\right)^d \left(\frac{x_{\bar{q}}}{x_{h_2}}\right)^d \delta(1-x_q-x_{\bar{q}}) \\*
& \times \int   d^d p_{2\perp}  \int d^d z_{1\perp}   \frac{e^{i z_{1\perp}\cdot \left(\frac{x_q}{2 x_{h_1}} p_{h_1\perp} + \frac{x_{\bar{q}}}{2 x_{h_2}} p_{h_2 \perp} -p_{2\perp} \right)} F(z_{1\perp})}{x_q x_{\bar{q}} Q^2 +\left(\frac{x_{\bar{q}}}{x_{h_2}} \vec{p}_{h_2}-\vec{p}_{2}\right)^2} \\
& \times \int   d^d p_{2'\perp}  \int d^d z_{2\perp}   \frac{e^{-i z_{2\perp}\cdot \left(\frac{x_q}{2 x_{h_1}} p_{h_1\perp} + \frac{x_{\bar{q}}}{2 x_{h_2}} p_{h_2 \perp} -p_{2'\perp} \right)} F^*(z_{2\perp})}{x_q x_{\bar{q}} Q^2 +\left(\frac{x_{\bar{q}}}{x_{h_2}} \vec{p}_{h_2}-\vec{p}_{2'}\right)^2} \\
& \times \frac{\alpha_s}{2\pi} \frac{1}{\hat{\epsilon}}  Q_q^2 \left[
\int_{\frac{x_{h_1}}{x_q}}^{1}\frac{d\beta_1}{\beta_1} C_F \frac{1+ \beta_1^2}{(1-\beta_1)_+} D_q^{h_1}\left(\frac{x_{h_1}}{\beta_1 x_q},\mu_F\right) D_{\bar{q}}^{h_2}\left(\frac{x_{h_2}}{x_{\bar{q}}},\mu_F \right) \right. \\*
& + \int_{\frac{x_{h_1}}{x_q}}^{1-\frac{\alpha}{x_q}} d \beta_1 C_F \frac{2}{1-\beta_1} \left(\frac{c_0^2}{\left(\frac{z_{1\perp}-z_{2\perp}}{2}\right)^2 \mu^2}\right)^\epsilon D_q^{h_1}\left(\frac{x_{h_1}}{x_q},\mu_F\right) D_{\bar{q}}^{h_2}\left(\frac{x_{h_2}}{x_{\bar{q}}},\mu_F \right) \\
& \left. - 2  C_F \ln \left(1-\frac{x_{h_1}}{x_q}\right)  D_q^{h_1}\left(\frac{x_{h_1}}{x_q},\mu_F\right) D_{\bar{q}}^{h_2}\left(\frac{x_{h_2}}{x_{\bar{q}}},\mu_F \right) \right] + (h_1 \leftrightarrow h_2 ) \,.  \numberthis[coll_div_q] 
\end{align*}
The first term in the bracket cancels part of the first term in the bracket in eq.~\eqref{eq:ct_LL}, i.e. the part involving the $+$ prescription in the spitting function $P_{qq}$, and the remaining part of the $P_{qq}$ term is cancelled by an analogous contribution in the virtual part. The second term in eq.~\eqref{eq:coll_div_q}  has to be removed to avoid double counting as it corresponds to the soft contribution and will be taken into account later in the paper. The third and last term, in eq.~\eqref{eq:coll_div_q}, will compensate an analogous term in the soft contribution.

The finite part for the $LL$ contribution takes the form 
\begin{align*}
&  \frac{d \sigma_{3LL}^{q \bar{q} \rightarrow h_1 h_2}}{ d x_{h_1} d x_{h_2} d p_{h_1 \perp} d^d p_{h_2\perp} } \bigg |_{\text{coll. qg fin}} \\
&= \frac{4  \alpha_{\mathrm{em}} Q^2}{(2\pi)^{4(d-1)} N_c} \sum_{q} \int_{x_{h_1}}^{1} d x_q  \int_{x_{h_2}}^{1} d x_{\bar{q}} \; x_q x_{\bar{q}} \delta(1-x_q -x_{\bar{q}}) \left(\frac{x_q}{x_{h_1}}\right)^d \left(\frac{x_{\bar{q}}}{x_{h_2}}\right)^d \\
& \times \int   d^d p_{2\perp}  \int d^d z_{1\perp}   \frac{e^{i z_{1\perp}\cdot \left(\frac{x_q}{2 x_{h_1}} p_{h_1\perp} + \frac{x_{\bar{q}}}{2 x_{h_2}} p_{h_2 \perp} -p_{2\perp} \right)} F(z_{1\perp})}{x_q x_{\bar{q}} Q^2 +\left(\frac{x_{\bar{q}}}{x_{h_2}} \vec{p}_{h_2}-\vec{p}_{2}\right)^2} \\
& \times \int   d^d p_{2'\perp}  \int d^d z_{2\perp}   \frac{e^{-i z_{2\perp}\cdot \left(\frac{x_q}{2 x_{h_1}} p_{h_1\perp} + \frac{x_{\bar{q}}}{2 x_{h_2}} p_{h_2\perp} -p_{2'\perp} \right)} F^*(z_{2\perp})}{x_q x_{\bar{q}} Q^2 +\left(\frac{x_{\bar{q}}}{x_{h_2}} \vec{p}_{h_2}-\vec{p}_{2'}\right)^2} \\
& \times  \frac{\alpha_s C_F}{2\pi}  \left \{ \int_{\frac{x_{h_1}}{x_q}}^{1} \frac{d\beta_1}{\beta_1}  Q_q^2 D_q^{h_1}\left(\frac{x_{h_1}}{\beta_1 x_q}, \mu_F\right) D_{\bar{q}}^{h_2}\left(\frac{x_{h_2}}{x_{\bar{q}}}, \mu_F\right) \right. \\
& \times  \left[ \ln \left( \frac{c_0^2}{\left(\frac{z_{1\perp} - z_{2\perp}}{2}\right)^2 \mu^2}\right)  \frac{1+ \beta_1^2}{(1-\beta_1)_+}  + \frac{(1-\beta_1)^2 + 2 (1+ \beta_1^2) \ln \beta_1 }{(1-\beta_1)} \right] \\
& \left. - 2 \ln \left( 1- \frac{x_{h_1}}{x_q}\right) \ln \left( \frac{c_0^2}{\left(\frac{z_{1\perp} - z_{2\perp}}{2}\right)^2 \mu^2}\right) D_{q}^{h_1} \left(\frac{x_{h_1}}{x_q},\mu_F\right) D_{\bar{q}}^{h_q} \left(\frac{x_{h_2}}{x_{\bar{q}}},\mu_F \right)\right\} + (h_1 \leftrightarrow h_2 )\,. \numberthis[collqg_LL_fin]
    \end{align*}

Similarly, one gets for the $TL$ case
\begin{align*}
& \frac{d \sigma_{3TL}^{q \bar{q} \rightarrow h_1 h_2}}{ d x_{h_1} d x_{h_2} d p_{h_1 \perp} d^d p_{h_2\perp} } \bigg |_{\text{coll. qg }} \\
& =  \frac{2  \alpha_{\mathrm{em}} Q}{(2\pi)^{4(d-1)} N_c} \sum_{q}  \int_{x_{h_1}}^{1} d x_q  \int_{x_{h_2}}^{1} d x_{\bar{q}} \; (x_{\bar{q}}-x_q)\delta(1-x_q -x_{\bar{q}}) \left(\frac{x_q}{x_{h_1}}\right)^d \left(\frac{x_{\bar{q}}}{x_{h_2}}\right)^d \\
& \times \int   d^d p_{2\perp}  \int d^d z_{1\perp}   \frac{e^{i z_{1\perp}\cdot \left(\frac{x_q}{2 x_{h_1}} p_{h_1\perp} + \frac{x_{\bar{q}}}{2 x_{h_2}} p_{h_2 \perp} -p_{2\perp} \right)} F(z_{1\perp})}{x_q x_{\bar{q}} Q^2 +\left(\frac{x_{\bar{q}}}{x_{h_2}} \vec{p}_{h_2}-\vec{p}_{2}\right)^2} \\
& \times \int   d^d p_{2'\perp}  \int d^d z_{2\perp}   \frac{e^{-i z_{2\perp}\cdot \left(\frac{x_q}{2 x_{h_1}} p_{h_1\perp} + \frac{x_{\bar{q}}}{2 x_{h_2}} p_{h_2 \perp} -p_{2'\perp} \right)} F^*(z_{2\perp})}{x_q x_{\bar{q}} Q^2 +\left(\frac{x_{\bar{q}}}{x_{h_2}} \vec{p}_{h_2}-\vec{p}_{2'}\right)^2}   \left( \frac{x_{\bar{q}}}{x_{h_2}} \vec{p}_{h_2}  - \vec{p}_{2'}\right) \cdot \vec{\varepsilon}^{\,*}_T  \\
& \times  \int_{\frac{x_{h_1}}{x_q}}^{1-\frac{\alpha}{x_q}}\frac{d\beta_1}{\beta_1} Q_q^2 D_q^{h_1}\left(\frac{x_{h_1}}{\beta_1 x_q}, \mu_F\right) D_{\bar{q}}^{h_2}\left(\frac{x_{h_2}}{x_{\bar{q}}}, \mu_F\right) \\
& \times \frac{\alpha_s C_F}{2\pi} \left[ \frac{1}{\hat{\epsilon}} \left(\frac{c_0^2}{\left(\frac{z_{1\perp}-z_{2\perp}}{2}\right)^2 \mu^2}\right)^\epsilon \frac{1+ \beta_1^2}{1-\beta_1}  + \frac{(1-\beta_1)^2 + 2 (1+ \beta_1^2) \ln \beta_1 }{(1-\beta_1)} \right] + (h_1 \leftrightarrow h_2 ) \\
&= \frac{d \sigma_{3TL}^{q \bar{q} \rightarrow h_1 h_2}}{ d x_{h_1} d x_{h_2} d p_{h_1 \perp} d^d p_{h_2\perp} } \bigg |_{\text{coll. qg div}}     + \frac{d \sigma_{3TL}^{q \bar{q} \rightarrow h_1 h_2}}{ d x_{h_1} d x_{h_2} d p_{h_1 \perp} d^d p_{h_2\perp} } \bigg |_{\text{coll. qg fin}}   ,  \numberthis
    \end{align*} %
where
\begin{align*}
& \frac{d \sigma_{3TL}^{q \bar{q} \rightarrow h_1 h_2}}{ d x_{h_1} d x_{h_2} d p_{h_1 \perp} d^d p_{h_2\perp} } \bigg |_{\text{coll. qg div}}       \\
&=  \frac{ 2 \alpha_{\mathrm{em}} Q}{(2\pi)^{4(d-1)} N_c} \sum_{q} \int_{x_{h_1}}^{1} d x_q \int_{x_{h_2}}^1 d x_{\bar{q}}  \; (x_{\bar{q}}-x_q) \left(\frac{x_q}{x_{h_1}}\right)^d \left(\frac{x_{\bar{q}}}{x_{h_2}}\right)^d \delta(1-x_q-x_{\bar{q}}) \\*
& \times \int   d^d p_{2\perp}  \int d^d z_{1\perp}   \frac{e^{i z_{1\perp}\cdot \left(\frac{x_q}{2 x_{h_1}} p_{h_1\perp} + \frac{x_{\bar{q}}}{2 x_{h_2}} p_{h_2 \perp} -p_{2\perp} \right)} F(z_{1\perp})}{x_q x_{\bar{q}} Q^2 +\left(\frac{x_{\bar{q}}}{x_{h_2}} \vec{p}_{h_2}-\vec{p}_{2}\right)^2} \\
& \times \int   d^d p_{2'\perp}  \int d^d z_{2\perp}   \frac{e^{-i z_{2\perp}\cdot \left(\frac{x_q}{2 x_{h_1}} p_{h_1\perp} + \frac{x_{\bar{q}}}{2 x_{h_2}} p_{h_2 \perp} -p_{2'\perp} \right)} F^*(z_{2\perp})}{x_q x_{\bar{q}} Q^2 +\left(\frac{x_{\bar{q}}}{x_{h_2}} \vec{p}_{h_2}-\vec{p}_{2'}\right)^2} \left( \frac{x_{\bar{q}}}{x_{h_2}} \vec{p}_{h_2}  - \vec{p}_{2'}\right) \cdot \vec{\varepsilon}^{\,*}_T \\
& \times \frac{\alpha_s}{2\pi} \frac{1}{\hat{\epsilon}}  Q_q^2 \left[
\int_{\frac{x_{h_1}}{x_q}}^{1}\frac{d\beta_1}{\beta_1} C_F \frac{1+ \beta_1^2}{(1-\beta_1)_+} D_q^{h_1}\left(\frac{x_{h_1}}{\beta_1 x_q},\mu_F\right) D_{\bar{q}}^{h_2}\left(\frac{x_{h_2}}{x_{\bar{q}}},\mu_F \right) \right. \\*
& + \int_{\frac{x_{h_1}}{x_q}}^{1-\frac{\alpha}{x_q}} d \beta_1 C_F \frac{2}{1-\beta_1} \left(\frac{c_0^2}{\left(\frac{z_{1\perp}-z_{2\perp}}{2}\right)^2 \mu^2}\right)^\epsilon D_q^{h_1}\left(\frac{x_{h_1}}{x_q},\mu_F\right) D_{\bar{q}}^{h_2}\left(\frac{x_{h_2}}{x_{\bar{q}}},\mu_F \right) \\
& \left. - 2  C_F \ln \left(1-\frac{x_{h_1}}{x_q}\right)  D_q^{h_1}\left(\frac{x_{h_1}}{x_q},\mu_F\right) D_{\bar{q}}^{h_2}\left(\frac{x_{h_2}}{x_{\bar{q}}},\mu_F \right) \right] + (h_1 \leftrightarrow h_2 ) \,,  
\end{align*}
and
\begin{align*}
&  \frac{d \sigma_{3TL}^{q \bar{q} \rightarrow h_1 h_2}}{ d x_{h_1} d x_{h_2} d p_{h_1 \perp} d^d p_{h_2\perp} } \bigg |_{\text{coll. qg fin}} \\
&= \frac{2 \alpha_{\mathrm{em}} Q}{(2\pi)^{4(d-1)} N_c} \sum_{q} \int_{x_{h_1}}^{1} d x_q  \int_{x_{h_2}}^{1} d x_{\bar{q}}  \; (x_{\bar{q}}-x_q) \delta(1-x_q -x_{\bar{q}}) \left(\frac{x_q}{x_{h_1}}\right)^d \left(\frac{x_{\bar{q}}}{x_{h_2}}\right)^d \\
& \times \int   d^d p_{2\perp}  \int d^d z_{1\perp}   \frac{e^{i z_{1\perp}\cdot \left(\frac{x_q}{2 x_{h_1}} p_{h_1\perp} + \frac{x_{\bar{q}}}{2 x_{h_2}} p_{h_2 \perp} -p_{2\perp} \right)} F(z_{1\perp})}{x_q x_{\bar{q}} Q^2 +\left(\frac{x_{\bar{q}}}{x_{h_2}} \vec{p}_{h_2}-\vec{p}_{2}\right)^2} \\
& \times \int   d^d p_{2'\perp}  \int d^d z_{2\perp}   \frac{e^{-i z_{2\perp}\cdot \left(\frac{x_q}{2 x_{h_1}} p_{h_1\perp} + \frac{x_{\bar{q}}}{2 x_{h_2}} p_{h_2\perp} -p_{2'\perp} \right)} F^*(z_{2\perp})}{x_q x_{\bar{q}} Q^2 +\left(\frac{x_{\bar{q}}}{x_{h_2}} \vec{p}_{h_2}-\vec{p}_{2'}\right)^2} \left( \frac{x_{\bar{q}}}{x_{h_2}} \vec{p}_{h_2}  - \vec{p}_{2'}\right) \cdot \vec{\varepsilon}^{\,*}_T\\
& \times  \frac{\alpha_s C_F}{2\pi}  \left \{ \int_{\frac{x_{h_1}}{x_q}}^{1} \frac{d\beta_1}{\beta_1}  Q_q^2 D_q^{h_1}\left(\frac{x_{h_1}}{\beta_1 x_q}, \mu_F\right) D_{\bar{q}}^{h_2}\left(\frac{x_{h_2}}{x_{\bar{q}}}, \mu_F\right) \right. \\
& \times  \left[ \ln \left( \frac{c_0^2}{\left(\frac{z_{1\perp} - z_{2\perp}}{2}\right)^2 \mu^2}\right)  \frac{1+ \beta_1^2}{(1-\beta_1)_+}  + \frac{(1-\beta_1)^2 + 2 (1+ \beta_1^2) \ln \beta_1 }{(1-\beta_1)} \right] \\
& \left. - 2 \ln \left( 1- \frac{x_{h_1}}{x_q}\right) \ln \left( \frac{c_0^2}{\left(\frac{z_{1\perp} - z_{2\perp}}{2}\right)^2 \mu^2}\right) D_{q}^{h_1} \left(\frac{x_{h_1}}{x_q},\mu_F\right) D_{\bar{q}}^{h_q} \left(\frac{x_{h_2}}{x_{\bar{q}}},\mu_F \right)\right\} + (h_1 \leftrightarrow h_2 )\,, \numberthis[TL_coll_qg_finite]
    \end{align*}% 
and finally, one obtains for the $TT$ case
\begin{align*}
& \frac{d \sigma_{3TT}^{q \bar{q} \rightarrow h_1 h_2}}{ d x_{h_1} d x_{h_2} d p_{h_1 \perp} d^d p_{h_2\perp} } \bigg |_{\text{coll. qg }} \\*
& =  \frac{  \alpha_{\mathrm{em}} }{(2\pi)^{4(d-1)} N_c} \sum_{q} \int_{x_{h_1}}^{1} \frac{d x_q}{x_q}  \int_{x_{h_2}}^{1} \frac{d x_{\bar{q}}}{x_{\bar{q}}} \delta(1-x_q -x_{\bar{q}}) \left(\frac{x_q}{x_{h_1}}\right)^d \left(\frac{x_{\bar{q}}}{x_{h_2}}\right)^d \\
& \times \left[ (x_{\bar{q}} -x_q)^2 g_{\perp}^{ri}g_{\perp}^{lk} - g_{\perp}^{rk}g_{\perp}^{li} + g_{\perp}^{rl}g_{\perp}^{ik} \right] \\
& \times \int   d^d p_{2\perp}  \int d^d z_{1\perp}   \frac{e^{i z_{1\perp}\cdot \left(\frac{x_q}{2 x_{h_1}} p_{h_1\perp} + \frac{x_{\bar{q}}}{2 x_{h_2}} p_{h_2 \perp} -p_{2\perp} \right)} F(z_{1\perp})}{x_q (1-x_q)Q^2 +\left(\frac{x_{\bar{q}}}{x_{h_2}} \vec{p}_{h_2}-\vec{p}_{2}\right)^2} \left(\frac{x_{\bar{q}}}{x_{h_2}} p_{h_2} - p_{2}\right)_r \varepsilon_{T i} \\
& \times \int   d^d p_{2'\perp}  \int d^d z_{2\perp}   \frac{e^{-i z_{2\perp}\cdot \left(\frac{x_q}{2 x_{h_1}} p_{h_1\perp} + \frac{x_{\bar{q}}}{2 x_{h_2}} p_{h_2 \perp} -p_{2'\perp} \right)} F^*(z_{2\perp})}{x_q x_{\bar{q}} Q^2 +\left(\frac{x_{\bar{q}}}{x_{h_2}} \vec{p}_{h_2}-\vec{p}_{2'}\right)^2} \left(\frac{x_{\bar{q}}}{x_{h_2}} p_{h_2 } - p_{2'}\right)_l \varepsilon_{T k }^*  \\
& \times  \int_{\frac{x_{h_1}}{x_q}}^{1-\frac{\alpha}{x_q}}\frac{d\beta_1}{\beta_1}  Q_q^2 D_q^{h_1}\left(\frac{x_{h_1}}{\beta_1 x_q}, \mu_F\right) D_{\bar{q}}^{h_2}\left(\frac{x_{h_2}}{x_{\bar{q}}}, \mu_F\right) \\
& \times \frac{\alpha_s C_F}{2\pi} \left[ \frac{1}{\hat{\epsilon}} \left(\frac{c_0^2}{\left(\frac{z_{1\perp}-z_{2\perp}}{2}\right)^2 \mu^2}\right)^\epsilon \frac{1+ \beta_1^2}{1-\beta_1}  + \frac{(1-\beta_1)^2 + 2 (1+ \beta_1^2) \ln \beta_1 }{(1-\beta_1)} \right] + (h_1 \leftrightarrow h_2 ) \\
&= \frac{d \sigma_{3TT}^{q \bar{q} \rightarrow h_1 h_2}}{ d x_{h_1} d x_{h_2} d p_{h_1 \perp} d^d p_{h_2\perp} } \bigg |_{\text{coll. qg div}}     + \frac{d \sigma_{3TT}^{q \bar{q} \rightarrow h_1 h_2}}{ d x_{h_1} d x_{h_2} d p_{h_1 \perp} d^d p_{h_2\perp} } \bigg |_{\text{coll. qg fin}} \; ,  \numberthis   
    \end{align*}
where
\begin{align*}
& \frac{d \sigma_{3TT}^{q \bar{q} \rightarrow h_1 h_2}}{ d x_{h_1} d x_{h_2} d p_{h_1 \perp} d^d p_{h_2\perp} } \bigg |_{\text{coll. qg div}}       \\
&=  \frac{  \alpha_{\mathrm{em}} }{(2\pi)^{4(d-1)} N_c} \sum_{q} \int_{x_{h_1}}^{1} \frac{d x_q}{x_q} \int_{x_{h_2}}^1 \frac{d x_{\bar{q}}}{x_{\bar{q}}} \left(\frac{x_q}{x_{h_1}}\right)^d \left(\frac{x_{\bar{q}}}{x_{h_2}}\right)^d \delta(1-x_q-x_{\bar{q}}) \\*
& \times \left[ (x_{\bar{q}} -x_q)^2 g_{\perp}^{ri}g_{\perp}^{lk} - g_{\perp}^{rk}g_{\perp}^{li} + g_{\perp}^{rl}g_{\perp}^{ik} \right] \\
& \times \int   d^d p_{2\perp}  \int d^d z_{1\perp}   \frac{e^{i z_{1\perp}\cdot \left(\frac{x_q}{2 x_{h_1}} p_{h_1\perp} + \frac{x_{\bar{q}}}{2 x_{h_2}} p_{h_2 \perp} -p_{2\perp} \right)} F(z_{1\perp})}{x_q x_{\bar{q}} Q^2 +\left(\frac{x_{\bar{q}}}{x_{h_2}} \vec{p}_{h_2}-\vec{p}_{2}\right)^2}  \left(\frac{x_{\bar{q}}}{x_{h_2}} p_{h_2 } - p_{2 }\right)_r \varepsilon_{T i}  \\
& \times \int   d^d p_{2'\perp}  \int d^d z_{2\perp}   \frac{e^{-i z_{2\perp}\cdot \left(\frac{x_q}{2 x_{h_1}} p_{h_1\perp} + \frac{x_{\bar{q}}}{2 x_{h_2}} p_{h_2 \perp} -p_{2'\perp} \right)} F^*(z_{2\perp})}{x_q x_{\bar{q}} Q^2 +\left(\frac{x_{\bar{q}}}{x_{h_2}} \vec{p}_{h_2}-\vec{p}_{2'}\right)^2} \left(\frac{x_{\bar{q}}}{x_{h_2}} p_{h_2} - p_{2'}\right)_l \varepsilon_{T k}^* \\
& \times \frac{\alpha_s}{2\pi} \frac{1}{\hat{\epsilon}}  Q_q^2 \left[
\int_{\frac{x_{h_1}}{x_q}}^{1}\frac{d\beta_1}{\beta_1} C_F \frac{1+ \beta_1^2}{(1-\beta_1)_+} D_q^{h_1}\left(\frac{x_{h_1}}{\beta_1 x_q},\mu_F\right) D_{\bar{q}}^{h_2}\left(\frac{x_{h_2}}{x_{\bar{q}}},\mu_F \right) \right. \\*
& + \int_{\frac{x_{h_1}}{x_q}}^{1-\frac{\alpha}{x_q}} d \beta_1 C_F \frac{2}{1-\beta_1} \left(\frac{c_0^2}{\left(\frac{z_{1\perp}-z_{2\perp}}{2}\right)^2 \mu^2}\right)^\epsilon D_q^{h_1}\left(\frac{x_{h_1}}{x_q},\mu_F\right) D_{\bar{q}}^{h_2}\left(\frac{x_{h_2}}{x_{\bar{q}}},\mu_F \right) \\
& \left. - 2  C_F \ln \left(1-\frac{x_{h_1}}{x_q}\right)  D_q^{h_1}\left(\frac{x_{h_1}}{x_q},\mu_F\right) D_{\bar{q}}^{h_2}\left(\frac{x_{h_2}}{x_{\bar{q}}},\mu_F \right) \right] + (h_1 \leftrightarrow h_2 ) \,,  
\end{align*}
and
\begin{align*}
&  \frac{d \sigma_{3TT}^{q \bar{q} \rightarrow h_1 h_2}}{ d x_{h_1} d x_{h_2} d p_{h_1 \perp} d^d p_{h_2\perp} } \bigg |_{\text{coll. qg fin}} \\
&= \frac{ \alpha_{\mathrm{em}} }{(2\pi)^{4(d-1)} N_c} \sum_{q} \int_{x_{h_1}}^{1} \frac{d x_q}{x_q}  \int_{x_{h_2}}^{1} \frac{d x_{\bar{q}}}{x_{\bar{q}}} \delta(1-x_q -x_{\bar{q}}) \left(\frac{x_q}{x_{h_1}}\right)^d \left(\frac{x_{\bar{q}}}{x_{h_2}}\right)^d \\
& \times \left[ (x_{\bar{q}} -x_q)^2 g_{\perp}^{ri}g_{\perp}^{lk} - g_{\perp}^{rk}g_{\perp}^{li} + g_{\perp}^{rl}g_{\perp}^{ik} \right] \\
& \times \int   d^d p_{2\perp}  \int d^d z_{1\perp}   \frac{e^{i z_{1\perp}\cdot \left(\frac{x_q}{2 x_{h_1}} p_{h_1\perp} + \frac{x_{\bar{q}}}{2 x_{h_2}} p_{h_2 \perp} -p_{2\perp} \right)} F(z_{1\perp})}{x_q x_{\bar{q}} Q^2 +\left(\frac{x_{\bar{q}}}{x_{h_2}} \vec{p}_{h_2}-\vec{p}_{2}\right)^2} \left(\frac{x_{\bar{q}}}{x_{h_2}} p_{h_2} - p_{2}\right)_r \varepsilon_{T i} \\
& \times \int   d^d p_{2'\perp}  \int d^d z_{2\perp}   \frac{e^{-i z_{2\perp}\cdot \left(\frac{x_q}{2 x_{h_1}} p_{h_1\perp} + \frac{x_{\bar{q}}}{2 x_{h_2}} p_{h_2\perp} -p_{2'\perp} \right)} F^*(z_{2\perp})}{x_q x_{\bar{q}} Q^2 +\left(\frac{x_{\bar{q}}}{x_{h_2}} \vec{p}_{h_2}-\vec{p}_{2'}\right)^2} \left(\frac{x_{\bar{q}}}{x_{h_2}} p_{h_2} - p_{2'}\right)_l \varepsilon_{T k}^*\\
& \times  \frac{\alpha_s C_F}{2\pi}  \left \{ \int_{\frac{x_{h_1}}{x_q}}^{1} \frac{d\beta_1}{\beta_1}  Q_q^2 D_q^{h_1}\left(\frac{x_{h_1}}{\beta_1 x_q}, \mu_F\right) D_{\bar{q}}^{h_2}\left(\frac{x_{h_2}}{x_{\bar{q}}}, \mu_F\right) \right. \\
& \times  \left[ \ln \left( \frac{c_0^2}{\left(\frac{z_{1\perp} - z_{2\perp}}{2}\right)^2 \mu^2}\right)  \frac{1+ \beta_1^2}{(1-\beta_1)_+}  + \frac{(1-\beta_1)^2 + 2 (1+ \beta_1^2) \ln \beta_1 }{(1-\beta_1)} \right] \\
& \left. - 2 \ln \left( 1- \frac{x_{h_1}}{x_q}\right) \ln \left( \frac{c_0^2}{\left(\frac{z_{1\perp} - z_{2\perp}}{2}\right)^2 \mu^2}\right) D_{q}^{h_1} \left(\frac{x_{h_1}}{x_q},\mu_F\right) D_{\bar{q}}^{h_q} \left(\frac{x_{h_2}}{x_{\bar{q}}},\mu_F \right)\right\} + (h_1 \leftrightarrow h_2 )\,. \numberthis[collqg_TT_fin]
    \end{align*}

\subsubsection{Collinear contributions: $\bar{q}$-$g$ splitting}
\label{sec:qqbarfragColl-qbarg}

Here the term in \eqref{eq: div real impact factor} to consider is the third one. The calculation proceeds in the same way as for the quark-gluon collinear contribution but this time the integration is performed over $p_{2 \perp}$ and $p_{2' \perp}$. To observe the cancellation of these collinear divergences, one has to use the different representations we gave for the LO cross-section, as explained before, see eqs.~(\ref{eq:LL-LO-minus}, \ref{Ftilde-LL}) with respect to eqs.~(\ref{eq:LL-LO}, \ref{F-LL}).

\begin{figure}[h!]
\begin{picture}(420,160)
\put(-50,0){\includegraphics[scale=0.5]{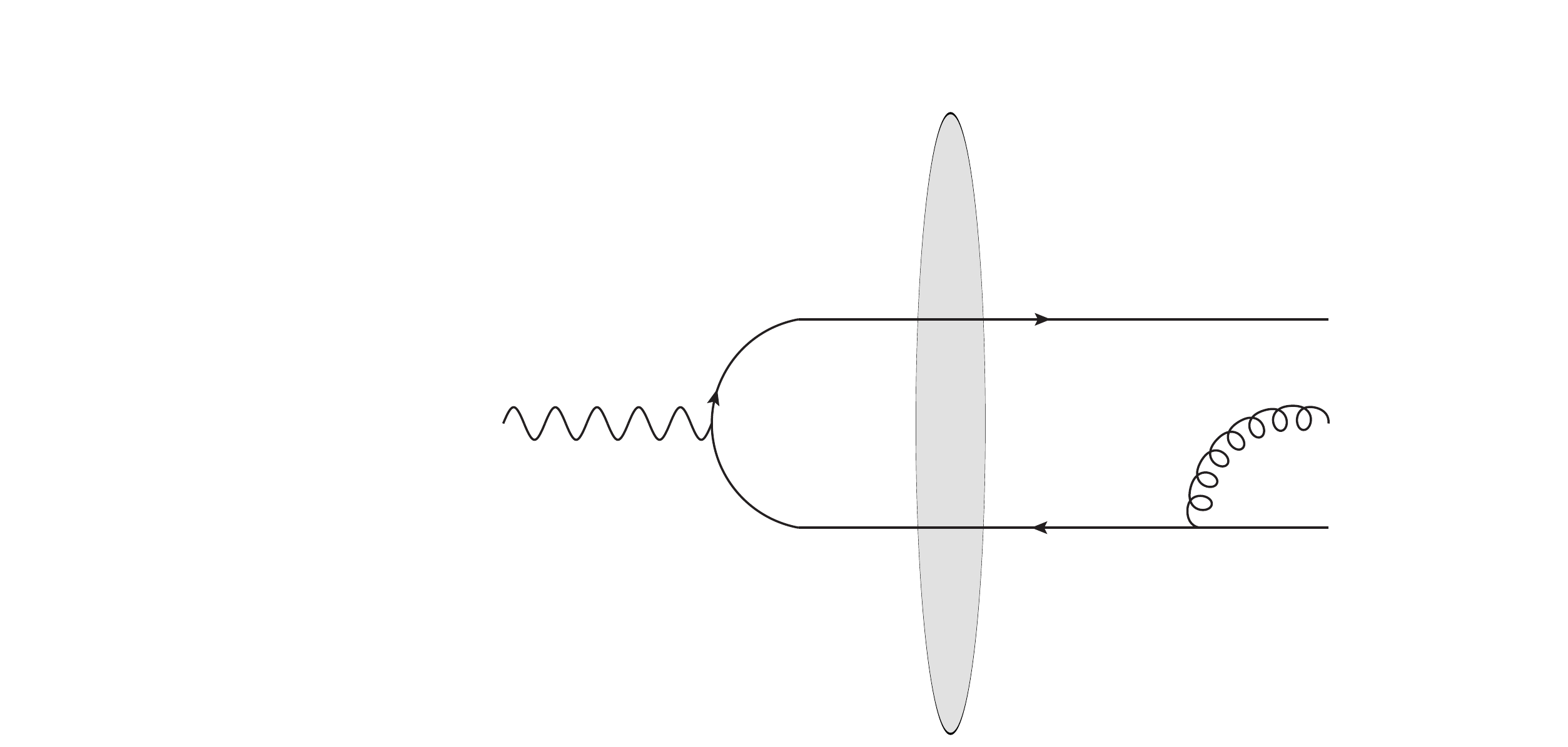}}
\put(180,35){$x_{\bar{q}}, \vec{p}_{\bar{q}}+\vec{p}_g$}
\put(270,100){$x_q, \vec{p}_q$}
\put(270,75){$x_g=(1-\beta_2)x_{\bar{q}}, \vec{p}_g$}
\put(270,50){$x'_{\bar{q}}=\beta_2 x_{\bar{q}}, \vec{p}_{\bar{q}}$}
\end{picture}
\caption{Kinematics for the $\bar{q}-g$ splitting contribution. We indicate the longitudinal fraction of momentum carried by the partons as well as their transverse momenta.}
\label{fig:kinematics_qbarg}
\end{figure}
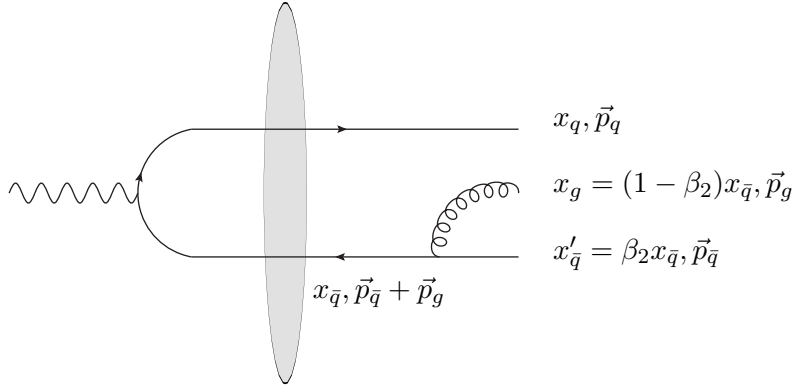

This time, the change of variable to be done is
\begin{align*}
    x_{\bar{q}}' & = \beta_2 x_{\bar{q}} , \\
    x_g & = (1-\beta_2) x_{\bar{q}} \,.\numberthis[transBeta2]
\end{align*}
This kinematics is illustrated in fig.~\ref{fig:kinematics_qbarg}.
The boundaries of integration for $(x_{\bar{q}},\beta_2)$ are calculated in the same spirit as in eq~\eqref{eq:change_variable_integral}. 

Thus, after changes of variable and integrations, the third term in \eqref{eq: div real impact factor} takes the form 

\begin{align*} 
& \frac{d \sigma_{3LL}^{q \bar{q} \rightarrow h_1 h_2}}{ d x_{h_1} d x_{h_2} d^d p_{h_1 \perp} d^d p_{h_2\perp} } \bigg |_{\text{coll. } \bar{q}g}  \\
& =  \frac{4  \alpha_{\mathrm{em}} Q^2}{(2\pi)^{4(d-1)} N_c} \sum_{q} \int_{x_{h_1}}^{1} d x_q  \int_{x_{h_2}}^{1} d x_{\bar{q}} \;  x_q x_{\bar{q}} \delta(1-x_q -x_{\bar{q}}) \left(\frac{x_q}{x_{h_1}}\right)^d \left(\frac{x_{\bar{q}}}{x_{h_2}}\right)^d \\
& \times \int   d^d p_{1\perp}  \int d^d z_{1\perp}   \frac{e^{i z_{1\perp}\cdot \left( -\frac{x_q}{2 x_{h_1}} p_{h_1\perp} - \frac{x_{\bar{q}}}{2 x_{h_2}} p_{h_2 \perp} + p_{1\perp} \right)} F(z_{1\perp})}{x_q x_{\bar{q}} Q^2 +\left(\frac{x_q}{x_{h_1}} \vec{p}_{h_1}-\vec{p}_{1}\right)^2} \\
& \times \int   d^d p_{1'\perp}  \int d^d z_{2\perp}   \frac{e^{-i z_{2\perp}\cdot \left(- \frac{x_q}{2 x_{h_1}} p_{h_1\perp} - \frac{x_{\bar{q}}}{2 x_{h_2}} p_{h_2 \perp} + p_{1'\perp} \right)} F^*(z_{2\perp})}{x_q x_{\bar{q}} Q^2 +\left(\frac{x_q}{x_{h_1}} \vec{p}_{h_1}-\vec{p}_{1'}\right)^2} \\
& \times  \int_{\frac{x_{h_2}}{x_{\bar{q}}}}^{1-\frac{\alpha}{x_{\bar{q}}}}\frac{d\beta_2}{\beta_2}  Q_q^2 D_q^{h_1}\left(\frac{x_{h_1}}{x_q}, \mu_F\right) D_{\bar{q}}^{h_2}\left(\frac{x_{h_2}}{\beta_2 x_{\bar{q}}}, \mu_F\right) \\
& \times \frac{\alpha_s C_F}{2\pi} \left[ \frac{1}{\hat{\epsilon}} \left(\frac{c_0^2}{\left(\frac{z_{2\perp}-z_{1\perp}}{2}\right)^2 \mu^2}\right)^\epsilon \frac{1+ \beta_2^2}{1-\beta_2}  + \frac{(1-\beta_2)^2 + 2 (1+ \beta_2^2) \ln \beta_2 }{(1-\beta_2)} \right] + (h_1 \leftrightarrow h_2 ) \\
& = \frac{d \sigma_{3LL}^{q \bar{q} \rightarrow h_1 h_2}}{ d x_{h_1} d x_{h_2} d p_{h_1 \perp} d^d p_{h_2\perp} } \bigg |_{\text{coll. } \bar{q}g  \text{ div}}     + \frac{d \sigma_{3LL}^{q \bar{q} \rightarrow h_1 h_2}}{ d x_{h_1} d x_{h_2} d p_{h_1 \perp} d^d p_{h_2\perp} } \bigg |_{\text{coll. } \bar{q}g \text{ fin}} \; . \numberthis[coll_qbar_g_qqbar_FF]
\end{align*}

Adding the $+$ prescription and expanding up to $\epsilon^0$, just like for the collinear $qg$ contribution,  one gets the finite and divergent part of \eqref{eq:coll_qbar_g_qqbar_FF}. 

The divergent part is 
\begin{align*}
&\frac{d \sigma_{3LL}^{q \bar{q} \rightarrow h_1 h_2}}{ d x_{h_1} d x_{h_2} d p_{h_1 \perp} d^d p_{h_2\perp} } \bigg |_{\text{coll. } \bar{q}g  \text{ div}} \\ 
&=  \frac{4  \alpha_{\mathrm{em}} Q^2}{(2\pi)^{4(d-1)} N_c} \sum_{q} \int_{x_{h_1}}^{1} d x_q \int_{x_{h_2}}^1 d x_{\bar{q}} \; x_q x_{\bar{q}} \left(\frac{x_q}{x_{h_1}}\right)^d \left(\frac{x_{\bar{q}}}{x_{h_2}}\right)^d \delta(1-x_q-x_{\bar{q}}) \\
& \times \int   d^d p_{1\perp}  \int d^d z_{1\perp}   \frac{e^{i z_{1\perp}\cdot \left( -\frac{x_q}{2 x_{h_1}} p_{h_1\perp} - \frac{x_{\bar{q}}}{2 x_{h_2}} p_{h_2 \perp} + p_{1\perp} \right)} F(z_{1\perp})}{x_q x_{\bar{q}} Q^2 +\left(\frac{x_q}{x_{h_1}} \vec{p}_{h_1}-\vec{p}_{1}\right)^2} \\
& \times \int   d^d p_{1'\perp}  \int d^d z_{2\perp}   \frac{e^{-i z_{2\perp}\cdot \left(- \frac{x_q}{2 x_{h_1}} p_{h_1\perp} - \frac{x_{\bar{q}}}{2 x_{h_2}} p_{h_2 \perp} + p_{1'\perp} \right)} F^*(z_{2\perp})}{x_q x_{\bar{q}} Q^2 +\left(\frac{x_q}{x_{h_1}} \vec{p}_{h_1}-\vec{p}_{1'}\right)^2} \\
& \times \frac{\alpha_s}{2\pi} \frac{1}{\hat{\epsilon}}  Q_q^2 \left[
\int_{\frac{x_{h_2}}{x_{\bar{q}}}}^{1}\frac{d\beta_2}{\beta_2} C_F \frac{1+ \beta_2^2}{(1-\beta_2)_+} D_q^{h_1}\left(\frac{x_{h_1}}{ x_q},\mu_F\right) D_{\bar{q}}^{h_2}\left(\frac{x_{h_2}}{\beta_2 x_{\bar{q}}},\mu_F \right) \right. \\
& + \int_{\frac{x_{h_2}}{x_{\bar{q}}}}^{1-\frac{\alpha}{x_q}} d \beta_2 C_F \frac{2}{1-\beta_2} \left(\frac{c_0^2}{\left(\frac{z_{1\perp}-z_{2\perp}}{2}\right)^2 \mu^2}\right)^\epsilon D_q^{h_1}\left(\frac{x_{h_1}}{x_q},\mu_F\right) D_{\bar{q}}^{h_2}\left(\frac{x_{h_2}}{x_{\bar{q}}},\mu_F \right) \\
& \left. - 2  C_F \ln \left(1-\frac{x_{h_2}}{x_{\bar{q}}}\right)  D_q^{h_1}\left(\frac{x_{h_1}}{x_q},\mu_F\right) D_{\bar{q}}^{h_2}\left(\frac{x_{h_2}}{x_{\bar{q}}},\mu_F \right) \right] + (h_1 \leftrightarrow h_2 ) \,.      \numberthis[coll_qbarg_div]
\end{align*}

The first term cancels with the + prescription term in the second $P_{qq}$ in \eqref{eq:ct_LL}.
We have to remove the second term to avoid double counting with the soft contribution. The third term is to be removed by the soft contribution. 

The finite part is 
\begin{align*}
&\frac{d \sigma_{3LL}^{q \bar{q} \rightarrow h_1 h_2}}{ d x_{h_1} d x_{h_2} d p_{h_1 \perp} d^d p_{h_2\perp} } \bigg |_{\text{coll. } \bar{q}g \text{ fin}} \\
&= \frac{4  \alpha_{\mathrm{em}} Q^2}{(2\pi)^{4(d-1)} N_c} \sum_{q} \int_{x_{h_1}}^{1} d x_q  \int_{x_{h_2}}^{1} d x_{\bar{q}} \;  x_q x_{\bar{q}} \delta(1-x_q -x_{\bar{q}}) \left(\frac{x_q}{x_{h_1}}\right)^d \left(\frac{x_{\bar{q}}}{x_{h_2}}\right)^d \\
& \times \int   d^d p_{1\perp}  \int d^d z_{1\perp}   \frac{e^{i z_{1\perp}\cdot \left( -\frac{x_q}{2 x_{h_1}} p_{h_1\perp} - \frac{x_{\bar{q}}}{2 x_{h_2}} p_{h_2 \perp} + p_{1\perp} \right)} F(z_{1\perp})}{x_q x_{\bar{q}} Q^2 +\left(\frac{x_q}{x_{h_1}} \vec{p}_{h_1}-\vec{p}_{1}\right)^2} \\
& \times \int   d^d p_{1'\perp}  \int d^d z_{2\perp}   \frac{e^{-i z_{2\perp}\cdot \left(- \frac{x_q}{2 x_{h_1}} p_{h_1\perp} - \frac{x_{\bar{q}}}{2 x_{h_2}} p_{h_2 \perp} + p_{1'\perp} \right)} F^*(z_{2\perp})}{x_q x_{\bar{q}} Q^2 +\left(\frac{x_q}{x_{h_1}} \vec{p}_{h_1}-\vec{p}_{1'}\right)^2} \\
& \times  \frac{\alpha_s C_F}{2\pi} \left\{ \int_{\frac{x_{h_2}}{x_{\bar{q}}}}^1\frac{d\beta_2}{\beta_2}  Q_q^2 D_q^{h_1}\left(\frac{x_{h_1}}{x_q}, \mu_F\right) D_{\bar{q}}^{h_2}\left(\frac{x_{h_2}}{\beta_2 x_{\bar{q}}}, \mu_F\right) \right. \\
& \times \left[ \ln \left( \frac{c_0^2}{\left(\frac{z_{1\perp} - z_{2\perp}}{2}\right)^2 \mu^2}\right)  \frac{1+ \beta_2^2}{(1-\beta_2)_+}  + \frac{(1-\beta_2)^2 + 2 (1+ \beta_2^2) \ln \beta_2}{(1-\beta_2)_+} \right] \\
& \left. - 2 \ln \left( 1- \frac{x_{h_2}}{x_{\bar{q}}}\right) \ln \left( \frac{c_0^2}{\left(\frac{z_{1\perp} - z_{2\perp}}{2}\right)^2 \mu^2}\right) D_{q}^{h_1} \left(\frac{x_{h_1}}{x_q},\mu_F\right) D_{\bar{q}}^{h_q} \left(\frac{x_{h_2}}{x_{\bar{q}}},\mu_F \right)\right\} + (h_1 \leftrightarrow h_2 )  \,. \numberthis[collqbarg_LL_fin]
\end{align*}

For the $TL$ transition, 
{\allowdisplaybreaks
\begin{align*} 
 & \frac{d \sigma_{3TL}^{q \bar{q} \rightarrow h_1 h_2}}{ d x_{h_1} d x_{h_2} d p_{h_1 \perp} d^d p_{h_2\perp} } \bigg |_{\text{coll. } \bar{q}g}  \\*
 & =  \frac{2 \alpha_{\mathrm{em}} Q}{(2\pi)^{4(d-1)} N_c} \sum_{q} \int_{x_{h_1}}^{1} d x_q  \int_{x_{h_2}}^{1} d x_{\bar{q}} \; (x_{\bar{q}} -x_q) \delta(1-x_q -x_{\bar{q}}) \left(\frac{x_q}{x_{h_1}}\right)^d \left(\frac{x_{\bar{q}}}{x_{h_2}}\right)^d \\
& \times \int   d^d p_{1\perp}  \int d^d z_{1\perp}   \frac{e^{i z_{1\perp}\cdot \left( -\frac{x_q}{2 x_{h_1}} p_{h_1\perp} - \frac{x_{\bar{q}}}{2 x_{h_2}} p_{h_2 \perp} + p_{1\perp} \right)} F(z_{1\perp})}{x_q x_{\bar{q}} Q^2 +\left(\frac{x_q}{x_{h_1}} \vec{p}_{h_1}-\vec{p}_{1}\right)^2} \\
& \times \int   d^d p_{1'\perp}  \int d^d z_{2\perp}   \frac{e^{-i z_{2\perp}\cdot \left(- \frac{x_q}{2 x_{h_1}} p_{h_1\perp} - \frac{x_{\bar{q}}}{2 x_{h_2}} p_{h_2 \perp} + p_{1'\perp} \right)} F^*(z_{2\perp})}{x_q x_{\bar{q}} Q^2 +\left(\frac{x_q}{x_{h_1}} \vec{p}_{h_1}-\vec{p}_{1'}\right)^2} \left( \frac{x_q}{x_{h_1}} p_{h_1 }  - p_{1'}\right) \cdot \varepsilon^* _{T}   \\
& \times  \int_{\frac{x_{h_2}}{x_{\bar{q}}}}^{1-\frac{\alpha}{x_{\bar{q}}}}\frac{d\beta_2}{\beta_2}  Q_q^2 D_q^{h_1}\left(\frac{x_{h_1}}{x_q}, \mu_F\right) D_{\bar{q}}^{h_2}\left(\frac{x_{h_2}}{ \beta_2 x_{\bar{q}}}, \mu_F\right) \\
& \times \frac{\alpha_s C_F}{2\pi} \left[ \frac{1}{\hat{\epsilon}} \left(\frac{c_0^2}{\left(\frac{z_{2\perp}-z_{1\perp}}{2}\right)^2 \mu^2}\right)^\epsilon \frac{1+ \beta_2^2}{1-\beta_2}  + \frac{(1-\beta_2)^2 + 2 (1+ \beta_2^2) \ln \beta_2 }{(1-\beta_2)} \right] + (h_1 \leftrightarrow h_2 ) \\
&= \frac{d \sigma_{3TL}^{q \bar{q} \rightarrow h_1 h_2}}{ d x_{h_1} d x_{h_2} d p_{h_1 \perp} d^d p_{h_2\perp} } \bigg |_{\text{coll. } \bar{q}g  \text{ div}}     + \frac{d \sigma_{3TL}^{q \bar{q} \rightarrow h_1 h_2}}{ d x_{h_1} d x_{h_2} d p_{h_1 \perp} d^d p_{h_2\perp} } \bigg |_{\text{coll. } \bar{q}g \text{ fin}}      \numberthis
\end{align*}
}
where 
\begin{align*}
&\frac{d \sigma_{3TL}^{q \bar{q} \rightarrow h_1 h_2}}{ d x_{h_1} d x_{h_2} d p_{h_1 \perp} d^d p_{h_2\perp} } \bigg |_{\text{coll. } \bar{q}g  \text{ div}} \\ 
&=  \frac{2  \alpha_{\mathrm{em}} Q}{(2\pi)^{4(d-1)} N_c} \sum_{q} \int_{x_{h_1}}^{1} d x_q \int_{x_{h_2}}^1 d x_{\bar{q}}  (x_{\bar{q}} -x_q)  \left(\frac{x_q}{x_{h_1}}\right)^d \left(\frac{x_{\bar{q}}}{x_{h_2}}\right)^d \delta(1-x_q-x_{\bar{q}}) \\
& \times \int   d^d p_{1\perp}  \int d^d z_{1\perp}   \frac{e^{i z_{1\perp}\cdot \left( -\frac{x_q}{2 x_{h_1}} p_{h_1\perp} - \frac{x_{\bar{q}}}{2 x_{h_2}} p_{h_2 \perp} + p_{1\perp} \right)} F(z_{1\perp})}{x_q x_{\bar{q}} Q^2 +\left(\frac{x_q}{x_{h_1}} \vec{p}_{h_1}-\vec{p}_{1}\right)^2} \\
& \times \int   d^d p_{1'\perp}  \int d^d z_{2\perp}   \frac{e^{-i z_{2\perp}\cdot \left(- \frac{x_q}{2 x_{h_1}} p_{h_1\perp} - \frac{x_{\bar{q}}}{2 x_{h_2}} p_{h_2 \perp} + p_{1'\perp} \right)} F^*(z_{2\perp})}{x_q x_{\bar{q}} Q^2 +\left(\frac{x_q}{x_{h_1}} \vec{p}_{h_1}-\vec{p}_{1'}\right)^2} \left( \frac{x_q}{x_{h_1}} p_{h_1}  - p_{1'}\right) \cdot \varepsilon^* _{T}  \\
& \times \frac{\alpha_s}{2\pi} \frac{1}{\hat{\epsilon}}  Q_q^2 \left[
\int_{\frac{x_{h_2}}{x_{\bar{q}}}}^{1}\frac{d\beta_2}{\beta_2} C_F \frac{1+ \beta_2^2}{(1-\beta_2)_+} D_q^{h_1}\left(\frac{x_{h_1}}{ x_q},\mu_F\right) D_{\bar{q}}^{h_2}\left(\frac{x_{h_2}}{\beta_2 x_{\bar{q}}},\mu_F \right) \right. \\
& + \int_{\frac{x_{h_2}}{x_{\bar{q}}}}^{1-\frac{\alpha}{x_{\bar{q}}}} d \beta_2 C_F \frac{2}{1-\beta_2} \left(\frac{c_0^2}{\left(\frac{z_{1\perp}-z_{2\perp}}{2}\right)^2 \mu^2}\right)^\epsilon D_q^{h_1}\left(\frac{x_{h_1}}{x_q},\mu_F\right) D_{\bar{q}}^{h_2}\left(\frac{x_{h_2}}{x_{\bar{q}}},\mu_F \right) \\
& \left. - 2  C_F \ln \left(1-\frac{x_{h_2}}{x_{\bar{q}}}\right)  D_q^{h_1}\left(\frac{x_{h_1}}{x_q},\mu_F\right) D_{\bar{q}}^{h_2}\left(\frac{x_{h_2}}{x_{\bar{q}}},\mu_F \right) \right] + (h_1 \leftrightarrow h_2 ) \,,   \numberthis
\end{align*}

and 
\begin{align*}
&\frac{d \sigma_{3TL}^{q \bar{q} \rightarrow h_1 h_2}}{ d x_{h_1} d x_{h_2} d p_{h_1 \perp} d^d p_{h_2\perp} } \bigg |_{\text{coll. } \bar{q}g \text{ fin}} \\
&= \frac{2  \alpha_{\mathrm{em}} Q}{(2\pi)^{4(d-1)} N_c} \sum_{q} \int_{x_{h_1}}^{1} d x_q  \int_{x_{h_2}}^{1} d x_{\bar{q}} \;  (x_{\bar{q}} -x_q)  \delta(1-x_q -x_{\bar{q}}) \left(\frac{x_q}{x_{h_1}}\right)^d \left(\frac{x_{\bar{q}}}{x_{h_2}}\right)^d \\
& \times \int   d^d p_{1\perp}  \int d^d z_{1\perp}   \frac{e^{i z_{1\perp}\cdot \left( -\frac{x_q}{2 x_{h_1}} p_{h_1\perp} - \frac{x_{\bar{q}}}{2 x_{h_2}} p_{h_2 \perp} + p_{1\perp} \right)} F(z_{1\perp})}{x_q x_{\bar{q}} Q^2 +\left(\frac{x_q}{x_{h_1}} \vec{p}_{h_1}-\vec{p}_{1}\right)^2} \\
& \times \int   d^d p_{1'\perp}  \int d^d z_{2\perp}   \frac{e^{-i z_{2\perp}\cdot \left(- \frac{x_q}{2 x_{h_1}} p_{h_1\perp} - \frac{x_{\bar{q}}}{2 x_{h_2}} p_{h_2 \perp} + p_{1'\perp} \right)} F^*(z_{2\perp})}{x_q x_{\bar{q}} Q^2 +\left(\frac{x_q}{x_{h_1}} \vec{p}_{h_1}-\vec{p}_{1'}\right)^2} \left( \frac{x_q}{x_{h_1}} p_{h_1}  - p_{1'}\right) \cdot \varepsilon^* _{T}  \\
& \times \frac{\alpha_s C_F}{2\pi} \left\{ \int_{\frac{x_{h_2}}{x_{\bar{q}}}}^1 \frac{d\beta_2}{\beta_2}  Q_q^2 D_q^{h_1}\left(\frac{x_{h_1}}{x_q}, \mu_F\right) D_{\bar{q}}^{h_2}\left(\frac{x_{h_2}}{\beta_2 x_{\bar{q}}}, \mu_F\right) \right. \\
& \times  \left[ \ln \left( \frac{c_0^2}{\left(\frac{z_{1\perp} - z_{2\perp}}{2}\right)^2 \mu^2}\right)  \frac{1+ \beta_2^2}{(1-\beta_2)_+}  + \frac{(1-\beta_2)^2 + 2 (1+ \beta_2^2) \ln \beta_2}{(1-\beta_2)_+} \right] \\
& \left. - 2 \ln \left( 1- \frac{x_{h_2}}{x_{\bar{q}}}\right) \ln \left( \frac{c_0^2}{\left(\frac{z_{1\perp} - z_{2\perp}}{2}\right)^2 \mu^2}\right) D_{q}^{h_1} \left(\frac{x_{h_1}}{x_q},\mu_F\right) D_{\bar{q}}^{h_q} \left(\frac{x_{h_2}}{x_{\bar{q}}},\mu_F \right)\right\} + (h_1 \leftrightarrow h_2 )  \,. \numberthis[collqbarg_TL_fin]
\end{align*}

For the TT case, we get 

{\allowdisplaybreaks
\begin{align*} 
& \frac{d \sigma_{3TT}^{q \bar{q} \rightarrow h_1 h_2}}{ d x_{h_1} d x_{h_2} d p_{h_1 \perp} d^d p_{h_2\perp} } \bigg |_{\text{coll. } \bar{q}g}  \\*
 & =  \frac{ \alpha_{\mathrm{em}} }{(2\pi)^{4(d-1)} N_c} \sum_{q} \int_{x_{h_1}}^{1} \frac{d x_q}{x_q}  \int_{x_{h_2}}^{1} \frac{d x_{\bar{q}} }{x_{\bar{q}}} \delta(1-x_q -x_{\bar{q}}) \left(\frac{x_q}{x_{h_1}}\right)^d \left(\frac{x_{\bar{q}}}{x_{h_2}}\right)^d  \\
 & \times \left[ (x_{\bar{q}} -x_q)^2 g_{\perp}^{ri}g_{\perp}^{lk} - g_{\perp}^{rk}g_{\perp}^{li} + g_{\perp}^{rl}g_{\perp}^{ik} \right] \\
& \times \int   d^d p_{1\perp}  \int d^d z_{1\perp}   \frac{e^{i z_{1\perp}\cdot \left( -\frac{x_q}{2 x_{h_1}} p_{h_1\perp} - \frac{x_{\bar{q}}}{2 x_{h_2}} p_{h_2 \perp} + p_{1\perp} \right)} F(z_{1\perp})}{x_q x_{\bar{q}} Q^2 +\left(\frac{x_q}{x_{h_1}} \vec{p}_{h_1}-\vec{p}_{1}\right)^2} \left(\frac{x_{q}}{x_{h_1}} p_{h_1} - p_{1}\right)_r \varepsilon_{T i}  \\
& \times \int   d^d p_{1'\perp}  \int d^d z_{2\perp}   \frac{e^{-i z_{2\perp}\cdot \left(- \frac{x_q}{2 x_{h_1}} p_{h_1\perp} - \frac{x_{\bar{q}}}{2 x_{h_2}} p_{h_2 \perp} + p_{1'\perp} \right)} F^*(z_{2\perp})}{x_q x_{\bar{q}} Q^2 +\left(\frac{x_q}{x_{h_1}} \vec{p}_{h_1}-\vec{p}_{1'}\right)^2}  \left(\frac{x_{q}}{x_{h_1}} p_{h_1} - p_{1'}\right)_l \varepsilon_{T k}^*  \\
& \times  \int_{\frac{x_{h_2}}{x_{\bar{q}}}}^{1-\frac{\alpha}{x_{\bar{q}}}}\frac{d\beta_2}{\beta_2}  Q_q^2 D_q^{h_1}\left(\frac{x_{h_1}}{x_q}, \mu_F\right) D_{\bar{q}}^{h_2}\left(\frac{x_{h_2}}{\beta_2 x_{\bar{q}}}, \mu_F\right)  \\
& \times \frac{\alpha_s C_F}{2\pi} \left[ \frac{1}{\hat{\epsilon}} \left(\frac{c_0^2}{\left(\frac{z_{2\perp}-z_{1\perp}}{2}\right)^2 \mu^2}\right)^\epsilon \frac{1+ \beta_2^2}{1-\beta_2}  + \frac{(1-\beta_2)^2 + 2 (1+ \beta_2^2)\ln \beta_2 }{(1-\beta_2)} \right] + (h_1 \leftrightarrow h_2 ) \\
&= \frac{d \sigma_{3TT}^{q \bar{q} \rightarrow h_1 h_2}}{ d x_{h_1} d x_{h_2} d p_{h_1 \perp} d^d p_{h_2\perp} } \bigg |_{\text{coll. } \bar{q}g  \text{ div}}     + \frac{d \sigma_{3TT}^{q \bar{q} \rightarrow h_1 h_2}}{ d x_{h_1} d x_{h_2} d p_{h_1 \perp} d^d p_{h_2\perp} } \bigg |_{\text{coll. } \bar{q}g \text{ fin}}  ,   \numberthis
\end{align*}}
where 
\begin{align*}
&\frac{d \sigma_{3TT}^{q \bar{q} \rightarrow h_1 h_2}}{ d x_{h_1} d x_{h_2} d p_{h_1 \perp} d^d p_{h_2\perp} } \bigg |_{\text{coll. } \bar{q}g  \text{ div}} \\ 
&=  \frac{  \alpha_{\mathrm{em}} }{(2\pi)^{4(d-1)} N_c} \sum_{q} \int_{x_{h_1}}^{1} \frac{d x_q}{x_q}\int_{x_{h_2}}^1 \frac{d x_{\bar{q}} }{x_{\bar{q}}} \left(\frac{x_q}{x_{h_1}}\right)^d \left(\frac{x_{\bar{q}}}{x_{h_2}}\right)^d \delta(1-x_q-x_{\bar{q}}) \\
& \times \left[ (x_{\bar{q}} -x_q)^2 g_{\perp}^{ri}g_{\perp}^{lk} - g_{\perp}^{rk}g_{\perp}^{li} + g_{\perp}^{rl}g_{\perp}^{ik} \right] \\
& \times \int   d^d p_{1\perp}  \int d^d z_{1\perp}   \frac{e^{i z_{1\perp}\cdot \left( -\frac{x_q}{2 x_{h_1}} p_{h_1\perp} - \frac{x_{\bar{q}}}{2 x_{h_2}} p_{h_2 \perp} + p_{1\perp} \right)} F(z_{1\perp})}{x_q x_{\bar{q}} Q^2 +\left(\frac{x_q}{x_{h_1}} \vec{p}_{h_1}-\vec{p}_{1}\right)^2} \left(\frac{x_{q}}{x_{h_1}} p_{h_1} - p_{1}\right)_r \varepsilon_{T i}  \\
& \times \int   d^d p_{1'\perp}  \int d^d z_{2\perp}   \frac{e^{-i z_{2\perp}\cdot \left(- \frac{x_q}{2 x_{h_1}} p_{h_1\perp} - \frac{x_{\bar{q}}}{2 x_{h_2}} p_{h_2 \perp} + p_{1'\perp} \right)} F^*(z_{2\perp})}{x_q x_{\bar{q}} Q^2 +\left(\frac{x_q}{x_{h_1}} \vec{p}_{h_1}-\vec{p}_{1'}\right)^2}  \left(\frac{x_{q}}{x_{h_1}} p_{h_1} - p_{1'}\right)_l \varepsilon_{T k}^*  \\
& \times \frac{\alpha_s}{2\pi} \frac{1}{\hat{\epsilon}}  Q_q^2 \left[
\int_{\frac{x_{h_2}}{x_{\bar{q}}}}^{1}\frac{d\beta_2}{\beta_2} C_F \frac{1+ \beta_2^2}{(1-\beta_2)_+} D_q^{h_1}\left(\frac{x_{h_1}}{ x_q},\mu_F\right) D_{\bar{q}}^{h_2}\left(\frac{x_{h_2}}{\beta_2 x_{\bar{q}}},\mu_F \right) \right. \\
& + \int_{\frac{x_{h_2}}{x_{\bar{q}}}}^{1-\frac{\alpha}{x_{\bar{q}}}} d \beta_2 C_F \frac{2}{1-\beta_2} \left(\frac{c_0^2}{\left(\frac{z_{1\perp}-z_{2\perp}}{2}\right)^2 \mu^2}\right)^\epsilon D_q^{h_1}\left(\frac{x_{h_1}}{x_q},\mu_F\right) D_{\bar{q}}^{h_2}\left(\frac{x_{h_2}}{x_{\bar{q}}},\mu_F \right) \\
& \left. - 2  C_F \ln \left(1-\frac{x_{h_2}}{x_{\bar{q}}}\right)  D_q^{h_1}\left(\frac{x_{h_1}}{x_q},\mu_F\right) D_{\bar{q}}^{h_2}\left(\frac{x_{h_2}}{x_{\bar{q}}},\mu_F \right) \right] + (h_1 \leftrightarrow h_2 ) \,,   \numberthis
\end{align*}
and
\begin{align*}
&\frac{d \sigma_{3TT}^{q \bar{q} \rightarrow h_1 h_2}}{ d x_{h_1} d x_{h_2} d p_{h_1 \perp} d^d p_{h_2\perp} } \bigg |_{\text{coll. } \bar{q}g \text{ fin}} \\
&= \frac{  \alpha_{\mathrm{em}} }{(2\pi)^{4(d-1)} N_c} \sum_{q} \int_{x_{h_1}}^{1} \frac{d x_q}{x_q}  \int_{x_{h_2}}^{1} \frac{d x_{\bar{q}}}{x_{\bar{q}}} \delta(1-x_q -x_{\bar{q}}) \left(\frac{x_q}{x_{h_1}}\right)^d \left(\frac{x_{\bar{q}}}{x_{h_2}}\right)^d \\
& \times \left[ (x_{\bar{q}} -x_q)^2 g_{\perp}^{ri}g_{\perp}^{lk} - g_{\perp}^{rk}g_{\perp}^{li} + g_{\perp}^{rl}g_{\perp}^{ik} \right] \\
& \times \int   d^d p_{1\perp}  \int d^d z_{1\perp}   \frac{e^{i z_{1\perp}\cdot \left( -\frac{x_q}{2 x_{h_1}} p_{h_1\perp} - \frac{x_{\bar{q}}}{2 x_{h_2}} p_{h_2 \perp} + p_{1\perp} \right)} F(z_{1\perp})}{x_q x_{\bar{q}} Q^2 +\left(\frac{x_q}{x_{h_1}} \vec{p}_{h_1}-\vec{p}_{1}\right)^2}  \left(\frac{x_{q}}{x_{h_1}} p_{h_1} - p_{1}\right)_r \varepsilon_{T i}  \\
& \times \int   d^d p_{1'\perp}  \int d^d z_{2\perp}   \frac{e^{-i z_{2\perp}\cdot \left(- \frac{x_q}{2 x_{h_1}} p_{h_1\perp} - \frac{x_{\bar{q}}}{2 x_{h_2}} p_{h_2 \perp} + p_{1'\perp} \right)} F^*(z_{2\perp})}{x_q x_{\bar{q}} Q^2 +\left(\frac{x_q}{x_{h_1}} \vec{p}_{h_1}-\vec{p}_{1'}\right)^2} \left(\frac{x_{q}}{x_{h_1}} p_{h_1} - p_{1'}\right)_l \varepsilon_{T k}^*   \\
& \times \frac{\alpha_s C_F}{2\pi} \left\{ \int_{\frac{x_{h_2}}{x_{\bar{q}}}}^{1}\frac{d\beta_2}{\beta_2}  Q_q^2 D_q^{h_1}\left(\frac{x_{h_1}}{x_q}, \mu_F\right) D_{\bar{q}}^{h_2}\left(\frac{x_{h_2}}{\beta_2 x_{\bar{q}}}, \mu_F\right) \right. \\
& \times  \left[ \ln \left( \frac{c_0^2}{\left(\frac{z_{1\perp} - z_{2\perp}}{2}\right)^2 \mu^2}\right)  \frac{1+ \beta_2^2}{(1-\beta_2)_+}  + \frac{(1-\beta_2)^2 + 2 (1+ \beta_2^2) \ln \beta_2}{(1-\beta_2)_+} \right] \\
& \left. - 2 \ln \left( 1- \frac{x_{h_2}}{x_{\bar{q}}}\right) \ln \left( \frac{c_0^2}{\left(\frac{z_{1\perp} - z_{2\perp}}{2}\right)^2 \mu^2}\right) D_{q}^{h_1} \left(\frac{x_{h_1}}{x_q},\mu_F\right) D_{\bar{q}}^{h_q} \left(\frac{x_{h_2}}{x_{\bar{q}}},\mu_F \right)\right\} + (h_1 \leftrightarrow h_2 )  \,. \numberthis[collqbarg_TT_fin]
\end{align*}

\subsubsection{Soft contribution}
\label{sec: soft contribution}

To calculate the soft contribution of the divergent part of the real emission cross-section, the soft limit of \eqref{eq:real_div_LL} is taken by setting  $\vec{p}_g = x_g \vec{u}$, where $|\vec{u}| \sim |\vec{p}_{h}|,$ which extracts the divergence on $x_g$.
\begin{align*}
& \frac{d \sigma_{3LL}^{q \bar{q} \rightarrow h_1 h_2 }}{d x_{h_1}d x_{h_2} d^d p_{h_1\perp} d p_{h_2 \perp}}\Bigg |_{\text{soft}} \\ 
&=  \frac{4  \alpha_{\mathrm{em}} Q^2}{(2\pi)^{4(d-1)} N_c} \sum_{q} \int_{x_{h_1}}^1 \frac{d x_q'}{x_q'} \int_{x_{h_2}}^1 \frac{d x_{\bar{q}}'}{ x_{\bar{q}}'}  Q_q^2 D_q^{h_1}\left(\frac{x_{h_1}}{x_q'},\mu_F\right)D_{\bar{q}}^{h_2}\left(\frac{x_{h_2}}{x_{\bar{q}}'}, \mu_F\right) \\
& \times \left(\frac{x_q'}{x_{h_1}}\right)^d \left(\frac{x_{\bar{q}}'}{x_{h_2}}\right)^d \int_{\alpha}^1 \frac{d x_g}{x_g^{3-d}} \delta(1-x_q'-x_{\bar{q}}'-x_g) \frac{\alpha_s C_F}{\mu^{2\epsilon}} \int \frac{d^d \vec{u}}{(2\pi)^d} \\
& \times \int d^d p_{1\perp} d^d p_{2\perp} \; \mathbf{F}\left(\frac{p_{12\perp}}{2}\right) \delta\left(\frac{x_q'}{x_{h_1}} p_{h_1\perp} - p_{1\perp} + \frac{x_{\bar{q}}'}{x_{h_2}}p_{h_2\perp} - p_{2\perp} + x_g u_{\perp}\right) \\
& \times \int d^d p_{1'\perp}  d^d p_{2'\perp} \; \mathbf{F}^*\left(\frac{p_{1'2'\perp}}{2}\right) \delta\left(\frac{x_q'}{x_{h_1}} p_{h_1\perp} - p_{1'\perp} + \frac{x_{\bar{q}}'}{x_{h_2}}p_{h_2\perp} - p_{2'\perp} + x_g u_{\perp}\right) \\
& \times \left \{ \frac{d x_g^2 + 4 x_q'(x_q'+x_g)}{\left(Q^2+ \frac{\left(\frac{x_{\bar{q}}'} {x_{h_2}}\vec{p}_{h_2}-\vec{p}_{2}\right)^2}{(1-x_{\bar{q}}')x_{\bar{q}}'}\right)\left(Q^2+ \frac{\left(\frac{x_{\bar{q}}'}{x_{h_2}}\vec{p}_{h_2}-\vec{p}_{2'}\right)^2}{(1-x_{\bar{q}}')x_{\bar{q}}'}\right) x_q'^2 \left( \vec{u}-\frac{\vec{p}_{h_1}}{x_{h_1}}\right)^2} \right.\\
& + \frac{d x_g^2 + 4x_{\bar{q}}'(x_{\bar{q}}' + x_g)}{\left(Q^2+ \frac{\left(\frac{x_{\bar{q}}'}{x_{h_2}}\vec{p}_{h_2}-\vec{p}_{2}+x_g\vec{u}\right)^2}{(1-x_q')x_q'}\right)\left(Q^2+ \frac{\left(\frac{x_{\bar{q}}'}{x_{h_2}}\vec{p}_{h_2}-\vec{p}_{2'}+x_g\vec{u}\right)^2}{(1-x_q')x_q'}\right)x_{\bar{q}}'^2 \left(\vec{u}-\frac{\vec{p}_{h_2}}{x_{h_2}}\right)^2} \\
& - \frac{[2x_g - d x_g^2 + 4x_q' x_{\bar{q}}'] \left(\vec{u}-\frac{\vec{p}_{h_1}}{x_{h_1}}\right)\cdot \left(\vec{u}-\frac{\vec{p}_{h_2}}{x_{h_2}}\right)}{\left(Q^2+ \frac{\left(\frac{x_{\bar{q}}'}{x_{h_2}}\vec{p}_{h_2}-\vec{p}_{2'}\right)^2}{(1-x_{\bar{q}}')x_{\bar{q}}'}\right)\left(Q^2+ \frac{\left(\frac{x_{\bar{q}}'}{x_{h_2}}\vec{p}_{h_2}-\vec{p}_{2}+x_g\vec{u}\right)^2}{(1-x_q')x_q'}\right) x_q' x_{\bar{q}}' \left(\vec{u}-\frac{\vec{p}_{h_1}}{x_{h_1}}\right)^2 \left(\vec{u}-\frac{\vec{p}_{h_2}}{x_{h_2}}\right)^2} \\
&  \left.  - \frac{[2x_g - d x_g^2 + 4x_q' x_{\bar{q}}'] \left(\vec{u}-\frac{\vec{p}_{h_1}}{x_{h_1}}\right)\cdot \left(\vec{u}-\frac{\vec{p}_{h_2}}{x_{h_2}}\right)}{\left(Q^2+ \frac{\left(\frac{x_{\bar{q}}'}{x_{h_2}}\vec{p}_{h_2}-\vec{p}_{2}\right)^2}{(1-x_{\bar{q}}')x_{\bar{q}}'}\right)\left(Q^2+ \frac{\left(\frac{x_{\bar{q}}'}{x_{h_2}}\vec{p}_{h_2}-\vec{p}_{2'}+x_g\vec{u}\right)^2}{(1-x_q')x_q'}\right) x_q' x_{\bar{q}}' \left(\vec{u}-\frac{\vec{p}_{h_1}}{x_{h_1}}\right)^2 \left(\vec{u}-\frac{\vec{p}_{h_2}}{x_{h_2}}\right)^2} \right\} \\
& + (h_1 \leftrightarrow h_2) \,. 
\end{align*}
The limit $x_g \rightarrow 0$ in the $F$ function and impact factor can be taken safely in the non- divergent terms of the cross-section, as $x_q'$ and $x_{\bar{q}}'$ are limited from below by $x_{h_1}, x_{h_2}$ and so cannot be arbitrary small (ie of order $x_g$). The cross-section in the soft limit becomes: 
\begin{align*}
& \frac{d \sigma_{3LL}^{q \bar{q} \rightarrow h_1 h_2 }}{d x_{h_1}d x_{h_2} d^d p_{h_1\perp} d p_{h_2 \perp}}\Bigg |_{\text{soft}} \\* 
&=  \frac{4  \alpha_{\mathrm{em}} Q^2}{(2\pi)^{4(d-1)} N_c} \sum_{q}   \int_{x_{h_1}}^1 \frac{d x_q'}{x_q'} \int_{x_{h_2}}^1 \frac{d x_{\bar{q}}'}{ x_{\bar{q}}'} Q_q^2 D_q^{h_1}\left(\frac{x_{h_1}}{x_q'},\mu_F\right)D_{\bar{q}}^{h_2}\left(\frac{x_{h_2}}{x_{\bar{q}}'}, \mu_F\right) \\
& \times \left(\frac{x_q'}{x_{h_1}}\right)^d \left(\frac{x_{\bar{q}}'}{x_{h_2}}\right)^d \int_{\alpha}^1 \frac{d x_g}{x_g^{3-d}} \delta(1-x_q'-x_{\bar{q}}'-x_g) \frac{\alpha_s C_F}{\mu^{2\epsilon}} \int \frac{d^d \vec{u}}{(2\pi)^d} \\
& \times \int d^d p_{1\perp} d^d p_{2\perp} \; \mathbf{F}\left(\frac{p_{12\perp}}{2}\right) \delta\left(\frac{x_q'}{x_{h_1}} p_{h_1\perp} - p_{1\perp} + \frac{x_{\bar{q}}'}{x_{h_2}}p_{h_2\perp} - p_{2\perp} \right) \\
& \times \int d^d p_{1'\perp} d^d p_{2'\perp} \; \mathbf{F}^*\left(\frac{p_{1'2'\perp}}{2}\right) \delta\left(\frac{x_q'}{x_{h_1}} p_{h_1\perp} - p_{1'\perp} + \frac{x_{\bar{q}}'}{x_{h_2}}p_{h_2\perp} - p_{2'\perp} \right) \\
& \times \left \{ \frac{ 4 }{\left(Q^2+ \frac{\left(\frac{x_{\bar{q}}'} {x_{h_2}}\vec{p}_{h_2}-\vec{p}_{2}\right)^2}{(1-x_{\bar{q}}')x_{\bar{q}}'}\right)\left(Q^2+ \frac{\left(\frac{x_{\bar{q}}'}{x_{h_2}}\vec{p}_{h_2}-\vec{p}_{2'}\right)^2}{(1-x_{\bar{q}}')x_{\bar{q}}'}\right) \left( \vec{u}-\frac{\vec{p}_{h_1}}{x_{h_1}}\right)^2} \right.\\
& + \frac{ 4 }{\left(Q^2+ \frac{\left(\frac{x_{\bar{q}}'}{x_{h_2}}\vec{p}_{h_2}-\vec{p}_{2}\right)^2}{(1-x_q')x_q'}\right)\left(Q^2+ \frac{\left(\frac{x_{\bar{q}}'}{x_{h_2}}\vec{p}_{h_2}-\vec{p}_{2'}\right)^2}{(1-x_q')x_q'}\right)\left(\vec{u}-\frac{\vec{p}_{h_2}}{x_{h_2}}\right)^2} \\
& - \frac{4 \left(\vec{u}-\frac{\vec{p}_{h_1}}{x_{h_1}}\right)\cdot \left(\vec{u}-\frac{\vec{p}_{h_2}}{x_{h_2}}\right)}{\left(Q^2+ \frac{\left(\frac{x_{\bar{q}}'}{x_{h_2}}\vec{p}_{h_2}-\vec{p}_{2'}\right)^2}{(1-x_{\bar{q}}')x_{\bar{q}}'}\right)\left(Q^2+ \frac{\left(\frac{x_{\bar{q}}'}{x_{h_2}}\vec{p}_{h_2}-\vec{p}_{2}\right)^2}{(1-x_q')x_q'}\right)  \left(\vec{u}-\frac{\vec{p}_{h_1}}{x_{h_1}}\right)^2 \left(\vec{u}-\frac{\vec{p}_{h_2}}{x_{h_2}}\right)^2} \\
&  \left.  - \frac{4 \left(\vec{u}-\frac{\vec{p}_{h_1}}{x_{h_1}}\right)\cdot \left(\vec{u}-\frac{\vec{p}_{h_2}}{x_{h_2}}\right)}{\left(Q^2+ \frac{\left(\frac{x_{\bar{q}}'}{x_{h_2}}\vec{p}_{h_2}-\vec{p}_{2}\right)^2}{(1-x_{\bar{q}}')x_{\bar{q}}'}\right)\left(Q^2+ \frac{\left(\frac{x_{\bar{q}}'}{x_{h_2}}\vec{p}_{h_2}-\vec{p}_{2}'\right)^2}{(1-x_q')x_q'}\right) \left(\vec{u}-\frac{\vec{p}_{h_1}}{x_{h_1}}\right)^2 \left(\vec{u}-\frac{\vec{p}_{h_2}}{x_{h_2}}\right)^2} 
\right\} + (h_1 \leftrightarrow h_2) \\ 
&=  \frac{4  \alpha_{\mathrm{em}} Q^2}{(2\pi)^{4(d-1)} N_c} \sum_{q} \int_{x_{h_1}}^1 \frac{d x_q'}{x_q'} \int_{x_{h_2}}^1 \frac{d x_{\bar{q}}'}{ x_{\bar{q}}'} Q_q^2 D_q^{h_1}\left(\frac{x_{h_1}}{x_q'},\mu_F\right)D_{\bar{q}}^{h_2}\left(\frac{x_{h_2}}{x_{\bar{q}}'}, \mu_F\right) \\
& \times \left(\frac{x_q'}{x_{h_1}}\right)^d \left(\frac{x_{\bar{q}}'}{x_{h_2}}\right)^d \int_{\alpha}^1 \frac{d x_g}{x_g^{3-d}} \delta(1-x_q'-x_{\bar{q}}'-x_g) \frac{\alpha_s C_F}{\mu^{2\epsilon}} \int \frac{d^d \vec{u}}{(2\pi)^d} \\
& \times \int d^d p_{1\perp} d^d p_{2\perp} \; \mathbf{F}\left(\frac{p_{12\perp}}{2}\right) \delta\left(\frac{x_q'}{x_{h_1}} p_{h_1\perp} - p_{1\perp} + \frac{x_{\bar{q}}'}{x_{h_2}}p_{h_2\perp} - p_{2\perp} \right) \\
& \times \int d^d p_{1'\perp}  d^d p_{2'\perp} \; \mathbf{F}^*\left(\frac{p_{1'2'\perp}}{2}\right) \delta\left(\frac{x_q'}{x_{h_1}} p_{h_1\perp} - p_{1'\perp} + \frac{x_{\bar{q}}'}{x_{h_2}}p_{h_2\perp} - p_{2'\perp} \right) \\
& \times \frac{ 4 }{\left(Q^2+ \frac{\left(\frac{x_{\bar{q}}'} {x_{h_2}}\vec{p}_{h_2}-\vec{p}_{2}\right)^2}{(1-x_{\bar{q}}')x_{\bar{q}}'}\right)\left(Q^2+ \frac{\left(\frac{x_{\bar{q}}'}{x_{h_2}}\vec{p}_{h_2}-\vec{p}_{2'}\right)^2}{(1-x_{\bar{q}}')x_{\bar{q}}'}\right) } \left \{ \frac{ 1 }{ \left( \vec{u}-\frac{\vec{p}_{h_1}}{x_h}\right)^2}  + \frac{ 1 }{\left(\vec{u}-\frac{\vec{p}_{h_2}}{x_{h_2}}\right)^2} \right. \\
& \left. - \frac{2 \left(\vec{u}-\frac{\vec{p}_{h_1}}{x_{h_1}}\right)\cdot \left(\vec{u}-\frac{\vec{p}_{h_2}}{x_{h_2}}\right)}{\left(\vec{u}-\frac{\vec{p}_{h_1}}{x_{h_1}}\right)^2 \left(\vec{u}-\frac{\vec{p}_{h_2}}{x_{h_2}}\right)^2} 
\right\} + (h_1 \leftrightarrow h_2) \; . 
\end{align*}

Then, the cross-section is divided into two parts in order to do two different changes of variables, which are the same changes as in eqs.~(\ref{eq:Transbeta}, \ref{eq:transBeta2}), in each part:
\begin{equation}
    x_q' = \beta_1 x_q \,,\hspace{2 cm}  x_g  = (1-\beta_1) x_q \,,
    \label{eq:Transbeta1}
\end{equation}
\begin{equation}
x_{\bar{q}}' = \beta_2 x_{\bar{q}}\,, \hspace{2 cm} x_g = (1-\beta_2) x_{\bar{q}}  \,,
   \label{eq:Transbeta2}
\end{equation}
 where the integration boundaries are calculated following the steps in eq.~(\ref{eq:change_variable_integral}). This division and changes of variable respect the symmetry between diagrams (1) + (2) on one side, and (3) + (4) on the other side in Fig.~\ref{fig:NLO-b-div}.
The limits $\beta_{1,2} \rightarrow 1 $, corresponding to $x_g \rightarrow 0$ are then taken. The choice of splitting the cross-section in this way, comes from 
the will to observe the cancellation of divergences at integrand level.
The first term, after the transformation (\ref{eq:Transbeta1}) and after taking the limit, gives 
 \begin{align*}
& \frac{d \sigma_{3LL}^{q \bar{q} \rightarrow h_1 h_2 }}{d x_{h_1}d x_{h_2} d^d p_{h_1\perp} d p_{h_2 \perp}}\Bigg |_{\text{soft } \beta_1} \\
&= \frac{2  \alpha_{\mathrm{em}} Q^2}{(2\pi)^{4(d-1)} N_c} \sum_{q} \int_{x_{h_1}}^{1-x_{h_2}} \frac{d x_q}{x_q} \frac{1}{1-x_q}\left(\frac{x_q}{x_{h_1}}\right)^d \left(\frac{1-x_q}{x_{h_2}}\right)^d \\
& \times   Q_q^2 D_q^{h_1}\left(\frac{x_{h_1}}{x_q},\mu_F\right)D_{\bar{q}}^{h_2}\left(\frac{x_{h_2}}{1-x_q}, \mu_F\right) \\
& \times \int  d^d p_{2\perp} \;  \mathbf{F}\left(\frac{x_q}{2x_{h_1}} p_{h_1\perp}  + \frac{1-x_q}{2x_{h_2}}p_{h_2\perp} - p_{2\perp}\right) \\
& \times  \int  d^d p_{2'\perp} \;  \mathbf{F}^*\left(\frac{x_q}{2x_{h_1}} p_{h_1\perp}  + \frac{1-x_q}{2x_{h_2}}p_{h_2\perp} - p_{2'\perp}\right) \\
& \times \int_{\frac{x_{h_1}}{x_q}}^{1-\frac{\alpha}{x_q}} \frac{d\beta_1}{ (1-\beta_1)^{1-2\epsilon} x_q^{1-2\epsilon}} x_q \\
& \times \frac{4 (1-x_q)^2 x_q^2 }{\left((1-x_q)x_q Q^2+ \left(\frac{1-x_q}{x_{h_2}}\vec{p}_{h_2}-\vec{p}_{2}\right)^2\right)\left((1-x_q)x_q Q^2+ \left(\frac{1-x_q}{x_{h_2}}\vec{p}_{h_2}-\vec{p}_{2'}\right)^2\right)} \\
& \times  \frac{\alpha_s C_F}{\mu^{2\epsilon}} \int \frac{d^d \vec{u}}{(2\pi)^d} \left \{ \frac{1}{\left( \vec{u}-\frac{\vec{p}_{h_1}}{x_{h_1}}\right)^2}  + \frac{1}{\left(\vec{u}-\frac{\vec{p}_{h_2}}{x_{h_2}}\right)^2}  - 2 \frac{ \left(\vec{u}-\frac{\vec{p}_{h_1}}{x_{h_1}}\right)\cdot \left(\vec{u}-\frac{\vec{p}_{h_2}}{x_{h_2}}\right)}{\left(\vec{u}-\frac{\vec{p}_{h_1}}{x_{h_1}}\right)^2 \left(\vec{u}-\frac{\vec{p}_{h_2}}{x_{h_2}}\right)^2} \right\} + (h_1 \leftrightarrow h_2) \,.
\numberthis[soft_beta_1]
\end{align*}
In a similar fashion, the transformation (\ref{eq:Transbeta2}) leads to second contribution, which reads
    \begin{align*}
& \frac{d \sigma_{3LL}^{q \bar{q} \rightarrow h_1 h_2 }}{d x_{h_1}d x_{h_2} d^d p_{h_1\perp} d p_{h_2 \perp}}\Bigg |_{\text{soft } \beta_2} \\
&= \frac{2 \alpha_{\mathrm{em}} Q^2}{(2\pi)^{4(d-1)} N_c} \sum_{q} \int_{x_{h_2}}^{1-x_{h_1}} \frac{d x_{\bar{q}}}{x_{\bar{q}}} \frac{1}{1-x_{\bar{q}}}\left(\frac{x_{\bar{q}}}{x_{h_2}}\right)^d \left(\frac{1-x_{\bar{q}}}{x_{h_1}}\right)^d \\
& \times   Q_q^2 D_q^{h_1}\left(\frac{x_{h_1}}{1-x_{\bar{q}}},\mu_F\right)D_{\bar{q}}^{h_2}\left(\frac{x_{h_2}}{x_{\bar{q}}}, \mu_F\right) \\
& \times  \int  d^d p_{2\perp} \; \mathbf{F}\left(\frac{1-x_{\bar{q}}}{2x_{h_1}} p_{h_1\perp}  + \frac{x_{\bar{q}}}{2x_{h_2}}p_{h_2\perp} - p_{2\perp}\right) \\
& \times  \int  d^d p_{2'\perp} \; \mathbf{F}^*\left(\frac{1-x_{\bar{q}}}{2x_{h_1}} p_{h_1\perp}  + \frac{x_{\bar{q}}}{2x_{h_2}}p_{h_2\perp} - p_{2'\perp}\right) \\
& \times \int_{\frac{x_{h_2}}{x_{\bar{q}}}}^{1-\frac{\alpha}{x_{\bar{q}}}} \frac{d\beta_2}{ (1-\beta_2)^{1-2\epsilon} x_{\bar{q}}^{1-2\epsilon}} x_{\bar{q}} \\
& \times \frac{4 (1-x_{\bar{q}})^2 x_{\bar{q}}^2 }{\left( (1-x_{\bar{q}}) x_{\bar{q}} Q^2+ \left(\frac{x_{\bar{q}}}{x_{h_2}}\vec{p}_{h_2}-\vec{p}_{2}\right)^2\right)\left((1-x_{\bar{q}}) x_{\bar{q}} Q^2+\left(\frac{x_{\bar{q}}}{x_{h_2}}\vec{p}_{h_2}-\vec{p}_{2'}\right)^2\right)} \\
& \times  \frac{\alpha_s C_F}{\mu^{2\epsilon}} \int \frac{d^d \vec{u}}{(2\pi)^d} \left \{ \frac{1}{\left( \vec{u}-\frac{\vec{p}_{h_1}}{x_{h_1}}\right)^2}  + \frac{1}{\left(\vec{u}-\frac{\vec{p}_{h_2}}{x_{h_2}}\right)^2}  - 2 \frac{ \left(\vec{u}-\frac{\vec{p}_{h_1}}{x_{h_1}}\right)\cdot \left(\vec{u}-\frac{\vec{p}_{h_2}}{x_{h_2}}\right)}{\left(\vec{u}-\frac{\vec{p}_{h_1}}{x_{h_1}}\right)^2 \left(\vec{u}-\frac{\vec{p}_{h_2}}{x_{h_2}}\right)^2} \right\} + (h_1 \leftrightarrow h_2) \,.\numberthis[soft_beta_2]
    \end{align*}
Next, we integrate over $\vec{u}$, which gives 
\begin{align*}
    I_u & = \frac{\alpha_s C_F}{\mu^{2\epsilon}} \int \frac{d^d \vec{u}}{(2\pi)^d} \left \{ \frac{1}{\left( \vec{u}-\frac{\vec{p}_{h_1}}{x_h}\right)^2}  + \frac{1}{\left(\vec{u}-\frac{\vec{p}_{h_2}}{x_{h_2}}\right)^2}  - 2 \frac{ \left(\vec{u}-\frac{\vec{p}_{h_1}}{x_{h_1}}\right)\cdot \left(\vec{u}-\frac{\vec{p}_{h_2}}{x_{h_2}}\right)}{\left(\vec{u}-\frac{\vec{p}_{h_1}}{x_{h_1}}\right)^2 \left(\vec{u}-\frac{\vec{p}_{h_2}}{x_{h_2}}\right)^2} \right\} \\
    & = \frac{\alpha_s C_F}{\mu^{2\epsilon}} \left(\frac{\vec{p}_{h_1}}{x_{h_1}} - \frac{\vec{p}_{h_2}}{x_{h_2}}\right)^2 \int \frac{d^d \vec{u}}{(2\pi)^d} \frac{1}{\left( \vec{u}-\frac{\vec{p}_{h_1}}{x_{h_1}}\right)^2 \left(\vec{u}-\frac{\vec{p}_{h_2}}{x_{h_2}}\right)^2 } \\ 
    &= \frac{\alpha_s C_F}{\mu^{2\epsilon}} \left(\frac{\vec{p}_{h_1}}{x_{h_1}} - \frac{\vec{p}_{h_2}}{x_{h_2}}\right)^2 \int \frac{d^d \vec{u}}{(2\pi)^d} \frac{1}{\vec{u}^2 \left(\vec{u}-\left( \frac{\vec{p}_{h_1}}{x_{h_1}} -\frac{\vec{p}_{h_2}}{x_{h_2}} \right)\right)^2 } \\
    &= \frac{\alpha_s C_F}{\mu^{2\epsilon}} \left(\frac{\vec{p}_{h_1}}{x_{h_1}} - \frac{\vec{p}_{h_2}}{x_{h_2}}\right)^2  \frac{1}{(2 \pi)^d} \pi^{1+\epsilon} \Gamma(1-\epsilon) \beta(\epsilon,\epsilon) \left[\left( \frac{\vec{p}_{h_1}}{x_{h_1}} -\frac{\vec{p}_{h_2}}{x_{h_2}} \right)^2\right]^{\epsilon-1} \\
    &= \frac{\alpha_s}{2\pi} C_F \frac{1}{\hat{\epsilon}} \left[ 1 + \epsilon \ln \left(\frac{\left(\frac{\vec{p}_{h_1}}{x_{h_1}} -\frac{\vec{p}_{h_2}}{x_{h_2}} \right)^2}{\mu^2}\right)\right]. \numberthis[integral_u]
    \end{align*}
Finally, the integral over $\beta$ leads to
\begin{align*}
    \int_{\frac{x_h}{x}}^{1-\frac{\alpha}{x}} \frac{d \beta}{(1-\beta)^{3-d}} 
    &=  \int_{\frac{x_h}{x}}^{1-\frac{\alpha}{x}} \frac{d \beta}{1-\beta} \left[1 + 2 \epsilon\ln  (1-\beta)\right] \\
    &= -\ln\left(\frac{\alpha}{x}\right)  + \ln\left(1-\frac{x_h}{x}\right) - \epsilon \ln^2\left(\frac{\alpha}{x}\right) + \epsilon \ln^2\left(1-\frac{x_h}{x}\right) \\
    &=  -\ln \alpha + \ln x + \ln\left(1-\frac{x_h}{x}\right) - \epsilon \left[\ln^2 \alpha -2\ln \alpha \ln x + \ln^2 x\right]  \\
    & + \epsilon \ln^2\left(1-\frac{x_h}{x}\right) \numberthis[integral_beta].
\end{align*}
Combining both integrals over $\beta$ \eqref{eq:integral_beta} and over $\vec{u}$ \eqref{eq:integral_u} and keeping only the divergent terms, eqs.~\eqref{eq:soft_beta_1} and \eqref{eq:soft_beta_2} become respectively
\begin{align*}
& \frac{d \sigma_{3LL}^{q \bar{q} \rightarrow h_1 h_2 }}{d x_{h_1}d x_{h_2} d^d p_{h_1\perp} d p_{h_2 \perp}}\Bigg |_{\text{soft } \beta_1} \\
&= \frac{2  \alpha_{\mathrm{em}} Q^2}{(2\pi)^{4(d-1)} N_c}  \sum_{q}  \int_{x_{h_1}}^{1} d x_q \int_{x_{h_2}}^{1} d x_{\bar{q}} \; x_q x_{\bar{q}} \left(\frac{x_q}{x_{h_1}}\right)^d \left(\frac{x_{\bar{q}}}{x_{h_2}}\right)^d \delta(1-x_q -x_{\bar{q}}) \\
&  \times Q_q^2 D_q^{h_1}\left(\frac{x_{h_1}}{x_q},\mu_F\right)D_{\bar{q}}^{h_2}\left(\frac{x_{h_2}}{x_{\bar{q}}}, \mu_F\right) \mathcal{F}_{LL} \frac{\alpha_S C_F}{2\pi} \frac{4}{\Hat{\epsilon}}  \left[ - \ln \alpha + \ln x_q + \ln\left(1- \frac{x_{h_1}}{x_q}\right) \right. \\
& \left. - \epsilon \ln^2 \alpha - \epsilon \ln \alpha \ln \left(\frac{\left(\frac{\vec{p}_{h_1}}{x_{h_1}} -\frac{\vec{p}_{h_2}}{x_{h_2}} \right)^2}{\mu^2}\right) + \epsilon \ln x_q \Bigg( \ln x_q + 2 \ln \left( 1 - \frac{x_{h_1}}{x_{q}} \right) \right. \\ & \left. \left. + \ln \left(\frac{\left(\frac{\vec{p}_{h_1}}{x_{h_1}} -\frac{\vec{p}_{h_2}}{x_{h_2}} \right)^2}{\mu^2}\right) \right) + \epsilon \ln \left( 1 - \frac{x_{h_1}}{x_q} \right) \left( \ln \left(\frac{\left(\frac{\vec{p}_{h_1}}{x_{h_1}} -\frac{\vec{p}_{h_2}}{x_{h_2}} \right)^2}{\mu^2}\right)  + \ln \left( 1 - \frac{x_{h_1}}{x_q} \right) \right) \right] \\ & + (h_1 \leftrightarrow h_2)\, ,
\end{align*}
and
\begin{align*}
& \frac{d \sigma_{3LL}^{q \bar{q} \rightarrow h_1 h_2 } }{d x_{h_1}d x_{h_2} d^d p_{h_1\perp} d p_{h_2 \perp}}\Bigg |_{\text{soft } \beta_2} \\
&=  \frac{2  \alpha_{\mathrm{em}} Q^2}{(2\pi)^{4(d-1)} N_c} \sum_{q} \int_{x_{h_1}}^{1} d x_q \int_{x_{h_2}}^{1} d x_{\bar{q}} \;  x_q x_{\bar{q}} \left(\frac{x_q}{x_{h_1}}\right)^d \left(\frac{x_{\bar{q}}}{x_{h_2}}\right)^d   \delta(1-x_q -x_{\bar{q}})    \\
&  \times Q_q^2 D_q^{h_1}\left(\frac{x_{h_1}}{x_q},\mu_F\right)D_{\bar{q}}^{h_2}\left(\frac{x_{h_2}}{x_{\bar{q}}}, \mu_F\right) \mathcal{F}_{LL} \frac{\alpha_S C_F}{2\pi} \frac{4}{\Hat{\epsilon}}  \left[ - \ln \alpha + \ln x_{\bar{q}} + \ln\left(1- \frac{x_{h_2}}{x_{\bar{q}}}\right) \right. \\
& \left. - \epsilon \ln^2 \alpha - \epsilon \ln \alpha \ln \left(\frac{\left(\frac{\vec{p}_{h_1}}{x_{h_1}} -\frac{\vec{p}_{h_2}}{x_{h_2}} \right)^2}{\mu^2}\right) + \epsilon \ln x_{\bar{q}} \Bigg( \ln x_{\bar{q}} + 2 \ln \left( 1 - \frac{x_{h_2}}{x_{\bar{q}}} \right) \right. \\ & \left. \left. + \ln \left(\frac{\left(\frac{\vec{p}_{h_1}}{x_{h_1}} -\frac{\vec{p}_{h_2}}{x_{h_2}} \right)^2}{\mu^2}\right) \right) + \epsilon \ln \left( 1 - \frac{x_{h_2}}{x_{\bar{q}}} \right) \left( \ln \left(\frac{\left(\frac{\vec{p}_{h_1}}{x_{h_1}} -\frac{\vec{p}_{h_2}}{x_{h_2}} \right)^2}{\mu^2}\right) + \ln \left( 1 - \frac{x_{h_2}}{x_{\bar{q}}} \right) \right) \right] \\ & + (h_1 \leftrightarrow h_2) \, . \\
\end{align*}

As said above, the total soft contribution expression is found by summing the two above equations. As usual, we split the final result into divergent and finite part. In the LL case, we obtain
\begin{align*}
\frac{d \sigma_{3LL}^{q \bar{q} \rightarrow h_1 h_2 
}}{d x_{h_1}d x_{h_2} d^d p_{h_1\perp} d p_{h_2 \perp}} & \Bigg |_{\text{soft div} } 
=  \frac{4  \alpha_{\mathrm{em}} Q^2}{(2\pi)^{4(d-1)} N_c} \sum_{q} \int_{x_{h_1}}^{1} d x_q \int_{x_{h_2}}^{1} d x_{\bar{q}} x_q x_{\bar{q}} \left(\frac{x_q}{x_{h_1}}\right)^d \left(\frac{x_{\bar{q}}}{x_{h_2}}\right)^d  \\
& \times \delta(1-x_q -x_{\bar{q}})   Q_q^2 D_q^{h_1}\left(\frac{x_{h_1}}{x_q},\mu_F\right)D_{\bar{q}}^{h_2}\left(\frac{x_{h_2}}{x_{\bar{q}}}, \mu_F\right) \mathcal{F}_{LL}\\
& \times \frac{\alpha_s C_F }{2\pi} \frac{1}{\hat{\epsilon}} \left[ - 4 \ln \alpha + 2 \ln x_q + 2 \ln\left(1-\frac{x_{h_1}}{x_q}\right) -4 \epsilon \ln^2 \alpha \right. \\
& \left. - 4 \epsilon \ln \alpha \ln \left(\frac{\left(\frac{\vec{p}_{h_1}}{x_{h_1}} -\frac{\vec{p}_{h_2}}{x_{h_2}} \right)^2}{\mu^2}\right)  + 2 \ln x_{\bar{q}} + 2 \ln \left(1-\frac{x_{h_2}}{x_{\bar{q}}}\right) 
\right] \\
& + (h_1 \leftrightarrow h_2)\, .
\numberthis[total_soft LL]
\end{align*}
and 
\begin{align*}
\frac{d \sigma_{3LL}^{q \bar{q} \rightarrow h_1 h_2 
}}{d x_{h_1}d x_{h_2} d^d p_{h_1\perp} d p_{h_2 \perp}} & \Bigg |_{\text{soft fin} } 
=  \frac{4  \alpha_{\mathrm{em}} Q^2}{(2\pi)^{4(d-1)} N_c} \sum_{q} \int_{x_{h_1}}^{1} d x_q \int_{x_{h_2}}^{1} d x_{\bar{q}} x_q x_{\bar{q}} \left(\frac{x_q}{x_{h_1}}\right)^d \left(\frac{x_{\bar{q}}}{x_{h_2}}\right)^d  \\
& \times \delta(1-x_q -x_{\bar{q}})   Q_q^2 D_q^{h_1}\left(\frac{x_{h_1}}{x_q},\mu_F\right)D_{\bar{q}}^{h_2}\left(\frac{x_{h_2}}{x_{\bar{q}}}, \mu_F\right) \mathcal{F}_{LL} \\
& \times \frac{\alpha_s C_F }{\pi} \left[ \ln x_q \left( \ln x_q + 2 \ln \left( 1 - \frac{x_{h_1}}{x_{q}} \right) + \ln \left(\frac{\left(\frac{\vec{p}_{h_1}}{x_{h_1}} -\frac{\vec{p}_{h_2}}{x_{h_2}} \right)^2}{\mu^2} \right) \right) \right. \\ & \left.  + \ln \left( 1 - \frac{x_{h_1}}{x_q} \right) \left( \ln \left(\frac{\left(\frac{\vec{p}_{h_1}}{x_{h_1}} -\frac{\vec{p}_{h_2}}{x_{h_2}} \right)^2}{\mu^2}\right)  + \ln \left( 1 - \frac{x_{h_1}}{x_q} \right) \right) \right. \\ & \left.
+ \ln x_{\bar{q}} \Bigg( \ln x_{\bar{q}} + 2 \ln \left( 1 - \frac{x_{h_2}}{x_{\bar{q}}} \right) \left. + \ln \left(\frac{\left(\frac{\vec{p}_{h_1}}{x_{h_1}} -\frac{\vec{p}_{h_2}}{x_{h_2}} \right)^2}{\mu^2}\right) \right) \right. \\ & \left. + \ln \left( 1 - \frac{x_{h_2}}{x_{\bar{q}}} \right) \left( \ln \left(\frac{\left(\frac{\vec{p}_{h_1}}{x_{h_1}} -\frac{\vec{p}_{h_2}}{x_{h_2}} \right)^2}{\mu^2}\right) + \ln \left( 1 - \frac{x_{h_2}}{x_{\bar{q}}} \right) \right) \right] \\ & + (h_1 \leftrightarrow h_2) \, .
\numberthis[total_soft_fin LL]
\end{align*}
For the TL and TT case, the calculation leads  respectively to
\begin{align*}
\frac{d \sigma_{3TL}^{q \bar{q} \rightarrow h_1 h_2 
}}{d x_{h_1}d x_{h_2} d^d p_{h_1\perp} d p_{h_2 \perp}} & \Bigg |_{\text{soft div} } 
=  \frac{2  \alpha_{\mathrm{em}} Q}{(2\pi)^{4(d-1)} N_c} \sum_{q} \int_{x_{h_1}}^{1} \hspace{-0.3 cm} d x_q \int_{x_{h_2}}^{1} \hspace{-0.3 cm} d x_{\bar{q}} (x_{\bar{q}}-x_q)  \left(\frac{x_q}{x_{h_1}}\right)^d \left(\frac{x_{\bar{q}}}{x_{h_2}}\right)^d  \\
& \times \delta(1-x_q -x_{\bar{q}})   Q_q^2 D_q^{h_1}\left(\frac{x_{h_1}}{x_q},\mu_F\right)D_{\bar{q}}^{h_2}\left(\frac{x_{h_2}}{x_{\bar{q}}}, \mu_F\right) \mathcal{F}_{TL}\\
& \times \frac{\alpha_s C_F }{2\pi} \frac{1}{\hat{\epsilon}} \left[ - 4 \ln \alpha + 2 \ln x_q + 2 \ln\left(1-\frac{x_{h_1}}{x_q}\right) -4 \epsilon \ln^2 \alpha \right. \\
& \left. - 4 \epsilon \ln \alpha \ln \left(\frac{\left(\frac{\vec{p}_{h_1}}{x_{h_1}} -\frac{\vec{p}_{h_2}}{x_{h_2}} \right)^2}{\mu^2}\right)  + 2 \ln x_{\bar{q}} + 2 \ln \left(1-\frac{x_{h_2}}{x_{\bar{q}}}\right) 
\right] \\
& + (h_1 \leftrightarrow h_2)\, ,
\numberthis[total_soft TL]
\end{align*}
\begin{align*}
\frac{d \sigma_{3TL}^{q \bar{q} \rightarrow h_1 h_2 
}}{d x_{h_1}d x_{h_2} d^d p_{h_1\perp} d p_{h_2 \perp}} & \Bigg |_{\text{soft fin} } 
=  \frac{2 \alpha_{\mathrm{em}} Q}{(2\pi)^{4(d-1)} N_c} \sum_{q} \int_{x_{h_1}}^{1} \hspace{-0.3 cm} d x_q \int_{x_{h_2}}^{1} \hspace{-0.3 cm} d x_{\bar{q}} (x_{\bar{q}} - x_q)  \left(\frac{x_q}{x_{h_1}}\right)^d \left(\frac{x_{\bar{q}}}{x_{h_2}}\right)^d  \\
& \times \delta(1-x_q -x_{\bar{q}})   Q_q^2 D_q^{h_1}\left(\frac{x_{h_1}}{x_q},\mu_F\right)D_{\bar{q}}^{h_2}\left(\frac{x_{h_2}}{x_{\bar{q}}}, \mu_F\right) \mathcal{F}_{TL} \\
& \times \frac{\alpha_s C_F }{\pi} \left[ \ln x_q \left( \ln x_q + 2 \ln \left( 1 - \frac{x_{h_1}}{x_{q}} \right) + \ln \left(\frac{\left(\frac{\vec{p}_{h_1}}{x_{h_1}} -\frac{\vec{p}_{h_2}}{x_{h_2}} \right)^2}{\mu^2} \right) \right) \right. \\ & \left.  + \ln \left( 1 - \frac{x_{h_1}}{x_q} \right) \left( \ln \left(\frac{\left(\frac{\vec{p}_{h_1}}{x_{h_1}} -\frac{\vec{p}_{h_2}}{x_{h_2}} \right)^2}{\mu^2}\right)  + \ln \left( 1 - \frac{x_{h_1}}{x_q} \right) \right) \right. \\ & \left.
+ \ln x_{\bar{q}} \Bigg( \ln x_{\bar{q}} + 2 \ln \left( 1 - \frac{x_{h_2}}{x_{\bar{q}}} \right) \left. + \ln \left(\frac{\left(\frac{\vec{p}_{h_1}}{x_{h_1}} -\frac{\vec{p}_{h_2}}{x_{h_2}} \right)^2}{\mu^2}\right) \right) \right. \\ & \left. + \ln \left( 1 - \frac{x_{h_2}}{x_{\bar{q}}} \right) \left( \ln \left(\frac{\left(\frac{\vec{p}_{h_1}}{x_{h_1}} -\frac{\vec{p}_{h_2}}{x_{h_2}} \right)^2}{\mu^2}\right) + \ln \left( 1 - \frac{x_{h_2}}{x_{\bar{q}}} \right) \right) \right] \\ & + (h_1 \leftrightarrow h_2) \, .
\numberthis[total_soft_fin TL]
\end{align*}
and
\begin{align*}
\frac{d \sigma_{3TT}^{q \bar{q} \rightarrow h_1 h_2 
}}{d x_{h_1}d x_{h_2} d^d p_{h_1\perp} d p_{h_2 \perp}} & \Bigg |_{\text{soft div} } 
=  \frac{  \alpha_{\mathrm{em}} }{(2\pi)^{4(d-1)} N_c} \sum_{q} \int_{x_{h_1}}^{1} \frac{d x_q}{x_q} \int_{x_{h_2}}^{1} \frac{d x_{\bar{q}}}{x_{\bar{q}}}  \left(\frac{x_q}{x_{h_1}}\right)^d \left(\frac{x_{\bar{q}}}{x_{h_2}}\right)^d  \\
& \times \delta(1-x_q -x_{\bar{q}})   Q_q^2 D_q^{h_1}\left(\frac{x_{h_1}}{x_q},\mu_F\right)D_{\bar{q}}^{h_2}\left(\frac{x_{h_2}}{x_{\bar{q}}}, \mu_F\right) \mathcal{F}_{TT}\\
& \times \frac{\alpha_s C_F }{2\pi} \frac{1}{\hat{\epsilon}} \left[ - 4 \ln \alpha + 2 \ln x_q + 2 \ln\left(1-\frac{x_{h_1}}{x_q}\right) -4 \epsilon \ln^2 \alpha \right. \\
& \left. - 4 \epsilon \ln \alpha \ln \left(\frac{\left(\frac{\vec{p}_{h_1}}{x_{h_1}} -\frac{\vec{p}_{h_2}}{x_{h_2}} \right)^2}{\mu^2}\right)  + 2 \ln x_{\bar{q}} + 2 \ln \left(1-\frac{x_{h_2}}{x_{\bar{q}}}\right) 
\right] \\
& + (h_1 \leftrightarrow h_2)\, ,
\numberthis[total_soft TT]
\end{align*}
\begin{align*}
\frac{d \sigma_{3TT}^{q \bar{q} \rightarrow h_1 h_2 
}}{d x_{h_1}d x_{h_2} d^d p_{h_1\perp} d p_{h_2 \perp}} & \Bigg |_{\text{soft fin} } 
=  \frac{\alpha_{\mathrm{em}} }{(2\pi)^{4(d-1)} N_c} \sum_{q} \int_{x_{h_1}}^{1}  \frac{d x_q}{x_q} \int_{x_{h_2}}^{1}  \frac{d x_{\bar{q}}}{x_{\bar{q}}}  \left(\frac{x_q}{x_{h_1}}\right)^d \left(\frac{x_{\bar{q}}}{x_{h_2}}\right)^d  \\
& \times \delta(1-x_q -x_{\bar{q}})   Q_q^2 D_q^{h_1}\left(\frac{x_{h_1}}{x_q},\mu_F\right)D_{\bar{q}}^{h_2}\left(\frac{x_{h_2}}{x_{\bar{q}}}, \mu_F\right) \mathcal{F}_{TT} \\
& \times \frac{\alpha_s C_F }{\pi} \left[ \ln x_q \left( \ln x_q + 2 \ln \left( 1 - \frac{x_{h_1}}{x_{q}} \right) + \ln \left(\frac{\left(\frac{\vec{p}_{h_1}}{x_{h_1}} -\frac{\vec{p}_{h_2}}{x_{h_2}} \right)^2}{\mu^2} \right) \right) \right. \\ & \left.  + \ln \left( 1 - \frac{x_{h_1}}{x_q} \right) \left( \ln \left(\frac{\left(\frac{\vec{p}_{h_1}}{x_{h_1}} -\frac{\vec{p}_{h_2}}{x_{h_2}} \right)^2}{\mu^2}\right)  + \ln \left( 1 - \frac{x_{h_1}}{x_q} \right) \right) \right. \\ & \left.
+ \ln x_{\bar{q}} \Bigg( \ln x_{\bar{q}} + 2 \ln \left( 1 - \frac{x_{h_2}}{x_{\bar{q}}} \right) \left. + \ln \left(\frac{\left(\frac{\vec{p}_{h_1}}{x_{h_1}} -\frac{\vec{p}_{h_2}}{x_{h_2}} \right)^2}{\mu^2}\right) \right) \right. \\ & \left. + \ln \left( 1 - \frac{x_{h_2}}{x_{\bar{q}}} \right) \left( \ln \left(\frac{\left(\frac{\vec{p}_{h_1}}{x_{h_1}} -\frac{\vec{p}_{h_2}}{x_{h_2}} \right)^2}{\mu^2}\right) + \ln \left( 1 - \frac{x_{h_2}}{x_{\bar{q}}} \right) \right) \right] \\ & + (h_1 \leftrightarrow h_2) \, .
\numberthis[total_soft_fin TT]
\end{align*}

At this level, we are already able to observe the full cancellation of soft divergences (and hence the disappearance of $\ln \alpha$-terms). Consider, for instance, the longitudinal cross-section.
Combining the divergent soft contribution, coming from the real part, see eq.~\eqref{eq:total_soft LL}, with the virtual contribution \eqref{eq:virtual div LL} we see the complete cancellation of these $\ln \alpha$-terms and also of $\frac{1}{\epsilon} \ln (x_q x_{\bar{q}})$-term. Moreover, surviving $\frac{1}{\epsilon}$ divergent terms cancel in combination with:
\begin{itemize}
    \item Terms proportional to $\frac{3}{2} \delta(1-\beta_i)$ appearing inside the splitting functions in \eqref{eq:ct_LL}
    \item Term proportional to $ \ln \left(1- \frac{x_{h_1}}{x_q} \right)$ in eq.~\eqref{eq:coll_div_q}
    \item Term proportional to $ \ln \left(1- \frac{x_{h_2}}{x_{\bar{q}}}\right)$ in \eqref{eq:coll_qbarg_div} 
\end{itemize}
Now, we are only left with collinearly divergent contributions related to the case of fragmentation from quark and gluon or from anti-quark and gluon. These should cancel the only two divergent contributions left in eq.~\eqref{eq:ct_LL}, i.e., the ones proportional to $P_{gq} (\beta_i)$.

\subsection{Fragmentation from anti-quark and gluon}
\label{sec:qgfrag}
In this section, we deal with extracting the collinear divergences associated with the contribution (d) in Fig. \ref{fig:sigma-NLO}. This contribution corresponds to the situation in which the anti-quark and the gluon fragment, while the quark plays the role of "spectator" emitted particle. This case is much simpler than before. We do not have to deal with any soft divergence and the only IR divergence that appears is when the fragmenting gluon is emitted by the quark line after the shockwave and the emitted quark and gluon become collinear.
Hence, we can directly compute the contribution due to the first term of eq. (\ref{eq:real_div_LL}). \\
We emphasize the difference with the contribution calculated in section \ref{sec:qqbarfragColl-qbarg}. Although at the level of hard computation the term that generates the present divergence is the same as the one that generates the collinear divergence in section \ref{sec:qqbarfragColl-qbarg}, the situation is completely different. In the present case, we integrate out the quark kinematic variables and remain differential in the variable of the emitted gluon, while, in section \ref{sec:qqbarfragColl-qbarg} it is exactly the opposite.

\subsubsection{Collinear contribution: $q$-$g$ splitting}

According to the above discussion, we should focus on the first term of
 eq.~\eqref{eq: div real impact factor}, which exhibits a collinear pole, namely
\begin{align*}
    & \left.\frac{d \sigma_{3LL}^{g \bar{q} \rightarrow h_1 h_2} }{d x_{h_1} d x_{h_2} d^d p_{h_1 \perp} d^d p_{h_2\perp}}\right|_{\rm coll\, qg} \\
    &= \frac{4  \alpha_{\mathrm{em}} Q^2}{(2\pi)^{4(d-1)} N_c} \sum_{q} \int_{x_{h_1}}^1 \frac{d x_g}{x_g^2} \int_0^1  d x_q' \int_{x_{h_2}}^1 \frac{d x_{\bar{q}}}{x_{\bar{q}}} \delta(1-x_q'-x_{\bar{q}}-x_g)\\
    & \times  \left(\frac{x_g}{x_{h_1}}\right)^d \left(\frac{x_{\bar{q}}}{x_{h_2}}\right)^d  Q_q^2 D_g^{h_1}\left(\frac{x_{h_1}}{x_g}, \mu_F\right) D_{\bar{q}}^{h_2}\left(\frac{x_{h_2}}{x_{\bar{q}}}, \mu_F\right) \frac{\alpha_s}{\mu^{2\epsilon}} C_F  \int \frac{d^d p_{q\perp}}{(2\pi)^d} \\
    & \times \int   d^d p_{2\perp}     \mathbf{F}  \left(\frac{p_{q\perp}}{2} +\frac{x_{\bar{q}}}{2 x_{h_2}}p_{h_2 \perp} -p_{2\perp} + \frac{x_g}{2 x_{h_1}} p_{h_1\perp}\right) \\
    & \times  \int   d^d p_{2'\perp}  \mathbf{F}^*\left(\frac{p_{q\perp}}{2} + \frac{x_{\bar{q}}}{2 x_{h_2}}p_{h_2 \perp} -p_{2'\perp} + \frac{x_g}{2 x_{h_1}} p_{h_1\perp}\right)  \\
    & \times \frac{1}{\left( x_{\bar{q}} (1-x_{\bar{q}})Q^2 +\left(\frac{x_{\bar{q}}}{x_{h_2}} \vec{p}_{h_2}-\vec{p}_{2}\right)^2\right) \left( x_{\bar{q}} (1-x_{\bar{q}})Q^2 +\left(\frac{x_{\bar{q}}}{x_{h_2}} \vec{p}_{h_2}-\vec{p}_{2'}\right)^2\right)}\\
    & \times \frac{(d x_g^2 + 4 x_q' (x_q' + x_g)) x_{\bar{q}}^2 (1-x_{\bar{q}})^2}{\left(x_q' \frac{x_g}{x_{h_1}} \vec{p}_{h_1} - x_g \vec{p}_q\right)^2} + (h_1 \leftrightarrow h_2)\,.
\end{align*}
Using a change of variable similar to \eqref{eq:Transbeta}, here
    \begin{align*}
        x_g & = \beta_1 x_q \\ 
        x_q'&= (1-\beta_1) x_q  
    \end{align*}
    with the Jacobian $d x_q' d x_g = d x_q d \beta_1 x_q$ and treating the integration over longitudinal fractions as follows 
\begin{align*}
    & \int_{x_{h_1}}^1 \frac{d x_g}{x_g^2}  \int_{x_{h_2}}^1 \frac{d x_{\bar{q}}}{x_{\bar{q}}} \int_{0}^1  d x_q' \delta(1-x_q'-x_{\bar{q}}-x_g) \\ 
    &= \int_{x_{h_1}}^1 \frac{d x_g}{x_g^2} \int_0^1  d x_q' \int_{- \infty}^{+\infty} \frac{d x_{\bar{q}}}{x_{\bar{q}}} \theta(x_{\bar{q}}-x_{h_2}) \theta(1- x_{\bar{q}}) \delta(1-x_q'-x_{\bar{q}}-x_g) \\
    &= \int_{x_{h_1}}^{1-x_{h_2}} d x_q \frac{1}{x_q(1-x_q)}  \int_{\frac{x_{h_1}}{x_q}}^1 \frac{d \beta_1}{\beta_1^2 } \; , 
\end{align*}
we get 

\begin{align*}
& \left. \frac{d \sigma_{3LL}^{g \bar{q} \rightarrow h_1 h_2}}{d x_{h_1} d x_{h_2} d p_{h_1 \perp} d^d p_{h_2\perp}} \right|_{\text{coll. qg}}   \\
&= \frac{4  \alpha_{\mathrm{em}} Q^2}{(2\pi)^{4(d-1)} N_c} \sum_{q} \int_{x_{h_1}}^{1-x_{h_2}} d x_q \int_{\frac{x_{h_1}}{x_q}}^1 \frac{d \beta_1}{\beta_1}  x_q (1-x_q)\left(\frac{x_q}{x_{h_1}}\right)^d \left(\frac{1-x_q}{x_{h_2}}\right)^d  \\
& \times   Q_q^2 D_g^{h_1}\left(\frac{x_{h_1}}{\beta_1 x_q}, \mu_F\right) D_{\bar{q}}^{h_2}\left(\frac{x_{h_2}}{1-x_q}, \mu_F\right) \\
& \times \int   d^d p_{2\perp}  \int d^d z_{1\perp} \frac{e^{i z_{1\perp} \cdot  \left(\frac{1-x_q}{2 x_{h_2}}p_{h_2 \perp} -p_{2\perp} + \frac{\beta_1 x_q}{2 x_{h_1}} p_{h_1\perp}\right)}}{x_q (1-x_q)Q^2 +\left(\frac{1-x_q}{x_{h_2}} \vec{p}_{h_2}-\vec{p}_{2}\right)^2} F(z_{1\perp}) \\
& \times \int   d^d p_{2'\perp}  \int d^d z_{2\perp} \frac{e^{-i z_{2\perp} \cdot  \left(\frac{1-x_q}{2 x_{h_2}}p_{h_2 \perp} -p_{2'\perp} + \frac{\beta_1 x_q}{2 x_{h_1}} p_{h_1\perp}\right)}}{x_q (1-x_q)Q^2 +\left(\frac{1-x_q}{x_{h_2}} \vec{p}_{h_2}-\vec{p}_{2'}\right)^2} F^*(z_{2\perp}) \\
&\times  \frac{2 ( (1-\beta_1)^2 +1 ) + 2 \epsilon \beta_1^2}{\beta_1}\beta_1^{d-2}\frac{\alpha_s}{\mu^{2\epsilon}} C_F \int  \frac{d^d p_{q\perp}}{(2\pi)^d} \frac{e^{i \left(\frac{z_{1\perp}-z_{2\perp}}{2}\right) \cdot p_{q\perp}}}{\left(\frac{(1-\beta_1)x_q}{x_{h_1}}\vec{p}_{h_1} - \vec{p}_q\right)^2} \\
& + (h_1 \leftrightarrow h_2) \,.
\numberthis[coll_qg_qbarg_FF_total]
\end{align*} 
Using eq.~\eqref{eq:expo} in eq.~\eqref{eq:coll_qg_qbarg_FF_total} to perform the integration over quark transverse momenta, we obtain
\begin{align*}
& \left. \frac{d \sigma_{3LL}^{g \bar{q} \rightarrow h_1 h_2}}{d x_{h_1} d x_{h_2} d p_{h_1 \perp} d^d p_{h_2\perp}} \right|_{\text{coll. qg}}  \\  
&= \frac{4  \alpha_{\mathrm{em}} Q^2}{(2\pi)^{4(d-1)} N_c} \sum_{q} \int_{x_{h_1}}^{1-x_{h_2}} d x_q \int_{\frac{x_{h_1}}{x_q}}^1 \frac{d \beta_1}{\beta_1}  x_q (1-x_q)\left(\frac{x_q}{x_{h_1}}\right)^d  \\
& \times \left(\frac{1-x_q}{x_{h_2}}\right)^d   Q_q^2 D_g^{h_1}\left(\frac{x_{h_1}}{\beta_1 x_q}, \mu_F\right) D_{\bar{q}}^{h_2}\left(\frac{x_{h_2}}{1-x_q}, \mu_F\right) \\
& \times \int   d^d p_{2\perp}  \int d^d z_{1\perp} \frac{e^{i z_{1\perp} \cdot  \left(\frac{1-x_q}{2 x_{h_2}}p_{h_2 \perp} -p_{2\perp} + \frac{\beta_1 x_q}{2 x_{h_1}} p_{h_1\perp}\right)}}{x_q (1-x_q)Q^2 +\left(\frac{1-x_q}{x_{h_2}} \vec{p}_{h_2}-\vec{p}_{2}\right)^2} F(z_{1\perp}) \\
& \times \int   d^d p_{2'\perp}  \int d^d z_{2\perp} \frac{e^{-i z_{2\perp} \cdot  \left(\frac{1-x_q}{2 x_{h_2}}p_{h_2 \perp} -p_{2'\perp} + \frac{\beta_1 x_q}{2 x_{h_1}} p_{h_1\perp}\right)}}{x_q (1-x_q)Q^2 +\left(\frac{1-x_q}{x_{h_2}} \vec{p}_{h_2}-\vec{p}_{2'}\right)^2} F^*(z_{2\perp}) \\
&\times  \frac{2 ( (1-\beta_1)^2 +1 ) + 2 \epsilon \beta_1^2 + 4 \epsilon ((1-\beta_1)^2 + 1) \ln \beta_1 }{\beta_1} \\ 
& \times e^{i \left(\frac{z_{1\perp}-z_{2\perp}}{2}\right) \cdot \frac{(1-\beta_1)x_q}{x_{h_1}} p_{h_1\perp}} \frac{\alpha_s C_F}{4\pi} \left( \frac{1}{\hat{\epsilon}} + \ln \left( \frac{c_0^2}{\left(\frac{z_{1\perp} - z_{2\perp}}{2}\right)^2 \mu^2}\right) \right) + (h_1 \leftrightarrow h_2) \\
&= \frac{4  \alpha_{\mathrm{em}} Q^2}{(2\pi)^{4(d-1)} N_c} \sum_{q} \int_{x_{h_1}}^{1-x_{h_2}} d x_q \int_{\frac{x_{h_1}}{x_q}}^1 \frac{d \beta_1}{\beta_1}  x_q (1-x_q)\left(\frac{x_q}{x_{h_1}}\right)^d \\
& \times   \left(\frac{1-x_q}{x_{h_2}}\right)^d  Q_q^2 D_g^{h_1}\left(\frac{x_{h_1}}{\beta_1 x_q}, \mu_F\right) D_{\bar{q}}^{h_2}\left(\frac{x_{h_2}}{1-x_q}, \mu_F\right) \\
& \times \int   d^d p_{2\perp}  \int d^d z_{1\perp} \frac{e^{i z_{1\perp} \cdot  \left(\frac{1-x_q}{2 x_{h_2}}p_{h_2 \perp} -p_{2\perp} + \frac{x_q}{2 x_{h_1}} p_{h_1\perp}\right)}}{x_q (1-x_q)Q^2 +\left(\frac{1-x_q}{x_{h_2}} \vec{p}_{h_2}-\vec{p}_{2}\right)^2} F(z_{1\perp}) \\
& \times \int   d^d p_{2'\perp}  \int d^d z_{2\perp} \frac{e^{-i z_{2\perp} \cdot  \left(\frac{1-x_q}{2 x_{h_2}}p_{h_2 \perp} -p_{2'\perp} + \frac{x_q}{2 x_{h_1}} p_{h_1\perp}\right)}}{x_q (1-x_q)Q^2 +\left(\frac{1-x_q}{x_{h_2}} \vec{p}_{h_2}-\vec{p}_{2'}\right)^2} F^*(z_{2\perp}) \\
& \frac{\alpha_s}{2\pi} C_F \Bigg [\frac{1}{\hat{\epsilon}}\frac{1 + (1-\beta_1)^2}{\beta_1} + \beta_1 + \frac{2  (1 + (1-\beta)^2) \ln \beta_1}{\beta_1} \\
& + \frac{1+ (1-\beta_1)^2}{\beta_1} \ln \left( \frac{c_0^2}{\left(\frac{z_{1\perp} - z_{2\perp}}{2}\right)^2 \mu^2}\right)\Bigg] + (h_1 \leftrightarrow h_2)  \\
&= \frac{d \sigma_{3LL}^{g \bar{q} \rightarrow h_1 h_2}}{ d x_{h_1} d x_{h_2} d p_{h_1 \perp} d^d p_{h_2\perp} } \bigg |_{\text{coll. qg div}}       +  \frac{d \sigma_{3LL}^{g \bar{q} \rightarrow h_1 h_2}}{ d x_{h_1} d x_{h_2} d p_{h_1 \perp} d^d p_{h_2\perp} } \bigg |_{\text{coll.qg fin}} \numberthis[coll_qg_qbar_g_beta]\; ,
\end{align*}
where the term labeled with "div" contains the first term of the square bracket. \\
Putting back $x_{\bar{q}}$ using eq.~\eqref{eq: xq xbarq }, this divergent term takes the form: 
\begin{align*}
& \left. \frac{d \sigma_{3LL}^{g \bar{q} \rightarrow h_1 h_2}}{d x_{h_1} d x_{h_2} d p_{h_1 \perp} d^d p_{h_2\perp}}\right|_{\text{coll. qg div}} \\
&= \frac{4  \alpha_{\mathrm{em}} Q^2}{(2\pi)^{4(d-1)} N_c}   \sum_{q}  \int_{x_{h_1}}^{1}  d x_q \int_{\frac{x_{h_2}}{x_q}}^1  d x_{\bar{q}} \;  x_q x_{\bar{q}}  \left(\frac{x_q}{x_{h_1}}\right)^d \left(\frac{x_{\bar{q}}}{x_{h_2}}\right)^d \\*
& \times  \delta(1-x_q -x_{\bar{q}}) \mathcal{F}_{LL} \frac{\alpha_s }{2\pi} \frac{1}{\hat{\epsilon}} \int_{\frac{x_{h_1}}{x_q}}^1 \frac{d \beta_1}{\beta_1}  \\
& \times   Q_q^2 D_g^{h_1}\left(\frac{x_{h_1}}{\beta_1 x_q},\mu_F\right) D_{\bar{q}}^{h_2}\left(\frac{x_{h_2}}{x_{\bar{q}}}, \mu_F\right) C_F \frac{1+(1-\beta_1)^2}{\beta_1} + (h_1 \leftrightarrow h_2) \,.\numberthis
\end{align*}
This is the term needed to cancel the divergent term proportional to $P_{gq} (\beta_1)$ in \eqref{eq:ct_LL}. Instead, the finite part in eq.~\eqref{eq:coll_qg_qbar_g_beta} reads
\begin{align*}
& \left. \frac{d \sigma_{3LL}^{g \bar{q} \rightarrow h_1 h_2}}{d x_{h_1} d x_{h_2} d p_{h_1 \perp} d^d p_{h_2\perp}}\right |_{\text{coll. qg fin}}  \\
& = \frac{4  \alpha_{\mathrm{em}} Q^2}{(2\pi)^{4(d-1)} N_c}   \sum_{q}  \int_{x_{h_1}}^{1}  d x_q \int_{\frac{x_{h_2}}{x_q}}^1  d x_{\bar{q}} \;  x_q x_{\bar{q}}  \left(\frac{x_q}{x_{h_1}}\right)^d \left(\frac{x_{\bar{q}}}{x_{h_2}}\right)^d  \\
& \times \delta(1-x_q -x_{\bar{q}})  \int_{\frac{x_{h_1}}{x_q}}^1 \frac{d \beta_1}{\beta_1}  Q_q^2 D_g^{h_1}\left(\frac{x_{h_1}}{\beta_1 x_q}, \mu_F\right) D_{\bar{q}}^{h_2}\left(\frac{x_{h_2}}{1-x_q}, \mu_F\right) \\
& \times \int   d^d p_{2\perp}  \int d^d z_{1\perp} \frac{e^{i z_{1\perp} \cdot  \left(\frac{1-x_q}{2 x_{h_2}}p_{h_2 \perp} -p_{2\perp} + \frac{x_q}{2 x_{h_1}} p_{h_1\perp}\right)}}{x_q (1-x_q)Q^2 +\left(\frac{1-x_q}{x_{h_2}} \vec{p}_{h_2}-\vec{p}_{2}\right)^2} F(z_{1\perp}) \\
& \times \int   d^d p_{2'\perp}  \int d^d z_{2\perp} \frac{e^{-i z_{2\perp} \cdot  \left(\frac{1-x_q}{2 x_{h_2}}p_{h_2 \perp} -p_{2'\perp} + \frac{x_q}{2 x_{h_1}} p_{h_1\perp}\right)}}{x_q (1-x_q)Q^2 +\left(\frac{1-x_q}{x_{h_2}} \vec{p}_{h_2}-\vec{p}_{2'}\right)^2} F^*(z_{2\perp}) \\
& \times \frac{\alpha_s}{2\pi} C_F \Bigg [\beta_1 + \frac{2 (1 + (1-\beta)^2) \ln \beta_1 }{\beta_1}  \\
& + \frac{1+ (1-\beta_1)^2}{\beta_1} \ln \left( \frac{c_0^2}{\left(\frac{z_{1\perp} - z_{2\perp}}{2}\right)^2 \mu^2}\right)\Bigg] + (h_1 \leftrightarrow h_2)\,. \numberthis
\end{align*}
In a similar way, in the TL case, we get
\begin{align*}
& \left. \frac{d \sigma_{3TL}^{g \bar{q} \rightarrow h_1 h_2}}{d x_{h_1} d x_{h_2} d p_{h_1 \perp} d^d p_{h_2\perp}}\right |_{\text{coll. qg }}  \\
&= \frac{2  \alpha_{\mathrm{em}} Q}{(2\pi)^{4(d-1)} N_c}  \sum_{q} \int_{x_{h_1}}^{1} d x_q  \int_{x_{h_2}}^1 d x_{\bar{q}}  \left(\frac{x_q}{x_{h_1}}\right)^d \left(\frac{x_{\bar{q}}}{x_{h_2}}\right)^d (x_{\bar{q}}-x_q)  \\
& \times \delta(1-x_q-x_{\bar{q}})  \int_{\frac{x_{h_1}}{x_q}}^1 \frac{d \beta_1}{\beta_1}   Q_q^2 D_g^{h_1}\left(\frac{x_{h_1}}{\beta_1 x_q}, \mu_F\right) D_{\bar{q}}^{h_2}\left(\frac{x_{h_2}}{x_{\bar{q}}}, \mu_F\right)  \\
& \times \int   d^d p_{2\perp}  \int d^d z_{1\perp} \frac{e^{i z_{1\perp} \cdot  \left(\frac{1-x_q}{2 x_{h_2}}p_{h_2 \perp} -p_{2\perp} + \frac{x_q}{2 x_{h_1}} p_{h_1\perp}\right)}}{x_q (1-x_q)Q^2 +\left(\frac{1-x_q}{x_{h_2}} \vec{p}_{h_2}-\vec{p}_{2}\right)^2} F(z_{1\perp}) \\
& \times \int   d^d p_{2'\perp}  \int d^d z_{2\perp} \frac{e^{-i z_{2\perp} \cdot  \left(\frac{1-x_q}{2 x_{h_2}}p_{h_2 \perp} -p_{2'\perp} + \frac{x_q}{2 x_{h_1}} p_{h_1\perp}\right)}}{x_q (1-x_q)Q^2 +\left(\frac{1-x_q}{x_{h_2}} \vec{p}_{h_2}-\vec{p}_{2'}\right)^2} F^*(z_{2\perp})  \left( \frac{x_{\bar{q}}}{x_{h_2}} \vec{p}_{h_2}  - \vec{p}_{2'}\right) \cdot \vec{\varepsilon}_T^{\; *}  \\
& \times \frac{\alpha_s}{2\pi} C_F \Bigg [\frac{1}{\hat{\epsilon}}\frac{1 + (1-\beta_1)^2}{\beta_1} + \beta_1 + \frac{2  (1 + (1-\beta)^2) \ln \beta_1}{\beta_1} \\
& + \frac{1+ (1-\beta_1)^2}{\beta_1} \ln \left( \frac{c_0^2}{\left(\frac{z_{1\perp} - z_{2\perp}}{2}\right)^2 \mu^2}\right)\Bigg] + (h_1 \leftrightarrow h_2)  \\
&= \frac{d \sigma_{3TL}^{g \bar{q} \rightarrow h_1 h_2}}{ d x_{h_1} d x_{h_2} d p_{h_1 \perp} d^d p_{h_2\perp} } \bigg |_{\text{coll. qg div}}       +  \frac{d \sigma_{3TL}^{g \bar{q} \rightarrow h_1 h_2}}{ d x_{h_1} d x_{h_2} d p_{h_1 \perp} d^d p_{h_2\perp} } \bigg |_{\text{coll. qg fin}} .
\numberthis
\end{align*}
Finally, in the TT case, we have
\begin{align*}
& \left.  \frac{d \sigma_{3TT}^{g \bar{q} \rightarrow h_1 h_2}}{d x_{h_1} d x_{h_2} d p_{h_1 \perp} d^d p_{h_2\perp}} \right |_{\text{coll. qg }}  \\
&= \frac{  \alpha_{\mathrm{em}} }{(2\pi)^{4(d-1)} N_c} \sum_{q}  \int_{x_{h_1}}^{1} \frac{d x_q }{x_q} \int_{x_{h_2}}^1 \frac{d x_{\bar{q}}}{x_{\bar{q}}} \left(\frac{x_q}{x_{h_1}}\right)^d \left(\frac{x_{\bar{q}}}{x_{h_2}}\right)^d  \delta(1-x_q-x_{\bar{q}}) \\
& \times \int_{\frac{x_{h_1}}{x_q}}^1 \frac{d \beta_1}{\beta_1}   Q_q^2 D_g^{h_1}\left(\frac{x_{h_1}}{\beta_1 x_q}, \mu_F\right) D_{\bar{q}}^{h_2}\left(\frac{x_{h_2}}{x_{\bar{q}}}, \mu_F\right) \left[ (x_{\bar{q}} -x_q)^2 g_{\perp}^{ri}g_{\perp}^{lk} - g_{\perp}^{rk}g_{\perp}^{li} + g_{\perp}^{rl}g_{\perp}^{ik} \right] \\
& \times \int   d^d p_{2\perp}  \int d^d z_{1\perp} \frac{e^{i z_{1\perp} \cdot  \left(\frac{1-x_q}{2 x_{h_2}}p_{h_2 \perp} -p_{2\perp} + \frac{x_q}{2 x_{h_1}} p_{h_1\perp}\right)}}{x_q (1-x_q)Q^2 +\left(\frac{1-x_q}{x_{h_2}} \vec{p}_{h_2}-\vec{p}_{2}\right)^2} F(z_{1\perp}) \left(\frac{x_{\bar{q}}}{x_{h_2}} p_{h_2} - p_{2}\right)_r \varepsilon_{T i}  \\
& \times \int   d^d p_{2'\perp}  \int d^d z_{2\perp} \frac{e^{-i z_{2\perp} \cdot  \left(\frac{1-x_q}{2 x_{h_2}}p_{h_2 \perp} -p_{2'\perp} + \frac{x_q}{2 x_{h_1}} p_{h_1\perp}\right)}}{x_q (1-x_q)Q^2 +\left(\frac{1-x_q}{x_{h_2}} \vec{p}_{h_2}-\vec{p}_{2'}\right)^2} F^*(z_{2\perp}) \left(\frac{x_{\bar{q}}}{x_{h_2}} p_{h_2} - p_{2'}\right)_l \varepsilon_{T k}^*\\
& \times \frac{\alpha_s}{2\pi} C_F \Bigg [\frac{1}{\hat{\epsilon}}\frac{1 + (1-\beta_1)^2}{\beta_1} + \beta_1 + \frac{2 (1 + (1-\beta)^2) \ln \beta_1 }{\beta_1} \\
& + \frac{1+ (1-\beta_1)^2}{\beta_1} \ln \left( \frac{c_0^2}{\left(\frac{z_{1\perp} - z_{2\perp}}{2}\right)^2 \mu^2}\right)\Bigg] + (h_1 \leftrightarrow h_2) \\
&= \frac{d \sigma_{3TT}^{g \bar{q} \rightarrow h_1 h_2}}{ d x_{h_1} d x_{h_2} d p_{h_1 \perp} d^d p_{h_2\perp} } \bigg |_{\text{coll. qg div}}       +  \frac{d \sigma_{3TT}^{g \bar{q} \rightarrow h_1 h_2}}{ d x_{h_1} d x_{h_2} d p_{h_1 \perp} d^d p_{h_2\perp} } \bigg |_{\text{coll. qg fin}} .\numberthis
\end{align*}
These results conclude the discussion of divergences in the case of fragmentation from antiquark and gluon.

\subsection{Fragmentation from quark and gluon}
In this section, we deal with extracting the collinear divergences associated with the contribution (c) in Fig. \ref{fig:sigma-NLO}. This contribution corresponds to the situation in which the quark and the gluon fragment, while the anti-quark plays the role of the "spectator" emitted particle.

\subsubsection{Collinear contribution: $\bar{q}$-$g$ splitting}

The term in eq.~\eqref{eq: div real impact factor} to consider is the third one. The calculation proceeds in the same way as for the anti-quark and gluon fragmentation, but this time the integration is over $p_{2,2' \perp}$ in the $F$ function. 

For the LL case, we get 
\begin{align*}
& \left. \frac{d \sigma_{3LL}^{ q g  \rightarrow h_1 h_2}}{d x_{h_1} d x_{h_2} d p_{h_1 \perp} d^d p_{h_2\perp}} \right|_{\text{coll. } \bar{q}g}  \\
&= \frac{4  \alpha_{\mathrm{em}} Q^2}{(2\pi)^{4(d-1)} N_c} \sum_{q}  \int_{x_{h_1}}^{1} d x_q \int_{x_{h_2}}^1 d x_{\bar{q}} \;  x_q x_{\bar{q}}  \delta(1-x_q -x_{\bar{q}})   \left(\frac{x_q}{x_{h_1}}\right)^d \left(\frac{x_{\bar{q}}}{x_{h_2}}\right)^d  \\*
& \times \int_{\frac{x_{h_2}}{x_{\bar{q}}}}^1 \frac{d \beta_2}{\beta_2}  Q_q^2  D_{q}^{h_2}\left(\frac{x_{h_1}}{x_q}, \mu_F\right)  D_g^{h_2}\left(\frac{x_{h_2}}{\beta_2 x_{\bar{q}}}, \mu_F\right) \\
& \times \int   d^d p_{1\perp}  \int d^d z_{1\perp} \frac{e^{i z_{1\perp} \cdot  \left(-\frac{x_{\bar{q}}}{2 x_{h_2}}p_{h_2 \perp} + p_{1\perp} - \frac{x_q}{2 x_{h_1}} p_{h_1\perp}\right)}}{x_q x_{\bar{q}} Q^2 +\left(\frac{x_q}{x_{h_1}} \vec{p}_{h_1}-\vec{p}_{1}\right)^2} F(z_{1\perp}) \\
& \times \int   d^d p_{1'\perp}  \int d^d z_{2\perp} \frac{e^{-i z_{2\perp} \cdot  \left(-\frac{x_{ \bar{q}}}{2 x_{h_2}}p_{h_2 \perp} -p_{1'\perp} -\frac{x_q}{2 x_{h_1}} p_{h_1\perp}\right)}}{x_q (1-x_q)Q^2 +\left(\frac{x_q}{x_{h_1}} \vec{p}_{h_1}-\vec{p}_{1'}\right)^2} F^*(z_{2\perp}) \\
& \times \frac{\alpha_s}{2\pi} C_F \Bigg [\frac{1}{\hat{\epsilon}}\frac{1 + (1-\beta_2)^2}{\beta_2} + \beta_2 + \frac{2 (1 + (1-\beta_2)^2)  \ln \beta_2 }{\beta_2} \\
& + \frac{1+ (1-\beta_2)^2}{\beta_2} \ln \left( \frac{c_0^2}{\left(\frac{z_{2\perp} - z_{1\perp}}{2}\right)^2 \mu^2}\right)\Bigg] + (h_1 \leftrightarrow h_2) \\
&= \frac{d \sigma_{3LL}^{q g\rightarrow h_1 h_2}}{ d x_{h_1} d x_{h_2} d p_{h_1 \perp} d^d p_{h_2\perp} } \bigg |_{\text{coll. } \bar{q}g \text{ div}}       +  \frac{d \sigma_{3LL}^{q g  \rightarrow h_1 h_2}}{ d x_{h_1} d x_{h_2} d p_{h_1 \perp} d^d p_{h_2\perp} } \bigg |_{\text{coll. } \bar{q}g\text{ fin}} .
\numberthis
\end{align*}
This term cancels the divergent term proportional to $P_{gq} (\beta_2)$ in \eqref{eq:ct_LL}. This is the last remaining cancellation of divergences, the rest of the cross-section is now completely finite.

For the TL case, we get 
{\allowdisplaybreaks
\begin{align*}
& \left. \frac{d \sigma_{3TL}^{ q g  \rightarrow h_1 h_2}}{d x_{h_1} d x_{h_2} d p_{h_1 \perp} d^d p_{h_2\perp}} \right|_{\text{coll. } \bar{q}g}   \\
&= \frac{2 \alpha_{\mathrm{em}} Q}{(2\pi)^{4(d-1)} N_c} \sum_{q} \int_{x_{h_1}}^{1} d x_q \int_{x_{h_2}}^1 d x_{\bar{q}}  \;  (x_{\bar{q}}-x_q) \delta(1-x_q -x_{\bar{q}})   \left(\frac{x_q}{x_{h_1}}\right)^d \left(\frac{x_{\bar{q}}}{x_{h_2}}\right)^d  \\
& \times \int_{\frac{x_{h_2}}{x_{\bar{q}}}}^1 \frac{d \beta_2}{\beta_2}   Q_q^2  D_{q}^{h_2}\left(\frac{x_{h_1}}{x_q}, \mu_F\right)  D_g^{h_2}\left(\frac{x_{h_2}}{\beta_2 x_{\bar{q}}}, \mu_F\right) \\
& \times \int   d^d p_{1\perp}  \int d^d z_{1\perp} \frac{e^{i z_{1\perp} \cdot  \left(-\frac{x_{\bar{q}}}{2 x_{h_2}}p_{h_2 \perp} + p_{1\perp} - \frac{x_q}{2 x_{h_1}} p_{h_1\perp}\right)}}{x_q x_{\bar{q}} Q^2 +\left(\frac{x_q}{x_{h_1}} \vec{p}_{h_1}-\vec{p}_{1}\right)^2} F(z_{1\perp}) \\
& \times \int   d^d p_{1'\perp}  \int d^d z_{2\perp} \frac{e^{-i z_{2\perp} \cdot  \left(-\frac{x_{ \bar{q}}}{2 x_{h_2}}p_{h_2 \perp} -p_{1'\perp} -\frac{x_q}{2 x_{h_1}} p_{h_1\perp}\right)}}{x_q (1-x_q)Q^2 +\left(\frac{x_q}{x_{h_1}} \vec{p}_{h_1}-\vec{p}_{1'}\right)^2} F^*(z_{2\perp})  \left( \frac{x_q}{x_{h_1}} p_{h_1 }  - p_{1'}\right) \cdot \varepsilon^* _{T}  \\
& \times \frac{\alpha_s}{2\pi} C_F \Bigg [\frac{1}{\hat{\epsilon}}\frac{1 + (1-\beta_2)^2}{\beta_2} + \beta_2 + \frac{2 \ln \beta_2 (1 + (1-\beta_2)^2)}{\beta_2}  \\
& + \frac{1+ (1-\beta_2)^2}{\beta_2} \ln \left( \frac{c_0^2}{\left(\frac{z_{2\perp} - z_{1\perp}}{2}\right)^2 \mu^2}\right)\Bigg] + (h_1 \leftrightarrow h_2) \\
&= \frac{d \sigma_{3TL}^{q g\rightarrow h_1 h_2}}{ d x_{h_1} d x_{h_2} d p_{h_1 \perp} d^d p_{h_2\perp} } \bigg |_{\text{coll. } \bar{q}g \text{ div}}       +  \frac{d \sigma_{3TL}^{q g  \rightarrow h_1 h_2}}{ d x_{h_1} d x_{h_2} d p_{h_1 \perp} d^d p_{h_2\perp} }  \bigg |_{\text{coll. } \bar{q}g\text{ fin}}
.
\numberthis
\end{align*}}

Finally, for the TT case, we get 

\begin{align*}
 & \left. \frac{d \sigma_{3TT}^{ q g  \rightarrow h_1 h_2}}{d x_{h_1} d x_{h_2} d p_{h_1 \perp} d^d p_{h_2\perp}} \right|_{\text{coll. } \bar{q}g}  \\
&= \frac{ \alpha_{\mathrm{em}} }{(2\pi)^{4(d-1)} N_c} \sum_{q} \int_{x_{h_1}}^{1} \frac{d x_q}{x_q} \int_{x_{h_2}}^1 \frac{d x_{\bar{q}} }{x_{\bar{q}}}  \delta(1-x_q -x_{\bar{q}})   \left(\frac{x_q}{x_{h_1}}\right)^d \left(\frac{x_{\bar{q}}}{x_{h_2}}\right)^d  \\
& \times \int_{\frac{x_{h_2}}{x_{\bar{q}}}}^1 \frac{d \beta_2}{\beta_2}   Q_q^2  D_{q}^{h_2}\left(\frac{x_{h_1}}{x_q}, \mu_F\right)  D_g^{h_2}\left(\frac{x_{h_2}}{\beta_2 x_{\bar{q}}}, \mu_F\right) \left[ (x_{\bar{q}} -x_q)^2 g_{\perp}^{ri}g_{\perp}^{lk} - g_{\perp}^{rk}g_{\perp}^{li} + g_{\perp}^{rl}g_{\perp}^{ik} \right]  \\
& \times \int   d^d p_{1\perp}  \int d^d z_{1\perp} \frac{e^{i z_{1\perp} \cdot  \left(-\frac{x_{\bar{q}}}{2 x_{h_2}}p_{h_2 \perp} + p_{1\perp} - \frac{x_q}{2 x_{h_1}} p_{h_1\perp}\right)}}{x_q x_{\bar{q}} Q^2 +\left(\frac{x_q}{x_{h_1}} \vec{p}_{h_1}-\vec{p}_{1}\right)^2} F(z_{1\perp}) \left(\frac{x_{q}}{x_{h_1}} p_{h_1} - p_{1}\right)_r \varepsilon_{T i}  \\
& \times \int   d^d p_{1'\perp}  \int d^d z_{2\perp} \frac{e^{-i z_{2\perp} \cdot  \left(-\frac{x_{ \bar{q}}}{2 x_{h_2}}p_{h_2 \perp} -p_{1'\perp} -\frac{x_q}{2 x_{h_1}} p_{h_1\perp}\right)}}{x_q (1-x_q)Q^2 +\left(\frac{x_q}{x_{h_1}} \vec{p}_{h_1}-\vec{p}_{1'}\right)^2} F^*(z_{2\perp}) \left(\frac{x_{q}}{x_{h_1}} p_{h_1} - p_{1'}\right)_l \varepsilon_{T k}^* \\
& \times \frac{\alpha_s}{2\pi} C_F \Bigg [\frac{1}{\hat{\epsilon}}\frac{1 + (1-\beta_2)^2}{\beta_2} + \beta_2 + \frac{2 \ln \beta_2 (1 + (1-\beta_2)^2)}{\beta_2} \\
& + \frac{1+ (1-\beta_2)^2}{\beta_2} \ln \left( \frac{c_0^2}{\left(\frac{z_{2\perp} - z_{1\perp}}{2}\right)^2 \mu^2}\right)\Bigg] + (h_1 \leftrightarrow h_2)  \\
&= \frac{d \sigma_{3TT}^{q g\rightarrow h_1 h_2}}{ d x_{h_1} d x_{h_2} d p_{h_1 \perp} d^d p_{h_2\perp} }\bigg |_{\text{coll. } \bar{q}g \text{ div}}      +  \frac{d \sigma_{3TT}^{q g  \rightarrow h_1 h_2}}{ d x_{h_1} d x_{h_2} d p_{h_1 \perp} d^d p_{h_2\perp} } \bigg |_{\text{coll. } \bar{q}g\text{ fin}}.
\numberthis
\end{align*}
\section{Additional finite terms}
\label{sec:AdditionalFin}

Some of the finite terms of our calculation are presented in previous sections. They come as a result of the extraction of divergences. There are many other terms, completely disconnected from divergences, which however contribute to the final result. We proceed to list them, also emphasizing again what their nature is.
\subsection{Virtual corrections: Dipole $\times$ double-dipole contribution}
The $1$-loop correction to the $\gamma^{*} \rightarrow q \bar{q}$ contains a dipole and double-dipole terms. The first one receives a contribution from all diagrams, while the second one gets contributions only from diagrams where the virtual gluon crosses the shockwave. At the cross-section level there will therefore be two contributions: 
\begin{itemize}
    \item[\textbullet] The one due to the interference between the dipole correction and the Born amplitude. This contains divergences and it is the one that we have completely computed in section \ref{sec: VirtualDiv}.
    \item[\textbullet] The one due to the interference between the double-dipole correction and the Born amplitude. Any  rapidity divergence present in this term is completely reabsorbed into the renormalized Wilson operator, at the amplitude level, with the help of the B-JIMWLK evolution. After this operation, this contribution is finite and can be taken in convolution with FFs without any additional manipulation.
\end{itemize}
Starting from eq.~(5.34) of \cite{Boussarie:2016ogo}, we get
\begin{align*}
& \frac{d\sigma_{2LL}^{q \bar{q} \rightarrow  h_1 h_2}}{d x_{h_1 } d^2 p_{h_1 \perp } d x_{h_2 } d^2 p_{h_2 \perp }  } \\
&= \hspace{-0.05 cm} \frac{\alpha_{\mathrm{em}} \alpha_s Q^2}{(2 \pi)^5 N_c x_{h_1}^2 x_{h_2}^2}  \sum_{q} \frac{Q_q^2}{2}  \int_{x_{h_1}}^1 \hspace{-0.3 cm} d x_q \int_{x_{h_2}}^1 \hspace{-0.3 cm} d x_{\bar{q}} \; x_q  x_{\bar{q}} \; \delta(1-x_q -x_{\bar{q}}) D_q^{h_1} \left(\frac{x_{h_1}}{x_q}, \mu_F\right) D_{\bar{q}}^{h_2} \left(\frac{x_{h_2}}{x_{\bar{q}}},\mu_F\right)  \\
& \times  \int d^2 p_{1 \perp} d^2 p_{2 \perp} d^2 p_{1' \perp} d^2 p_{2' \perp} \int \frac{d^2 p_{3 \perp}}{(2\pi)^2} \frac{\tilde{\mathbf{F}}\left(\frac{p_{12\perp}}{2}, p_{3\perp}\right) 
\mathbf{F}^*\left(\frac{p_{1'2'\perp}}{2}\right)}{\left(\frac{x_q}{x_{h_1}} \vec{p}_{h_1}-\vec{p}_{1'} \right)^2 + x_q x_{\bar{q}}Q^2 }\\ 
& \times \delta \left( \frac{x_q}{x_{h_1}} p_{h_1 \perp}- p_{1\perp} + \frac{x_{\bar{q}}}{x_{h_2}} p_{h_2 \perp} - p_{2 \perp} - p_{3\perp} \right)  \delta(p_{11' \perp} + p_{22' \perp}+ p_{3\perp})  \\
& \times \left \{ 4 x_q x_{\bar{q}} \left[  \frac{x_q x_{\bar{q}} \left(\vec{p}_3^2 - \left(\frac{x_{\bar{q}}}{x_{h_2}} \vec{p}_{h_2} - \vec{p}_2\right)^2 - \left(\frac{x_q}{x_{h_1}} \vec{p}_{h_1} - \vec{p}_1 \right)^2 -2 x_q x_{\bar{q}}Q^2  \right)}{\left(\left(\frac{x_{\bar{q}}}{x_{h_2}}\vec{p}_{h_2} - \vec{p}_2 \right)^2 + x_q x_{\bar{q}} Q^2 \right) \left(\left(\frac{x_q}{x_{h_1}} \vec{p}_{h_1} - \vec{p}_1 \right)^2 + x_q x_{\bar{q}} Q^2 \right) - x_q x_{\bar{q}} Q^2 \vec{p}_3^2 } \right. \right. \\
& \times \ln \left(\frac{x_q x_{\bar{q}}}{e^{2 \eta}}\right) \ln \left( \frac{\left(\left(\frac{x_{\bar{q}}}{x_{h_2}}\vec{p}_{h_2} - \vec{p}_2 \right)^2 + x_q x_{\bar{q}} Q^2 \right) \left(\left(\frac{x_q}{x_{h_1}} \vec{p}_{h_1} - \vec{p}_1 \right)^2 + x_q x_{\bar{q}} Q^2 \right)}{x_q x_{\bar{q}} Q^2 \vec{p}_3^2 }\right) \\
& - \left. \left( \frac{2 x_q x_{\bar{q}}}{Q^2 x_q x_{\bar{q}}  + \left(\frac{x_q}{x_{h_1}} \vec{p}_{h_1} - \vec{p}_1\right)^2 }  \ln \left(\frac{x_q}{e^\eta}\right) \ln \left( \frac{\vec{p}_3^2}{\mu^2}\right) + (q \leftrightarrow \bar{q}) \right) \right] \\
& + \left[ Q^2 \int_0^{x_q} d z \left[\left(\phi_5 + \phi_6\right)_{LL}\right]_+ + (q \leftrightarrow \bar{q}) \right] \Bigg \} + h.c. +  (h_{1} \leftrightarrow h_2)  \numberthis \; . 
\end{align*}

Concerning other transitions, we have 
\begin{align*}
& \frac{d\sigma_{2TL}^{q \bar{q} \rightarrow h_1 h_2}}{d x_{h_1 } d^2 p_{h_1 \perp } d x_{h_2 } d^2 p_{h_2 \perp }  } \\
& = \hspace{-0.05 cm} \frac{\alpha_{\mathrm{em}} \alpha_s Q}{(2 \pi)^5 N_c x_{h_1}^2 x_{h_2}^2}  \sum_{q} \frac{Q_q^2}{2}  \int_{x_{h_1}}^1 \hspace{-0.3 cm} d x_q \int_{x_{h_2}}^1 \hspace{-0.3 cm} d x_{\bar{q}} \; x_q  x_{\bar{q}} \; \delta(1-x_q -x_{\bar{q}}) \\
& \times D_q^{h_1} \left(\frac{x_{h_1}}{x_q}, \mu_F\right) D_{\bar{q}}^{h_2} \left(\frac{x_{h_2}}{x_{\bar{q}}},\mu_F\right) \int d^2 p_{1 \perp} d^2 p_{2 \perp} d^2 p_{1' \perp} d^2 p_{2' \perp} \frac{d^2 p_{3 \perp} d^2 p_{3' \perp} }{(2\pi)^2} \varepsilon_{T i}^*  \\
& \times \delta \left( \frac{x_q}{x_{h_1}} p_{h_1 \perp }- p_{1\perp} + \frac{x_{\bar{q}}}{x_{h_2}} p_{h_2 \perp} - p_{2 \perp} - p_{3\perp} \right) \delta \left( p_{11'\perp} + p_{22'\perp} + p_{33'\perp} \right) \\
& \times   \left\{ \delta(p_{3' \perp}) \frac{  \tilde{\mathbf{F}}\left(\frac{p_{12\perp}}{2}, p_{3\perp}\right) \mathbf{F}^*\left(\frac{p_{1'2'\perp}}{2}\right)}{\left(\frac{x_q}{x_{h_1}}\vec{p}_{h_1}-\vec{p}_{1'}\right)^2 + x_q x_{\bar{q}} Q^2 }  \Bigg \{  2 (x_{\bar{q}} - x_q )  \left(\frac{x_q}{x_{h_1}} p_{h_1} - p_{1'}\right)^i \right. \\
& \times \left[  \frac{x_q x_{\bar{q}} \left(\vec{p}_3^2 - \left(\frac{x_{\bar{q}}}{x_{h_2}} \vec{p}_{h_2} - \vec{p}_2\right)^2 - \left(\frac{x_q}{x_{h_1}} \vec{p}_{h_1} - \vec{p}_1 \right)^2 -2 x_q x_{\bar{q}}Q^2  \right)}{\left(\left(\frac{x_{\bar{q}}}{x_{h_2}}\vec{p}_{h_2} - \vec{p}_2 \right)^2 + x_q x_{\bar{q}} Q^2 \right) \left(\left(\frac{x_q}{x_{h_1}} \vec{p}_{h_1} - \vec{p}_1 \right)^2 + x_q x_{\bar{q}} Q^2 \right) - x_q x_{\bar{q}} Q^2 \vec{p}_3^2 } \ln\left(\frac{x_q x_{\bar{q}}}{e^{2\eta}}\right) \right. \\
& \times \ln \left( \frac{\left(\left(\frac{x_{\bar{q}}}{x_{h_2}}\vec{p}_{h_2} - \vec{p}_2 \right)^2 + x_q x_{\bar{q}} Q^2 \right) \left(\left(\frac{x_q}{x_{h_1}} \vec{p}_{h_1} - \vec{p}_1 \right)^2 + x_q x_{\bar{q}} Q^2 \right)}{x_q x_{\bar{q}} Q^2 \vec{p}_3^2 }\right) \\
& - \left. \left( \frac{2 x_q x_{\bar{q}}}{Q^2 x_q x_{\bar{q}} + \left(\frac{x_q}{x_{h_1}} \vec{p}_{h_1} - \vec{p}_1\right)^2 }  \ln \left(\frac{x_q}{e^\eta}\right) \ln \left( \frac{\vec{p}_3^2}{\mu^2}\right) + (q \leftrightarrow \bar{q}) \right) \right] \\
&  + \left[ \frac{1}{2 x_q x_{\bar{q}}} \int_0^{x_q} d z \left[\left( \phi^i_5 + \phi^i_6 \right)_{TL}\right]_+  + (q \leftrightarrow \bar{q }) \right] \Bigg \} \\
& + \delta(p_{3 \perp}) \frac{ \mathbf{F}\left(\frac{p_{12\perp}}{2}\right) \tilde{\mathbf{F} }^*\left(\frac{p_{1'2'\perp}}{2}, p_{3'\perp}\right) }{\left(\frac{x_q}{x_{h_1}}\vec{p}_{h_1}-\vec{p}_{1}\right)^2 + x_q x_{\bar{q}} Q^2 }  \Bigg \{  \Bigg [ 2 x_q x_{\bar{q}} (x_{\bar{q}} - x_q )  \left(\frac{x_q}{x_{h_1}} p_{h_1} - p_{1'}\right)^i \\
& \times \left( \frac{-2 }{Q^2  x_q x_{\bar{q}} + \left(\frac{x_q}{x_{h_1}} \vec{p}_{h_1} - \vec{p}_{1'}\right)^2} \ln\left(\frac{x}{e^\eta}\right) \ln\left(\frac{\vec{p}_{3'}^2}{\mu^2}\right) \right. \\
& - \ln \left(\frac{x_q x_{\bar{q}}}{e^{2 \eta }}\right)  \frac{ \left(\frac{x_{\bar{q}}}{x_{h_2}} \vec{p}_{h_2} - \vec{p}_{2'}\right)^2 + x_q x_{\bar{q}} Q^2 }{ \left(\left(\frac{x_{\bar{q}}}{x_{h_2}} \vec{p}_{h_2} - \vec{p}_{2'}\right)^2 + x_q x_{\bar{q}} Q^2\right) \left(\left(\frac{x_q}{x_{h_1}} \vec{p}_{h_1} - \vec{p}_{1'}\right)^2 + x_q x_{\bar{q}} Q^2\right) - x_q x_{\bar{q}} Q^2 \vec{p}_{3'}^2  } \\
& \times \ln \left(\frac{ \left(\left(\frac{x_{\bar{q}}}{x_{h_2}} \vec{p}_{h_2} - \vec{p}_{2'}\right)^2 + x_q x_{\bar{q}} Q^2\right) \left(\left(\frac{x_q}{x_{h_1}} \vec{p}_{h_1} - \vec{p}_{1'}\right)^2 + x_q x_{\bar{q}} Q^2\right)}{x_q x_{\bar{q}} Q^2 \vec{p}_{3'}^2 }\right) \\
& + \left. \left.  \frac{1}{ \left( \frac{x_q}{x_{h_1}} \vec{p}_{h_1} -\vec{p}_{1'}\right)^2 } \ln \left(\frac{x_q x_{\bar{q}}}{e^{2 \eta}}\right) \ln\left(  \frac{ \left( \frac{x_q}{x_{h_1}} \vec{p}_{h_1} -\vec{p}_{1'}\right)^2 + x_q x_{\bar{q}} Q^2 }{x_q x_{\bar{q}} Q^2 } \right) \right) + (q \leftrightarrow \bar{q}) \right] \\
& + \left. \left[ \int_0^{x_q} d z \left[\left(\phi_5^{i*} + \phi_6^{i*} \right)_{LT}\right]_+ + (q \leftrightarrow \bar{q})    \right]  \right. \Bigg \} \Bigg \} \\
& + ( h_1 \leftrightarrow h_2 ) \; , \numberthis
\end{align*} 
and
\begin{align*}
& \frac{d\sigma_{2TT}^{q \bar{q} \rightarrow h_1 h_2}}{d x_{h_1 } d^2 p_{h_1 \perp } d x_{h_2 } d^2 p_{h_2 \perp }  } \\
&= \hspace{-0.05 cm} \frac{\alpha_{\mathrm{em}} \alpha_s}{(2 \pi)^5 N_c x_{h_1}^2 x_{h_2}^2}  \sum_{q} \frac{Q_q^2}{2}  \int_{x_{h_1}}^1 \hspace{-0.3 cm} d x_q \int_{x_{h_2}}^1 \hspace{-0.3 cm} d x_{\bar{q}} \; x_q  x_{\bar{q}} \; \delta(1-x_q -x_{\bar{q}}) D_q^{h_1} \left(\frac{x_{h_1}}{x_q}, \mu_F\right) D_{\bar{q}}^{h_2} \left(\frac{x_{h_2}}{x_{\bar{q}}},\mu_F\right) \\
& \times \int d^2 p_{1 \perp} d^2 p_{2 \perp} \int \frac{d^2 p_{3 \perp}}{(2\pi)^2} \int d^2 p_{1' \perp} d^2 p_{2' \perp} \frac{\varepsilon_{T i} \varepsilon_{T j}^* }{\left( \frac{x_q}{x_{h_1}} \vec{p}_{h_1}-\vec{p}_{1'}\right)^2 + x_q x_{\bar{q}} Q^2} \\
& \times  \left[ \delta \left( \frac{x_q}{x_{h_1}} p_{h_1 \perp} - p_{1\perp} + \frac{x_{\bar{q}}}{x_{h_2}} p_{h_2 \perp} - p_{2 \perp} - p_{3\perp} \right)  \delta \left( \frac{x_q}{x_{h_1}} p_{h_1 \perp} - p_{1'\perp} + \frac{x_{\bar{q}}}{x_{h_2}} p_{h_2 \perp} - p_{2' \perp}  \right)  \right.  \\
& \times \tilde{\mathbf{F}}\left(\frac{p_{12\perp}}{2}, p_{3\perp}\right) \mathbf{F}^*\left( \frac{p_{1'2'\perp}}{2}\right)  \Bigg \{  \Bigg [ \left( \frac{x_q}{x_{h_1}} p_{h_1} - p_{1'}\right)_l \left( \frac{x_q}{x_{h_1}} p_{h_1} - p_{1}\right)_k   \\
& \times \left( (x_{\bar{q}} -x_q)^2 g_{\perp}^{ki} g_{\perp}^{lj} - g_{\perp}^{kj} g_{\perp}^{li} + g_{\perp}^{kl} g_{\perp}^{ij} \right) \left( \frac{-2}{Q^2  x_q x_{\bar{q}} +  \left( \frac{x_q}{x_{h_1}} \vec{p}_{h_1} - \vec{p}_{1}\right)^2 }  \ln \left( \frac{x_q}{e^\eta} \right)  \ln \left(\frac{\vec{p}_{3}^2 }{\mu^2 }\right) \right. \\
& + \frac{1}{ \left( \frac{x_q}{x_{h_1}} \vec{p}_{h_1} -\vec{p}_{1}\right)^2 } \ln \left(\frac{x_q x_{\bar{q}}}{e^{2 \eta}}\right) \ln\left(  \frac{ \left( \frac{x_q}{x_{h_1}} \vec{p}_{h_1} -\vec{p}_{1}\right)^2 + x_q x_{\bar{q}} Q^2 }{x_q x_{\bar{q}} Q^2 } \right) \\
& - \ln \left(\frac{x_q x_{\bar{q}}}{e^{2 \eta }}\right) \frac{ \left(\frac{x_{\bar{q}}}{x_{h_2}} \vec{p}_{h_2} - \vec{p}_{2}\right)^2 + x_q x_{\bar{q}} Q^2 }{ \left(\left(\frac{x_{\bar{q}}}{x_{h_2}} \vec{p}_{h_2} - \vec{p}_{2}\right)^2 + x_q x_{\bar{q}} Q^2\right) \left(\left(\frac{x_q}{x_{h_1}} \vec{p}_{h_1} - \vec{p}_{1}\right)^2 + x_q x_{\bar{q}} Q^2\right) - x_q x_{\bar{q}} Q^2 \vec{p}_{3}^2  } \\
& \left. \left. \times \ln \left(\frac{ \left(\left(\frac{x_{\bar{q}}}{x_{h_2}} \vec{p}_{h_2} - \vec{p}_{2}\right)^2 + x_q x_{\bar{q}} Q^2\right) \left(\left(\frac{x_q}{x_{h_1}} \vec{p}_{h_1} - \vec{p}_{1}\right)^2 + x_q x_{\bar{q}} Q^2\right)}{x_q x_{\bar{q}} Q^2 \vec{p}_{3}^2 }\right) \right) + (q \leftrightarrow \bar{q}) \right] \\
& \left. +  \left. \frac{1}{x_q x_{\bar{q}}} \left[ \int_{0}^{x_q} d z\left[\left(\phi_{5}^{i j} + \phi_{6}^{i j} \right)_{T T}\right]_{+} d z+ (q \leftrightarrow \bar{q}) \right]  \right \} +h.c \big |_{(p_1, p_3 \leftrightarrow p_{1'}, p_{3'}) ,  (i \leftrightarrow j)} \right]  \\
& + (h_1 \leftrightarrow h_2 )  \; .\numberthis
\end{align*} 

The $\phi$ function are defined in \ref{AppendixB}.

\subsection{Real corrections: Fragmentation from quark and anti-quark}
In this case, we refer to the finite terms related to the contribution (b) of Fig. \ref{fig:sigma-NLO}. In order to better understand what these contributions are, referring to Ref. \cite{Boussarie:2014lxa}, we recall that the impact factor for the transition $\gamma^{*} \rightarrow q  \bar{q} g$ has a double dipole contribution ($\Phi_4^{(+,i)}$) and a single dipole ($\Phi_3^{(+,i)}$) contribution. The finite contributions which we obtain are
\begin{itemize}
    \item Finite terms related to the dipole $\times$ dipole contribution. 
    \item Dipole $\times$ double dipole contribution. 
    \item Double dipole $\times$ double dipole contribution. 
\end{itemize}

\subsubsection{Finite part of dipole $\times$ dipole contribution}
When we square the dipole contribution, using shorthand notation introduced in (\ref{eq:ShortHand}), we obtain the following structure:
\begin{align*}
\Phi_3^\alpha(\vec{p}_1,\vec{p}_2) \Phi_3^{\beta*}(\vec{p}_{1'},\vec{p}_{2'}) & = \tilde{\Phi}_3^\alpha(\vec{p}_1,\vec{p}_2) \tilde{\Phi}_3^{\beta*}(\vec{p}_{1'},\vec{p}_{2'}) \\
& + \left( \tilde{\Phi}_3^\alpha(\vec{p}_1, \vec{p}_2) \Phi_4^{\beta *}(\vec{p}_{1'}, \vec{p}_{2'},\vec{0})+\Phi_4^\alpha(\vec{p}_1, \vec{p}_2,\vec{0}) \Phi_3^{\beta *}(\vec{p}_{1'},\vec{p}_{2'})\right)  \\ 
& + \Phi_4^\alpha(\vec{p}_1,\vec{p}_2,\vec{0}) \Phi_4^{\beta*}(\vec{p}_{1'},\vec{p}_{2'},\vec{0}) \,.
\numberthis[phi3_squared]
\end{align*}
The first term in the RHS is the one containing divergences that we have considered in previous sections. After isolating soft and collinear divergences, finite terms remain. The finite contributions for
$\tilde{\sigma}_{(b)div,1}$ and $\tilde{\sigma}_{(b)div,3}$ have been computed respectively in
sections \ref{sec:qqbarfragColl-qg} and \ref{sec:qqbarfragColl-qbarg}, see eqs.~\eqref{eq:collqg_LL_fin},
\eqref{eq:TL_coll_qg_finite},
\eqref{eq:collqg_TT_fin} and
\eqref{eq:collqbarg_LL_fin}
\eqref{eq:collqbarg_TL_fin}
\eqref{eq:collqbarg_TT_fin}.
Besides,
the terms $(\tilde{\sigma}_{(b)div,2}-\tilde{\sigma}_{(b)div,2}^{soft})$ and $(\tilde{\sigma}_{(b)div,4}-\tilde{\sigma}_{(b)div,4}^{soft})$ in eq. \eqref{sigmatilde_q-qbar} are finite. Their contribution read
\begin{align*}
& \frac{d \sigma_{3LL}^{q \bar{q} \rightarrow h_1 h_2}}{d x_{h_1} d x_{h_2} d^d p_{h_1\perp} d^d p_{h_2 \perp}}\Bigg |_{\text{finite, (b) 2,4}} \\*
&= \frac{\alpha_s C_F}{\mu^{2\epsilon}} \frac{4  \alpha_{\mathrm{em}} Q^2}{(2\pi)^{4(d-1)} N_c \; x_{h_1}^d x_{h_2}^d} \sum_{q} \int_{x_{h_1}}^1 \frac{d x_q}{x_q} \int_{x_{h_2}}^1 \frac{d x_{\bar{q}}}{ x_{\bar{q}}} \int_{\alpha}^1 \frac{d x_g}{x_g^{3-d}}  \left(x_qx_{\bar{q}} \right)^{d-1}  \delta(1-x_q-x_{\bar{q}}-x_g)  \\
& \times Q_q^2 D_q^{h_1}\left(\frac{x_{h_1}}{x_q},\mu_F\right)D_{\bar{q}}^{h_2}\left(\frac{x_{h_2}}{x_{\bar{q}}}, \mu_F\right)  \int \frac{d^d \vec{u}}{(2\pi)^d} \int d^d p_{1\perp} d^d p_{2\perp} \mathbf{F} \left(\frac{p_{12\perp}}{2}\right)  \\
& \times \int d^d p_{1'\perp}  d^d p_{2'\perp} \mathbf{F}^*\left( \frac{p_{1'2'\perp}}{2}\right) \delta\left(p_{1 1' \perp} + p_{2 2'\perp} \right) \\
& \times \left \{ \frac{8 x_q x_{\bar{q}} \; \delta \left(\frac{x_q}{x_{h_1}} p_{h_1\perp} - p_{1\perp} + \frac{x_{\bar{q}}}{x_{h_2}} p_{h_2\perp} - p_{2\perp} \right)}{\left(Q^2+ \frac{\left(\frac{x_{\bar{q}}}{x_{h_2}}\vec{p}_{h_2}-\vec{p}_{2'}\right)^2}{x_{\bar{q}} (1-x_{\bar{q}})}\right)\left(Q^2+ \frac{\left(\frac{x_q}{x_{h_1}} \vec{p}_{h_1}-\vec{p}_{1}\right)^2}{x_q (1-x_q)}\right)} \frac{\left(\vec{u}-\frac{\vec{p}_{h_1}}{x_{h_1}}\right)\cdot \left(\vec{u}-\frac{\vec{p}_{h_2}}{x_{h_2}}\right)}{\left(\vec{u}-\frac{\vec{p}_{h_1}}{x_{h_1}}\right)^2 \left(\vec{u}-\frac{\vec{p}_{h_2}}{x_{h_2}}\right)^2} \right. \\ 
& - \frac{(2 x_g - d x_g^2 + 4 x_q x_{\bar{q}}) \delta\left(\frac{x_q}{x_{h_1}} p_{h_1\perp} - p_{1\perp} + \frac{x_{\bar{q}}}{x_{h_2}} p_{h_2\perp} - p_{2\perp} + x_g u_{ \perp} \right)}{\left(Q^2+ \frac{\left(\frac{x_{\bar{q}}}{x_{h_2}}\vec{p}_{h_2}-\vec{p}_{2'}\right)^2}{x_{\bar{q}} (1-x_{\bar{q}})}\right)\left(Q^2+ \frac{\left(\frac{x_q}{x_{h_1}} \vec{p}_{h_1}-\vec{p}_{1}\right)^2}{x_q (1-x_q)}\right)}  \frac{\left(\vec{u}-\frac{\vec{p}_{h_1}}{x_{h_1}}\right)\cdot \left(\vec{u}-\frac{\vec{p}_{h_2}}{x_{h_2}}\right)}{\left(\vec{u}-\frac{\vec{p}_{h_1}}{x_{h_1}}\right)^2 \left(\vec{u}-\frac{\vec{p}_{h_2}}{x_{h_2}}\right)^2} \\
&   - \frac{(2 x_g - d x_g^2 + 4 x_q x_{\bar{q}}) \delta\left(\frac{x_q}{x_{h_1}} p_{h_1\perp} - p_{1\perp} + \frac{x_{\bar{q}}}{x_{h_2}}p_{h_2\perp} - p_{2\perp} + x_g u_{ \perp} \right) }{\left(Q^2+ \frac{\left(\frac{x_{\bar{q}}}{x_{h_2}}\vec{p}_{h_2}-\vec{p}_{2}\right)^2}{x_{\bar{q}} (1-x_{\bar{q}})}\right)\left(Q^2+ \frac{\left(\frac{x_q}{x_{h_1}} \vec{p}_{h_1}-\vec{p}_{1'}\right)^2}{x_q (1-x_q)}\right)} \\
& \times \left. \frac{\left(\vec{u}-\frac{\vec{p}_{h_1}}{x_{h_1}}\right)\cdot \left(\vec{u}-\frac{\vec{p}_{h_2}}{x_{h_2}}\right)}{\left(\vec{u}-\frac{\vec{p}_{h_1}}{x_{h_1}}\right)^2 \left(\vec{u}-\frac{\vec{p}_{h_2}}{x_{h_2}}\right)^2}
\right\} + (h_1 \leftrightarrow h_2) \; , \numberthis \\ 
\end{align*}
in the LL case. The same contribution in the TL and TT cases is, respectively,
\begin{align*}
& \frac{d \sigma_{3TL}^{q \bar{q} \rightarrow h_1 h_2}}{d x_{h_1} d x_{h_2} d^d p_{h_1\perp} d^d p_{h_2 \perp}}\Bigg |_{\text{finite, (b) 2,4}} \\*
&= \frac{\alpha_s C_F}{\mu^{2\epsilon}} \frac{2 \alpha_{\mathrm{em}}Q}{(2\pi)^{4(d-1)}N_c x_{h_1}^d x_{h_2}^d} \sum_{q} \int_{x_{h_1}}^1 \frac{d x_q}{x_q} \int_{x_{h_2}}^1 \frac{x_{\bar{q}}}{x_{\bar{q}}} \int_{\alpha}^1 \frac{d x_g}{x_g^{3-d}} (x_q x_{\bar{q}})^{d-1} \delta(1-x_q -x_{\bar{q}}-x_g) \\
&  \times  Q_q^2  D_q^{h_1}\left(\frac{x_{h_1}}{x_q}, \mu_F\right) D_{\bar{q}}^{h_2}\left(\frac{x_{h_2}}{ x_{\bar{q}}},\mu_F \right) \int \frac{d^d \vec{u}}{(2\pi)^d }   \int d^d p_{1 \perp} d^d p_{2 \perp} \mathbf{F}\left(\frac{p_{12 \perp}}{2}\right) \\
& \times  \int d^d p_{1' \perp } d^d p_{2' \perp} \mathbf{F}^*\left(\frac{p_{1'2' \perp}}{2}\right)   \delta \left(p_{11' \perp} + p_{22' \perp} \right) \varepsilon_{T i}^* \\
& \times \left\{   \frac{\delta \left( \frac{x_q}{x_{h_1}} p_{h_1 \perp} - p_{1 \perp} + \frac{x_{\bar{q}}}{x_{h_2}} p_{h_2 \perp} - p_{2 \perp} + x_g u_\perp \right)}{  \left(Q^2 + \frac{\left( \frac{x_{\bar{q}}}{x_{h_2}} \vec{p}_{h_2} - \vec{p}_{2'} \right)^2}{x_{\bar{q}} (1-x_{\bar{q}} )}\right) \left(Q^2 + \frac{\left(\frac{x_q}{x_{h_1}} \vec{p}_{h_1} -\vec{p}_1\right)^2}{x_q (1-x_q) } \right)}  \frac{\left(u_\perp - \frac{p_{h_1 \perp}}{x_{h_1}} \right)_\mu \left(u_\perp - \frac{p_{h_2 \perp}}{x_{h_2}} \right)_\nu}{\left(\vec{u}-\frac{\vec{p}_{h_1}}{x_{h_1}}\right)^2 \left(\vec{u}-\frac{\vec{p}_{h_2}}{x_{h_2}}\right)^2}\right.   \\
& \times \frac{1}{x_{\bar{q}} (x_q + x_g)} \left[ x_g (4 x_{\bar{q}} + d x_g -2) \left( \left(\frac{x_{\bar{q}}}{x_{h_2}} p_{h_2 \perp} - p_{2' \perp} \right)^\mu g_\perp^{i \nu} -  \left(\frac{x_{\bar{q}}}{x_{h_2}} p_{h_2 \perp} - p_{2' \perp} \right)^\nu g_\perp^{i \mu}   \right) \right. \\
& \left. - (2 x_{\bar{q}} -1 ) (4 x_{\bar{q}} x_q + x_g(2-x_g d)) g_\perp^{\mu \nu} \left( \frac{x_{\bar{q}}}{x_{h_2}} p_{h_2 \perp} - p_{2' \perp} \right)^i \right]  \\ 
& +   \frac{\delta \left( \frac{x_q}{x_{h_1}} p_{h_1 \perp} - p_{1 \perp} + \frac{x_{\bar{q}}}{x_{h_2}} p_{h_2 \perp} - p_{2 \perp} + x_g u_\perp \right)}{  \left(Q^2 + \frac{\left( \frac{x_q}{x_{h_1}} \vec{p}_{h_1} - \vec{p}_{1'} \right)^2}{x_q (1-x_q )}\right) \left(Q^2 + \frac{\left(\frac{x_{\bar{q}}}{x_{h_2}} \vec{p}_{h_2} -\vec{p}_2\right)^2}{x_{\bar{q}} (1-x_{\bar{q}}) } \right)}  \frac{\left(u_\perp - \frac{p_{h_2 \perp}}{x_{h_2}} \right)_\mu \left(u_\perp - \frac{p_{h_1 \perp}}{x_{h_1}} \right)_\nu}{\left(\vec{u}-\frac{\vec{p}_{h_1}}{x_{h_1}}\right)^2 \left(\vec{u}-\frac{\vec{p}_{h_2}}{x_{h_2}}\right)^2} \\
& \times \frac{1}{x_q (x_{\bar{q}} + x_g)} \left[ x_g (4 x_q + d x_g -2) \left( \left(\frac{x_q}{x_{h_1}} p_{h_1 \perp} - p_{1' \perp} \right)^\mu g_\perp^{i \nu} -  \left(\frac{x_q}{x_{h_1}} p_{h_1 \perp} - p_{1' \perp} \right)^\nu g_\perp^{i \mu}   \right) \right. \\
& \left. - (2 x_q -1 ) (4 x_{\bar{q}} x_q + x_g(2-x_g d)) g_\perp^{\mu \nu} \left( \frac{x_q}{x_{h_2}} p_{h_1 \perp} - p_{1' \perp} \right)^i \right]  \\ 
& - \frac{ 8 \; \delta \left(\frac{x_q }{x_{h_1}} p_{h_1 \perp} - p_{1 \perp} + \frac{x_{\bar{q}}}{x_{h_2}} p_{h_2 \perp} - p_{2 \perp}  \right) }{\left(Q^2 + \frac{\left( \frac{x_{\bar{q}}}{x_{h_2}} \vec{p}_{h_2} - \vec{p}_{2'} \right)^2}{x_{\bar{q}} (1-x_{\bar{q}} )}\right) \left(Q^2 + \frac{\left(\frac{x_q}{x_{h_1}} \vec{p}_{h_1} -\vec{p}_1\right)^2}{x_q (1-x_q) } \right)} \frac{\left( \vec{u} - \frac{\vec{p}_{h_1}}{x_{h_1}} \right) \cdot \left( \vec{u} - \frac{\vec{p}_{h_2}}{x_{h_2}} \right) }{\left( \vec{u} - \frac{\vec{p}_{h_1}}{x_{h_1}} \right)^2  \left( \vec{u} - \frac{\vec{p}_{h_2}}{x_{h_2}} \right)^2}  \\ 
& \times  \left. \left( \frac{x_{\bar{q}} }{x_{h_2}} p_{h_2 \perp} - p_{2' \perp} \right)^i (x_{\bar{q}} -x_q) \right\} + (h_1 \leftrightarrow h_2) \; ,
\end{align*}
and
\begin{align*}
& \frac{d \sigma_{3TT}^{q \bar{q} \rightarrow h_1 h_2}}{d x_{h_1} d x_{h_2} d^d p_{h_1\perp} d^d p_{h_2 \perp}}\Bigg |_{\text{finite, (b) 2,4}} \\*
&= \frac{\alpha_s C_F}{\mu^{2\epsilon}} \frac{\alpha_{\mathrm{em}} }{(2 \pi)^{4(d-1)} N_c x_{h_1}^d x_{h_2}^d } \sum_q \int_{x_{h_1}}^1 \frac{d x_q}{x_q} \int_{x_{h_2}}^1 \frac{d x_{\bar{q}}}{x_{\bar{q}}} \int_\alpha^1 \frac{d x_g}{x_g^{3-d}} \delta(1-x_q -x_{\bar{q}} -x_g) (x_q x_{\bar{q}})^{d-1} \\
& \times Q_q ^2  D_q^{h_1}\left(\frac{x_{h_1}}{x_q}, \mu_F\right) D_{\bar{q}}^{h_2}\left(\frac{x_{h_2} }{ x_{\bar{q}}},\mu_F \right) \int \frac{d^d \vec{u}}{(2\pi)^d} \int d^d p_{1 \perp} d^d p_{2 \perp}  \mathbf{F}\left(\frac{p_{12 \perp}}{2}\right) \\ 
& \times \int d^d p_{1' \perp} d^d p_{2' \perp}  \mathbf{F}^*\left(\frac{p_{1'2' \perp}}{2}\right) \delta (p_{11' \perp} + p_{22' \perp} ) \varepsilon_{T i} \varepsilon_{T k}^* \\ 
&  \times \left\{ \left[ \left( - \frac{ \delta \left( p_{q1\perp} + p_{\bar{q}2 \perp} + x_g u_\perp \right) }{\left(Q^2+\frac{\vec{p}_{\bar{q} 2}^2}{x_{\bar{q}}\left(1-x_{\bar{q}}\right)}\right)\left(Q^2+\frac{\vec{p}_{q 1^{\prime}}^2}{x_q\left(1-x_q\right)}\right)} \frac{\left(u_\perp - \frac{p_{q\perp}}{x_q} \right)_\mu \left(u_\perp - \frac{p_{\bar{q} \perp}}{x_{\bar{q}}} \right)_\nu}{\left(\vec{u} - \frac{\vec{p}_{q}}{x_q} \right)^2 \left(\vec{u} - \frac{p_{\bar{q}}}{x_{\bar{q}} } \right)^2 }   \right. \right.  \right. \\ 
& \times \frac{1}{(x_q + x_g)(x_{\bar{q}} + x_g) x_q x_{\bar{q}} } \left\{ x_g ((d-4)) x_g -2) \left[p_{q 1^{\prime} \perp}^\nu\left(p_{\bar{q} 2 \perp}^\mu g_{\perp}^{i k}+p_{\bar{q} 2 \perp}^k g_{\perp}^{\mu i}\right) \right. \right. \\
& \left. + g_{\perp}^{\mu \nu}\left(\left(\vec{p}_{q 1^{\prime}} \cdot \vec{p}_{\bar{q} 2}\right) g_{\perp}^{i k}+p_{q 1^{\prime} \perp}^i p_{\bar{q} 2 \perp}^k\right) -g_{\perp}^{\nu k} p_{q 1^{\prime} \perp}^i p_{\bar{q} 2 \perp}^\mu -g_{\perp}^{\mu i} g_{\perp}^{\nu k}\left(\vec{p}_{q 1^{\prime}} \cdot \vec{p}_{\bar{q} 2}\right) \right] -g_{\perp}^{\mu \nu} \\
& \times \hspace{-0.15 cm} \left[ \left(2x_q -1 \right) \left(2 x_{\bar{q}} - 1\right) p_{q1'\perp}^k p_{\bar{q}2\perp}^i \left( 4 x_q x_{\bar{q}} + x_g (2 - x_g d)\right)  + 4 x_q x_{\bar{q}} ((\vec{p}_{q1'} \cdot \vec{p}_{\bar{q}2})g_\perp^{ik} + p_{q1'\perp}^i p_{\bar{q}2\perp}^k  )\right] \\
& + \left( p_{q1'\perp}^\mu p_{\bar{q}2\perp}^\nu g_\perp^{ik} - p_{q1'\perp}^\mu p_{\bar{q}2\perp}^k g_\perp^{\nu i } - p_{q1'\perp}^i p_{\bar{q}2\perp}^\nu g_\perp^{\mu k } - g_\perp^{\mu k } g_\perp^{\nu i } (\vec{p}_{q1'} \cdot \vec{p}_{\bar{q}2} ) \right) \\ 
& \times x_g ((d-4)x_g + 2) + x_g (2x_{\bar{q}} - 1 ) (x_g d + 4 x_q -2 ) \left( g_\perp^{\mu k } p_{q1'\perp}^\nu - g_\perp^{\nu k} p_{q1'\perp}^\mu \right) p_{\bar{q}2\perp}^i \\
&   + \left. \left. \left. x_g (2 x_q -1 ) p_{q1'\perp}^k (4 x_{\bar{q}} + x_g d -2) \left( g_\perp^{\nu i } p_{\bar{q}2\perp}^\mu -g_\perp^{\nu k } p_{q1'\perp}^\nu \right) \right\}  \right) + (q \leftrightarrow \bar{q}) \right] \\
& + \frac{8 \; \delta(p_{q1 \perp} + p_{\bar{q}2\perp} ) }{\left(Q^2+\frac{\vec{p}_{\bar{q} 2}^2}{x_{\bar{q}}\left(1-x_{\bar{q}}\right)}\right)\left(Q^2+\frac{\vec{p}_{q 1^{\prime}}^2}{x_q\left(1-x_q\right)}\right)} \frac{\left( \vec{u} - \frac{\vec{p}_{q}}{x_q} \right) \cdot \left( \vec{u} - \frac{\vec{p}_{\bar{q}}}{x_{\bar{q}}} \right) }{\left( \vec{u} - \frac{\vec{p}_{q}}{x_q} \right)^2 \left( \vec{u} - \frac{\vec{p}_{\bar{q}}}{x_{\bar{q}}} \right)^2  } \\
& \times  \left.  \frac{1}{x_q x_{\bar{q}} } \left[ -(x_{\bar{q}} - x_q )^2 g_\perp^{ri} g_\perp^{kl} +  g_\perp^{il} g_\perp^{rk} -  g_\perp^{rl} g_\perp^{ik}\right] p_{\bar{q}2 \perp r} p_{q1' \perp l } 
\right\} + (h_1 \leftrightarrow h_2 ) \; .
\end{align*}
In the above expression, the following replacement needs to be done: 
\begin{equation}
    p_{q \perp} = \frac{x_q}{x_{h_1}} p_{h_1 \perp} \; , \hspace{0.5cm}  p_{\bar{q} \perp} = \frac{x_{\bar{q}}}{x_{h_2}} p_{h_2 \perp} \; .
\end{equation}
The remaining term in eq. \eqref{eq:phi3_squared}, for arbitrary polarization, is
\begin{align*}
   & \frac{d {\sigma}_{3JI}^{q \bar{q} \rightarrow h_1 h_2}}{d x_{h_1} d x_{h_2} d^d p_{h_1} d^d p_{h_2}} \\ & =  \frac{2 \alpha_s \alpha_{\mathrm{em}} C_F }{\mu^{2\epsilon}(2\pi)^{4(d-1)}N_c} \frac{(p_0^-)^2}{s^2 x_{h_1}^d x_{h_2}^d} \sum_q Q_q^2 \int_{x_{h_1}}^1 \frac{d x_q}{x_q} \int_{x_{h_2}}^{1}  \frac{d x_{\bar{q}}}{x_{\bar{q}}} \; (x_q x_{\bar{q}})^{d-1} D_q^{h_1} \left( \frac{x_{h_1}}{x_q} , \mu_F \right)  \\
    & \times D_{\bar{q}}^{h_2} \left( \frac{x_{h_2}}{x_{\bar{q}}}, \mu_F \right) \int_0^1 \frac{d x_g  }{x_g } \int \frac{d^d p_{g\perp}}{(2\pi)^d} \delta (1-x_q-x_{\bar{q}}-x_g) \int  d^d p_{1\perp} d^d p_{2\perp}   d^d p_{1'\perp} d^d p_{2'\perp}  \\
    & \times  \mathbf{F}\left(\frac{p_{12\perp}}{2}\right) \mathbf{F}^*\left(\frac{p_{1'2'\perp}}{2}\right)  \delta \left(\frac{x_q}{x_{h_1} } p_{h_1 \perp}-p_{1\perp} + \frac{x_{\bar{q}}}{x_{h_2}} p_{h_2 \perp} -  p_{2\perp} + p_{g\perp}\right) \delta (p_{11'\perp} + p_{22'\perp} ) \\
   &  \times \varepsilon_{I\alpha} \; \varepsilon_{J\beta}^* \left[ \Phi_3^\alpha (p_{1\perp}, p_{2\perp}) \Phi_3^{\beta*} (p_{1'\perp}, p_{2'\perp}) - \tilde{\Phi}_3^{\alpha} (p_{1\perp}, p_{2\perp}) \tilde{\Phi}_3^{\beta*} (p_{1'\perp}, p_{2'\perp}) \right] \\
   & + (h_1 \leftrightarrow h_2) \; . \numberthis
\end{align*}

\subsubsection{Dipole $\times$ double-dipole contribution and double-dipole $\times$ double-dipole contribution}

The dipole $\times$ double dipole contribution, for arbitrary polarization, is given by
\begin{align*}
& \frac{d\sigma_{4JI}^{q \bar{q} \rightarrow h_1 h_2}}{d x_{h_1} d x_{h_2} d^d p_{h_1 \perp} d^d p_{h_2 \perp}} \\  & = \frac{\alpha_{s}\alpha_{\mathrm{em}} }{\mu^{2\epsilon}(2\pi)^{4(d-1)}N_{c}}\frac{(p_{0}%
^{-})^{2}}{s^{2}x_{h_1}^d x_{h_2}^d} \sum_q Q_{q}^{2} \hspace{-0.1 cm} \int_{x_{h_1}}^1  \hspace{-0.25 cm} \frac{d x_q}{x_q} \hspace{-0.15 cm} \int_{x_{h_2}}^1 \hspace{-0.25 cm} \frac{d x_{\bar{q}}}{x_{\bar{q}}} \; (x_q x_{\bar{q}})^{d-1} D_q^{h_1}\left(\frac{x_{h_1}}{x_q},\mu_F\right)D_{\bar{q}}^{h_2}\left(\frac{x_{h_2}}{x_{\bar{q}}}, \mu_F\right) \\ 
& \times \int_0^1 \frac{d x_g }{x_g } \int \frac{d^{d} p_{g\bot}}{(2\pi)^{d}} \delta(1-x_q -x_{\bar{q}}-x_g) \int d^dp_{1\bot}d^dp_{2\bot} d^dp_{1\bot}^\prime d^dp_{2\bot}^\prime \frac{d^dp_{3\bot}d^dp_{3\bot}^\prime}{\left( 2\pi \right)^d} \\ & \times \delta\left(\frac{x_q}{x_{h_1}}p_{h_1\perp} - p_{1\bot}+ \frac{x_{\bar{q}}}{x_{h_2}} p_{h_2 \perp} -p_{2\bot}+p_{g3\bot}\right) \delta(p_{11^\prime \bot}+p_{22^\prime \bot}+p_{33^\prime \bot}) (\varepsilon_{I\alpha} \varepsilon_{J\beta}^\ast) \\
& \times \left[ \Phi_3^\alpha(p_{1\bot},p_{2\bot}) \Phi_4^{\beta\ast}(p_{1\bot}^\prime, p_{2\bot}^\prime, p_{3\bot}^\prime) \mathbf{F}\left(\frac{p_{12\bot}}{2}\right) \mathbf{\tilde{F}}^\ast \left( \frac{p_{1^\prime 2^\prime \bot}}{2}, p_{3\bot}^\prime \right) \delta(p_{3\bot}) \right. \numberthis \\
& + \left. \Phi_4^\alpha (p_{1\bot}, p_{2\bot}, p_{3\bot}) \Phi_3^{\beta\ast} \left(p_{1'\perp},p_{2'\perp} \right) \mathbf{\tilde{F}}\left(\frac{p_{12\bot}}{2}, p_{3\bot}\right) \mathbf{F}^\ast\left(\frac{p_{1^\prime 2^\prime\bot}}{2} \right) \delta(p_{3\bot}^\prime) \right] + (h_1 \leftrightarrow h_2) \; .
\end{align*}
The double-dipole $\times$ double-dipole contribution, for arbitrary polarization, is given by
\begin{align*}
& \frac{d\sigma_{5JI}^{q \bar{q} \rightarrow h_1 h_2}}{d x_{h_1} d x_{h_2} d^d p_{h_1} d^d p_{h_2}} \\ & = \frac{\alpha_{s} \alpha_{\mathrm{em}} (\varepsilon_{I\alpha} \varepsilon_{J\beta}^*) }{\mu^{2\epsilon}(2\pi)^{4(d-1)} (N_{c}^{2}-1)}\frac{(p_{0}^{-})^{2}}{s^{2}x_{h_1}^d x_{h_2}^d} \sum_q Q_{q}^{2} \int_{x_{h_1}}^1 \hspace{-0.15 cm} \frac{d x_q}{x_q} \hspace{-0.15 cm} \int_{x_{h_2}}^1 \hspace{-0.15 cm} \frac{d x_{\bar{q}}}{x_{\bar{q}}} \; (x_q x_{\bar{q}})^{d-1} D_q^{h_1}\left(\frac{x_{h_1}}{x_q},\mu_F\right) \\ 
& \times D_{\bar{q}}^{h_2}\left(\frac{x_{h_2}}{x_{\bar{q}}}, \mu_F\right) \int_0^1 \frac{d x_g }{x_g } \int \frac{d^{d}p_{g\bot}}{(2\pi)^{d}}\delta(1-x_q -x_{\bar{q}}-x_g) \int d^dp_{1\bot}d^dp_{2\bot} d^dp_{1\bot}^\prime d^dp_{2\bot}^\prime \\
& \times \int \frac{d^dp_{3\bot}d^dp_{3\bot}^\prime}{\left( 2\pi \right)^{2d}} \delta\left(\frac{x_q}{x_{h_1}}p_{h_1\perp} - p_{1\bot}+ \frac{x_{\bar{q}}}{x_{h_2}} p_{h_2 \perp} -p_{2\bot}+p_{g3\bot}\right) \delta(p_{11^\prime \bot}+p_{22^\prime \bot}+p_{33^\prime \bot})  \\
& \times \Phi_4^\alpha(p_{1\bot},p_{2\bot},p_{3\bot}) \Phi_4^{\beta\ast}(p_{1\bot}^\prime,p_{2\bot}^\prime, p_{3\bot}^\prime) \mathbf{\tilde{F}}\left( \frac{p_{12\bot}}{2}, p_{3\bot} \right) \mathbf{\tilde{F}}^\ast \left(\frac{p_{1^\prime 2^\prime \bot}}{2},p_{3\bot}^\prime\right)  \\
& + (h_1 \leftrightarrow h_2)   \; . \numberthis 
\end{align*}
Expression for the squared impact factors can be found in Appendix \ref{sec: appendixC}. They are written in terms of $p_q, p_{\bar{q}}, p_{g}, z$ and the following identification should be done:
\begin{equation}
    p_{q \perp} = \frac{x_q}{x_{h_1}} p_{h_1 \perp} \; , \hspace{0.5cm}  p_{\bar{q} \perp} = \frac{x_{\bar{q}}}{x_{h_2}} p_{h_2 \perp} \; , \hspace{0.5 cm} z=x_g \; .
\end{equation}

\subsection{Real corrections: Fragmentation from anti-quark and gluon}

\subsubsection{Finite part of dipole $\times$ dipole contribution}

When squaring the dipole contribution, we have also finite terms. This time we write separately for each polarization transition. In the LL case, we have
\begin{align*}
   & \frac{d {\sigma}_{3LL}^{g \bar{q} \rightarrow h_1 h_2}}{d x_{h_1} d x_{h_2} d^d p_{h_1} d^d p_{h_2}} \\ & =  \frac{2 \alpha_s \alpha_{\mathrm{em}} C_F }{\mu^{2\epsilon}(2\pi)^{4(d-1)}N_c} \frac{(p_0^-)^2}{s^2 x_{h_1}^d x_{h_2}^d} \sum_q Q_q^2 \int_{x_{h_1}}^{1} \frac{d x_g}{x_g} \int_{x_{h_2}}^{1} \frac{d x_{\bar{q}} }{x_{\bar{q}}} \; (x_g x_{\bar{q}})^{d-1} D_g^{h_1} \left( \frac{x_{h_1}}{x_g} , \mu_F \right) \\
    & \times  D_{\bar{q}}^{h_2} \left( \frac{x_{h_2}}{x_{\bar{q}}}, \mu_F \right) \int_0^1 \frac{d x_q  }{x_q } \int \frac{d^d p_{q\perp}}{(2\pi)^d}\delta (1-x_q-x_{\bar{q}}-x_g) \int d^d p_{1\perp} d^d p_{2\perp}  d^d p_{1'\perp} d^d p_{2'\perp}  \\
    & \times \mathbf{F}\left(\frac{p_{12\perp}}{2}\right)  \mathbf{F}^*\left(\frac{p_{1'2'\perp}}{2}\right)  \delta \left(p_{q1\perp} + \frac{x_{\bar{q}}}{x_{h_2}} p_{h_2 \perp} - p_{2\perp} + \frac{x_g}{x_{h_1}}p_{h_1 \perp}\right)    \delta (p_{11'\perp} + p_{22'\perp} ) \frac{Q^2}{(p_\gamma^+)^2}  \\
    & \times \left[ \Phi_3^{+} (p_{1\perp}, p_{2\perp}) \Phi_3^{+*} (p_{1'\perp}, p_{2'\perp})  - \frac{8 x_q x_{\bar{q}} (p_\gamma^+)^4 \left( d x_g^2 + 4 x_q (x_q + x_g) \right)}{\left(Q^2 + \frac{\left(\frac{x_{\bar{q}}}{x_{h_2}}\vec{p}_{h_2}- \vec{p}_{2}\right)^2}{x_{\bar{q}}(1-x_{\bar{q}})} \right) \left(Q^2 + \frac{\left(\frac{x_{\bar{q}}}{x_{h_2}}\vec{p}_{h_2}- \vec{p}_{2'}\right)^2}{x_{\bar{q}}(1-x_{\bar{q}})} \right)  } \right. \\
   &  \times \left.  \frac{1}{\left(x_q \frac{x_{g}}{x_{h_1}} \vec{p}_{h_1} -x_g \vec{p}_q \right)^2} \right] + (h_1 \leftrightarrow h_2) \; . \numberthis \\ 
\end{align*}
For TL case, we have %
\begin{align*}
   & \frac{d {\sigma}_{3TL}^{g \bar{q} \rightarrow h_1 h_2}}{d x_{h_1} d x_{h_2} d^d p_{h_1} d^d p_{h_2}} \\ & =  \frac{2\alpha_s \alpha_{\mathrm{em}} C_F }{\mu^{2\epsilon}(2\pi)^{4(d-1)}N_c} \frac{(p_0^-)^2}{s^2 x_{h_1}^d x_{h_2}^d} \sum_q Q_q^2 \int_{x_{h_1}}^{1}   \frac{d x_g}{x_g} \int_{x_{h_2}}^{1}  \frac{d x_{\bar{q}}}{x_{\bar{q}}} \; (x_g x_{\bar{q}})^{d-1} D_g^{h_1} \left( \frac{x_{h_1}}{x_g} , \mu_F \right)  \\
    & \times D_{\bar{q}}^{h_2} \left( \frac{x_{h_2}}{x_{\bar{q}}}, \mu_F \right) \int_0^1 \frac{d x_q}{x_q } \int \frac{d^d p_{q\perp}}{(2\pi)^d} \delta (1-x_q-x_{\bar{q}}-x_g) \int d^d p_{1\perp} d^d p_{2\perp}  d^d p_{1'\perp} d^d p_{2'\perp}\\
    & \times  \mathbf{F}\left(\frac{p_{12\perp}}{2}\right)  \mathbf{F}^*\left(\frac{p_{1'2'\perp}}{2}\right) \delta \left(p_{q1\perp} + \frac{x_{\bar{q}}}{x_{h_2}}p_{h_2 \perp} - p_{2\perp} + \frac{x_{g}}{x_{h_1}} p_{h_1\perp}\right) \delta (p_{11'\perp} + p_{22'\perp}) \varepsilon_{T i}^* \frac{Q}{p_\gamma^+} \\
   & \times \left[ \Phi_3^{+} (p_{1\perp}, p_{2\perp}) \Phi_3^{ i *} (p_{1'\perp}, p_{2'\perp})  + \frac{4 x_q \left(p_\gamma^{+}\right)^3\left(2 x_{\bar{q}} -1\right)\left(x_g^2 d+4 x_q \left(x_q +x_g\right)\right) }{\left(Q^2+\frac{\left(\frac{x_{\bar{q}}}{x_{h_2}}\vec{p}_{h_2}- \vec{p}_{2}\right)^2}{x_{\bar{q}}\left(1-x_{\bar{q}}\right)}\right)\left(Q^2+\frac{\left(\frac{x_{\bar{q}}}{x_{h_2}}\vec{p}_{h_2}- \vec{p}_{2'}\right)^2}{x_{\bar{q}}\left(1-x_{\bar{q}}\right)}\right)} \right.\\
   & \left. \times  \frac{\left( \frac{x_{\bar{q}}}{x_{h_2}} p_{h_2 \perp}-p_{ 2' \perp}\right)^i}{\left(x_q+x_g\right) \left(x_q \frac{x_{g}}{x_{h_1}} \vec{p}_{h_1}  - x_g \vec{p}_q\right)^2}  \right]   + (h_1 \leftrightarrow h_2) \; . \numberthis \\
\end{align*}
Finally, for TT case, we obtain %
\begin{align*}
   & \frac{d {\sigma}_{3TT}^{g \bar{q} \rightarrow h_1 h_2}}{d x_{h_1} d x_{h_2} d^d p_{h_1} d^d p_{h_2}} \\ & =  \frac{2 \alpha_s \alpha_{\mathrm{em}} C_F }{\mu^{2 \epsilon}(2\pi)^{4(d-1)}N_c} \frac{(p_0^-)^2}{s^2 x_{h_1}^d x_{h_2}^d} \sum_q Q_q^2 \int_{x_{h_1}}^{1}  \frac{d x_g}{x_g}\int_{x_{h_2}}^{1}  \frac{d x_{\bar{q}} }{ x_{\bar{q}}} \; (x_g x_{\bar{q}})^{d-1} D_g^{h_1} \left( \frac{x_{h_1}}{x_g} , \mu_F \right)  \\
    & \times D_{\bar{q}}^{h_2} \left( \frac{x_{h_2}}{x_{\bar{q}}}, \mu_F \right) \int_0^1 \frac{d x_q }{x_q} \int \frac{ d^d p_{q\perp}}{ (2\pi)^d} \delta (1-x_q-x_{\bar{q}}-x_g) \int d^d p_{1\perp} d^d p_{2\perp} \int d^d p_{1'\perp} d^d p_{2'\perp}  \\
    & \times \mathbf{F}\left(\frac{p_{12\perp}}{2}\right)   \mathbf{F}^*\left(\frac{p_{1'2'\perp}}{2}\right)  \delta \left(p_{q1\perp} + \frac{x_{\bar{q}}}{x_{h_2}} p_{h_2}-p_{2\perp} +\frac{x_g}{x_{h_1}} p_{h_1 \perp} \right) \delta (p_{11'\perp} + p_{22'\perp} ) \varepsilon_{T i} \; \varepsilon_{T k}^*  \\
   &  \times \left[  \Phi_3^{i} (p_{1\perp}, p_{2\perp}) \Phi_3^{k*} (p_{1'\perp}, p_{2'\perp}) - \frac{2 x_q (p_\gamma^+)^2 (x_g^2 d + 4x_q (x_q + x_g))}{\left(Q^2+\frac{\left(\frac{x_{\bar{q}}}{x_{h_2}}\vec{p}_{h_2}- \vec{p}_{2}\right)^2}{x_{\bar{q}}\left(1-x_{\bar{q}}\right)}\right) \left(Q^2+\frac{\left(\frac{x_{\bar{q}}}{x_{h_2}}\vec{p}_{h_2}- \vec{p}_{2'}\right)^2}{x_{\bar{q}}\left(1-x_{\bar{q}}\right)}\right)  } \right. \\
   & \times \left. \frac{\left( (1-2x_{\bar{q}})^2 g_{\perp}^{ri}  g_{\perp}^{lk} - g_{\perp}^{li} g_{\perp}^{rk} + g_{\perp}^{rl} g_{\perp}^{ik}  \right)\left( \frac{x_{\bar{q}}}{x_{h_2}} p_{h_2 } - p_{2}\right)_r \left( \frac{x_{\bar{q}}}{x_{h_2}}p_{h_2 } - p_{2'} \right)_l}{x_{\bar{q}} (x_q + x_g)^2\left(x_q\frac{x_{g}}{x_{h_1}} \vec{p}_{h_1}  - x_g \vec{p}_q \right)^2}  \right] \\
   & + (h_1 \leftrightarrow h_2) \; . \numberthis
\end{align*}

\subsubsection{Dipole $\times$ double-dipole contribution and double-dipole $\times$ double-dipole contribution}

The dipole $\times$ double dipole contribution, for arbitrary polarization, is given by
\begin{align*}
& \frac{d\sigma_{4JI}^{g \bar{q} \rightarrow h_1 h_2}}{d x_{h_1} d x_{h_2} d^d p_{h_1 \perp} d^d p_{h_2 \perp}} \\  & = \frac{\alpha_{s}\alpha_{\mathrm{em}} }{\mu^{2 \epsilon} (2\pi)^{4(d-1)}N_{c}}\frac{(p_{0}^{-})^{2}}{s^{2}x_{h_1}^d x_{h_2}^d} \sum_q Q_{q}^{2} \int_{x_{h_1}}^1 \frac{d x_g}{x_g} \int_{x_{h_2}}^1 \frac{d x_{\bar{q}} }{x_{\bar{q}} }\; (x_g x_{\bar{q}})^{d-1} D_g^{h_1}\left(\frac{x_{h_1}}{x_g},\mu_F\right)\\ 
& \times D_{\bar{q}}^{h_2}\left(\frac{x_{h_2}}{x_{\bar{q}}}, \mu_F\right)  \int_0^1 \frac{d x_q }{x_q } \int \frac{d^{d} p_{q\bot}}{(2\pi)^{d}} \delta(1-x_q -x_{\bar{q}}-x_g) \int d^dp_{1\bot}d^dp_{2\bot} d^dp_{1\bot}^\prime d^dp_{2\bot}^\prime \\ 
& \times \frac{d^dp_{3\bot}d^dp_{3\bot}^\prime}{\left( 2\pi \right)^d}  \delta\left(p_{q1\bot}+\frac{x_{\bar{q}}}{x_{h_2}} p_{h_2 \perp}-p_{2\bot}+\frac{x_g}{x_{h_1}}p_{h_1 \perp}-p_{3\bot}\right) \delta(p_{11^\prime \bot}+p_{22^\prime \bot}+p_{33^\prime \bot})  \\
& \times (\varepsilon_{I\alpha} \varepsilon_{J\beta}^\ast) \left[ \Phi_3^\alpha(p_{1\bot},p_{2\bot}) \Phi_4^{\beta\ast}(p_{1\bot}^\prime, p_{2\bot}^\prime, p_{3\bot}^\prime) \mathbf{F}\left(\frac{p_{12\bot}}{2}\right) \mathbf{\tilde{F}}^\ast \left( \frac{p_{1^\prime 2^\prime \bot}}{2}, p_{3\bot}^\prime \right) \delta(p_{3\bot}) \right. \numberthis \\
& + \left. \Phi_4^\alpha (p_{1\bot}, p_{2\bot}, p_{3\bot}) \Phi_3^{\beta\ast} \left( p_{1^\prime \perp}, p_{ 2^\prime \bot} \right) \mathbf{\tilde{F}}\left(\frac{p_{12\bot}}{2}, p_{3\bot}\right) \mathbf{F}^\ast\left(\frac{p_{1^\prime 2^\prime\bot}}{2} \right) \delta(p_{3\bot}^\prime) \right]  + (h_1 \leftrightarrow h_2) \; .
\end{align*}

The double-dipole $\times$ double-dipole contribution, for arbitrary polarization, is given by
\begin{align*}
& \frac{d\sigma_{5JI}^{g \bar{q} \rightarrow h_1 h_2}}{d x_{h_1} d x_{h_2} d^d p_{h_1} d^d p_{h_2}} \\ 
& = \frac{\alpha_{s} \alpha_{\mathrm{em}}}{\mu^{2\epsilon}(2\pi)^{4(d-1)} (N_{c}^{2}-1)}\frac{(p_{0}^{-})^{2}}{s^{2}x_{h_1}^d x_{h_2}^d} \sum_q Q_{q}^{2} \int_{x_{h_1}}^1  \frac{d x_g}{x_g} \int_{x_{h_2}}^{1}  \frac{d x_{\bar{q}} }{x_{\bar{q}}}\; (x_g x_{\bar{q}} )^{d-1} D_g^{h_1}\left(\frac{x_{h_1}}{x_q},\mu_F\right) \\
& \times D_{\bar{q}}^{h_2}\left(\frac{x_{h_2}}{x_{\bar{q}}}, \mu_F\right) \int_0^1 \frac{d x_q}{x_q } \int \frac{ d^{d}p_{q \bot}}{(2\pi)^{d}} \delta(1-x_q -x_{\bar{q}}-x_g) \int d^dp_{1\bot} d^dp_{2\bot} d^dp_{1\bot}^\prime d^dp_{2\bot}^\prime  \\
& \times  \int \frac{d^dp_{3\bot}d^dp_{3\bot}^\prime}{\left( 2\pi \right)^{2d}} \delta\left(p_{q1\bot}+\frac{x_{\bar{q}}}{x_{h_2}}p_{h_2 \perp}-p_{2\bot}+\frac{x_{g}}{x_{h_1}} p_{h_1 \perp}-p_{3\bot}\right) \delta(p_{11^\prime \bot}+p_{22^\prime \bot}+p_{33^\prime \bot}) \\
& \times  (\varepsilon_{I\alpha} \varepsilon_{J\beta}^*)  \Phi_4^\alpha(p_{1\bot},p_{2\bot},p_{3\bot}) \Phi_4^{\beta\ast}(p_{1\bot}^\prime,p_{2\bot}^\prime, p_{3\bot}^\prime) \mathbf{\tilde{F}}\left( \frac{p_{12\bot}}{2}, p_{3\bot} \right) \mathbf{\tilde{F}}^\ast \left(\frac{p_{1^\prime 2^\prime \bot}}{2},p_{3\bot}^\prime\right) \\
& + (h_1 \leftrightarrow h_2) \; .  \numberthis
\end{align*}

Expression for the squared impact factors can be found in Appendix \ref{sec: appendixC}. They are written in terms of $p_q, p_{\bar{q}}, p_{g}, z$ and the following identification should be done:
\begin{equation}
    p_{g \perp} = \frac{x_g}{x_{h_1}} p_{h_1 \perp} \; , \hspace{0.5cm}  p_{\bar{q} \perp} = \frac{x_{\bar{q}}}{x_{h_2}} p_{h_2 \perp} \; , \hspace{0.5 cm} z=x_g \; .
\end{equation}

\subsection{Real corrections: Fragmentation from quark and gluon}

\subsubsection{Finite part of dipole $\times$ dipole contribution}

When squaring the dipole contribution, one also gets finite terms. This time we write separately for each polarization transition. In the LL case, we have
\begin{align*}
   & \frac{d {\sigma}_{3LL}^{q g \rightarrow h_1 h_2}}{d x_{h_1} d x_{h_2} d^d p_{h_1} d^d p_{h_2}} \\ & =  \frac{2 \alpha_s \alpha_{\mathrm{em}} C_F }{\mu^{2 \epsilon}(2\pi)^{4(d-1)}N_c} \frac{(p_0^-)^2}{s^2 x_{h_1}^d x_{h_2}^d} \sum_q Q_q^2 \int_{x_{h_1}}^{1}   \frac{d x_q}{x_q} \int_{x_{h_2}}^{1}  \frac{d x_{g}}{x_g} \; (x_q x_{g})^{d-1} D_q^{h_1} \left( \frac{x_{h_1}}{x_q} , \mu_F \right) \\
    & \times D_{g}^{h_2} \left( \frac{x_{h_2}}{x_{\bar{q}}}, \mu_F \right)  \int_0^1 \frac{d x_{\bar{q}}  }{x_{\bar{q}}} \int \frac{d^d p_{\bar{q}\perp}}{ (2\pi)^d} \delta (1-x_q-x_{\bar{q}}-x_g) \int d^d p_{1\perp} d^d p_{2\perp} \int d^d p_{1'\perp} d^d p_{2'\perp}    \\
    & \times \mathbf{F}\left(\frac{p_{12\perp}}{2}\right) \mathbf{F}^*\left(\frac{p_{1'2'\perp}}{2}\right) \delta \left( \frac{x_q}{x_{h_1}} p_{h_1 \perp}-p_{1\perp} + p_{\bar{q}2\perp} + \frac{x_g}{x_{h_2}} p_{h_2 \perp}\right) \delta (p_{11'\perp} + p_{22'\perp}) \frac{Q^2}{(p_{\gamma}^+)^2} \\
   & \times \left[ \Phi_3^{+} (p_{1\perp}, p_{2\perp}) \Phi_3^{+*} (p_{1'\perp}, p_{2'\perp}) - \frac{8 x_q x_{\bar{q}} (p_\gamma^+)^4 \left( d x_g^2 + 4 x_{\bar{q}} (x_{\bar{q}} + x_g) \right)}{\left(Q^2 + \frac{\left(\frac{x_q}{x_{h_1}} \vec{p}_{h_1}-\vec{p}_{ 1}\right)^2}{x_q(1-x_{q})} \right) \left(Q^2 + \frac{\left( \frac{x_q}{x_{h_1}}\vec{p}_{h_1}-\vec{p}_{1'}\right)^2}{x_{q}(1-x_{q})} \right) } \right. \\
   & \times \left. \frac{1}{\left(x_{\bar{q}} \frac{x_g}{x_{h_2}} \vec{p}_{h_2}-x_g \vec{p}_{\bar{q}}\right)^2 }  \right]+ (h_1 \leftrightarrow h_2) \; . \numberthis
\end{align*}
For TL case, we have  %
\begin{align*}
   & \frac{d {\sigma}_{3TL}^{q g \rightarrow h_1 h_2}}{d x_{h_1} d x_{h_2} d^d p_{h_1} d^d p_{h_2}} \\ & =  \frac{2 \alpha_s \alpha_{\mathrm{em}} C_F }{\mu^{2\epsilon}(2\pi)^{4(d-1)}N_c} \frac{(p_0^-)^2}{s^2 x_{h_1}^d x_{h_2}^d} \sum_q Q_q^2 \int_{x_{h_1}}^{1}   \frac{d x_q }{x_q} \int_{x_{h_2}}^{1}  \frac{d x_{g}}{x_g} \; (x_q x_g)^{d-1} D_q^{h_1} \left( \frac{x_{h_1}}{x_q} , \mu_F \right)  \\
    & \times D_{g}^{h_2} \left( \frac{x_{h_2}}{x_{g}}, \mu_F \right) \int_0^1 \frac{d x_{\bar{q}}  }{x_{\bar{q}} } \int \frac{d^d p_{\bar{q}\perp}}{(2\pi)^d}\delta (1-x_q-x_{\bar{q}}-x_g) \int d^d p_{1\perp} d^d p_{2\perp}  d^d p_{1'\perp} d^d p_{2'\perp}   \\
    & \times \mathbf{F}\left(\frac{p_{12\perp}}{2}\right) \mathbf{F}^*\left(\frac{p_{1'2'\perp}}{2}\right) \delta \left( \frac{x_q}{x_{h_1}} p_{h_1 \perp}-p_{1\perp} + p_{\bar{q}2\perp} + \frac{x_g}{x_{h_2}} p_{h_2 \perp}\right) \delta (p_{11'\perp} + p_{22'\perp}) \varepsilon_{T i}^* \frac{Q}{p_{\gamma}^+}  \\
   & \times \left[  \Phi_3^{+} (p_{1\perp}, p_{2\perp}) \Phi_3^{i*} (p_{1'\perp}, p_{2'\perp})  + \frac{4 x_{\bar{q}} \left(p_\gamma^{+}\right)^3\left(2 x_{q} -1\right)\left(x_g^2 d+4 x_{\bar{q}} \left(x_{\bar{q}} +x_g\right)\right) }{\left(Q^2+\frac{\left( \frac{x_q}{x_{h_1}}\vec{p}_{h_1}-\vec{p}_{ 1}\right)^2}{x_{q}\left(1-x_{q}\right)}\right)\left(Q^2+\frac{\left( \frac{x_q}{x_{h_1}}\vec{p}_{h_1}-\vec{p}_{ 1'}\right)^2}{x_{q}\left(1-x_{q}\right)}\right) }  \right.\\
   & \times \left. \frac{\left(\frac{x_q}{x_{h_1}}p_{h_1}- p_{ 1' }\right)^i}{\left(x_{\bar{q}}+x_g\right)\left(x_{\bar{q}} \frac{x_g}{x_{h_2}} \vec{p}_{h_2}- x_g \vec{p}_{\bar{q}}\right)^2} \right] + (h_1 \leftrightarrow h_2) \; . \numberthis
\end{align*}
Finally, for TT case, we obtain
\begin{align*}
   & \frac{d {\sigma}_{3TT}^{q g \rightarrow h_1 h_2}}{d x_{h_1} d x_{h_2} d^d p_{h_1} d^d p_{h_2}} \\*
   & =  \frac{2 \alpha_s \alpha_{\mathrm{em}} C_F }{\mu^{2 \epsilon} (2\pi)^{4(d-1)}N_c} \frac{(p_0^-)^2}{s^2 x_{h_1}^d x_{h_2}^d} \sum_q Q_q^2 \int_{x_{h_1}}^{1}   \frac{d x_q}{x_q} \int_{x_{h_2}}^{1}  \frac{d x_{g}}{x_g} \; (x_q x_{g})^{d-1} D_q^{h_1} \left( \frac{x_{h_1}}{x_q} , \mu_F \right)  \\
    & \times D_{g}^{h_2} \left( \frac{x_{h_2}}{x_{g}}, \mu_F \right) \int_0^1 \frac{d x_{\bar{q}}  }{x_{\bar{q}} } \int \frac{d^d p_{\bar{q}\perp}}{(2\pi)^d} \delta (1-x_q-x_{\bar{q}}-x_g) \int d^d p_{1\perp} d^d p_{2\perp} \int d^d p_{1'\perp} d^d p_{2'\perp} \\
    & \times   \mathbf{F}\left(\frac{p_{12\perp}}{2}\right)  \mathbf{F}^*\left(\frac{p_{1'2'\perp}}{2}\right) \delta \left( \frac{x_q}{x_{h_1}} p_{h_1 \perp}-p_{1\perp} + p_{\bar{q}2\perp} + \frac{x_g}{x_{h_2}} p_{h_2 \perp}\right)  \delta (p_{11'\perp} + p_{22'\perp} ) \varepsilon_{T i} \; \varepsilon_{T k}^*  \\
   &\times \left[ \Phi_3^{i} (p_{1\perp}, p_{2\perp}) \Phi_3^{k*} (p_{1'\perp}, p_{2'\perp}) - \frac{2 x_{\bar{q}} (p_\gamma^+)^2 (x_g^2 d + 4x_{\bar{q}} (x_{\bar{q}} + x_g))  }{ \left(Q^2 + \frac{\left( \frac{x_q}{x_{h_1}} \vec{p}_{h_1}- \vec{p}_{1}\right)^2 }{x_{q} (1 - x_{q})} \right) \left(Q^2 + \frac{\left( \frac{x_q}{x_{h_1}} \vec{p}_{h_1}- \vec{p}_{1'}\right)^2}{x_{q} (1 - x_{q})} \right)  }  \right. \\
   & \left. \times \frac{\left((1-2 x_q)^2 g_\perp^{ir} g_\perp^{lk} -  g_\perp^{il} g_\perp^{rl} +  g_\perp^{ik} g_\perp^{lr}\right) \left( \frac{x_q}{x_{h_1}} p_{h_1} - p_{1}\right)_r \left( \frac{x_q}{x_{h_1}} p_{h_1} - p_{1'}\right)_l}{x_{q} (x_{\bar{q}} + x_g)^2 \left(x_{\bar{q}} \frac{x_g}{x_{h_2}} \vec{p}_{h_2}- x_g \vec{p}_{\bar{q}} \right)^2}\right] \\
   & + (h_1 \leftrightarrow h_2) \; . \numberthis
\end{align*}

\subsubsection{Dipole $\times$ double-dipole contribution and double-dipole $\times$ double-dipole contribution}

The dipole $\times$ double dipole contribution is given, for arbitrary polarization, by
\begin{align*}
& \frac{d\sigma_{4JI}^{q \bar{q} \rightarrow h_1 h_2}}{d x_{h_1} d x_{h_2} d^d p_{h_1 \perp} d^d p_{h_2 \perp}} \\  & = \frac{\alpha_{s}\alpha_{\mathrm{em}} }{\mu^{2 \epsilon}(2\pi)^{4(d-1)}N_{c}}\frac{(p_{0}^{-})^{2}}{s^{2}x_{h_1}^d x_{h_2}^d} \sum_q Q_{q}^{2} \int_{x_{h_1}}^1  \frac{d x_q}{x_q}  \int_{x_{h_2}}^1  \frac{d x_g }{x_g} \; (x_q x_{g})^{d-1} D_q^{h_1}\left(\frac{x_{h_1}}{x_q},\mu_F\right) \\ 
& \times D_{g}^{h_2}\left(\frac{x_{h_2}}{x_g}, \mu_F\right) \int_0^1 \frac{d x_{\bar{q}} }{x_{\bar{q}}} \int \frac{d^{d} p_{ \bar{q} \bot}}{ (2\pi)^{d}} \delta(1-x_q -x_{\bar{q}}-x_g) \int d^dp_{1\bot}d^dp_{2\bot} d^dp_{1\bot}^\prime d^dp_{2\bot}^\prime  \\ 
& \times \frac{d^dp_{3\bot}d^dp_{3\bot}^\prime}{\left( 2\pi \right)^d} \delta\left(\frac{x_q}{x_{h_1}} p_{h_1 \perp}-p_{1\bot}+p_{\bar{q}2\bot}+ \frac{x_{g}}{x_{h_2}}p_{h_2 \perp}-p_{3\bot}\right) \delta(p_{11^\prime \bot}+p_{22^\prime \bot}+p_{33^\prime \bot})  \\
& \times (\varepsilon_{I\alpha} \varepsilon_{J\beta}^\ast) \left[ \Phi_3^\alpha(p_{1\bot},p_{2\bot}) \Phi_4^{\beta\ast}(p_{1\bot}^\prime, p_{2\bot}^\prime, p_{3\bot}^\prime) \mathbf{F}\left(\frac{p_{12\bot}}{2}\right) \mathbf{\tilde{F}}^\ast \left( \frac{p_{1^\prime 2^\prime \bot}}{2}, p_{3\bot}^\prime \right) \delta(p_{3\bot}) \right. \numberthis \\
& + \left. \Phi_4^\alpha (p_{1\bot}, p_{2\bot}, p_{3\bot}) \Phi_3^{\beta\ast} \left( p_{1'\perp}, p_{2'\perp}\right) \mathbf{\tilde{F}}\left(\frac{p_{12\bot}}{2}, p_{3\bot}\right) \mathbf{F}^\ast\left(\frac{p_{1^\prime 2^\prime\bot}}{2} \right) \delta(p_{3\bot}^\prime) \right]  + (h_1 \leftrightarrow h_2) \; .
\end{align*}
The double-dipole $\times$ double-dipole contribution is given, for arbitrary polarization, by
\begin{align*}
& \frac{d\sigma_{5JI}^{q \bar{q} \rightarrow h_1 h_2}}{d x_{h_1} d x_{h_2} d^d p_{h_1} d^d p_{h_2}} \\* 
& = \frac{\alpha_{s} \alpha_{\mathrm{em}}}{\mu^{2\epsilon}(2\pi)^{4(d-1)} (N_{c}^{2}-1)}\frac{(p_{0}^{-})^{2}}{s^{2}x_{h_1}^d x_{h_2}^d} \sum_q Q_{q}^{2} \int_{x_{h_1}}^1 \frac{d x_q } {x_q} \int_{x_{h_2}}^1  \frac{d x_{g}}{x_g} \; (x_q x_{g})^{d-1} D_q^{h_1}\left(\frac{x_{h_1}}{x_q},\mu_F\right) \\
& \times D_{g}^{h_2}\left(\frac{x_{h_2}}{x_{g}}, \mu_F\right) \int_0^1 \frac{d x_{\bar{q}} }{x_{\bar{q}} } \int \frac{d^{d}p_{ \bar{q} \bot}}{(2\pi)^{d}} \delta(1-x_q -x_{\bar{q}}-x_g) \int d^dp_{1\bot}d^dp_{2\bot} d^dp_{1\bot}^\prime d^dp_{2\bot}^\prime \\
& \times  \int \frac{d^dp_{3\bot}d^dp_{3\bot}^\prime}{\left( 2\pi \right)^{2d}}  \delta \left( \frac{x_q}{x_{h_1}} p_{h_1 \perp}-p_{1\bot}+p_{\bar{q}2\bot}+ \frac{x_g}{x_{h_2}} p_{h_2 \perp}-p_{3\bot}\right) \delta(p_{11^\prime \bot}+p_{22^\prime \bot}+p_{33^\prime \bot})  \\
& \times  (\varepsilon_{I\alpha} \varepsilon_{J\beta}^*)  \Phi_4^\alpha(p_{1\bot},p_{2\bot},p_{3\bot}) \Phi_4^{\beta\ast}(p_{1\bot}^\prime,p_{2\bot}^\prime, p_{3\bot}^\prime) \mathbf{\tilde{F}}\left( \frac{p_{12\bot}}{2}, p_{3\bot} \right) \mathbf{\tilde{F}}^\ast \left(\frac{p_{1^\prime 2^\prime \bot}}{2},p_{3\bot}^\prime\right) \\ 
& + (h_1 \leftrightarrow h_2) \numberthis \; .
\end{align*}
Expression for the squared impact factors can be found in Appendix \ref{sec: appendixC}. They are written in terms of $p_q, p_{\bar{q}}, p_{g}, z$ and the following identification should be done:
\begin{equation}
    p_{q \perp} = \frac{x_q}{x_{h_1}} p_{h_1 \perp} \; , \hspace{0.5cm}  p_{g \perp} = \frac{x_g}{x_{h_2}} p_{h_2 \perp} \; , \hspace{0.5 cm} z=x_g \; .
\end{equation}

\section{Summary and Conclusion}

In this work, we have considered, for the first time at NLO, the diffractive production of a pair of hadrons at large $p_T$, in $\gamma^{(*)}$ nucleon/nucleus scattering, in the most general kinematics. 

This new class of processes provides an  access to precision physics of gluon saturation dynamics, with very promising future phenomenological studies both at the LHC in UPC (in photoproduction) and at the future EIC (both in photoproduction and leptoproduction). Our main result is the explicit finite result for the cross-section at NLO, obtained after showing explicitly the cancellation of rapidity divergences (through the B-JIMWLK equation), soft divergences and collinear divergences between real, virtual contributions, and  DGLAP evolution equation governing fragmentation functions.
Finite contributions and purely divergent contributions have been separated and the sum of the latter has been shown to be zero. Hence, the collection of all terms labeled with "fin" in sections~\ref{sec:CounterTerms}, \ref{sec: VirtualDiv}, \ref{sec: RealDiv}, plus all the formulas in section~\ref{sec:AdditionalFin} give the final and main result of this paper.

This full NLO result adds a new piece in the list of processes which are very promising to probe gluonic saturation in nucleons and nuclei at NLO, including inclusive DIS~\cite{Beuf:2022ndu}, inclusive photoproduction of dijets~\cite{Altinoluk:2020qet,Taels:2022tza},  
photon-dijet production in DIS~\cite{Roy:2019hwr}, dijets in DIS~\cite{Caucal:2021ent,Caucal:2022ulg}, single hadron~\cite{Bergabo:2022zhe}  and dihadrons production in DIS~\cite{Bergabo:2022tcu,Iancu:2022gpw}, diffractive exclusive dijets~\cite{Boussarie:2014lxa,Boussarie:2016ogo,Boussarie:2019ero} and exclusive light meson production~\cite{Boussarie:2016bkq,Mantysaari:2022bsp}, exclusive quarkonium production~\cite{Mantysaari:2021ryb,Mantysaari:2022kdm}, and inclusive DDIS~\cite{Beuf:2022kyp}.

\acknowledgments

E.~L. and S.~W. thank Charlotte Van Hulse and Ronan McNulty for early discussions which motivated the present work.
We thank Renaud Boussarie, Michel Fontannaz, Saad Nabeebaccus, Maxim A.~Nefedov, Alessandro Papa and Farid Salazar for many useful discussions.

This  project  has  received  funding  from  the  European  Union’s  Horizon  2020  research  and  innovation program under grant agreement STRONG–2020 (WP 13 "NA-Small-x").

The  work of  L.~S. is  supported  by  the  grant  2019/33/B/ST2/02588  of  the  National  Science Center  in  Poland. L.~S. thanks the P2IO Laboratory
of Excellence (Programme Investissements d'Avenir ANR-10-LABEX-0038) and the P2I - Graduate School of Physics of Paris-Saclay University for support.
This work was also partly supported by the French CNRS via the GDR QCD.

M.~F.~thanks IJCLab for support during the time this work has been done.

\appendix

\section{LO impact factor squared }
The impact factors in the LL, TL and TT cases are respectively given by
\begin{align}
\sum_{\lambda_q,\lambda_{\bar{q}}} \Phi_0^+(\vec{p}_1,\vec{p}_2)\Phi_0^{+*}(\vec{p}_{1'},\vec{p}_{2'}) = \frac{32 (p_\gamma^+)^4 x_q^3 x_{\bar{q}}^3 }{\left(\vec{p}_{q1}^2 + x_q x_{\bar{q}} Q^2 \right) \left(\vec{p}_{q1'}^2 + x_q x_{\bar{q}} Q^2 \right)} \; ,
\end{align}

\begin{align}
\sum_{\lambda_q,\lambda_{\bar{q}}} \Phi_0^+(\vec{p}_1,\vec{p}_2) \Phi_0^{i*}(\vec{p}_{1'},\vec{p}_{2'}) = \frac{16 (p_\gamma^+)^3 x_q^2 x_{\bar{q}}^2 p_{q1\perp}^i(1-2x_q)}{\left(\vec{p}_{q1}^2 + x_q x_{\bar{q}} Q^2 \right) \left(\vec{p}_{q1'}^2 + x_q x_{\bar{q}} Q^2 \right)} \; ,
\end{align}

\begin{align}
 \sum_{\lambda_q,\lambda_{\bar{q}}} \Phi_0^{i} (\vec{p}_1,\vec{p}_2)\Phi_0^{k*}(\vec{p}_{1'},\vec{p}_{2'}) = \frac{8 (p_\gamma^+)^2 x_q x_{\bar{q}} \left[(1-2x_q)^2 g_\perp^{ri} g_\perp^{lk} - g_\perp^{rk} g_\perp^{li}+ g_\perp^{rl} g_\perp^{rl}g_\perp^{ik}\right] p_{q1\perp r} p_{q1'\perp l}}{\left(\vec{p}_{q1}^2 + x_q x_{\bar{q}} Q^2 \right) \left(\vec{p}_{q1'}^2 + x_q x_{\bar{q}} Q^2 \right)} \; .
\end{align}
The LT case is immediately obtained from TL by complex conjugation and $1,2 \leftrightarrow 1',2'$ substitution.

\section{Finite parts of virtual corrections}
\label{AppendixB}

\subsection{Building blocks integrals}
\label{sec:building_block}

\begin{eqnarray}
I_{1}^{k}(\vec{q}_1,\, \vec{q}_2,\, \Delta_1,\, \Delta_2) & \equiv & \frac{1}{\pi}\int\frac{d^{d}\vec{l}\left(l_{\perp}^{k}\right)}{\left[(\vec{l}-\vec{q}_{1})^{2}+\Delta_{1}\right]\left[(\vec{l}-\vec{q}_{2})^{2}+\Delta_{2}\right]\vec{l}^{^{\, \, 2}}} \label{I1k}, \\
I_2(\vec{q}_1,\, \vec{q}_2,\, \Delta_1,\, \Delta_2) & \equiv & \frac{1}{\pi}\int \frac{d^d \vec{l}}{\left[ (\vec{l}-\vec{q}_1)^2+\Delta_1 \right] \left[ (\vec{l}-\vec{q}_2)^2 +\Delta_2 \right]} \label{I2}, \\
I_3^k(\vec{q}_1,\, \vec{q}_2,\, \Delta_1,\, \Delta_2) & \equiv & \frac{1}{\pi}\int \frac{d^d \vec{l}\left( l_\bot^k \right)}{\left[ (\vec{l}-\vec{q}_1)^2+\Delta_1 \right] \left[ (\vec{l}-\vec{q}_2)^2 +\Delta_2 \right]} \label{I3k}, \\
I^{jk}(\vec{q}_1,\, \vec{q}_2,\, \Delta_1,\, \Delta_2) & \equiv & \frac{1}{\pi}\int\frac{d^{d}\vec{l}\left( l_{\perp}^j l_{\perp}^{k}\right)}{\left[(\vec{l}-\vec{q}_{1})^{2}+\Delta_{1}\right]\left[(\vec{l}-\vec{q}_{2})^{2}+\Delta_{2}\right]\vec{l}^{^{\, \, 2}}} \label{Ijk} \, .
\end{eqnarray}
The arguments of these integrals will be different for each diagram so we will write them explicitly before giving the expression of each diagram, but we will omit them in the equations for the reader's convenience. \\
Explicit results for the first 3 integrals in (\ref{I1k}-\ref{Ijk}) are obtained by a straightforward Feynman parameter integration. We will express them using the following variables :

\begin{eqnarray}
\rho_{1} & \equiv & \frac{\left(\vec{q}_{12}^{\, \, 2}+\Delta_{12}\right)-\sqrt{\left(\vec{q}_{12}^{\, \, 2}+\Delta_{12}\right)^{2}+4\vec{q}_{12}^{\, \, 2}\Delta_{2}}}{2\vec{q}_{12}^{\, \, 2}} ,\\
\rho_{2} & \equiv & \frac{\left(\vec{q}_{12}^{\, \, 2}+\Delta_{12}\right)+\sqrt{\left(\vec{q}_{12}^{\, \, 2}+\Delta_{12}\right)^{2}+4\vec{q}_{12}^{\, \, 2}\Delta_{2}}}{2\vec{q}_{12}^{\, \, 2}} \, , \label{rhovar}
\end{eqnarray}
where $\Delta_{ij} = \Delta_i - \Delta_j$ . \\
One gets :

\begin{eqnarray}
I_{1}^{k} & = & \frac{q_{1\perp}^{k}}{2\left[\vec{q}_{12}^{\, \, 2}\left(\vec{q}_{1}^{\, \, 2}+\Delta_{1}\right)\left(\vec{q}_{2}^{\, \, 2}+\Delta_{2}\right)-\left(\vec{q}_{1}^{\, \, 2}-\vec{q}_{2}^{\, \, 2}+\Delta_{12}\right)\left(\vec{q}_{1}^{\, \, 2}\Delta_{2}-\vec{q}_{2}^{\, \, 2}\Delta_{1}\right)\right]}\\ \nonumber
 & \times & \left\{ \frac{\left(\vec{q}_{2}^{\, \, 2}+\Delta_{2}\right)\vec{q}_{12}^{\, \, 2}+\vec{q}_{2}^{\, \, 2}\left(\Delta_{1}+\Delta_{2}\right)+\Delta_{2}\left(\Delta_{21}-2\vec{q}_{1}^{\, \, 2}\right)}{\left(\rho_{1}-\rho_{2}\right)\vec{q}_{12}^{\, \, 2}}\ln\left[\left(\frac{-\rho_{1}}{1-\rho_{1}}\right)\left(\frac{1-\rho_{2}}{-\rho_{2}}\right)\right]\right.\\ \nonumber
 & \times & \left.\left(\vec{q}_{2}^{\, \, 2}+\Delta_{2}\right)\ln\left[\frac{\Delta_{2}\left(\vec{q}_{1}^{\, \, 2}+\Delta_{1}\right)^{2}}{\Delta_{1}\left(\vec{q}_{2}^{\, \, 2}+\Delta_{2}\right)^{2}}\right]+\left(1\leftrightarrow2\right)\right\} \, ,
\end{eqnarray}

\begin{eqnarray}
I_{2} & = & \frac{1}{\vec{q}_{12}^{\, \, 2}\left(\rho_{1}-\rho_{2}\right)}\ln\left[\left(\frac{-\rho_{1}}{1-\rho_{1}}\right)\left(\frac{1-\rho_{2}}{-\rho_{2}}\right)\right] \, ,
\end{eqnarray}
and

\begin{eqnarray}
I_{3}^{k} & = & \frac{\left(\vec{q}_{12}^{\, \, 2}+\Delta_{12}\right)q_{1}^{k}+\left(\vec{q}_{21}^{\, \, 2}+\Delta_{21}\right)q_{2}^{k}}{2\left(\rho_{1}-\rho_{2}\right)(\vec{q}_{12}^{\, \, 2})^2}\ln\left[\left(\frac{-\rho_{1}}{1-\rho_{1}}\right)\left(\frac{1-\rho_{2}}{-\rho_{2}}\right)\right] \nonumber \\ &-& \frac{q_{12}^{k}}{2\vec{q}_{12}^{\, \, 2}}\ln\left(\frac{\Delta_{1}}{\Delta_{2}}\right) \, .
\end{eqnarray}
Please note that in some cases the real part of $\Delta_1$ or $\Delta_2$ will be negative so the previous results can acquire an imaginary part from the imaginary part $\pm \, i0$ of the arguments. \\ 
The last integral in (\ref{Ijk}) can be expressed in terms of the other ones by writing 
\begin{equation}
I^{jk} = I_{11}\left(q_{1\perp}^{j}q_{1\perp}^{k}\right)+I_{12}\left(q_{1\perp}^{j}q_{2\perp}^{k}+q_{2\perp}^{j}q_{1\perp}^{k}\right)+I_{22}\left(q_{2\perp}^{j}q_{2\perp}^{k}\right) \, ,
\end{equation}
with
\begin{align}
I_{11} & = -\frac{1}{2}\frac{\left[\vec{q}_{2}^{\, \, 2}q_{1\perp k}-\left(\vec{q}_{1}\cdot\vec{q}_{2}\right)q_{2\perp k}\right]}{\left[\vec{q}_{1}^{\, \, 2}\vec{q}_{2}^{\, \, 2}-\left(\vec{q}_{1} \cdot \vec{q}_{2}\right)^{2}\right]^{2}} \\ \nonumber
& \hspace{-0.25 cm} \times  \left[\left(\frac{\vec{q}_{1}\cdot\vec{q}_{2}}{\vec{q}_{1}^{\, \, 2}}\right)\ln\left(\frac{\vec{q}_{1}^{\, \, 2}+\Delta_{1}}{\Delta_{1}}\right)q_{1\perp}^{k}+\left(\vec{q}_{2}\cdot\vec{q}_{21}\right)I_{3}^{k}+\left\{ \vec{q}_{2}^{\, \, 2}\left(\vec{q}_{1}\cdot\vec{q}_{12}\right)+\Delta_{1}\vec{q}_{2}^{\, \, 2}-\Delta_{2}\left(\vec{q}_{1}\cdot\vec{q}_{2}\right)\right\} I_{1}^{k}\right]\\
I_{12} & = \frac{-1}{4\left[\vec{q}_{1}^{\, \, 2} \vec{q}_{2}^{\, \, 2} -\left(\vec{q}_1 \cdot \vec{q}_2\right)^2\right]} \ln \left(\frac{\vec{q}_{1}^{\, \, 2}+\Delta_1}{\Delta_1}\right)  \nonumber \\
&+ \frac{\vec{q}_{2}^{\, \, 2} \left(\vec{q}_1 \cdot \vec{q}_2\right)}{2\left[\vec{q}_{1}^{\, \, 2} \vec{q}_{2}^{\, \, 2} -\left(\vec{q}_1 \cdot \vec{q}_2\right)^2\right]^2}\left[\left(\vec{q}_{1}^{\, \, 2} +\Delta_1\right)\left(q_{1 \perp k} I_1^k\right)+\left(q_{1 \perp k} I_3^k\right)\right]  \nonumber\\
& -\frac{\left(\vec{q}_{1}^{\, \, 2} \vec{q}_{2}^{\, \, 2} \right)+\left(\vec{q}_1 \cdot \vec{q}_2\right)^2}{4\left[\vec{q}_{1}^{\, \, 2} \vec{q}_{2}^{\, \, 2} -\left(\vec{q}_1 \cdot \vec{q}_2\right)^2\right]^2}\left[\left(\vec{q}_{2}^{\, \, 2} +\Delta_2\right)\left(q_{1 \perp k} I_1^k\right)+\left(q_{1 \perp k} I_3^k\right)\right]+(1 \leftrightarrow 2), \\
I_{22}&  = I_{11}|_{1 \leftrightarrow 2} \, .
\end{align}

In what follows, for the $\phi$ function, $x=x_q$, $\bar{x} = x_{\bar{q}}$.

\subsection{$\phi_4$}

The arguments in the integrals of \ref{sec:building_block} are 
\begin{eqnarray*}
\vec{q}_{1} & = & \vec{p}_{1}-\left(\frac{x-z}{x}\right)\vec{p}_{q}, \quad \, \, \, \, \,
\vec{q}_{2}  =  \left(\frac{x-z}{x}\right)\left(x\vec{p}_{\bar{q}}-\bar{x}\vec{p}_{q}\right) \, ,\\
\Delta_{1} & = & \left(x-z\right)\left(\bar{x}+z\right)Q^{2}, \quad
\Delta_{2}  =  -\frac{x\left(\bar{x}+z\right)}{\bar{x}\left(x-z\right)}\vec{q}^{2}-i0\,.
\end{eqnarray*}
Let us write the impact factors in terms of these variables. \\They read: \vspace{0.2 cm} \\
(longitudinal NLO) $\times$ (longitudinal LO) contribution :
\begin{equation}
\left(\phi_{4}\right)_{LL}=-\frac{4(x-z)(\bar{x}+z)}{z}[-\bar{x}(x-z)(z+1)I_{2}+q_{2\bot k}(2x^{2}-(2x-z)(z+1))I_{1}^{k}] \, ,
\end{equation}
(longitudinal NLO) $\times$ (transverse LO) contribution :
\begin{equation}
\left(\phi_{4}\right)_{LT}^{j}=(1-2x)p_{q1^{\prime}}{}_{\bot}^{j}\left(\phi_{4}\right)_{LL}-4(x-z)(\bar{x}+z)(1-2x+z)[(\vec{q}\cdot\vec{p}_{q1^{\prime}})g_{\bot k}^{j}+q_{2\bot}^{j}p_{q1^{\prime}\bot k}]I_{1}^{k} \, ,
\end{equation}
(transverse NLO) $\times$ (longitudinal LO) contribution :
\begin{align}
\left(\phi_{4}\right)_{TL}^{i} & =2\{[(x-\bar{x}-z)q_{2\bot}^{i}q_{1\bot k}+(-8x\bar{x}-6xz+2z^{2}+3z+1)q_{1\perp}^{j}q_{2\bot k}]I_{1}^{k}\nonumber \\
 & -2[4x^{2}-x(3z+5)+(z+1)^{2}]q_{2\bot k}I^{ik}+(x-\bar{x}-z)\left(\vec{q}_{2}\cdot\vec{q}_{1}\right)I^{i}\nonumber \\
 & +I_{2}[(x-\bar{x}-z)q_{2\bot}^{i}+\bar{x}(2(x-z)^{2}-5x+3z+1)q_{1\perp}^{i}]\nonumber \\
 & -\bar{x}[2(x-z)^{2}-5x+3z+1]I_{3}^{i}\nonumber \\
 & +\frac{x\bar{x}(1-2x)}{z}[2q_{2\bot k}I^{ik}+I_{3}^{i}-q_{1\perp}^{i}(2q_{2\bot k}I_{1}^{k}+I_{2})]\} \, ,
\end{align}
(transverse NLO) $\times$ (transverse LO) contribution :
\begin{eqnarray} \nonumber
\left(\phi_{4}\right)_{TT}^{ij} & = & \left[(x-\bar{x}-2z)(x-\bar{x}-z)(\vec{q}_{2}\cdot\vec{p}_{q1^{\prime}})q_{1\perp}^{i}+(z+1)(\left(\vec{q}_{1}\cdot\vec{q}_{2}\right)p_{q1^{\prime}\perp}^{i}-(\vec{q}_{1}\cdot\vec{p}_{q1^{\prime}})q_{2\bot}^{i})\right]I_{1}^{j}\\ \nonumber
 &+& 2\bar{x}[q_{2\bot k}-(x-z)q_{1\perp k}](p_{q1^{\prime}\bot}^{i}I^{jk}-g_{\bot}^{ij}p_{q1^{\prime}\bot l}I^{kl}) \\ \nonumber
 &+& 2(x-z)[(2\bar{x}+z)(\vec{q}_{2}\cdot\vec{p}_{q1^{\prime}})-\bar{x}(\vec{q}_{1}\cdot\vec{p}_{q1^{\prime}})]I^{ij}\\ \nonumber
 &+& [(1-z)((\vec{q}_{1}\cdot\vec{p}_{q1^{\prime}})q_{2\bot}^{j}-(\vec{q}_{2}\cdot\vec{p}_{q1^{\prime}})q_{1\perp}^{j})-(1-2x)(\bar{x}-x+z)\left(\vec{q}_{1}\cdot\vec{q}_{2}\right)p_{q1^{\prime}\perp}^{j}]I_{1}^{i}\\ \nonumber
 &-& 2\left[(x-z)(\bar{x}q_{1\perp}^{j}-(2\bar{x}+z)q_{2\bot}^{j})p_{q1^{\prime}\perp k} \right. \\ \nonumber 
 &+& \left. (1-2x)\left(4x^{2}-(3z+5)x+(z+1)^{2}\right)q_{2\bot k}p_{q1^{\prime}}{}_{\bot}^{j}\right]I^{ik}\\ \nonumber
 &-& \bar{x}\left(\bar{x}-x\right)\left(2(x-z)^{2}-5x+3z+1\right)p_{q1^{\prime}\perp}^{j}I_{3}^{i} \\ \nonumber
 &+& \bar{x}\left(\bar{x}+z\right)(p_{q1^{\prime}\perp}^{i}I_{3}^{j}-g_{\bot}^{ij}p_{q1^{\prime}\perp k}I_{3}^{k})\\ \nonumber
 &+& I_{2}\left[g_{\bot}^{ij}\left((1-z)(\vec{q}_{2}\cdot\vec{p}_{q1^{\prime}})-\bar{x}(1+x-z)(\vec{q}_{1}\cdot\vec{p}_{q1^{\prime}})\right) \right. \\ \nonumber 
 &+& \left.((1-z)q_{2\bot}^{j}-\bar{x}(1+x-z)q_{1\perp}^{j})p_{q1^{\prime}}{}_{\bot}^{i}\right.\\ \nonumber
 &-& \left.(\bar{x}-x)\left((\bar{x}-x+z)q_{2\bot}^{i}-\bar{x}\left(2(x-z)^{2}-5x+3z+1\right)q_{1\perp}^{i}\right)p_{q1^{\prime}}{}_{\bot}^{j}\right]\\ \nonumber
 &+& I_{1}^{k}\left[g_{\bot}^{ij}\left((x-\bar{x}+z)(\vec{q}_{1}\cdot\vec{p}_{q1^{\prime}})q_{2\bot k}+(1-z)(\vec{q}_{2}\cdot\vec{p}_{q1^{\prime}})q_{1\bot k}-(z+1)\left(\vec{q}_{1}\cdot\vec{q}_{2}\right)p_{q1^{\prime}}{}_{\bot k}\right)\right.\\ \nonumber
 &+& q_{1\perp}^{j}((x-\bar{x}+z)q_{2\bot k}p_{q1^{\prime}\perp}^{i}-(z+1)q_{2\bot}^{i}p_{q1^{\prime}}{}_{\bot k})\\ \nonumber
 &+& q_{2\bot}^{j}((x-\bar{x}-2z)(x-\bar{x}-z)q_{1\perp}^{i}p_{q1^{\prime}\perp k}+(1-z)q_{1\perp k}p_{q1^{\prime}}{}_{\bot}^{i})\\ \nonumber
&-&\left.(1-2x)((1-2x+z)q_{2\bot}^{i}q_{1\perp k}-(2z^{2}+3z-x(8\bar{x}+6z)+1)q_{1\perp}^{i}q_{2\bot k})p_{q1^{\prime}}{}_{\bot}^{j}\right]\\ \nonumber
 &+& \frac{x\bar{x}}{z}\left[(x-\bar{x})^{2}p_{q1^{\prime}\perp}^{j}(2q_{2\bot k}I^{ik}+I_{3}^{i}-q_{1\perp}^{i}(I_{2}+2q_{2\bot k}I_{1}^{k}))\right.\\ \nonumber
 &+& p_{q1^{\prime}\perp}^{i}(q_{1\perp}^{j}(I_{2}+2q_{2\bot k}I_{1}^{k})-2q_{2\bot k}I^{jk}-I_{3}^{j})\\
 &+& \left.g_{\bot}^{ij}((\vec{q}_{1}\cdot\vec{p}_{q1^{\prime}})(I_{2}+2q_{2\bot k}I_{1}^{k})+p_{q1^{\prime}\perp k}(2q_{2\bot l}I^{kl}+I_{3}^{k}))\right]\, .
\end{eqnarray}

\subsection{$\phi_5$}

Here the integrals from \ref{sec:building_block} will have the following arguments :

\begin{equation}
\vec{q}_1 = \left( \frac{x-z}{x} \right) \vec{p}_3 -\frac{z}{x}\vec{p}_1, \quad \vec{q}_2 = \vec{p}_{q1} - \frac{z}{x}\vec{p}_q \, ,\label{var1D5}
\end{equation}
\begin{equation}
 \Delta_1 = \frac{z(x-z)}{x^2\bar{x}} (\vec{p}_{\bar{q}2}^{\, \, 2}+ x\bar{x}Q^2), \quad  \Delta_2 = (x-z)(\bar{x}+z)Q^2 \label{var2D5}\, ,
\end{equation}
With such variables, it is easy to see that the argument in the square roots in (\ref{rhovar}) are full squares.
In terms of the variables in (\ref{var1D5}), the impact factors read: \vspace{0.2 cm} \\
(longitudinal NLO) $\times$ (longitudinal LO) : 
\begin{eqnarray}
\left(\phi_{5}\right)_{LL}=\frac{4(x-z)(-2x(\bar{x}+z)+z^{2}+z)}{xz}\left[\bar{x}(x-z)I_{2}-\left(zq_{1\perp k}-x\left(\bar{x}+z\right)q_{2\bot k}\right)I_{1}^{k}\right] \, ,
\end{eqnarray}
(longitudinal NLO) $\times$ (transverse LO) : 
\begin{align}
\left(\phi_{5}\right)_{LT}^{j} & =(\bar{x}-x)p_{q1^{\prime}\bot}^{j}\left(\phi_{5}\right)_{LL} \\ \nonumber &+\frac{4(x-z)(x-\bar{x}-z)}{x}\left(zq_{1\perp}^{k}-x(\bar{x}+z)q_{2\perp}^{k}\right)p_{q1^{\prime}\perp l}\left(g_{\perp k}^{j}I_{1}^{l}+I_{1}^{j}\right)\, ,
\end{align}
(transverse NLO) $\times$ (longitudinal LO) : 
\begin{align}
\left(\phi_{5}\right)_{TL}^{i} & =2\left[(x-\bar{x}-z)\left(\vec{q}_{1}\cdot\vec{q}_{2}\right)-\bar{x}(x-z)^{2}Q^{2}+(\frac{z}{x}-x)\vec{q}_{1}^{\,\,2}\right]I_{1}^{i}\nonumber \\
 & +\frac{2}{x}\left[xq_{2\bot k}(-8x\bar{x}-6xz+2z^{2}+3z+1)+2q_{1\bot k}(2xz-2x^{2}+x-z^{2})\right]q_{1\perp}^{i}I_{1}^{k}\nonumber \\
 & +2q_{2\bot}^{i}q_{1\perp k}(x-\bar{x}-z)I_{1}^{k}+2\frac{\bar{x}}{x}(x(8x-3)-6xz+2z^{2}+z)I_{1}^{i}\nonumber \\
 & +\frac{2}{x}\left[xq_{2\bot}^{i}(x-\bar{x}-z)+q_{1\perp}^{i}(8x^{3}-6x^{2}(z+2)+x(z+3)(2z+1)-2z^{2})\right]I_{2}\nonumber \\
 & -\frac{4}{x}\left[(x-z)(\bar{x}+z)q_{1\perp k}+x(4x^{2}-x(3z+5)+(z+1)^{2})q_{2\perp k}\right]I^{ik}\nonumber \\
 & -\frac{4}{z}x\bar{x}(x-\bar{x})\left[q_{2\perp k}I^{ik}+I_{3}^{i}-q_{1\perp}^{i}\left(q_{2\perp k}I_{1}^{k}+I_{2}\right)\right] \, ,
\end{align}
(transverse NLO) $\times$ (transverse LO) : 

\begin{align*}
& \left(\phi_{5}\right)_{TT}^{ij}  =  -2(x-z)\left[\frac{z}{x}(\vec{q}_{1}\cdot\vec{p}_{q1^{\prime}})-(2\bar{x}+z)(\vec{q}_{2}\cdot\vec{p}_{q1^{\prime}})\right]I^{ij}\\ \nonumber
 & +  \left[-\bar{x}(x-z)^{2}Q^{2}p_{q1^{\prime}\perp}^{i}+(\bar{x}-x+2z)(\bar{x}-x+z)(\vec{q}_{2}\cdot\vec{p}_{q1^{\prime}})q_{1\perp}^{i}\right.\\
 & - \left.(\vec{q}_{1}\cdot\vec{p}_{q1^{\prime}})((z+1)q_{2\bot}^{i}-2\frac{z}{x}(2x-z)q_{1\perp}^{i}) \right. \\ \nonumber
 &+ \left. ((z+1)\left(\vec{q}_{1}\cdot\vec{q}_{2}\right)-\left(x+\frac{z}{x}\right)\vec{r}^{\,\,2})p_{q1^{\prime}\bot}^{i}\right]I_{1}^{j}\\ \nonumber
 & -  2\frac{\bar{x}}{x}(xq_{2\perp k}+(x-z)q_{1\perp k})\left(g_{\bot}^{ij}p_{q1^{\prime}\perp l}I^{kl}-p_{q1^{\prime}}{}_{\bot}^{i}I^{jk}\right)\\ \nonumber
 & + \left[\bar{x}\left(x-\bar{x}\right)(x-z)^{2}Q^{2}p_{q1^{\prime}\perp}^{j}-(z-1)(\vec{q}_{1}\cdot\vec{p}_{q1^{\prime}})q_{2\bot}^{j}\right.\\  \nonumber
 & +  \left.(z-1)(\vec{q}_{2}\cdot\vec{p}_{q1^{\prime}})q_{1\perp}^{j}+\frac{x-\bar{x}}{x}\left((x^{2}-z)\vec{q}_{1}^{\,\,2}+x(\bar{x}-x+z)(\vec{q}_{1}\cdot\vec{q}_{2})\right)p_{q1^{\prime}\perp}^{k}\right]I_{1}^{i}\\ \nonumber
 & +  2\left[\frac{x-\bar{x}}{x}\left(x(4x^{2}-(3z+5)x+(z+1)^{2})q_{2\bot k}+(x-z)(\bar{x}+z)q_{1\perp k}\right)p_{q1^{\prime}\perp}^{j}\right.\\
 & -  \left.\frac{x-z}{x}\left(x(2x-z-2)q_{2\bot}^{j}+zq_{1\perp}^{j}\right)p_{q1^{\prime}\perp k}\right]I^{ik} \\ \nonumber 
 & +  \frac{\bar{x}\left(\bar{x}-x\right)}{x}\left(2z^{2}-6xz+z+x(8x-3)\right)p_{q1^{\prime}\perp}^{j}I_{3}^{i}\\ \nonumber
 & + \left[(x-\bar{x})\left((\bar{x}-x+z)q_{2\bot}^{i}+\left(6(z+2)x-8x^{2}-(z+3)(2z+1)+2\frac{z^{2}}{x}\right)q_{1\perp}^{i}r_{\bot}^{i}\right)p_{q1^{\prime}\perp}^{j}\right.\\ \nonumber
 & +  \left.(1-z)(g_{\bot}^{ij}(\vec{q}_{2}\cdot\vec{p}_{q1^{\prime}})+q_{2\bot}^{k}p_{q1^{\prime}\perp}^{i})+(2x+z-3)(g_{\bot}^{ik}(\vec{q}_{1}\cdot\vec{p}_{q1^{\prime}})+q_{1\perp}^{k}p_{q1^{\prime}\perp}^{i})\right]I_{2}\\ \nonumber
 & +  \left(3\bar{x}+z-\frac{z}{x}\right)p_{q1^{\prime}\perp}^{i}I_{3}^{k}-\frac{\bar{x}}{x}(3x-z)g_{\bot}^{ij}p_{q1^{\prime}\perp k}I_{3}^{k}\\
 & +  \left[(x-\bar{x})p_{q1^{\prime}\perp}^{j}\left\{ (\bar{x}-x+z)q_{2\bot}^{i}q_{1\perp k}-(2z^{2}-6xz+3z-8x\bar{x}+1)q_{2\perp k}q_{1\perp}^{i}\right.\right.\\ \nonumber
 & -  \left. 2(\bar{x}-x+2z-\frac{z^{2}}{x})q_{1\perp k}q_{1\perp}^{i}\right\} +\bar{x}(x-z)^{2}Q^{2}g_{\bot}^{ij}p_{q1^{\prime}\perp k} \\ \nonumber 
 &+ (1-z)q_{1\perp k}(g_{\bot}^{ij}(\vec{q}_{2}\cdot\vec{p}_{q1^{\prime}})+q_{2\bot}^{j}p_{q1^{\prime}\perp}^{i})\\ \nonumber
 & +  \left((x-\bar{x}+z)q_{2\bot k}-2q_{1\perp k}\right)(g_{\bot}^{ij}(\vec{q}_{1}\cdot\vec{p}_{q1^{\prime}})+q_{1\perp}^{j}p_{q1^{\prime}\perp}^{i}) \\ \nonumber 
 &+ g_{\bot}^{ij}\left(\left(x+\frac{z}{x}\right)\vec{q}_{1}^{\,\,2}-(z+1)(\vec{q}_{1}\cdot\vec{q}_{2})\right)p_{q1^{\prime}\perp k}\\ \nonumber
 & + \left.\left((x-\bar{x}-2z)(x-\bar{x}-z)q_{1\perp}^{i}q_{2\perp}^{j}-(z+1)q_{2\perp}^{i}q_{1\perp}^{j}+2(2x-z)\frac{z}{x}q_{1\perp}^{i}q_{1\perp}^{j}\right)p_{q1^{\prime}\perp k}\right]I_{1}^{k}\\ \nonumber
 & +  \frac{2x\bar{x}}{z}\left[(x-\bar{x})^{2}p_{q1^{\prime}\perp}^{j}(q_{2\bot k}I^{ik}+I_{3}^{i})-p_{q1^{\prime}\perp}^{i}(q_{2\bot k}I^{jk}+I_{3}^{k})+g_{\bot}^{ij}p_{q1^{\prime}\bot k}(q_{2\bot l}I^{kl}+I_{3}^{k})\right.\\ \nonumber
 & +  \left.(I_{2}+q_{2\perp k}I_{1}^{k})\left(g_{\bot}^{ij}(\vec{q}_{1}\cdot\vec{p}_{q1^{\prime}})+q_{1\perp}^{j}p_{q1^{\prime}\perp}^{i}-(1-2x)^{2}q_{1\perp}^{i}p_{q1^{\prime}\perp}^{j}\right)\right]. \numberthis
\end{align*}

\subsection{$\phi_6$}
We will use the variable
\begin{equation}
\vec{q}=\left( \frac{x-z}{x} \right) \vec{p}_{3}-\frac{z}{x}\vec{p}_1 \, .
\end{equation}
(longitudinal NLO) $\times $ (longitudinal NLO) :
\begin{equation}
(\phi_6)_{LL}=-4x\bar{x}^2 J_0 \, ,
\end{equation}
(longitudinal NLO) $\times $ (transverse NLO) :
\begin{equation}
(\phi_6)_{LT}^j = (1-2x)p_{q1^\prime\bot}^j(\phi_6)_{LL}\, ,
\end{equation}
(transverse NLO) $\times $ (longitudinal NLO) :
\begin{equation}
(\phi_6)_{TL}^i = 2\bar{x}\left[ (1-2x) p_{\bar{q}2\bot}^{i} J_0 - J_{1\bot}^i \right] \, ,
\end{equation}
(transverse NLO) $\times $ (transverse NLO) :
\begin{align}
(\phi_6)_{TT}^{ij} &  =\bar{x}\left[(x-\bar{x})^{2}p_{\bar{q}2\bot}^{i}%
p_{q1^{\prime}\bot}^{j}-g_{\bot}^{ij}(\vec{p}_{\bar{q}2}
\cdot\vec{p}_{q1^{\prime}})-p_{q1^{\prime}\bot}^{i}p_{\bar{q}2\bot}^{j}\right]J_0 \nonumber\\
&  +\bar{x} \left[(x-\bar{x})p_{q1^{\prime}\bot}^{j}g_{\bot k}^i - p_{q1^\prime\bot k}g_{\bot}^{ij}+p_{q1^{\prime
}\bot}^{i}g_{\bot k}^j\right]J_{1\bot}^k \, .
\end{align}
We introduced
\begin{align}
J_{1\bot}^{k}  &  =\frac{(x-z)^{2}}{x^{2}}\frac{q_{\bot}^{k}}{\vec
{q}^{\,\,2}}\ln\left(  \frac{\vec{p}_{\bar{q}2}^{\,\,2}+x\bar{x}Q^{2}}{\vec{p}_{\bar{q}2}^{\,\,2}+x\bar{x}Q^{2}+	\frac{x^2 \bar{x}}{z(x-z)}\vec{q}^{\,\,2}%
}\right)  ,\\ \nonumber
\mathrm{and} \\
J_0 &  =\frac{z}{x(\vec{p}_{\bar{q}2}^{\,\,2}+x\bar{x}Q^{2})}  -\frac{2x(x-z)+z^{2}}{xz(\vec{p}_{\bar{q}2}^{\,\,2}+x\bar{x}Q^{2})}
\ln\left(  \frac{x^2\bar{x}\mu^{2}}{z(x-z)(\vec{p}_{\bar{q}2}^{\,\,2}+x\bar{x}Q^{2})+x^{2}\bar{x}\vec{q}^{\,\,2}}\right)  .
\end{align}

\section{Finite part of the squared impact factors for real corrections  }
\label{sec: appendixC}
 \subsection{LL transition}
The double-dipole $\times$ double-dipole contribution is
\begin{align}
\label{phi4plus_squared}
\Phi &  _{4}^{+}(p_{1\bot},p_{2\bot},p_{3\bot})\Phi_{4}^{+*}(p_{1\bot}^{\prime
},p_{2\bot}^{\prime},p_{3\bot}^{\prime})=\frac{8p_{\gamma}^{+}{}^{4}%
}{z^{2}\left(  \frac{\vec{p}_{\bar{q}2^\prime}^{\, \, 2}%
}{x_{\bar{q}}\left( 1 - x_{\bar{q}} \right) }+Q^{2}\right)  \left( Q^2+\frac{\vec{p}_{q1^{\prime}}^{\,\,2}}{x_{q}%
} + \frac{\vec{p}_{\bar{q}2^\prime}^{\, \, 2}}{x_{\bar{q}}}+\frac
{\vec{p}_{g3^{\prime}}^{\,\,2}}{z}\right)  }\nonumber\\
\times &  \left[  \frac{x_{\bar{q}}\left(  dz^{2}+4x_{q}\left(  x_{q}%
+z\right)  \right)  \left(  x_{q}\vec{p}_{g3}-z\vec{p}_{q1})(x_{q}\vec
{p}_{g3^{\prime}}-z\vec{p}_{q1^{\prime}}\right)  }{x_{q}\left(  x_{q}%
+z\right)  ^{2}{}\left(  \frac{(\vec{p}_{g3}+\vec{p}_{q1}){}^{2}}{x_{\bar{q}%
}\left(  x_{q}+z\right)  }+Q^{2}\right)  \left(  \frac{(\vec{p}_{g3}+\vec
{p}_{q1}){}^{2}}{x_{\bar{q}}}+\frac{\vec{p}_{g3}^{\,\,2}}{z}+\frac{\vec{p}_{q1}%
^{\,\,2}}{x_{q}}+Q^{2}\right)  }\right. \nonumber\\
-  &  \left.  \frac{(4x_{q}x_{\bar{q}}+2z-dz^{2})(x_{\bar{q}}\vec{p}%
_{g3}-z\vec{p}_{\bar{q}2})(x_{q}\vec{p}_{g3^{\prime}}-z\vec{p}_{q1^{\prime}}%
)}{\left(  x_{\bar{q}}+z\right)  \left(  x_{q}+z\right)  \left(  \frac
{(\vec{p}_{\bar{q}2}+\vec{p}_{g3}){}^{2}}{x_{q}\left(  x_{\bar{q}}+z\right)
}+Q^{2}\right)  \left(  \frac{(\vec{p}_{\bar{q}2}+\vec{p}_{g3}){}^{2}}{x_{q}%
}+\frac{\vec{p}_{g3}^{\,\,2}}{z}+\frac{\vec{p}_{\bar{q}2}^{\,\,2}}{x_{\bar{q}}%
}+Q^{2}\right)  }\right]  +(q\leftrightarrow\bar{q}).
\end{align}
The interference term in the dipole $\times$ dipole contribution reads
{\allowdisplaybreaks
\begin{align*}
& \left( \tilde{\Phi}_3^+(\vec{p}_1, \vec{p}_2) \Phi_4^{+*}(\vec{p}_{1'}, \vec{p}_{2'},\vec{0}) +\Phi_4^+(\vec{p}_1, \vec{p}_2,\vec{0}) \tilde{\Phi}_3^{+*}(\vec{p}_{1'},\vec{p}_{2'})\right) \\
&  =\left[  \frac{8p_{\gamma}^{+}{}^{4}}{z\left(  x_{q}+z\right) \left(  \frac{\vec{p}{}_{\bar{q}2^{\prime}}^{\,\,2}}{x_{\bar{q}}\left(x_{q}+z\right)  }+Q^{2}\right)  \left(  \frac{\vec{p}{}_{q1^{\prime}}^{\,\,2}%
}{x_{q}}+\frac{\vec{p}{}_{\bar{q}2^{\prime}}^{\,\,2}}{x_{\bar{q}}}+\frac
{\vec{p}_{g}{}^{2}}{z}+Q^{2}\right)  }\right. \nonumber\\
&  \times\left\{  \frac{\left(  4x_{q}x_{\bar{q}}+z(2-dz)\right)  (\vec{p}_{g}%
-\frac{z}{x_{\bar{q}}}\vec{p}_{\bar{q}})(x_{q}\vec{p}_{g}-z\vec
{p}_{q1^{\prime}})}{(\vec{p}_{g}-\frac{z\vec{p}_{\bar{q}}}{x_{\bar{q}}}){}%
^{2}\left(  \frac{\vec{p}{}_{q1'}^{\,\,2}}{x_{q}\left(  x_{\bar{q}}+z\right)
}+Q^{2}\right)  }\right. \nonumber\\
&  -\left.  \left.  \frac{x_{\bar{q}}\left(  dz^{2}+4x_{q}\left(
x_{q}+z\right)  \right)  ({}\vec{p}_{g}-\frac{z}{x_{q}}\vec{p}_{q})(\vec{p}_{g}%
-\frac{z}{x_{q}}\vec{p}_{q1^{\prime}})}{(\vec{p}_{g}-\frac{z\vec{p}_{q}%
}{x_{q}}){}^{2}\left(  \frac{\vec{p}{}_{\bar{q}2}^{\,\,2}}{x_{\bar{q}}\left(
x_{q}+z\right)  }+Q^{2}\right)  }\right\}  +(q\leftrightarrow\bar{q})\right]
\nonumber\\
&  +(1\leftrightarrow1^{\prime},2\leftrightarrow2^{\prime}). \numberthis
\end{align*}}
The double-dipole $\times$ dipole contribution has the form 

\begin{equation}
\Phi_{4}^{+}( \vec{p}_1, \vec{p}_2, \vec{p}_3 )\,\Phi_{3}^{+*}(\vec{p}_{1'}, \vec{p}_{2'})=\Phi_{4}^{+} ( \vec{p}_1, \vec{p}_2, \vec{p}_3 ) \Phi_{4}^{+*}(\vec{p}_{1'}, \vec{p}_{2'}, \vec{0})+\Phi_4^+(\vec{p}_1, \vec{p}_2, \vec{p}_3) \tilde{\Phi}_3^{+*}(\vec{p}_{1'}, \vec{p}_{2'}) ,
\end{equation}
where
\begin{align*}
\Phi_4^+(\vec{p}_1, \vec{p}_2, \vec{p}_3) & \tilde{\Phi}_3^{+*}(\vec{p}_{1'}, \vec{p}_{2'})  =\frac{8p_{\gamma}^{+}{}^{4}}{z\left(  x_{q}+z\right)  \left(
\frac{\vec{p}{}_{\bar{q}2}^{\,\,2}}{x_{\bar{q}}\left(  x_{q}+z\right)  }%
+Q^{2}\right)  \left(  \frac{\vec{p}{}_{q1}^{\,\,2}}{x_{q}}+\frac{\vec{p}%
{}_{\bar{q}2}^{\,\,2}}{x_{\bar{q}}}+\frac{\vec{p}_{g3}^{\,\,2}}{z}+Q^{2}\right)
}\nonumber\\
&  \times\left\{  \frac{\left(  4x_{q}x_{\bar{q}}+z(2-dz)\right)  (\vec{p}_{g}%
-\frac{z}{x_{\bar{q}}}\vec{p}_{\bar{q}})(x_{q}\vec{p}_{g3}-z\vec{p}_{q1}%
)}{(\vec{p}_{g}-\frac{z\vec{p}_{\bar{q}}}{x_{\bar{q}}}){}^{2}\left(
\frac{\vec{p}{}_{q1^{\prime}}^{\,\,2}}{x_{q}\left(  x_{\bar{q}}+z\right)
}+Q^{2}\right)  }\right. \nonumber\\
&  -\left.  \frac{x_{\bar{q}}\left(  dz^{2}+4x_{q}\left(  x_{q}+z\right)
\right)  (\vec{p}_{g}-\frac{z}{x_{q}}\vec{p}_{q})(\vec{p}_{g3}-\frac{z}{x_{q}}%
\vec{p}_{q1})}{(\vec{p}_{g}-\frac{z\vec{p}_{q}}{x_{q}}){}^{2}\left(
\frac{\vec{p}{}_{\bar{q}2^{\prime}}^{\,\,2}}{x_{\bar{q}}\left(  x_{q}%
+z\right)  }+Q^{2}\right)  }\right\}  +( q \leftrightarrow \bar{q} ). \numberthis[finite_double_dipole_dipole_LL]
\end{align*}
For the dipole $\times$ double-dipole contribution, one just has to complex conjugate \eqref{eq:finite_double_dipole_dipole_LL} and also invert the name of the momenta i.e. $1',2' \leftrightarrow 1,2$. 

\subsection{LT/TL transition}
The double-dipole $\times$ double-dipole contribution is 
\begin{align*}
&  \Phi_{4}^{i}(p_{1\bot},p_{2\bot},p_{3\bot})\Phi_{4}^{+*}(p_{1\bot}^{\prime
},p_{2\bot}^{\prime},p_{3\bot}^{\prime}) \\
& =\frac{-4p_{\gamma}^{+}{}^{3}%
}{\left(  Q^{2}+\frac{\vec{p}{}_{g3}^{\,\,2}}{z}+\frac{\vec{p}{}_{q1}^{\,\,2}%
}{x_{q}}+\frac{\vec{p}{}_{{\bar{q}}2}^{\,\,2}}{x_{\bar{q}}}\right)  \left(
Q^{2}+\frac{\vec{p}{}_{g3^{\prime}}^{\,\,2}}{z}+\frac{\vec{p}{}_{q1^{\prime}%
}^{\,\,2}}{x_{q}}+\frac{\vec{p}{}_{{\bar{q}}2^{\prime}}^{\,\,2}}{x_{\bar{q}}%
}\right)  }\nonumber\\
& \hspace{-0.1 cm} \times \hspace{-0.1 cm} \left(  \frac{z\left(  (\vec{P} \hspace{-0.1 cm}  \cdot \hspace{-0.1 cm}  \vec{p}_{q1})G_{\bot}^{i} \hspace{-0.1 cm}  - \hspace{-0.1 cm} (\vec
{G} \hspace{-0.1 cm}  \cdot  \hspace{-0.1 cm} \vec{p}_{q1})P_{\bot}^{i}\right)  \left(  dz+4x_{q}-4\right)  -(\vec{G} \cdot 
\vec{P})p_{q1}^{i}{}_{\bot}\left(  2x_{q}-1\right)  \left(  4\left(
x_{q}-1\right)  x_{\bar{q}}-dz^{2}\right)  }{z^{2}x_{\bar{q}}\left(
z+x_{\bar{q}}\right)  {}^{3}\left(  Q^{2}+\frac{\vec{p}{}_{q1}^{\,\,2}}%
{x_{q}\left(  z+x_{\bar{q}}\right)  }\right)  \left(  Q^{2}+\frac{\vec{p}%
{}_{q1^{\prime}}^{\,\,2}}{x_{q}\left(  z+x_{\bar{q}}\right)  }\right)
}\right. \nonumber\\
 &  + \hspace{-0.05 cm}  \frac{z\left(  (\vec{P} \hspace{-0.1 cm} \cdot \hspace{-0.1 cm}  \vec{p}_{q1})H_{\bot}^{i} \hspace{-0.1 cm} -(\vec{H} \cdot \vec{p}_{q1})P_{\bot}^{i}\right)  \left(  dz+4x_{q}-2\right)  -(\vec{H} \cdot \vec{P}%
)p_{q1}^{i}{}_{\bot}\left(  2x_{q}-1\right)  \left(  z(2-dz)+4x_{q}x_{\bar{q}%
}\right)  }{z^{2}x_{q}\left(  z+x_{q}\right)  \left(  z+x_{\bar{q}}\right)
{}^{2}\left(  Q^{2}+\frac{\vec{p}{}_{\bar{q}2^{\prime}}^{\,\,2}}{\left(
z+x_{q}\right)  x_{\bar{q}}}\right)  \left(  Q^{2}+\frac{\vec{p}{}%
_{q1}^{\,\,2}}{x_{q}\left(  z+x_{\bar{q}}\right)  }\right)  }\nonumber\\
&  +   \left.  \frac{H_{\bot}^{i}\left(  z(zd+d-2)+x_{q}\left(  2-4x_{\bar{q}%
}\right)  \right)  x_{\bar{q}}}{z\left(  z+x_{q}\right)  {}^{2}\left(
z+x_{\bar{q}}\right)  \left(  Q^{2}+\frac{\vec{p}{}_{\bar{q}2^{\prime}%
}^{\,\,2}}{\left(  z+x_{q}\right)  x_{\bar{q}}}\right)  }\right)
+(q\leftrightarrow\bar{q}). \numberthis[phi_4_phi_4_LT]
\end{align*}
Here, 
\begin{equation}
G_{\bot}^{i}=x_{\bar{q}}p_{g3^{\prime}\bot}^{i}-zp_{\bar{q}2^{\prime}\bot}%
^{i},\quad H_{\bot}^{i}=x_{q}p_{g3^{\prime}\bot}^{i}-zp_{q1^{\prime}\bot}%
^{i},\quad P_{\bot}^{i}=x_{\bar{q}}p_{g3\bot}^{i}-zp_{\bar{q}2\bot}^{i}.
\end{equation}
The interference term in the dipole $\times$ dipole contribution reads
\begin{align*}
&  \left( \Phi_4^{i}(\vec{p}_{1}, \vec{p}_2, \vec{0}) \tilde{\Phi}_3^{+*}(\vec{p}_{1'}, \vec{p}_{2'}) + \tilde{\Phi}_3^{i}(\vec{p}_{1}, \vec{p}_{2}) \Phi_4^{+*} (\vec{p}_{1'}, \vec{p}_{2'}, \vec{0})\right) \\
& =4p_{\gamma}^{+}{}^{3}\left(  \frac{\Delta_{q}{}_{\bot}^{i}%
x_{q}x_{\bar{q}}\left(  dz^{2}+dz-2z+2x_{q}-4x_{q}x_{\bar{q}}\right)  }%
{\vec{\Delta}{}_{q}^{2}\left(  z+x_{q}\right)  {}^{2}\left(  z+x_{\bar{q}%
}\right)  \left(  Q^{2}+\frac{\vec{p}_{g}^{\,\,2}}{z}+\frac{\vec{p}{}%
_{q1}^{\,\,2}}{x_{q}}+\frac{\vec{p}{}_{\bar{q}2}^{\,\,2}}{x_{\bar{q}}}\right)
\left(  Q^{2}+\frac{\vec{p}{}_{\bar{q}2^{\prime}}^{\,\,2}}{\left(
z+x_{q}\right)  x_{\bar{q}}}\right)  }\right. \nonumber\\
&  - \frac{(\vec{J} \cdot \vec{\Delta}_{q})p_{\bar{q}2}^{i}{}_{\bot}\left(
dz^{2}+4x_{q}\left(  z+x_{q}\right)  \right)  \left(  1-2x_{\bar{q}}\right)
+z\left(  (\vec{J} \cdot \vec{p}_{\bar{q}2})\Delta_{q}^{i}{}_{\bot}-(\vec{p}_{\bar
{q}2} \cdot \vec{\Delta}_{q})J_{\bot}^{i}\right)  \left(  dz+4x_{\bar{q}}-4\right)
}{z\left(  z+x_{q}\right)  {}^{3}\vec{\Delta}{}_{q}^{2}\left(  Q^{2}%
+\frac{\vec{p}_{g}^{\,\,2}}{z}+\frac{\vec{p}{}_{q1^{\prime}}^{\,\,2}}{x_{q}%
}+\frac{\vec{p}{}_{\bar{q}2^{\prime}}^{\,\,2}}{x_{\bar{q}}}\right)  \left(
Q^{2}+\frac{\vec{p}{}_{\bar{q}2}^{\,\,2}}{\left(  z+x_{q}\right)  x_{\bar{q}}%
}\right)  \left(  Q^{2}+\frac{\vec{p}{}_{\bar{q}2^{\prime}}^{\,\,2}}{\left(
z+x_{q}\right)  x_{\bar{q}}}\right)  }\nonumber\\
&  -\frac{x_{q}\left(  z\left(  (\vec{K} \cdot \vec{p}_{\bar{q}2})\Delta_{q}^{i}%
{}_{\bot}-(\vec{p}_{\bar{q}2} \cdot \vec{\Delta}_{q})K_{\bot}^{i}\right)  \left(
dz+4x_{\bar{q}}-2\right)  +(\vec{K} \cdot \vec{\Delta}_{q})p_{\bar{q}2}^{i}{}_{\bot
}\left(  1-2x_{\bar{q}}\right)  \left(  z(dz-2)-4x_{q}x_{\bar{q}}\right)
\right)  }{z\left(  z+x_{q}\right)  {}^{2}x_{\bar{q}}\left(  z+x_{\bar{q}%
}\right)  \vec{\Delta}{}_{q}^{2}\left(  Q^{2}+\frac{\vec{p}_{g}^{\,\,2}}%
{z}+\frac{\vec{p}{}_{q1^{\prime}}^{\,\,2}}{x_{q}}+\frac{\vec{p}{}_{\bar
{q}2^{\prime}}^{\,\,2}}{x_{\bar{q}}}\right)  \left(  Q^{2}+\frac{\vec{p}%
{}_{\bar{q}2}^{\,\,2}}{\left(  z+x_{q}\right)  x_{\bar{q}}}\right)  \left(
Q^{2}+\frac{\vec{p}{}_{q1^{\prime}}^{\,\,2}}{x_{q}\left(  z+x_{\bar{q}%
}\right)  }\right)  }\nonumber\\
&  -\frac{z\left(  (\vec{p}_{q1} \cdot \vec{\Delta}_{q})X_{\bot}^{i}-(\vec{X} \cdot \vec{p}_{q1}) \Delta_{q}^{i}{}_{\bot}\right)  \left(  dz+4x_{q}-2\right)  +(\vec
{X} \cdot \vec{\Delta}_{q})p_{q1}^{i}{}_{\bot}\left(  1-2x_{q}\right)  \left(
z(dz-2)-4x_{q}x_{\bar{q}}\right)  }{z\vec{\Delta}{}_{q}^{2}\left(
z+x_{q}\right)  \left(  z+x_{\bar{q}}\right)  {}^{2}\left(  Q^{2}+\frac
{\vec{p}_{g}^{\,\,2}}{z}+\frac{\vec{p}{}_{q1}^{\,\,2}}{x_{q}}+\frac{\vec{p}%
{}_{\bar{q}2}^{\,\,2}}{x_{\bar{q}}}\right)  \left(  Q^{2}+\frac{\vec{p}%
{}_{\bar{q}2^{\prime}}^{\,\,2}}{\left(  z+x_{q}\right)  x_{\bar{q}}}\right)
\left(  Q^{2}+\frac{\vec{p}{}_{q1}^{\,\,2}}{x_{q}\left(  z+x_{\bar{q}}\right)
}\right)  }\nonumber\\
&  +\left.  \frac{z\left(  (\vec{X} \cdot \vec{p}_{q1})\Delta_{\bar{q}}^{i}{}_{\bot
}-(\vec{p}_{q1} \cdot \vec{\Delta}_{\bar{q}})X_{\bot}^{i}\right)  \left(
dz+4x_{q}-4\right)  -(\vec{X} \cdot \vec{\Delta}_{\bar{q}})p_{q1}^{i}{}_{\bot}\left(
2x_{q}-1\right)  \left(  4\left(  x_{q}-1\right)  x_{\bar{q}}-dz^{2}\right)
}{z\left(  z+x_{\bar{q}}\right)  {}^{3}\vec{\Delta}{}_{\bar{q}}^{2}\left(
Q^{2}+\frac{\vec{p}_{g}^{\,\,2}}{z}+\frac{\vec{p}{}_{q1}^{\,\,2}}{x_{q}}%
+\frac{\vec{p}{}_{\bar{q}2}^{\,\,2}}{x_{\bar{q}}}\right)  \left(  Q^{2}%
+\frac{\vec{p}{}_{q1}^{\,\,2}}{x_{q}\left(  z+x_{\bar{q}}\right)  }\right)
\left(  Q^{2}+\frac{\vec{p}{}_{q1^{\prime}}^{\,\,2}}{x_{q}\left(  z+x_{\bar
{q}}\right)  }\right)  }\right) \nonumber\\
&  +(q\leftrightarrow\bar{q}) \; , \numberthis[phi_tilde_phi_4_dipole_dipole_LT]
\end{align*}
where 

\begin{equation}
\vec{\Delta}_{q} = \frac{x_q \vec{p}_g - x_g \vec{p}_q}{x_q + x_g}
\vec{\Delta}_{\bar{q}} = \frac{x_{\bar{q}} \vec{p}_g - x_g \vec{p}_{\bar{q}}}{x_q + x_g}
\end{equation}
\begin{align}
X_{\bot}^{i}   =x_{\bar{q}}p_{g\bot}^{i}-zp_{\bar{q}2\bot}^{i} = & P_{\bot}%
^{i}|_{p_{3}=0},\quad J_{\bot}^{i}=x_{q}p_{g\bot}^{i}-zp_{q1^{\prime}\bot}%
^{i}=H_{\bot}^{i}|_{p_{3}^{\prime}=0},\nonumber\\
K_{\bot}^{i}  &  =x_{\bar{q}}p_{g\bot}^{i}-zp_{\bar{q}2^{\prime}\bot}%
^{i}=G_{\bot}^{i}|_{p_{3}^{\prime}=0}.
\end{align}
The TL transition is obtained from above by complex conjugation and inverting the naming of the different momenta in \eqref{eq:phi_tilde_phi_4_dipole_dipole_LT} and \eqref{eq:phi_4_phi_4_LT}. \\
The double-dipole $\times$ dipole have, respectively, the form 
\begin{equation}
\Phi_4^{i}(\vec{p}_{1}, \vec{p}_{2}, \vec{p}_{3}) \Phi_3^{+*}(\vec{p}_{1'}, \vec{p}_{2'}) = \Phi_4^i(\vec{p}_1, \vec{p}_2, \vec{p}_3) \Phi_4^{+*}(\vec{p}_{1'}, \vec{p}_{2'}, 0) + \Phi_4^i(\vec{p}_1, \vec{p}_2, \vec{p}_3) \tilde{\Phi}_3^{+*}(\vec{p}_{1'}, \vec{p}_{2'})  \; ,
\end{equation}
\begin{equation}
   \Phi_4^{+} (\vec{p}_{1}, \vec{p}_{2}, \vec{p}_{3}) \Phi_3 ^{i*}(\vec{p}_{1'}, \vec{p}_{2'}) = \Phi_4^{+} (\vec{p}_{1}, \vec{p}_{2}, \vec{p}_{3}) \Phi_4^{i*}(\vec{p}_{1'}, \vec{p}_{2'}, \vec{0}) + \Phi_4^{+} (\vec{p}_{1}, \vec{p}_{2}, \vec{p}_{3}) \tilde{\Phi}_3^{i*}(\vec{p}_{1'}, \vec{p}_{2'}) \; ,
\end{equation}
where 
\begin{align}
& \Phi_4^i(\vec{p}_1, \vec{p}_2, \vec{p}_3) \tilde{\Phi}_3^{+*}(\vec{p}_{1'}, \vec{p}_{2'})  =\frac{4p_{\gamma}^{+}{}^{3}}{\left(  x_{q}+z\right)  \vec{\Delta}_{q}^{2}\left(  \frac{\vec{p}{}_{\bar{q}2^{\prime}}^{\,\,2}}%
{x_{\bar{q}}\left(  x_{q}+z\right)  }+Q^{2}\right)  \left(  \frac{\vec{p}%
{}_{q1}^{\,\,2}}{x_{q}}+\frac{\vec{p}{}_{\bar{q}2}^{\,\,2}}{x_{\bar{q}}}%
+\frac{\vec{p}_{g3}^{\; 2}}{z}+Q^{2}\right)  } \nonumber\\
&  \times\left\{  \frac{x_{q}x_{\bar{q}}\Delta_{q}^{i}\left(
dz(z+1)-2\left(  1-2x_{q}\right)  \left(  x_{q}+z\right)  \right)  }{\left(
x_{q}+z\right)  {}\left(  x_{\bar{q}}+z\right) } + \frac{\left(  dz+4x_{q}-2\right)  \left(  \Delta_{q}^{i}
\vec{P} \cdot \vec{p}_{q1} -P^{i}   \vec{p}_{q1} \cdot
\vec{\Delta}_{q}  \right)  }{\left(  x_{\bar{q}}+z\right)  {}%
^{2}\left(  \frac{\vec{p}_{q1}^{ \; 2}}{x_{q}\left(  x_{\bar{q}}+z\right)
}+Q^{2}\right)  } \right. \nonumber\\
&  +\frac{\left(  2x_{q}-1\right)  p_{q1}^{i} \vec{P} \cdot
\vec{\Delta}_{q} \left(  z(dz-2)-4x_{q}x_{\bar{q}}\right)
}{z\left(  x_{\bar{q}}+z\right)  {}^{2}\left(  \frac{\vec{p}_{q1}^{\; 2}%
}{x_{q}\left(  x_{\bar{q}}+z\right)  }+Q^{2}\right)} - \frac{\left(  (d-4)z-4x_{q}\right)  \left( W^{i}  \vec{p}_{\bar{q}2} \cdot \vec{\Delta}_{q}  -\Delta_{q}^{i} 
\vec{W} \cdot \vec{p}_{\bar{q}2} \right) }{\left(  x_{q}+z\right)
{}^{2}\left(  \frac{\vec{p}_{\bar{q}2}^{ \; 2}}{x_{\bar{q}}\left(
x_{q}+z\right)  }+Q^{2}\right)  } \nonumber\\
&  +\left.  \frac{\left(  2x_{\bar{q}}-1\right)  \left(  dz^{2}+4x_{q}\left(
x_{q}+z\right)  \right)  p_{\bar{q}2}^{i}  \vec{W} \cdot \vec{\Delta}_{q} }{z\left(  x_{q}+z\right)  {}^{2}\left(  \frac
{\vec{p}_{\bar{q}2}^{\; 2}}{x_{\bar{q}}\left(  x_{q}+z\right)  }%
+Q^{2}\right)  }\right\} +(q \leftrightarrow \bar{q}) \; ,
\end{align}%
and
\begin{align}
\Phi_4^{+} (\vec{p}_{1}, \vec{p}_{2}, \vec{p}_{3}) & \tilde{\Phi}_3^{i*}(\vec{p}_{1'}, \vec{p}_{2'}) =\frac{4p_{\gamma}^{+}{}^{3}}{z\vec{\Delta}{}_{q}^{2}\left(
x_{q}+z\right)  {}^{2}\left(  Q^{2}+\frac{\vec{p}_{g3}^{\,\,2}}{z}+\frac
{\vec{p}{}_{q1}^{\,\,2}}{x_{q}}+\frac{\vec{p}{}_{\bar{q}2}^{\,\,2}}{x_{\bar
{q}}}\right)  \left(  Q^{2}+\frac{\vec{p}{}_{\bar{q}2^{\prime}}^{\,\,2}%
}{\left(  z+x_{q}\right)  x_{\bar{q}}}\right)  }\nonumber\\
&  \times\left[  \frac{x_{q}z\left(  (d-4)z-4x_{q}+2\right)  \left(
P^{i}\left(  \vec{p}_{\bar{q}2^{\prime}} \cdot \vec{\Delta}_{q}\right)
-\Delta_{q}^{i}\left( \vec{P} \cdot \vec{p}_{\bar{q}2^{\prime}}\right)
\right)  }{x_{\bar{q}}\left(  x_{\bar{q}}+z\right)  \left(  \frac
{\vec{p}_{q1}^{\; 2}}{x_{q}\left(  x_{\bar{q}}+z\right)  }+Q^{2}\right)
}\right.  \nonumber\\
&  -\frac{x_{q}\left(  x_{q}-x_{\bar{q}}+z\right)  p_{\bar{q}2^{\prime}}^{i}\left(  \vec{P} \cdot \vec{\Delta}_{q} \right)  \left(  z(dz-2)-4x_{q}%
x_{\bar{q}}\right)}{x_{\bar{q}}\left(  x_{\bar{q}}+z\right)  \left(
\frac{\vec{p}_{q1}^{\; 2}}{x_{q}\left(  x_{\bar{q}}+z\right)  }%
+Q^{2}\right)  }\nonumber\\
&  -\frac{\left(  x_{q}-x_{\bar{q}}+z\right)  \left(  dz^{2}+4x_{q}\left(
x_{q}+z\right)  \right)  p_{\bar{q}2^{\prime}}^{i}\left(  \vec{W} \cdot \vec{\Delta}_{q} \right)  }{\left(  x_{q}+z\right)  {}\left(
\frac{\vec{p}_{\bar{q}2}^{\; 2}}{x_{\bar{q}}\left(  x_{q}+z\right)
}+Q^{2}\right)  }\nonumber\\
&  -\left.  \frac{z\left(  (d-4)z-4x_{q}\right)  \left(  \Delta_{q}^{i} \left( \vec{W} \cdot \vec{p}_{\bar{q}2^{\prime}} \right) - W^{i} \left(
\vec{p}_{\bar{q}2^{\prime}} \cdot \vec{\Delta}_{q} \right)  \right)
}{\left(  x_{q}+z\right)  {}\left(  \frac{\vec{p}_{\bar{q}2}^{\,\, 2}%
}{x_{\bar{q}}\left(  x_{q}+z\right)  }+Q^{2}\right)  }\right] +(q \leftrightarrow \bar{q}) .
\end{align}
Here, we introduced 
\begin{equation}
W_{\bot}^{i}=x_{q}p_{g3\bot}^{i}-zp_{q1\bot}^{i}.
\end{equation}

\subsection{TT transition}
The double-dipole $\times$ double-dipole contribution is 

\begin{align}
&  \Phi_{4}^{i}(p_{1\bot},p_{2\bot},p_{3\bot})\Phi_{4}^{k}(p_{1\bot}^{\prime
},p_{2\bot}^{\prime},p_{3\bot}^{\prime})^{\ast}=\left(  \frac{p_{\gamma}^{+}%
{}^{2}}{\left(  Q^{2}+\frac{\vec{p}_{g3}^{\;2}}{z}+\frac{\vec{p}_{q1}^{\;2}%
}{x_{q}}+\frac{\vec{p}_{\bar{q}2}^{\,\,2}}{x_{\bar{q}}}\right)  \left(
Q^{2}+\frac{\vec{p}_{g3^{\prime}}^{\;2}}{z}+\frac{\vec{p}_{q1^{\prime}}^{\;2}%
}{x_{q}}+\frac{\vec{p}_{\bar{q}2^{\prime}}^{\,\,2}}{x_{\bar{q}}}\right)
}\right. \nonumber\\
&  \times\left[  -\frac{g_{\bot}^{ik}x_{q}x_{\bar{q}}\left(zd+d-2+2x_{\bar
{q}}\right)  }{\left(  z+x_{q}\right)^{2}\left(z+x_{\bar{q}}\right)
}-\frac{2P_{\bot}^{k} p_{q1\bot}^{i}\left( 1-2x_{q}\right)  }{z\left(
z+x_{\bar{q}}\right)^{2}\left( Q^{2}+\frac{\vec{p}_{q1}^{\;2}}%
{x_{q}\left(  z+x_{\bar{q}}\right)  }\right)  }\left(  \frac{(d-2)z-2x_{\bar
{q}}}{z+x_{\bar{q}}}+\frac{dz+2x_{\bar{q}}}{z+x_{q}}\right)  \right.
\nonumber\\
&  -\frac{2\left(  g_{\bot}^{ik}(\vec{P} \cdot \vec{p}_{q1})+P_{\bot}^{i}p_{q1\bot}%
{}^{k}\right)  }{z\left(  z+x_{\bar{q}}\right)^{2}\left(  Q^{2}+\frac
{\vec{p}_{q1}^{\;2}}{x_{q}\left(  z+x_{\bar{q}}\right)  }\right)  }\left(
\frac{(d-4)z-2x_{\bar{q}}}{z+x_{q}}+\frac{(d-2)z-2x_{\bar{q}}}{z+x_{\bar{q}}%
}\right) \nonumber
\end{align}%
\begin{align}
&  -\frac{1}{z^{2} x_{q}\left(  z+x_{q}\right)^{2} x_{\bar{q}}\left(
z+x_{\bar{q}}\right)^{2}\left(  Q^{2}+\frac{\vec{p}_{\bar{q}2^{\prime}%
}^{\,\,2}}{\left(  z+x_{q}\right)  x_{\bar{q}}}\right)  \left(  Q^{2}%
+\frac{\vec{p}_{q1}^{\;2}}{x_{q}\left(  z+x_{\bar{q}}\right)  }\right)
}\left\{  (\vec{H}  \cdot \vec{P})\left[  p_{q1\bot}^{i}{}p_{\bar{q}2^{\prime}\bot
}^{k}{}\left(  1-2x_{q}\right)  \right.  \right. \nonumber\\
&  \times\left.  \left(  1-2x_{\bar{q}}\right)  \left(  z(2-dz)+4x_{q}%
x_{\bar{q}}\right)  +(g_{\bot}^{ik}(\vec{p}_{q1}  \cdot \vec{p}_{\bar{q}2^{\prime}%
})+p_{q1\bot}^{k} p_{\bar{q}2^{\prime}\bot}^{i})\left(  z(2-(d-4)z)+4x_{q}%
x_{\bar{q}}\right)  \right] \nonumber\\
&  +((d-4)z-2)\left[  z(\vec{H}  \cdot  \vec{p}_{\bar{q}2^{\prime}})(g_{\bot}^{ik}%
(\vec{P}  \cdot  \vec{p}_{q1})+P_{\bot}^{i}p_{q1\bot}^{k})+z H_{\bot}^{k}\left(
(\vec{P}  \cdot \vec{p}_{q1})p_{\bar{q}2^{\prime}\bot}^{i}-(\vec{p}_{q1}  \cdot \vec{p}_{\bar{q}2^{\prime}})P_{\bot}^{i}\right)  \right] \nonumber\\
&  +((d-4)z+2)\left[  zH^{i}\left(  (\vec{P}  \cdot  \vec{p}_{\bar{q}2^{\prime}%
})p_{q1\bot}^{k}-(\vec{p}_{q1}  \cdot \vec{p}_{\bar{q}2^{\prime}}) P_{\bot}%
^{k}\right)  +z(\vec{H}  \cdot  \vec{p}_{q1})(g_{\bot}^{ik}(\vec{P}  \cdot \vec{p}_{\bar{q}2^{\prime}})+P_{\bot}^{k}p_{\bar{q}2^{\prime}\bot}^{i})\right]
\nonumber\\
&  +\left.  2z\left(  (\vec{H}  \cdot  \vec{p}_{\bar{q}2^{\prime}})P_{\bot}^{k}%
-(\vec{P}  \cdot  \vec{p}_{\bar{q}2^{\prime}}) H_{\bot}^{k}\right)  p_{q1\bot}{}%
^{i}\left(  1-2x_{q}\right)  \left(  dz+4x_{\bar{q}}-2\right)  \right\}
\nonumber
\end{align}%
\begin{align}
&  -\frac{1}{z^{2}x_{q}x_{\bar{q}}\left(  z+x_{\bar{q}}\right)  {}^{4}\left(
Q^{2}+\frac{\vec{p}_{q1}^{\;2}}{x_{q}\left(  z+x_{\bar{q}}\right)  }\right)
\left(  Q^{2}+\frac{\vec{p}_{q1^{\prime}}^{\;2}}{x_{q}\left(  z+x_{\bar{q}%
}\right)  }\right)  }\left\{  z\left(  (d-4)z-4x_{\bar{q}}\right)  \frac{{}%
}{{}}\right. \nonumber\\
&  \times\left[  g_{\bot}^{ik}\left(  (\vec{G}  \cdot \vec{p}_{q1^{\prime}}) (\vec{P}  \cdot  \vec{p}_{q1})-(\vec{G}  \cdot  \vec{p}_{q1})(\vec{P}  \cdot  \vec{p}_{q1^{\prime}})\right) +(\vec{p}_{q1}  \cdot  \vec{p}_{q1^{\prime}})\left(  G_{\bot}^{i}P_{\bot}^{k}-G_{\bot}^{k}P_{\bot}^{i}\right)  \right. \nonumber\\
&  +\left.  2(\vec{G}  \cdot  \vec{p}_{q1^{\prime}})\left(  P_{\bot}^{i}p_{q1\bot}^{k}+P_{\bot}^{k} p_{q1\bot}^{i}\left(  1-2x_{q}\right)  \right)
-2(\vec{G}  \cdot  \vec{p}_{q1}) \left(  P_{\bot}^{k} p_{q1^{\prime}\bot}^{i}+P_{\bot}^{i} p_{q1^{\prime}\bot}^{k}\left(  1-2x_{q}\right)  \right)  \right] \nonumber\\
&  +\left.  \left.  (\vec{G}  \cdot  \vec{P})\left[  p_{q1\bot}^{k} p_{q1^{\prime}\bot}^{i}-p_{q1\bot}^{i} p_{q1^{\prime}\bot}^{k}\left(  1-2x_{q}\right)^{2}+g_{\bot}^{ik}\left(  \vec{p}_{q1}  \cdot  \vec{p}_{q1^{\prime}}\right)  \right] \left(  dz^{2}+4x_{\bar{q}}\left(  z+x_{\bar{q}}\right)  \right)  \right\}
\right] \nonumber\\
&  +\left.  \frac{{}}{{}}(1\leftrightarrow1^{\prime},2\leftrightarrow
2^{\prime},3\leftrightarrow3^{\prime},i\leftrightarrow k)\right)
+(q\leftrightarrow\bar{q}).
\end{align}
The interference term in the dipole $\times$ dipole contribution reads
\begin{align*}
& \left( \tilde{\Phi}_3^i(\vec{p}_1,\vec{p}_2)\Phi_4^{k*}(\vec{p}_{1'}, \vec{p}_{2'}, \vec{0}) + \Phi_4^i(\vec{p}_1, \vec{p}_{2}, \vec{0}) \tilde{\Phi}_3^{k*}(\vec{p}_{1'}, \vec{p}_{2'}) \right) \\
&  =\left(  \frac{2p_{\gamma}^{+}{}^{2}}{\vec{\Delta}{}_{q}^{2}\left(
Q^{2}+\frac{\vec{p}_{g}^{\,\,2}}{z}+\frac{\vec{p}{}_{q1}^{\,\,2}}{x_{q}}%
+\frac{\vec{p}{}_{\bar{q}2}^{\,\,2}}{x_{\bar{q}}}\right)  \left(  Q^{2}%
+\frac{\vec{p}{}_{\bar{q}2^{\prime}}^{\,\,2}}{\left(  z+x_{q}\right)
x_{\bar{q}}}\right)  }\right. \\
& \times \left[  \frac{\left(  (d-2)z-2x_{q}\right)  x_{q}}{\left(
z+x_{q}\right)  {}^{3}}\left(  g_{\bot}^{ik}(\vec{p}_{\bar{q}2^{\prime}} \cdot \vec{\Delta}_{q})+p_{\bar{q}2^{\prime}}{}_{\bot}^{i}\Delta_{q\bot}^{k} +p_{\bar{q}2^{\prime}\bot}^{k}\Delta_{q\bot}^{i}\left(  1-2x_{\bar{q}}\right)  \right)  \right. \\
&  +\frac{x_{q}\left(  \left(  (d-4)z-2x_{q}\right)  \left(  g_{\bot}^{ik}(\vec{p}_{\bar{q}2^{\prime}} \cdot \vec{\Delta}_{q})+p_{\bar{q}2^{\prime}\bot}^{i}{}\Delta_{q\bot}^{k}{}\right)  +p_{\bar{q}2^{\prime}\bot}^{k} \Delta_{q\bot}^{i} \left(  dz+2x_{q}\right)  \left(  1-2x_{\bar{q}}\right) \right)  }{\left(  z+x_{q}\right)  {}^{2}\left(  z+x_{\bar{q}}\right)  } \\
& -\frac{1}{z\left(  z+x_{q}\right)  {}^{2}x_{\bar{q}}\left(  z+x_{\bar{q}%
}\right)  {}^{2}\left(  Q^{2}+\frac{\vec{p}_{q1}^{\;2}}{x_{q}\left(
z+x_{\bar{q}}\right)  }\right)  }\left\{  z((d-4)z+2)\frac{{}}{{}}\right. \\
& \hspace{-0.1 cm} \times \hspace{-0.1 cm} \left[  p_{q1}{}_{\bot}^{i} \left(  (\vec{p}_{\bar{q}2^{\prime}}\cdot \vec{\Delta}_{q})X_{\bot}^{k}-(\vec{X} \cdot \vec{p}_{\bar{q}2^{\prime}}) \Delta_{q \bot}^{k}\right)  \left( 2x_{q}-1\right)  -(\vec{X} \cdot \vec{p}_{\bar{q}2^{\prime}})\left(  g_{\bot}^{ik}(\vec{p}_{q1} \cdot \vec{\Delta}_{q})+p_{q1\bot}^{k} \Delta_{q\bot}^{i}\right)  \right. \\
&  \left. - \hspace{-0.05 cm}  X_{\bot}^{k} \hspace{-0.05 cm} \left(  (\vec{p}_{q1} \hspace{-0.05 cm} \cdot \hspace{-0.05 cm} \vec{\Delta}_{q}) p_{\bar
{q}2^{\prime} \bot}^{i} \hspace{-0.1 cm} -(\vec{p}_{q1} \hspace{-0.05 cm} \cdot \vec{p}_{\bar{q}2^{\prime}})\Delta_{q}{}_{\bot}^{i}\right)  \right] \hspace{-0.05 cm} + \hspace{-0.05 cm} 4x_{q}z \left(  1 \hspace{-0.05 cm} - \hspace{-0.05 cm} 2x_{q}\right)
p_{q1\bot}^{i} \hspace{-0.1 cm} \left(  (\vec{p}_{\bar{q}2^{\prime}} \hspace{-0.05 cm} \cdot \hspace{-0.05 cm} \vec{\Delta}%
_{q})X_{\bot}^{k} \hspace{-0.1 cm} - \hspace{-0.1 cm}(\vec{X} \hspace{-0.05 cm} \cdot \vec{p}_{\bar{q}2^{\prime}}) \Delta_{q\bot}%
^{k}\right)  \\
&  +z\left(  1-2x_{\bar{q}}\right)  \left(  dz+4x_{q}-2\right)  p_{\bar
{q}2^{\prime}\bot}^{k}\left(  (\vec{p}_{q1} \cdot \vec{\Delta}_{q}) X_{\bot}%
^{i}-(\vec{X} \cdot \vec{p}_{q1}) \Delta_{q\bot}^{i}\right)  -z((d-4)z-2)\\
&  \times \left[  \left(  g_{\bot}^{ik} (\vec{X} \cdot \vec{p}_{q1})+X_{\bot}^{i} 
p_{q1\bot}^{k}\right)  (\vec{p}_{\bar{q}2^{\prime}} \cdot \vec{\Delta}%
_{q})+ \left(  (\vec{X}\vec{p}_{q1})p_{\bar{q}2^{\prime}}{}_{\bot}^{i}-(\vec
{p}_{q1}\vec{p}_{\bar{q}2^{\prime}})X_{\bot}^{i}\right)  \Delta_{q\bot
}^{k}\right] \\
&  +(\vec{X} \cdot \vec{\Delta}_{q})p_{q1\bot}^{i}p_{\bar{q}2^{\prime} \bot
}^{k}\left( 1-2x_{q}\right)  \left( 1-2x_{\bar{q}}\right)  \left(
z(d z-2)-4x_{q}x_{\bar{q}}\right) \\
&  - \left.  (\vec{X} \cdot \vec{\Delta}_{q})\left(  g_{\bot}^{ik}(\vec{p}_{q1} \cdot \vec{p}_{\bar{q}2^{\prime}})+p_{q1\bot}^{k} p_{\bar{q}2^{\prime}\bot}^{i}\right)  \left(  z(2-(d-4)z)+4x_{q}x_{\bar{q}}\right)  \right\} \\
& -    \frac{1}{z\left(  z+x_{q}\right)  {}^{4}\left(  Q^{2}+\frac{\vec{p}%
{}_{\bar{q}2}^{\,\,2}}{\left(  z+x_{q}\right)  x_{\bar{q}}}\right)  x_{\bar
{q}}}\left\{  z\left(  dz+4x_{\bar{q}}-4\right)  \left[  \left(  1-2x_{\bar
{q}}\right)  \frac{{}}{{}}\right.  \right. \nonumber\\
&  \times  \left(  p_{\bar{q}2^{\prime}\bot}^{k} \left(  (\vec{p}_{\bar
{q}2} \cdot \vec{\Delta}_{q}) V_{\bot}^{i}-(\vec{V} \cdot \vec{p}_{\bar{q}2})\Delta_{q\bot}^{i} \right)  +p_{\bar{q}2\bot}^{i}\left(  (\vec{V}\vec{p}_{\bar{q}2^{\prime}})\Delta_{q\bot}^{k}-(\vec{p}_{\bar{q}2^{\prime}} \cdot \vec{\Delta}_{q}) V_{\bot}^{k}\right)  \right) \nonumber\\
& +    V_{\bot}^{k} \left(  (\vec{p}_{\bar{q}2} \cdot \vec{\Delta}_{q}) p_{\bar
{q}2^{\prime}\bot}^{i}-(\vec{p}_{\bar{q}2} \cdot \vec{p}_{\bar{q}2^{\prime}}) \Delta_{q\bot}^{i}\right)  + \left(  (\vec{p}_{\bar{q}2} \cdot \vec{p}_{\bar
{q}2^{\prime}}) V_{\bot}^{i}-(\vec{V} \cdot \vec{p}_{\bar{q}2}) p_{\bar{q}2^{\prime}\bot}^{i}\right)  \Delta_{q \bot}^{k}\nonumber\\
& +    \left.  g_{\bot}^{ik} \left(  (\vec{V} \cdot \vec{p}_{\bar{q}2^{\prime}}%
)(\vec{p}_{\bar{q}2} \cdot \vec{\Delta}_{q})-(\vec{V} \cdot \vec{p}_{\bar{q}2} )(\vec
{p}_{\bar{q}2^{\prime}} \cdot  \vec{\Delta}_{q}) \right)  + p_{\bar{q}2\bot}^{k}\left(  (\vec{V} \cdot \vec{p}_{\bar{q}2^{\prime}}) \Delta_{q\bot}^{i} 
- (\vec{p}_{\bar{q}2^{\prime}} \cdot \vec{\Delta}_{q}) V_{\bot}^{i}\right)  \right]
\nonumber\\
& +    \left.  \left.  (\vec{V} \cdot \vec{\Delta}_{q})\left(  p_{\bar{q}2\bot
}^{i}p_{\bar{q}2^{\prime}\bot}^{k}\left( 1-2x_{\bar{q}}\right)^{2}-g_{\bot}^{ik}(\vec{p}_{\bar{q}2} \cdot \vec{p}_{\bar{q}2^{\prime}})-p_{\bar{q}2\bot}^{k} p_{\bar{q}2^{\prime} \bot}^{i}\right)  \left(d z^{2}-4x_{q}\left( x_{\bar{q}}-1\right)  \right)  \right\}  \frac{{}}{{}%
}\right] \nonumber \\
&  +   \left.  \frac{{}}{{}}(1\leftrightarrow1^{\prime},2\leftrightarrow
2^{\prime},i\leftrightarrow k)\right)  +(q\leftrightarrow\bar{q}). \numberthis
\end{align*}
Here,
\begin{equation}
V_{\bot}^{i}=x_{q}p_{g\bot}^{i}-zp_{q1\bot}^{i}.
\end{equation}
The double-dipole $\times$ dipole contribution has the form  
\begin{equation}
\Phi_{4}^{i}(\vec{p}_1, \vec{p}_2, \vec{p}_3)\,\Phi_{3}^{k*}(\vec{p}_{1'}, \vec{p}_{2'})=\Phi_{4}^{i}(\vec{p}_1, \vec{p}_2, \vec{p}_3)\Phi_{4}^{k*}(\vec{p}_{1'}, \vec{p}_{2'}, \vec{0})+\Phi_4^i(\vec{p}_1, \vec{p}_2, \vec{p}_3) \tilde{\Phi}_3^{k*}(\vec{p}_{1'}, \vec{p}_{2'}),
\end{equation}
where
\begin{align*}
& \Phi_4^i(\vec{p}_1, \vec{p}_2, \vec{p}_3) \tilde{\Phi}_3^{k*}(\vec{p}_{1'}, \vec{p}_{2'})  \\
&  =\frac{2p_{\gamma}^{+}{}^{2}}{\vec{\Delta}{}_{q}^{2}\left(
Q^{2}+\frac{\vec{p}_{g3}^{\,\,2}}{z}+\frac{\vec{p}{}_{q1}^{\,\,2}}{x_{q}%
}+\frac{\vec{p}{}_{\bar{q}2}^{\,\,2}}{x_{\bar{q}}}\right)  \left(  Q^{2}%
+\frac{\vec{p}{}_{\bar{q}2^{\prime}}^{\,\,2}}{\left(  z+x_{q}\right)
x_{\bar{q}}}\right)  }\nonumber\\
&  \times\left[  \frac{\left(  (d-2)z-2x_{q}\right)  x_{q}}{\left(
z+x_{q}\right)  {}^{3}}\left(  g_{\bot}^{ik}(\vec{p}_{\bar{q}2^{\prime}}%
\vec{\Delta}_{q})+p_{\bar{q}2^{\prime}}{}_{\bot}^{i}\Delta_{q\bot}{}%
^{k}+p_{\bar{q}2^{\prime}\bot}{}^{k}\Delta_{q\bot}{}^{i}\left(  1-2x_{\bar{q}%
}\right)  \right)  \right. \nonumber\\
&  +\frac{x_{q}\left(  \left(  (d-4)z-2x_{q}\right)  \left(  g_{\bot}%
^{ik}(\vec{p}_{\bar{q}2^{\prime}}\vec{\Delta}_{q})+p_{\bar{q}2^{\prime}\bot
}^{i}{}\Delta_{q\bot}^{k}{}\right)  +p_{\bar{q}2^{\prime}\bot}^{k}{}%
\Delta_{q\bot}^{i}{}\left(  dz+2x_{q}\right)  \left(  1-2x_{\bar{q}}\right)
\right)  }{\left(  z+x_{q}\right)  {}^{2}\left(  z+x_{\bar{q}}\right)
}\nonumber\\
&  -\frac{1}{z\left(  z+x_{q}\right)  {}^{2}x_{\bar{q}}\left(  z+x_{\bar{q}%
}\right)  {}^{2}\left(  Q^{2}+\frac{\vec{p}_{q1}^{\;2}}{x_{q}\left(
z+x_{\bar{q}}\right)  }\right)  }\left\{  z((d-4)z+2)\frac{{}}{{}}\right.
\nonumber\\
&  \times\left[  p_{q1}{}_{\bot}^{i}\left(  (\vec{p}_{\bar{q}2^{\prime}}%
\vec{\Delta}_{q})P_{\bot}^{k}-(\vec{P}\vec{p}_{\bar{q}2^{\prime}})\Delta_{q}%
{}_{\bot}^{k}\right)  \left(  2x_{q}-1\right)  -(\vec{P}\vec{p}_{\bar
{q}2^{\prime}})\left(  g_{\bot}^{ik}(\vec{p}_{q1}\vec{\Delta}_{q})+p_{q1}%
{}_{\bot}^{k}\Delta_{q}{}_{\bot}^{i}\right)  \right. \nonumber\\
&  -\!\left.  P_{\bot}^{k}\!\left(  \!(\vec{p}_{q1}\vec{\Delta}_{q})p_{\bar
{q}2^{\prime}}{}_{\bot}^{i}\!-\!(\vec{p}_{q1}\vec{p}_{\bar{q}2^{\prime}%
})\Delta_{q}{}_{\bot}^{i}\!\right)  \!\right]  +4x_{q}z\left(  1-2x_{q}%
\right)  p_{q1}{}_{\bot}^{i}\!\left(  \!(\vec{p}_{\bar{q}2^{\prime}}%
\vec{\Delta}_{q})P_{\bot}^{k}-(\vec{P}\vec{p}_{\bar{q}2^{\prime}})\Delta_{q}%
{}_{\bot}^{k}\!\right) \nonumber\\
&  +z\left(  1-2x_{\bar{q}}\right)  \left(  dz+4x_{q}-2\right)  p_{\bar
{q}2^{\prime}}{}_{\bot}^{k}\left(  (\vec{p}_{q1}\vec{\Delta}_{q})P_{\bot}%
^{i}-(\vec{P}\vec{p}_{q1})\Delta_{q}{}_{\bot}^{i}\right)
-z((d-4)z-2)\nonumber\\
&  \times\left[  \left(  g_{\bot}^{ik}(\vec{P}\vec{p}_{q1})+P_{\bot}^{i}%
p_{q1}{}_{\bot}^{k}\right)  (\vec{p}_{\bar{q}2^{\prime}}\vec{\Delta}%
_{q})+\left(  (\vec{P}\vec{p}_{q1})p_{\bar{q}2^{\prime}}{}_{\bot}^{i}-(\vec
{p}_{q1}\vec{p}_{\bar{q}2^{\prime}})P_{\bot}^{i}\right)  \Delta_{q}{}_{\bot
}^{k}\right] \nonumber\\
&  +(\vec{P}\vec{\Delta}_{q})p_{q1}{}_{\bot}^{i}p_{\bar{q}2^{\prime}}{}_{\bot
}^{k}\left(  1-2x_{q}\right)  \left(  1-2x_{\bar{q}}\right)  \left(
z(dz-2)-4x_{q}x_{\bar{q}}\right) \nonumber\\
&  -\left.  (\vec{P}\vec{\Delta}_{q})\left(  g_{\bot}^{ik}(\vec{p}_{q1}\vec
{p}_{\bar{q}2^{\prime}})+p_{q1}{}_{\bot}^{k}p_{\bar{q}2^{\prime}}{}_{\bot}%
^{i}\right)  \left(  z(2-(d-4)z)+4x_{q}x_{\bar{q}}\right)  \right\} \; . \numberthis
\end{align*}%
As above, the dipole $\times$ double-dipole contribution is obtained by complex conjugation and changing the momenta. 

\providecommand{\href}[2]{#2}\begingroup\raggedright\endgroup


\begin{thebibliography}{10}

\bibitem{Wusthoff:1999cr}
M.~W{\"u}sthoff and A.~D. Martin, {\it {The QCD description of diffractive
  processes}},  {\em J. Phys.} {\bf G25} (1999) R309--R344,
  [\href{http://xxx.lanl.gov/abs/hep-ph/9909362}{{\tt hep-ph/9909362}}].

\bibitem{Wolf:2009jm}
G.~Wolf, {\it {Review of High Energy Diffraction in Real and Virtual Photon
  Proton scattering at HERA}},  {\em Rept. Prog. Phys.} {\bf 73} (2010) 116202,
  [\href{http://xxx.lanl.gov/abs/0907.1217}{{\tt arXiv:0907.1217}}].

\bibitem{Aktas:2006hx}
{\bf H1} Collaboration, A.~Aktas {\em et.~al.}, {\it {Diffractive
  deep-inelastic scattering with a leading proton at HERA}},  {\em Eur. Phys.
  J.} {\bf C48} (2006) 749--766,
  [\href{http://xxx.lanl.gov/abs/hep-ex/0606003}{{\tt hep-ex/0606003}}].

\bibitem{Aktas:2006hy}
{\bf H1} Collaboration, A.~Aktas {\em et.~al.}, {\it {Measurement and {QCD}
  analysis of the diffractive deep- inelastic scattering cross-section at
  HERA}},  {\em Eur. Phys. J.} {\bf C48} (2006) 715--748,
  [\href{http://xxx.lanl.gov/abs/hep-ex/0606004}{{\tt hep-ex/0606004}}].

\bibitem{Chekanov:2004hy}
{\bf ZEUS Collaboration} Collaboration, S.~Chekanov {\em et.~al.}, {\it
  {Dissociation of virtual photons in events with a leading proton at HERA}},
  {\em Eur. Phys. J.} {\bf C38} (2004) 43--67,
  [\href{http://xxx.lanl.gov/abs/hep-ex/0408009}{{\tt hep-ex/0408009}}].

\bibitem{Chekanov:2005vv}
{\bf ZEUS} Collaboration, S.~Chekanov {\em et.~al.}, {\it {Study of deep
  inelastic inclusive and diffractive scattering with the ZEUS forward plug
  calorimeter}},  {\em Nucl. Phys.} {\bf B713} (2005) 3--80,
  [\href{http://xxx.lanl.gov/abs/hep-ex/0501060}{{\tt hep-ex/0501060}}].

\bibitem{Aaron:2010aa}
F.~Aaron, C.~Alexa, V.~Andreev, S.~Backovic, A.~Baghdasaryan, {\em et.~al.},
  {\it {Measurement of the cross section for diffractive deep-inelastic
  scattering with a leading proton at HERA}},  {\em Eur. Phys. J.} {\bf C71}
  (2011) 1578, [\href{http://xxx.lanl.gov/abs/1010.1476}{{\tt
  arXiv:1010.1476}}].

\bibitem{Aaron:2012ad}
{\bf H1 Collaboration} Collaboration, F.~Aaron {\em et.~al.}, {\it {Inclusive
  Measurement of Diffractive Deep-Inelastic Scattering at HERA}},  {\em Eur.
  Phys. J.} {\bf C72} (2012) 2074,
  [\href{http://xxx.lanl.gov/abs/1203.4495}{{\tt arXiv:1203.4495}}].

\bibitem{Chekanov:2008fh}
{\bf ZEUS Collaboration} Collaboration, S.~Chekanov {\em et.~al.}, {\it {Deep
  inelastic scattering with leading protons or large rapidity gaps at HERA}},
  {\em Nucl. Phys.} {\bf B816} (2009) 1--61,
  [\href{http://xxx.lanl.gov/abs/0812.2003}{{\tt arXiv:0812.2003}}].

\bibitem{Aaron:2012hua}
{\bf H1 Collaboration, ZEUS Collaboration} Collaboration, F.~Aaron {\em
  et.~al.}, {\it {Combined inclusive diffractive cross sections measured with
  forward proton spectrometers in deep inelastic $ep$ scattering at HERA}},
  {\em Eur. Phys. J.} {\bf C72} (2012) 2175,
  [\href{http://xxx.lanl.gov/abs/1207.4864}{{\tt arXiv:1207.4864}}].

\bibitem{Collins:1997sr}
J.~C. Collins, {\it {Proof of factorization for diffractive hard scattering}},
  {\em Phys. Rev.} {\bf D57} (1998) 3051--3056,
  [\href{http://xxx.lanl.gov/abs/hep-ph/9709499}{{\tt hep-ph/9709499}}].

\bibitem{Boussarie:2014lxa}
R.~Boussarie, A.~Grabovsky, L.~Szymanowski, and S.~Wallon, {\it {Impact factor
  for high-energy two and three jets diffractive production}},  {\em JHEP} {\bf
  1409} (2014) 026, [\href{http://xxx.lanl.gov/abs/1405.7676}{{\tt
  arXiv:1405.7676}}].

\bibitem{Boussarie:2016ogo}
R.~Boussarie, A.~V. Grabovsky, L.~Szymanowski, and S.~Wallon, {\it {On the one
  loop $ {\gamma}^{\left(\ast \right)}\to q\overline{q} $ impact factor and the
  exclusive diffractive cross sections for the production of two or three
  jets}},  {\em JHEP} {\bf 11} (2016) 149,
  [\href{http://xxx.lanl.gov/abs/1606.0041}{{\tt arXiv:1606.0041}}].

\bibitem{Boussarie:2019ero}
R.~Boussarie, A.~V. Grabovsky, L.~Szymanowski, and S.~Wallon, {\it {Towards a
  complete next-to-logarithmic description of forward exclusive diffractive
  dijet electroproduction at HERA: real corrections}},  {\em Phys. Rev.} {\bf
  D100} (2019), no.~7 074020, [\href{http://xxx.lanl.gov/abs/1905.0737}{{\tt
  arXiv:1905.0737}}].

\bibitem{Boussarie:2016bkq}
R.~Boussarie, A.~V. Grabovsky, D.~{\relax Yu}. Ivanov, L.~Szymanowski, and
  S.~Wallon, {\it {Next-to-Leading Order Computation of Exclusive Diffractive
  Light Vector Meson Production in a Saturation Framework}},  {\em Phys. Rev.
  Lett.} {\bf 119} (2017), no.~7 072002,
  [\href{http://xxx.lanl.gov/abs/1612.0802}{{\tt arXiv:1612.0802}}].

\bibitem{Balitsky:1995ub}
I.~Balitsky, {\it Operator expansion for high-energy scattering},  {\em Nucl.
  Phys.} {\bf B463} (1996) 99--160,
  [\href{http://xxx.lanl.gov/abs/hep-ph/9509348}{{\tt hep-ph/9509348}}].

\bibitem{Balitsky:1998kc}
I.~Balitsky, {\it Factorization for high-energy scattering},  {\em Phys. Rev.
  Lett.} {\bf 81} (1998) 2024--2027,
  [\href{http://xxx.lanl.gov/abs/hep-ph/9807434}{{\tt hep-ph/9807434}}].

\bibitem{Balitsky:1998ya}
I.~Balitsky, {\it Factorization and high-energy effective action},  {\em Phys.
  Rev.} {\bf D60} (1999) 014020,
  [\href{http://xxx.lanl.gov/abs/hep-ph/9812311}{{\tt hep-ph/9812311}}].

\bibitem{Balitsky:2001re}
I.~Balitsky, {\it Effective field theory for the small-x evolution},  {\em
  Phys. Lett.} {\bf B518} (2001) 235--242,
  [\href{http://xxx.lanl.gov/abs/hep-ph/0105334}{{\tt hep-ph/0105334}}].

\bibitem{JalilianMarian:1997jx}
J.~Jalilian-Marian, A.~Kovner, A.~Leonidov, and H.~Weigert, {\it The {BFKL}
  equation from the {W}ilson renormalization group},  {\em Nucl. Phys.} {\bf
  B504} (1997) 415--431, [\href{http://xxx.lanl.gov/abs/hep-ph/9701284}{{\tt
  hep-ph/9701284}}].

\bibitem{JalilianMarian:1997gr}
J.~Jalilian-Marian, A.~Kovner, A.~Leonidov, and H.~Weigert, {\it {The {W}ilson
  renormalization group for low x physics: Towards the high density regime}},
  {\em Phys. Rev.} {\bf D59} (1999) 014014,
  [\href{http://xxx.lanl.gov/abs/hep-ph/9706377}{{\tt hep-ph/9706377}}].

\bibitem{JalilianMarian:1997dw}
J.~Jalilian-Marian, A.~Kovner, and H.~Weigert, {\it {The {W}ilson
  renormalization group for low x physics: Gluon evolution at finite parton
  density}},  {\em Phys. Rev.} {\bf D59} (1999) 014015,
  [\href{http://xxx.lanl.gov/abs/hep-ph/9709432}{{\tt hep-ph/9709432}}].

\bibitem{JalilianMarian:1998cb}
J.~Jalilian-Marian, A.~Kovner, A.~Leonidov, and H.~Weigert, {\it Unitarization
  of gluon distribution in the doubly logarithmic regime at high density},
  {\em Phys. Rev.} {\bf D59} (1999) 034007,
  [\href{http://xxx.lanl.gov/abs/hep-ph/9807462}{{\tt hep-ph/9807462}}].

\bibitem{Kovner:2000pt}
A.~Kovner, J.~G. Milhano, and H.~Weigert, {\it Relating different approaches to
  nonlinear {QCD} evolution at finite gluon density},  {\em Phys. Rev.} {\bf
  D62} (2000) 114005, [\href{http://xxx.lanl.gov/abs/hep-ph/0004014}{{\tt
  hep-ph/0004014}}].

\bibitem{Weigert:2000gi}
H.~Weigert, {\it Unitarity at small {B}jorken x},  {\em Nucl. Phys.} {\bf A703}
  (2002) 823--860, [\href{http://xxx.lanl.gov/abs/hep-ph/0004044}{{\tt
  hep-ph/0004044}}].

\bibitem{Iancu:2000hn}
E.~Iancu, A.~Leonidov, and L.~D. McLerran, {\it {Nonlinear gluon evolution in
  the color glass condensate. I}},  {\em Nucl. Phys.} {\bf A692} (2001)
  583--645, [\href{http://xxx.lanl.gov/abs/hep-ph/0011241}{{\tt
  hep-ph/0011241}}].

\bibitem{Iancu:2001ad}
E.~Iancu, A.~Leonidov, and L.~D. McLerran, {\it The renormalization group
  equation for the color glass condensate},  {\em Phys. Lett.} {\bf B510}
  (2001) 133--144, [\href{http://xxx.lanl.gov/abs/hep-ph/0102009}{{\tt
  hep-ph/0102009}}].

\bibitem{Ferreiro:2001qy}
E.~Ferreiro, E.~Iancu, A.~Leonidov, and L.~McLerran, {\it {Nonlinear gluon
  evolution in the color glass condensate. II}},  {\em Nucl. Phys.} {\bf A703}
  (2002) 489--538, [\href{http://xxx.lanl.gov/abs/hep-ph/0109115}{{\tt
  hep-ph/0109115}}].

\bibitem{Fadin:1975cb}
V.~S. Fadin, E.~A. Kuraev, and L.~N. Lipatov, {\it {On the {P}omeranchuk
  Singularity in Asymptotically Free Theories}},  {\em Phys. Lett.} {\bf B60}
  (1975) 50--52.

\bibitem{Kuraev:1976ge}
E.~A. Kuraev, L.~N. Lipatov, and V.~S. Fadin, {\it {Multi - Reggeon Processes
  in the {Y}ang-{M}ills Theory}},  {\em Sov. Phys. JETP} {\bf 44} (1976)
  443--450.

\bibitem{Kuraev:1977fs}
E.~A. Kuraev, L.~N. Lipatov, and V.~S. Fadin, {\it {The Pomeranchuk Singularity
  in Nonabelian Gauge Theories}},  {\em Sov. Phys. JETP} {\bf 45} (1977)
  199--204.

\bibitem{Balitsky:1978ic}
I.~I. Balitsky and L.~N. Lipatov, {\it {The Pomeranchuk Singularity in Quantum
  Chromodynamics}},  {\em Sov. J. Nucl. Phys.} {\bf 28} (1978) 822--829.

\bibitem{Fadin:1998py}
V.~S. Fadin and L.~N. Lipatov, {\it {BFKL pomeron in the next-to-leading
  approximation}},  {\em Phys. Lett.} {\bf B429} (1998) 127--134,
  [\href{http://xxx.lanl.gov/abs/hep-ph/9802290}{{\tt hep-ph/9802290}}].

\bibitem{Ciafaloni:1998gs}
M.~Ciafaloni and G.~Camici, {\it {Energy scale(s) and next-to-leading BFKL
  equation}},  {\em Phys. Lett.} {\bf B430} (1998) 349--354,
  [\href{http://xxx.lanl.gov/abs/hep-ph/9803389}{{\tt hep-ph/9803389}}].

\bibitem{Fadin:2004zq}
V.~Fadin and R.~Fiore, {\it {Non-forward BFKL pomeron at next-to-leading
  order}},  {\em Phys. Lett.} {\bf B610} (2005) 61--66,
  [\href{http://xxx.lanl.gov/abs/hep-ph/0412386}{{\tt hep-ph/0412386}}].

\bibitem{Fadin:2005zj}
V.~S. Fadin and R.~Fiore, {\it {Non-forward NLO BFKL kernel}},  {\em Phys.
  Rev.} {\bf D72} (2005) 014018,
  [\href{http://xxx.lanl.gov/abs/hep-ph/0502045}{{\tt hep-ph/0502045}}].

\bibitem{Celiberto:2020wpk}
F.~G. Celiberto, {\it {Hunting BFKL in semi-hard reactions at the LHC}},  {\em
  Eur. Phys. J. C} {\bf 81} (2021), no.~8 691,
  [\href{http://xxx.lanl.gov/abs/2008.0737}{{\tt arXiv:2008.0737}}].  

\bibitem{Kovchegov:1999yj}
Y.~V. Kovchegov, {\it {Small-$x$ $F_2$ structure function of a nucleus
  including multiple pomeron exchanges}},  {\em Phys. Rev.} {\bf D60} (1999)
  034008, [\href{http://xxx.lanl.gov/abs/hep-ph/9901281}{{\tt
  hep-ph/9901281}}].

\bibitem{Kovchegov:1999ua}
Y.~V. Kovchegov, {\it {Unitarization of the BFKL pomeron on a nucleus}},  {\em
  Phys. Rev.} {\bf D61} (2000) 074018,
  [\href{http://xxx.lanl.gov/abs/hep-ph/9905214}{{\tt hep-ph/9905214}}].

\bibitem{Chirilli:2013kca}
G.~A. Chirilli and Y.~V. Kovchegov, {\it {Solution of the NLO BFKL Equation and
  a Strategy for Solving the All-Order BFKL Equation}},  {\em JHEP} {\bf 06}
  (2013) 055, [\href{http://xxx.lanl.gov/abs/1305.1924}{{\tt
  arXiv:1305.1924}}].

\bibitem{Grabovsky:2013gta}
A.~V. Grabovsky, {\it {On the solution to the NLO forward BFKL equation}},
  {\em JHEP} {\bf 09} (2013) 098,
  [\href{http://xxx.lanl.gov/abs/1307.3152}{{\tt arXiv:1307.3152}}].

\bibitem{Collins:2011zzd}
J.~Collins, {\em {Foundations of perturbative QCD}}, vol.~32.
\newblock Cambridge University Press, 11, 2013.

\bibitem{Altarelli:1979kv}
G.~Altarelli, R.~K. Ellis, G.~Martinelli, and S.-Y. Pi, {\it {Processes
  Involving Fragmentation Functions Beyond the Leading Order in QCD}},  {\em
  Nucl. Phys. B} {\bf 160} (1979) 301--329.

\bibitem{Gribov:1972ri}
V.~N. Gribov and L.~N. Lipatov, {\it Deep inelastic e p scattering in
  perturbation theory},  {\em Sov. J. Nucl. Phys.} {\bf 15} (1972) 438--450.

\bibitem{Lipatov:1974qm}
L.~N. Lipatov, {\it The parton model and perturbation theory},  {\em Sov. J.
  Nucl. Phys.} {\bf 20} (1975) 94--102.

\bibitem{Altarelli:1977zs}
G.~Altarelli and G.~Parisi, {\it Asymptotic freedom in parton language},  {\em
  Nucl. Phys.} {\bf B126} (1977) 298.

\bibitem{Dokshitzer:1977sg}
Y.~L. Dokshitzer, {\it {Calculation of the Structure Functions for Deep
  Inelastic Scattering and $e^+ e^-$ Annihilation by Perturbation Theory in
  {Q}uantum {C}hromodynamics}},  {\em Sov. Phys. JETP} {\bf 46} (1977)
  641--653.

\bibitem{Ivanov:2012iv}
D.~{\relax Yu}. Ivanov and A.~Papa, {\it {Inclusive production of a pair of
  hadrons separated by a large interval of rapidity in proton collisions}},
  {\em JHEP} {\bf 07} (2012) 045,
  [\href{http://xxx.lanl.gov/abs/1205.6068}{{\tt arXiv:1205.6068}}].

\bibitem{Chirilli:2012jd}
G.~A. Chirilli, B.-W. Xiao, and F.~Yuan, {\it {Inclusive Hadron Productions in
  pA Collisions}},  {\em Phys. Rev.} {\bf D86} (2012) 054005,
  [\href{http://xxx.lanl.gov/abs/1203.6139}{{\tt arXiv:1203.6139}}].

\bibitem{Beuf:2022ndu}
G.~Beuf, T.~Lappi, and R.~Paatelainen, {\it {Massive quarks in NLO dipole
  factorization for DIS: Transverse photon}},
  \href{http://xxx.lanl.gov/abs/2204.0248}{{\tt arXiv:2204.0248}}.
   
\bibitem{Altinoluk:2020qet}
T.~Altinoluk, R.~Boussarie, C.~Marquet, and P.~Taels, {\it {Photoproduction of
  three jets in the CGC: gluon TMDs and dilute limit}},  {\em JHEP} {\bf 07}
  (2020) 143, [\href{http://xxx.lanl.gov/abs/2001.0076}{{\tt
  arXiv:2001.0076}}].   
   
\bibitem{Taels:2022tza}
P.~Taels, T.~Altinoluk, G.~Beuf, and C.~Marquet, {\it {Dijet photoproduction at low x at next-to-leading order and its back-to-back limit}},  {\em JHEP} {\bf 10} (2022) 184, [\href{http://xxx.lanl.gov/abs/2204.1165}{{\tt
  arXiv:2204.1165}}].  

\bibitem{Roy:2019hwr}
K.~Roy and R.~Venugopalan, {\it {NLO impact factor for inclusive photon$+$dijet
  production in $e+A$ DIS at small $x$}},  {\em Phys. Rev. D} {\bf 101} (2020),
  no.~3 034028, [\href{http://xxx.lanl.gov/abs/1911.0453}{{\tt
  arXiv:1911.0453}}].

\bibitem{Caucal:2021ent}
P.~Caucal, F.~Salazar, and R.~Venugopalan, {\it {Dijet impact factor in DIS at
  next-to-leading order in the Color Glass Condensate}},  {\em JHEP} {\bf 11}
  (2021) 222, [\href{http://xxx.lanl.gov/abs/2108.0634}{{\tt
  arXiv:2108.0634}}].
  
\bibitem{Caucal:2022ulg}
P.~Caucal, F.~Salazar, B.~Schenke, and R.~Venugopalan, {\it {Back-to-back
  inclusive dijets in DIS at small $x$: Sudakov suppression and gluon
  saturation at NLO}},  \href{http://xxx.lanl.gov/abs/2208.1387}{{\tt
  arXiv:2208.1387}}.

\bibitem{Bergabo:2022zhe}
F.~Bergabo and J.~Jalilian-Marian, {\it {Single Inclusive Hadron Production in
  DIS at Small $x$: Next to Leading Order Corrections}},
  \href{http://xxx.lanl.gov/abs/2210.0320}{{\tt arXiv:2210.0320}}.

\bibitem{Bergabo:2022tcu}
F.~Bergabo and J.~Jalilian-Marian, {\it {One-loop corrections to dihadron
  production in DIS at small x}},  {\em Phys. Rev. D} {\bf 106} (2022), no.~5
  054035, [\href{http://xxx.lanl.gov/abs/2207.0360}{{\tt arXiv:2207.0360}}].

\bibitem{Iancu:2022gpw}
E.~Iancu and Y.~Mulian, {\it {Dihadron production in DIS at NLO: the real
  corrections}},  \href{http://xxx.lanl.gov/abs/2211.0483}{{\tt
  arXiv:2211.0483}}.

\bibitem{Mantysaari:2022bsp}
H.~M\"antysaari and J.~Penttala, {\it {Exclusive production of light vector
  mesons at next-to-leading order in the dipole picture}},  {\em Phys. Rev. D}
  {\bf 105} (2022), no.~11 114038,
  [\href{http://xxx.lanl.gov/abs/2203.1691}{{\tt arXiv:2203.1691}}].

\bibitem{Mantysaari:2021ryb}
H.~M\"antysaari and J.~Penttala, {\it {Exclusive heavy vector meson production
  at next-to-leading order in the dipole picture}},  {\em Phys. Lett. B} {\bf
  823} (2021) 136723, [\href{http://xxx.lanl.gov/abs/2104.0234}{{\tt
  arXiv:2104.0234}}].

\bibitem{Mantysaari:2022kdm}
H.~M\"antysaari and J.~Penttala, {\it {Complete calculation of exclusive heavy
  vector meson production at next-to-leading order in the dipole picture}},
  \href{http://xxx.lanl.gov/abs/2204.1403}{{\tt arXiv:2204.1403}}.

\bibitem{Beuf:2022kyp}
G.~Beuf, H.~H\"anninen, T.~Lappi, Y.~Mulian, and H.~M\"antysaari, {\it
  {Diffractive deep inelastic scattering at NLO in the dipole picture: the $q
  \bar q g$ contribution}},  \href{http://xxx.lanl.gov/abs/2206.1316}{{\tt
  arXiv:2206.1316}}.



\end{thebibliography}
\end{document}